\documentclass[twocolumn, preprint, times]{aastex62}

\usepackage{amsmath, amssymb}
\usepackage{soul}
\usepackage{amsmath}
\usepackage{upgreek}
\usepackage{natbib}

\usepackage{subfigure}
\usepackage{longtable}
\usepackage{threeparttablex}
\usepackage{booktabs}
\usepackage{scalerel}
\usepackage{tikz}
\usetikzlibrary{svg.path}

\usepackage{tikz}
\usetikzlibrary{svg.path}
\definecolor{orcidlogocol}{HTML}{A6CE39}
\tikzset{
  orcidlogo/.pic={
    \fill[orcidlogocol] svg{M256,128c0,70.7-57.3,128-128,128C57.3,256,0,198.7,0,128C0,57.3,57.3,0,128,0C198.7,0,256,57.3,256,128z};
    \fill[white] svg{M86.3,186.2H70.9V79.1h15.4v48.4V186.2z}
                 svg{M108.9,79.1h41.6c39.6,0,57,28.3,57,53.6c0,27.5-21.5,53.6-56.8,53.6h-41.8V79.1z M124.3,172.4h24.5c34.9,0,42.9-26.5,42.9-39.7c0-21.5-13.7-39.7-43.7-39.7h-23.7V172.4z}
                 svg{M88.7,56.8c0,5.5-4.5,10.1-10.1,10.1c-5.6,0-10.1-4.6-10.1-10.1c0-5.6,4.5-10.1,10.1-10.1C84.2,46.7,88.7,51.3,88.7,56.8z};
  }
}
\newcommand\orcidicon[1]{\href{https://orcid.org/#1}{\mbox{\scalerel*{
\begin{tikzpicture}[yscale=-1,transform shape]
\pic{orcidlogo};
\end{tikzpicture}
}{|}}}}

\title{published_repeaters_polarization}
\begin{document}

\title{Revealing the Dynamic Magneto-ionic Environments of Repeating Fast Radio Burst Sources through Multi-year Polarimetric Monitoring with CHIME/FRB}

%affiliations
% \newcommand{\mcgillphysics}{Department of Physics, McGill University, 3600 rue University, Montr\'eal, QC H3A 2T8, Canada}
% \newcommand{\msi}{McGill Space Institute, McGill University, 3550 rue University, Montr\'eal, QC H3A 2A7, Canada}
% \newcommand{\cita}{Canadian Institute for Theoretical Astrophysics, 60 St. George Street, Toronto, ON M5S 3H8, Canada}
% \newcommand{\dunlapinstitute}{Dunlap Institute for Astronomy \& Astrophysics, University of Toronto, 50 St. George Street, Toronto, ON M5S 3H4, Canada}
% \newcommand{\dunlapdep}{David A. Dunlap Department of Astronomy \& Astrophysics, University of Toronto, 50 St. George Street, Toronto, ON M5S 3H4, Canada}
% \newcommand{\mitkavli}{MIT Kavli Institute for Astrophysics and Space Research, Massachusetts Institute of Technology, 77 Massachusetts Ave, Cambridge, MA 02139, USA}
% \newcommand{\mitphysics}{Department of Physics, Massachusetts Institute of Technology, 77 Massachusetts Ave, Cambridge, MA 02139, USA}
% \newcommand{\ubc}{Department of Physics and Astronomy, University of British Columbia, 6224 Agricultural Road, Vancouver, BC V6T
% 23 1Z1 Canada}
% \newcommand{\sidrat}{Sidrat Research, PO Box 73527 RPO Wychwood, Toronto, Ontario, M6C 4A7, Canada}
\newcommand{\mcgillphysics}{Department of Physics, McGill University, 3600 rue University, Montr\'eal, QC H3A 2T8, Canada}
\newcommand{\msi}{McGill Space Institute, McGill University, 3550 rue University, Montr\'eal, QC H3A 2A7, Canada}
\newcommand{\wvuphysics}{Department of Physics and Astronomy, West Virginia University, P.O. Box 6315, Morgantown, WV 26506, USA}
\newcommand{\wvugws}{Center for Gravitational Waves and Cosmology, West Virginia University, Chestnut Ridge Research Building, Morgantown, WV 26505, USA}
\newcommand{\wvucs}{Lane Department of Computer Science and Electrical Engineering, 1220 Evansdale Drive, PO Box 6109  Morgantown, WV 26506, USA}
\newcommand{\uoftphysics}{Department of Physics, University of Toronto, 60 St. George Street, Toronto, ON M5S 1A7, Canada}
\newcommand{\cita}{Canadian Institute for Theoretical Astrophysics, 60 St. George Street, Toronto, ON M5S 3H8, Canada}
\newcommand{\dunlapinstitute}{Dunlap Institute for Astronomy \& Astrophysics, University of Toronto, 50 St. George Street, Toronto, ON M5S 3H4, Canada}
\newcommand{\dunlapdep}{David A. Dunlap Department of Astronomy \& Astrophysics, University of Toronto, 50 St. George Street, Toronto, ON M5S 3H4, Canada}
\newcommand{\nrao}{National Radio Astronomy Observatory, 520 Edgemont Rd, Charlottesville, VA 22903, USA}
\newcommand{\haa}{National Research Council Canada, Herzberg Astronomy and Astrophysics Research Centre, Dominion Radio Astrophysical Observatory, PO Box 248, Penticton, British Columbia, V2A 6J9 Canada}
\newcommand{\mitkavli}{MIT Kavli Institute for Astrophysics and Space Research, Massachusetts Institute of Technology, 77 Massachusetts Ave, Cambridge, MA 02139, USA}
\newcommand{\mitphysics}{Department of Physics, Massachusetts Institute of Technology, 77 Massachusetts Ave, Cambridge, MA 02139, USA}
\newcommand{\ubc}{Dept. of Physics and Astronomy, 6224 Agricultural Road, Vancouver, BC V6T 1Z1 Canada}
\newcommand{\sidrat}{Sidrat Research, PO Box 73527 RPO Wychwood, Toronto, Ontario, M6C 4A7, Canada}
\newcommand{\perimeter}{Perimeter Institute for Theoretical Physics, 31 Caroline Street N, Waterloo ON N2L 2Y5 Canada}
\newcommand{\waterloo}{Department of Physics and Astronomy, University of Waterloo, Waterloo, ON N2L 3G1, Canada}
\newcommand{\tata}{Department of Astronomy and Astrophysics, Tata Institute of Fundamental Research, Mumbai, 400005, India}
\newcommand{\ncra}{National Centre for Radio Astrophysics, Post Bag 3, Ganeshkhind, Pune, 411007, India}
\newcommand{\uva}{Anton Pannekoek Institute for Astronomy, University of Amsterdam, Science Park 904, 1098 XH Amsterdam, The Netherlands}
\newcommand{\caltech}{Cahill Center for Astronomy and Astrophysics, California Institute of Technology, 1216 E California Boulevard, Pasadena, CA 91125, USA}

\author{R. Mckinven \orcidicon{0000-0001-7348-6900}}
\affiliation{\mcgillphysics}
\affiliation{\msi}
\affiliation{\dunlapinstitute}
\affiliation{\dunlapdep}
\author{B.M. Gaensler \orcidicon{0000-0002-3382-9558}}
\affiliation{\dunlapinstitute}
\affiliation{\dunlapdep}
\author{D. Michilli \orcidicon{0000-0002-2551-7554}}
\affiliation{\mitphysics}
\affiliation{\mitkavli}
\author{K. Masui \orcidicon{0000-0002-4279-6946}}
\affiliation{\mitkavli}
\affiliation{\mitphysics}
\author{V.M. Kaspi \orcidicon{0000-0001-9345-0307}}
\affiliation{\mcgillphysics}
\affiliation{\msi}
\author{J. Su \orcidicon{0000-0003-0607-8194}}
\affiliation{\mcgillphysics}
% \author{D. Cubranic \orcidicon{0000-0003-2319-9676}}
% \affiliation{\ubc}
\author{M. Bhardwaj \orcidicon{0000-0002-3615-3514}}
\affiliation{\mcgillphysics}
\affiliation{\msi}
\author{T. Cassanelli \orcidicon{0000-0003-2047-5276}}
\affiliation{\dunlapdep}
\affiliation{\dunlapinstitute}
\author{P. Chawla \orcidicon{0000-0002-3426-7606}}
\affiliation{\uva}
\author{F. (Adam) Dong \orcidicon{0000-0003-4098-5222}}
\affiliation{\ubc}
\author{E. Fonseca \orcidicon{0000-0001-8384-5049}}
\affiliation{\wvucs}
\affiliation{\wvugws}
\affiliation{\wvuphysics}
\author{C. Leung \orcidicon{0000-0002-4209-7408}}
\affiliation{\mitkavli}
\affiliation{\mitphysics}
\author{D.Z. Li \orcidicon{0000-0001-7931-0607}}
\affiliation{\caltech}
\author{C. Ng \orcidicon{0000-0002-3616-5160}}
\affiliation{\dunlapinstitute}
\author{C. Patel \orcidicon{0000-0003-3367-1073}}
\affiliation{\mcgillphysics}
\affiliation{\dunlapinstitute}
\author{A.B. Pearlman \orcidicon{0000-0002-8912-0732}}
\affiliation{\mcgillphysics}
\affiliation{\msi}
\author{E. Petroff \orcidicon{0000-0002-9822-8008}}
\affiliation{\mcgillphysics}
\affiliation{\msi}
\affiliation{\uva}
\author{Z. Pleunis, \orcidicon{0000-0002-4795-697X}}
\affiliation{\dunlapinstitute}
\author{M. Rafiei-Ravandi \orcidicon{0000-0001-7694-6650}}
\affiliation{\mcgillphysics}
\affiliation{\msi}
\author{M. Rahman \orcidicon{0000-0003-1842-6096}}
\affiliation{\sidrat}
\author{K.R. Sand \orcidicon{0000-0003-3154-3676}}
\affiliation{\mcgillphysics}
\affiliation{\msi}
\author{K. Shin \orcidicon{0000-0002-6823-2073}}
\affiliation{\mitkavli}
\affiliation{\mitphysics}
\author{I.H. Stairs \orcidicon{0000-0001-9784-8670}}
\affiliation{\ubc}
\author{S. Tendulkar \orcidicon{0000-0003-2548-2926}}
\affiliation{\tata}
\affiliation{\ncra}

% \author{C. Leung \orcidicon{0000-0002-4209-7408}}
% \affiliation{\mitkavli}
% \affiliation{\mitphysics}
% \author{P. J. Boyle \orcidicon{0000-0001-8537-9299}}
% \affiliation{\mcgillphysics}
% \affiliation{\msi}
% \author{C. Brar \orcidicon{0000-0002-1800-8233}}
% \affiliation{\mcgillphysics}
% \affiliation{\msi}

% \author{J. Mena-Parra \orcidicon{0000-0002-0772-9326}}
% \affiliation{\mitkavli}

\correspondingauthor{R. Mckinven}
\email{ryan.mckinven@mcgill.ca}

\begin{abstract}
Fast radio bursts (FRBs) display a confounding variety of burst properties and host galaxy associations. Repeating FRBs offer insight into the FRB population by enabling spectral, temporal and polarimetric properties to be tracked over time. Here, we report on the polarized observations of 12 repeating sources using multi-year monitoring with the Canadian Hydrogen Intensity Mapping Experiment (CHIME) over 400-800 MHz. We observe significant RM variations from many sources in our sample, including RM changes of several hundred $\rm{rad\, m^{-2}}$ over month timescales from FRBs 20181119A, 20190303A and 20190417A, and more modest RM variability ($\rm{\Delta RM \lesssim}$ few tens rad m$^{-2}$) from FRBs 20181030A, 20190208A, 20190213B and 20190117A over equivalent timescales. Several repeaters display a frequency dependent degree of linear polarization that is consistent with depolarization via scattering. Combining our measurements of RM variations with equivalent constraints on DM variability, we estimate the average line-of-sight magnetic field strength in the local environment of each repeater. In general, repeating FRBs display RM variations that are more prevalent/extreme than those seen from radio pulsars in the Milky Way and the Magellanic Clouds, suggesting repeating FRBs and pulsars occupy distinct magneto-ionic environments.
\end{abstract}

\section{Introduction}
\label{sec:intro}

Fast radio bursts \citep[FRBs;][]{Lorimer2007,Petroff2022} are short (microsecond to millisecond) bursts of radio emission with uncertain origins. FRB polarization measurements can characterize several properties that contain information on the emission physics of the source and/or its environment. These properties can be summarized as a handful of quantities: the position angle of the linear polarization vector (PA), the linear/circular polarized fractions ($L/I$, $V/I$), and the Faraday rotation (quantified through the rotation measure, RM). 

Evolution in the PA can be tracked both within a burst and, for a repeating source, between bursts. Such information can greatly improve our understanding of the geometry of the FRB emission mechanism in a manner similar to the rotating vector model \citep[RVM;][]{Radhakrishnan1969} of pulsar radio emission. To date, a wide variety of PA behavior has been observed from FRBs. Large PA swings have been observed from repeating source FRB 20180301A \citep{Luo2020}. Meanwhile, other sources displaying apparent flat PA curves have since been shown to exhibit subtle intraburst PA variability when observed at high temporal resolution \citep[FRB 20180916B;][]{Nimmo2021}. Some repeating sources also display significant PA variations across bursts. A recent example is FRB 20201124A \citep{Hilmarsson2021b, Kumar2021, Lee2021}, which also exhibits significant circular polarization \citep{Hilmarsson2021b, Lee2021} that in many cases exceeds what can be produced from Faraday conversion between linear and circular polarized modes \citep{Vedantham2019, Levin2020}.

Recently, \citet{Wang2021} have demonstrated that other mechanisms may produce anomalous circular polarization seen from FRB 20201124A, including mechanisms such as inverse Compton scattering or cyclotron resonance absorption. Polarized observations are, therefore, tremendously valuable for both constraining FRB emission models and possibly the magneto-ionic properties of the local environment. For example, both the PA evolution and substantial $V/I$ fractions observed in some FRB sources are features that are naturally favorable to models invoking magnetospheric origins \citep[e.g.,][]{Zhang2020}. However, synchrotron maser emission models \citep{Metzger2019} may also be capable of explaining such features under certain conditions. Indeed, recent simulations of a flaring magnetar \citep{Yuan2020} have demonstrated the existence of strong variations in the orientation of the magnetar wind's magnetic field that would be required to explain the PA variability observed in some FRBs.

Meanwhile, the $\rm{RM}$ probes the intervening magnetized plasma between the source and the observer. As such, this quantity can be decomposed in terms of the $\rm{RM}$ contributions of the various media encountered along the line-of-sight (LoS)\footnote{For simplicity, we have absorbed the diluting effect of cosmological expansion on $\rm{RM}$ by representing each $\rm{RM}$ component in the observer's frame. As such, both the $\rm{RM_{IGM}}$ and $\rm{RM_{host}}$ contributions are diluted by a factor of $(1+z)^2$, $z$ being the redshift of the Faraday rotating medium.}, 
\begin{equation}
    \rm{RM = RM_{MW} + RM_{IGM} + RM_{host}}.
\end{equation}                          
Here, $\rm{RM_{MW}}$ represents the $\rm{RM}$ contribution of the Milky Way and $\rm{RM_{IGM}}$ is the modest contribution ($\rm{|RM_{IGM}|\ll 10 \; rad \, m^{-2}}$) expected from the intergalactic medium \citep[IGM;][]{Akahori2016}. $\rm{RM_{host}}$ is the RM contribution of the FRB host galaxy, which represents the shared contribution from its interstellar medium (ISM) and the local magneto-ionic environment of the FRB source. FRB RMs often significantly exceed $\rm{RM_{MW}}$ estimates from Galactic RM maps (see Section~\ref{sec:galdmrm}). In such cases, FRB RMs offer a window into the magneto-ionic properties of the local environments and hosts galaxies of FRB sources \citep{Masui2015, Petroff2017, Michilli2018}. 

When applying this analysis to repeaters, any temporal changes in the RM can be used to constrain the magnetization, configuration and dynamics of the medium producing those changes. If the changes are substantial, then these variations are a direct probe of a changing magneto-ionic environment that is, very likely, in the vicinity of the FRB source. This analysis was recently done for FRB 20121102A, whose non-monotonically decreasing RM evolution was found to be consistent with theoretical models describing it in terms of an evolving supernova remnant \citep[SNR;][]{Hilmarsson2021}. A similar interpretation has been put forth to describe observations of the repeater, FRB 20190520B \citep{Niu2021}, which along with displaying an association with a compact persistent radio source (PRS), also shares a number of other similarities with FRB 20121102A. Based on these similarities, \citet{Zhao2021} argue for a young ($\sim 14-22$ years) magnetar embedded in a wind nebula and SNR. While young SNR systems remain a popular model for explaining these observations, other possibilities exist. Indeed, interesting similarities have been drawn with the RM behavior of the radio-loud Galactic center magnetar, whose substantial and evolving RM is very likely related to its shared environment with the nearby supermassive black hole, Sagittarius A$^{*}$ \citep{Desvignes2018}. Meanwhile, recent observations by the FAST telescope of nearly two thousand polarized bursts from FRB 20201124A \citep{Lee2021} have shown a number of interesting properties that include significant and irregular RM variations on week-long timescales. Such behavior is discrepant with the secular decrease expected if the RM is dominated by a shocked, uniform medium \citep[e.g.][]{Piro2018} and instead suggests a high degree of non-uniformity in the local environment of the source. These observations, however, are insufficient to rule out the existence of an associated SNR since the observed non-monotonic RM evolution could be produced by the SNR's small-scale turbulent magnetic field \citep{Dickel1976,Milne1987}. In such a scenario, the observed RM behavior would manifest either through the relative motion between the FRB source and the small-scale field or through the dynamic evolution of turbulences in the SNR itself.     

This paper reports on polarimetric observations of a sample of 12 previously published repeating FRB sources \citep{chime/frb2019b, chime/frb2019, Fonseca2020}. This sample greatly expands the number of repeaters with corresponding polarization information, revealing interesting behavior in individual sources as well as enabling the aggregate properties of this sample to be compared to the (apparently) non-repeating FRB sample and other sources. The paper is organized as follows: Section~\ref{sec:methods} reviews the observations and analysis methods, Section~\ref{sec:results} summarizes results and is followed by Section~\ref{sec:discussion}, which explores the implications of these results for different sources and how these results may be used in aggregate to characterize the local environments of repeaters. 

\section{Observations, Data Reduction \& Analysis}
\label{sec:methods}

\begin{figure}
% 	\centering
% \begin{center}
    \includegraphics[width=0.45\textwidth]{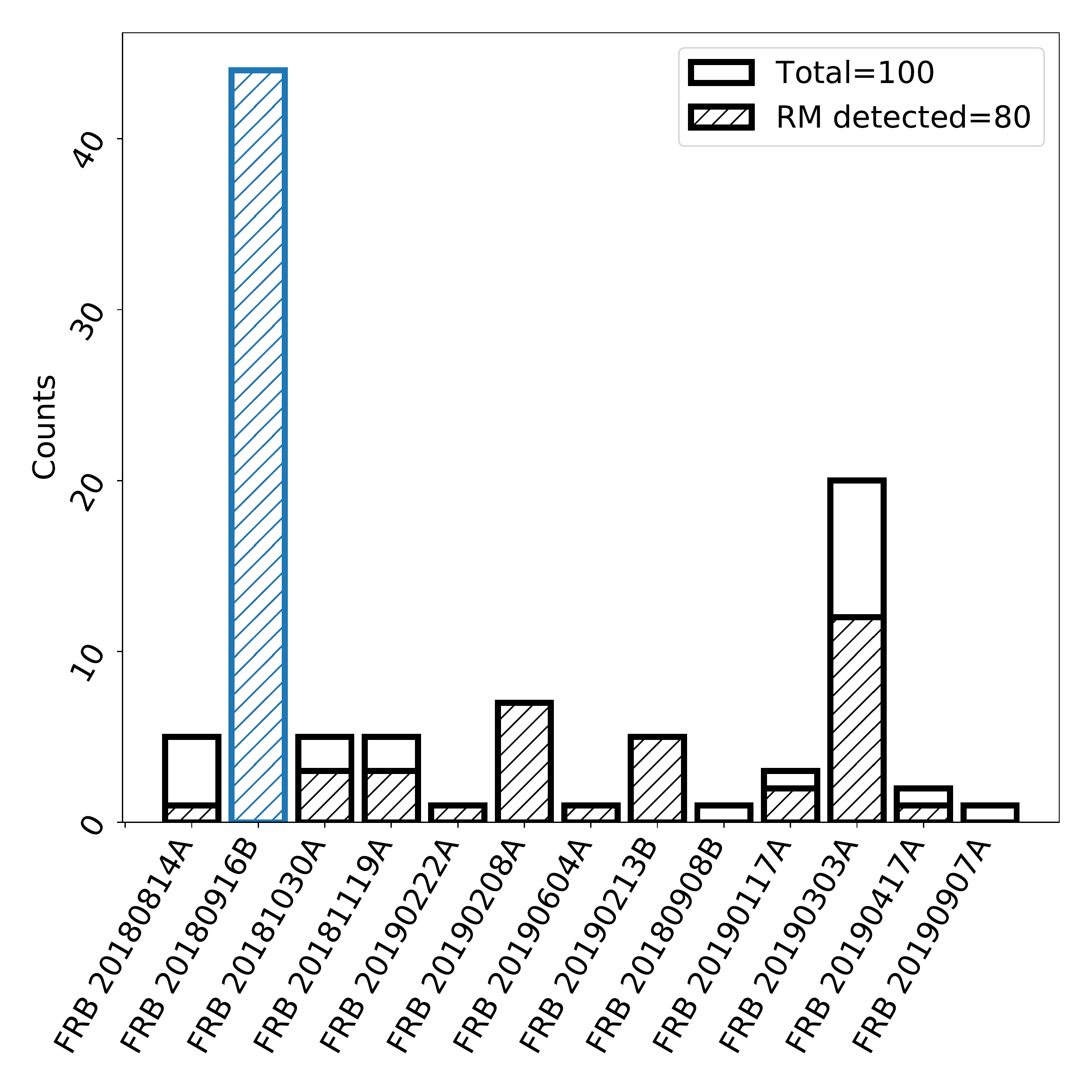}
    \caption{The number of bursts observed in polarization from each repeating source in our sample and the number of RM detections (hatched). The 44-burst sample from FRB 20180916B is highlighted in blue and is reported on in a companion paper \citep{Mckinven2022}.}
    \label{fig:summary_hist}
% \end{center}
\end{figure}

Observations reported in this paper correspond to bursts from previously published repeating FRB sources \citep{chime/frb2019,Fonseca2020} that successfully triggered baseband callback data \citep{Michilli2021} over a three year period between 2018 December and 2021 December. Figure~\ref{fig:summary_hist} summarizes the number of bursts from each repeater of this sample along with the number of significant RM detections (hatched). A significant fraction of these bursts are associated with the periodically active source, FRB 20180916B (44 bursts). Results of the polarization analysis of this repeater can be found in a companion paper \citep{Mckinven2022} to which we refer the reader for a summary of the baseband data format and a description of the methods used to determine relevant quantities like the RM, polarized fraction ($L/I$) and polarization position angle (PA) curve \citep[see also;][]{Mckinven2021}. This paper studies the polarized properties of the remaining 12 repeaters of this sample, which, with the exception of FRB 20190303A and FRB 20190604A \citep[FRB 190303.J1353+48 \& FRB 190604.J1435+53, respectively;][]{Fonseca2020}, are the first polarization measurements for these sources. Due to resource limitations and the threat posed by false positives, a higher S/N threshold has been historically set for callbacks of baseband versus intensity data \citep{chime/frb2018}. Hence, the detections reported here are not entirely reflective of the total number of detections from each source, which will be reported elsewhere. 

\subsection{Galactic Dispersion and Rotation Measure Estimates}
\label{sec:galdmrm}

 Galactic RM contributions and associated uncertainties are estimated from the updated all-sky Faraday Sky map of \citet{Hutschenreuter2022}. This map reconstructs the Galactic RM contribution, $\rm{RM_{MW}}$, using RM measurements of polarized extragalactic sources. Galactic DM contributions are estimated using the {\tt PyGEDM} package\footnote{https://github.com/FRBs/pygedm}, a Python interface to the YMW16 \citep{Yao2017} and NE2001 \citep{Cordes2002,Cordes2003} electron density models. Table~\ref{ta:dmrmgal} summarizes estimates of the Galactic dispersion and rotation measure contributions for the sample of repeating FRB sources studied here. Sources are referred to by their transient name server (TNS)\footnote{https://www.wis-tns.org/} names. Previous names along with the discovery paper of each source are indicated here as a reference.

% \begin{table}[!htbp]
\begin{table*}
% \centering
\begin{center}
\caption{Galactic DM$^{a}$ and RM estimates for our target sources}
%\begin{tabular}{*5c}
\begin{tabular}{l l |c c| c }
\toprule
\multicolumn{2}{c}{Source} & \multicolumn{2}{c}{$\rm{DM_{MW}}\; \rm{[pc \, cm^{-3}]}$} & $\rm{RM_{MW}}\; \rm{[rad \, m^{-2}]}$ \\
TNS Name & Previous Name  & NE2001 & YMW16 &  \\
% \toprule
\hline
FRB 20180814A & FRB 180814.J0422+73$^{b}$ & $\sim 88$ & $\sim 108$ & $-45(18)$ \\
% FRB 20180916B & FRB 180916.J0158+65$^{c}$ & $\sim 199$ & $\sim 325$ & $-94(45)$   \\
FRB 20181030A & FRB 181030.J1054+73$^{c}$ & $\sim 41$ & $\sim 33$ & $-19(9)$ \\
FRB 20181119A & FRB 181119.J12+65$^{c}$ & $\sim 34$ & $\sim 26$ & $+22(5)$ \\
% FRB 20190116A & FRB 190116.J1249+27$^{c}$ & $\sim 20$ & $\sim 20$ & $+2.8(1.1)$ \\ 
FRB 20190222A & FRB 190222.J2052+69$^{c}$ & $\sim 87$ & $\sim 101$ & $+10(7)$ \\
FRB 20190208A & FRB 190208.J1855+46$^{d}$ & $\sim 65$ & $\sim 71$ & $+8(13)$ \\
FRB 20190604A & FRB 190604.J1435+53$^{d}$ & $\sim 32$ & $\sim 24$ & $+12(2)$ \\
FRB 20190213B & FRB 190212.J18+81$^{d}$ & $\sim 53$ & $\sim 49$ & $+9(13)$\\
FRB 20180908B & FRB 180908.J1232+74$^{d}$ & $\sim 38$ & $\sim 31$ & $-7(7)$ \\
FRB 20190117A & FRB 190117.J2207+17$^{d}$ & $\sim 48$ & $\sim 40$ & $-26(8)$\\
FRB 20190303A & FRB 190303.J1353+48$^{d}$ & $\sim30$ & $\sim22$ & $+21(5)$  \\
FRB 20190417A & FRB 190417.J1939+59$^{d}$ & $\sim 80$ & $\sim 83$ & $+36(13)$\\
FRB 20190907A & FRB 190907.J08+46$^{d}$ & $\sim 53$ & $\sim 50$ & $+7(4)$ \\
%\bottomrule
 \hline
\end{tabular}
\label{ta:dmrmgal}
\end{center}
$^a$ These estimates do not include the DM contribution of the Galactic halo, which is expected to contribute an average value of $\sim 30-50 \; \rm{rad\, m^{-2}}$, depending on the halo model assumed \citep[e.g.,][]{Dolag2015, Yamasaki2020}. \\
$^{b}$ \citet{chime/frb2019b} \\
$^{c}$ \citet{chime/frb2019} \\
$^{d}$ \citet{Fonseca2020}
\end{table*}

\subsection{Ionospheric Rotation Measure Estimates}

Earth's ionosphere imparts a small but measurable contribution to the observed Faraday rotation of any linear polarized source. This contribution, $\rm{RM_{iono}}$, must be corrected to determine the significance of any RM variations. $\rm{RM_{iono}}$ can vary significantly depending on time of day, solar cycle and pointing \citep{Mevius2018a}. However, as described in \citet{Mckinven2022}, most FRB detections with CHIME occur very near (a few degrees) of zenith, limiting the variation in $\rm{RM_{iono}}$ that would otherwise be introduced from extreme changes in pointings through the ionosphere. As was the case for our observations of FRB 20180916B, $\rm{RM_{iono}}$ estimates are obtained from the {\tt RMextract} package\footnote{https://github.com/lofar-astron/RMextract} \citep{Mevius2018b} whose values are provided in Table~\ref{ta:bursts}.

\section{Results}
\label{sec:results}

Table~\ref{ta:bursts} reports the results of our polarimetric analysis of our sample of known repeating FRB sources \citep{chime/frb2019b, chime/frb2019,Fonseca2020}. The sample contains several sources with burst separations of multiple years. Bursts from the prolific CHIME/FRB source, FRB 20190303A, are summarized in Figure~\ref{fig:waterfalls_R17} in the form of dynamic spectra (referred to as `waterfall plots'), displaying each burst as a function of frequency and time. Here, data have been rebinned to highlight substructure of individual bursts. Bursts have been dedispersed by their individual structure optimizing DM \citep[$\rm{DM_{struct}}$;][]{Seymour2019} 
which was determined by applying the {\tt DM\_phase} package\footnote{https://github.com/danielemichilli/DM\_phase} on rebinned data. Waterfall plots of bursts from other repeaters can be found in Appendix~\ref{appendix:waterfall} (see Figure set 1 for waterfall plots of remaining sample).

Burst profiles are shown directly above waterfall plots and display the \textit{subband} averaged signal in total (black), linear polarized (red) and circular polarized intensities (blue) as a function of time. Each burst's subband has been determined by eye and is indicated by an orange line along the left vertical axis and reported in Table~\ref{ta:bursts} under the ``Emitting Band" column. This quantity is not corrected for the non-uniform bandpass of the CHIME, including the well-known bias for detecting bursts in the lower part of the band where FFT beams of the realtime detection pipeline are wider \citep{chime/frb2018}. Adjoining the burst profiles in Figure~\ref{fig:waterfalls_R17} (see also Figure set 1) is an additional panel showing PA behavior as a function of time. CHIME/FRB observations cannot robustly calibrate for the PA value \citep{Mckinven2021}; thus, PA profiles shown are only informative for the evolution over the burst duration and have been determined from Equation 3 of \citet{Mckinven2022} which centers PA values relative to $\sim 0$ degrees. \footnote{The handful of events with PA curves that are significantly offset from PA$\sim 0$ degrees is an imprint of instrumental polarization, namely, a differential response between the primary beam of CHIME for the two X, Y polarizations.} Linear polarization, $L = \sqrt{Q^2+U^2}$, has been de-biased by applying Equation 11 of \citet{Everett2001} (see \citet{Mckinven2022} for further details). We do not display corresponding linear/circular/PA profiles for bursts without a significant RM detection since these events are disproportionately affected by residual instrumental polarization, which is more problematic when the polarized astrophysical signal is marginal or non-existent. These events are not so much unpolarized as polarized below the sensitivity threshold of CHIME, and are indicated in Figure~\ref{fig:waterfalls_R17} and elsewhere as events where only total intensity profiles are shown. 

The remainder of this section studies burst properties as a function of time of one of the more prolific repeating CHME/FRB sources, FRB 20190303A (\S\ref{sec:res_R17}). This is followed by a summary of equivalent analysis for the other repeaters that are more sparsely sampled in time (\S\ref{sec:other_rep}). 

 \pagebreak
\begin{ThreePartTable}
\begin{TableNotes}
\item[$^*$]{Bursts whose intensity data were previously reported by \citet{chime/frb2019}.} 
\item[$^{\dagger}$]{Bursts whose intensity data were previously reported by \citet{chime/frb2020c}.}
\item[$^a$]{Unconstrained parameters are listed as ``$-$''. Uncertainties are reported at the $1\sigma$ confidence level.} 
\item[$^b$]{Topocentric TOAs provided here are in Modified Julian Date (MJD) format, referenced at 400 MHz with $\sim 1$ second precision.} 
\item[$^c$]{From structure-optimization (see Section 2.2.1 of \citet{Mckinven2022}.}
\item[$^d$]{The downsampling factor that determines the time resolution (i.e. $n_{\rm{down}} \times 2.56 \, \mu s $) of the waterfalls plots displayed in Figure~\ref{fig:waterfalls_R17} and Figure set 1.}
\item[$^e$]{The linear polarization fraction determined from integrating $L$ over the burst profile. Upper limits of unpolarized bursts are indicated by a `$<$' prefix.}
\item[$^f$]{The portion of the 400$-$800 MHz CHIME band over which emission is observed, uncorrected for the non-uniform bandpass of the instrument and corresponding to vertical orange lines in burst waterfalls.}
\item[$^g$]{Quoted uncertainties refer to the statistical error, $\sigma_{\rm{stat}}$, and do not account for additional systematic errors, which we estimate as $\sigma_{\rm{sys}} = 0.85$ rad m$^{-2}$ based on multi-epoch polarimertic monitoring of giant pulse data from the Crab pulsar \citep[see][for details]{Mckinven2022}.}  
\end{TableNotes} 
{\small\tabcolsep=1.5pt  % hold it local
\begin{longtable*}[l]{cccccccccc}
\bottomrule
\insertTableNotes
\endlastfoot
\caption{Individual burst properties from a sample of repeating CHIME/FRB sources$^{a}$.} 
\endfirsthead
\caption{\textit{continued}} \endhead
\hline
Burst \# \hspace{0.5pt} & Arrival Time$^{b}$ & S/N & $\rm{DM_{struct}}^{c}$ & $\rm{n_{down}}^{d}$ & $\rm{\langle L/I \rangle}^{e}$ & Emitting Band$^{f}$ & $\rm{RM_{FDF}}$ & $\rm{RM_{QU}}$ & $\rm{RM_{iono}}$ \\
  & (MJD) & & (pc~cm$^{-3}$) & & & (MHz) & (rad m$^{-2}$) & (rad m$^{-2}$) & (rad m$^{-2}$) \\
\hline
 \multicolumn{10}{c}{FRB20180814A} \\
 \hline
1	&	58645.78660	&	43.8	&	$190.13(84)$	&	128	&	$<0.11$	&	600-730	&	$-$	&	$-$	&	+0.94\\
2	&	58659.73874	&	22.4	&	$188.743(70)$	&	256	&	$<0.22$	&	400-470	&	$-$	&	$-$	&	+0.89\\
3	&	58660.77631	&	26.8	&	$191.62(73)$	&	512	&	$<0.19$	&	400-470	&	$-$	&	$-$	&	+0.73\\
4	&	58785.40415	&	46.3	&	$189.02(35)$	&	128	&	$0.691(21)$	&	670-800	&	$+699.8(1.0)$	&	$+700.1(1.3)$	&	+0.17\\
5	&	58798.37171	&	33.0	&	$188.552(35)$	&	256	&	$<0.15$	&	520-650	&	$-$	&	$-$	&	+0.21\\
\hline
 \multicolumn{10}{c}{FRB 20181030A} \\
 \hline
1	&	58870.43093	&	43.9	&	$103.500(13)$	&	32	&	$0.728(23)$	&	400-560	&	$+36.39(29)$	&	$+36.52(30)$	&	+0.24\\
2	&	58870.43542	&	31.3	&	$103.504(11)$	&	32	&	$0.601(27)$	&	400-500	&	$+37.67(54)$	&	$+38.53(64)$	&	+0.23\\
3	&	58870.92327	&	22.5	&	$103.586(55)$	&	32	&	$<0.22$	&	400-520	&	$-$	&	$-$	&	+0.53\\
4	&	58870.92352	&	26.8	&	$103.473(35)$	&	128	&	$<0.19$	&	400-550	&	$-$	&	$-$	&	+0.53\\
5	&	58870.93287	&	30.6	&	$103.493(83)$	&	256	&	$0.928(43)$	&	400-600	&	$+37.08(51)$	&	$+36.60(58)$	&	+0.51\\
  \hline
 \multicolumn{10}{c}{FRB 20181119A} \\
 \hline
1	&	59021.11044	&	22.3	&	$364.03(16)$	&	128	&	$1.024(65)$	&	660-790	&	$+606.2(1.2)$	&	$+608.8(1.6)$	&	+0.77\\
2	&	59187.65926	&	18.1	&	$364.33(15)$	&	256	&	$0.473(37)$	&	630-730	&	$+1343.7(2.0)$	&	$+1339.3(2.7)$	&	+0.53\\
3	&	59379.13757	&	12.7	&	$364.009(29)$	&	512	&	$<0.39$	&	420-550	&	$-$	&	$-$	&	+0.88\\
4	&	59389.10515	&	36.9	&	$363.59(33)$	&	64	&	$0.679(26)$	&	590-720	&	$+480.95(59)$	&	$+479.20(81)$	&	+0.98\\
5	&	59529.72050	&	28.5	&	$364.78(67)$	&	256	&	$<0.18$	&	580-700	&	$-$	&	$-$	&	+0.96\\
  \hline
 \multicolumn{10}{c}{FRB 20190222A} \\
 \hline
 1$^*$	&	58543.75211	&	59.8	&	$459.58(10)$	&	16	&	$0.2461(58)$	&	530-660	&	$+571.2(1.7)$	&	$+567.7(2.3)$	&	+0.65\\
  \hline
 \multicolumn{10}{c}{FRB 20190208A} \\
 \hline
1	&	58872.77984	&	32.6	&	$579.889(55)$	&	64	&	$0.885(38)$	&	410-670	&	$+32.99(36)$	&	$+31.81(53)$	&	+0.74\\
2	&	58982.47538	&	23.0	&	$579.33(16)$	&	128	&	$0.884(54)$	&	650-800	&	$+17.5(1.0)$	&	$+17.4(1.1)$	&	+0.43\\
3	&	59248.74874	&	113.6	&	$579.858(19)$	&	8	&	$0.903(11)$	&	630-800	&	$+13.19(25)$	&	$+13.09(38)$	&	+0.86\\
4	&	59377.39380	&	19.7	&	$579.836(25)$	&	128	&	$0.965(69)$	&	485-605	&	$+10.9(1.1)$	&	$+10.1(1.5)$	&	+0.32\\
5	&	59496.07040	&	42.1	&	$579.859(11)$	&	32	&	$0.860(29)$	&	400-490	&	$+24.28(34)$	&	$+25.75(18)$	&	+0.77\\
6	&	59576.85105	&	43.1	&	$579.718(41)$	&	64	&	$0.913(30)$	&	400-570	&	$+27.90(23)$	&	$+29.00(19)$	&	+1.70\\
7	&	59577.84674	&	39.6	&	$579.97(11)$	&	64	&	$0.686(25)$	&	440-800	&	$+36.29(74)$	&	$+32.13(81)$	&	+1.66\\
\hline
\multicolumn{10}{c}{FRB 20190604A} \\
\hline
1$^\dagger$ &	58640.23231	& 21.7 & $552.53(24)$	&	128	&	$0.912(81)$	&	500-625	&	$-17.8(8)$	&	$-17.8(1.2)$	& +0.59\\
\hline
\multicolumn{10}{c}{FRB 20190213B} \\
\hline
1	&	58834.36037	&	21.3	&	$301.64(12)$	&	256	&	$0.742(49)$	&	400-670	&	$-3.63(41)$	&	$-5.28(63)$	&	+0.11\\
2	&	59055.23537	&	18.1	&	$301.38(22)$	&	256	&	$0.951(74)$	&	400-500	&	$+1.04(42)$	&	$-0.41(56)$	&	+0.34\\
3	&	59261.71552	&	35.8	&	$301.400(48)$	&	128	&	$0.921(36)$	&	420-500	&	$+1.06(52)$	&	$-0.2(1.0)$	&	+0.69\\
4	&	59314.04279	&	26.2	&	$301.720(64)$	&	256	&	$0.934(50)$	&	510-620	&	$+0.97(52)$	&	$-0.25(66)$	&	+0.54\\
5	&	59431.25656	&	22.6	&	$301.430(19)$	&	128	&	$1.009(63)$	&	400-510	&	$-0.28(42)$	&	$-0.05(64)$	&	+0.38\\
% \pagebreak
\hline
 \multicolumn{10}{c}{FRB 20180908B} \\
 \hline
1	&	58655.09819	&	28.0	&	$195.33(38)$	&	256	&	$-$	&	400-450	&	$<0.18$	&	$-$	&	+0.80\\
 \hline
 \multicolumn{10}{c}{FRB 20190117A} \\
 \hline
1	&	58500.92954	&	118.9	&	$393.062(74)$	&	8	&	$0.1945(23)$	&	460-570	&	$+74.30(68)$	&	$+70.97(80)$	&	+0.84\\
2	&	58840.00494	&	23.2	&	$392.1123(19)$	&	256	&	$<0.22$             &   460-590	&	$-$	&	$-$	&	+0.57\\
3	&	59532.10691	&	222.5	&	$395.804(36)$	&	8	&	$0.14333(91)$	&	420-590	&	$+76.31(40)$	&	$+79.57(58)$	&	+0.44 \\
\hline
\multicolumn{10}{c}{FRB 20190303A} \\
 \hline
1$^{\dagger}$	&	58666.13522	&	76.8	&	$221.43(21)$	&	8	&	$0.3248(60)$	&	640-800	&	$-489.3(6.3)$	&	$-499.5(3.4)$	&	+0.95\\
2	&	58769.85517	&	45.0	&	$221.233(11)$	&	256	&	$<0.11$	&	425-550	&	$-$	&	$-$	&	+0.86\\
3	&	58776.82559	&	82.1	&	$221.579(97)$	&	4	&	$0.5413(93)$	&	580-730	&	$-498.52(75)$	&	$-501.14(73)$	&	+0.90\\
4	&	58797.77293	&	113.7	&	$221.354(59)$	&	32	&	$0.1861(23)$	&	475-620	&	$-601.2(4.5)$	&	$-594.8(1.7)$	&	+0.80\\
5	&	58800.75865	&	54.8	&	$221.353(82)$	&	32	&	$<0.09$	&	510-610	&	$-$	&	$-$	&	+0.68\\
6	&	58803.75684	&	208.9	&	$221.473(62)$	&	1	&	$0.1810(12)$	&	575-715	&	$-701.31(43)$	&	$-703.40(58)$	&	+0.93\\
7	&	58804.76286	&	33.5	&	$221.264(61)$	&	256	&	$<0.15$	&	400-480	&	$-$	&	$-$	&	+0.80\\
8	&	58832.67612	&	46.8	&	$221.55(16)$	&	32	&	$<0.11$	&	450-610	&	$-$	&	$-$	&	+0.44\\
9	&	58848.63738	&	27.2	&	$221.64(57)$	&	64	&	$0.487(25)$	&	580-690	&	$-556.7(2.4)$	&	$-563.9(3.5)$	&	+0.36\\
10	&	58860.60178	&	38.8	&	$221.70(25)$	&	128	&	$<0.13$	&	425-505	&	$-$	&	$-$	&	+0.27\\
11	&	59022.16102	&	36.8	&	$222.04(29)$	&	128	&	$0.716(27)$	&	700-800	&	$-559.3(3.9)$	&	$-560.1(4.3)$	&	+0.91\\
12	&	59070.02806	&	57.8	&	$221.39(38)$	&	4	&	$0.723(18)$	&	550-685	&	$-483.72(74)$	&	$-485.57(83)$	&	+0.96\\
13	&	59101.93754	&	68.7	&	$221.71(15)$	&	128	&	$<0.07$	&	410-460	&	$-$	&	$-$	&	+0.82\\
14	&	59248.53543	&	57.7	&	$221.333(43)$	&	16	&	$0.768(19)$	&	660-800	&	$-362.65(73)$	&	$-363.74(83)$	&	+0.26\\
15	&	59252.52666	&	80.2	&	$221.78(24)$	&	8	&	$0.841(15)$	&	635-800	&	$-408.76(33)$	&	$-409.11(34)$	&	+0.23\\
16	&	59254.52323	&	102.2	&	$221.72(12)$	&	8	&	$0.4339(60)$	&	600-800	&	$-421.33(37)$	&	$-421.54(45)$	&	+0.18\\
17	&	59275.46633	&	31.3	&	$221.387(90)$	&	128	&	$<0.16$	&	480-615	&	$-$	&	$-$	&	+0.23\\
18	&	59501.84489	&	29.9	&	$221.36(50)$	&	128	&	$0.611(29)$	&	520-690	&	$-270.2(1.2)$	&	$-269.0(1.7)$	&	+1.40\\
19	&	59525.77798	&	60.1	&	$221.49(48)$	&	16	&	$0.555(13)$	&	400-560	&	$-200.34(30)$	&	$-205.41(42)$	&	+1.26\\
20	&	59543.73255	&	37.5	&	$221.91(31)$	&	128	&	$<0.13$	&	400-500	&	$-$	&	$-$	&	+0.93\\
\hline
 \multicolumn{10}{c}{FRB 20190417A} \\
  \hline 
1	&	59056.30674	&	56.2	&	$1378.58(13)$	&	16	&	$0.1466(37)$	&	620-710	&	$+4429.8(3.5)$	&	$+4429.3(4.1)$	&	+0.26\\
2	&	59277.69667	&	36.0	&	$1378.58(22)$	&	64	&	$<0.14$	&	500-620	&	$-$	&	$-$	&	+0.70\\
\hline
 \multicolumn{10}{c}{FRB 20190907A} \\
 \hline
1	&	59059.82090	&	18.1	&	$309.30(31)$	&	512	&	$<0.28$	&	400-535	&	$-$	&	$-$	&	+0.90\\
\hline
\label{ta:bursts}
\end{longtable*}
}
\end{ThreePartTable}

%%%%% R17 %%%%%%

\subsection{FRB 20190303A} % R17
\label{sec:res_R17}

\begin{figure*}[!htb]
\vspace{0.7in}
	\centering
\begin{center}
    \includegraphics[width=0.19\textwidth]{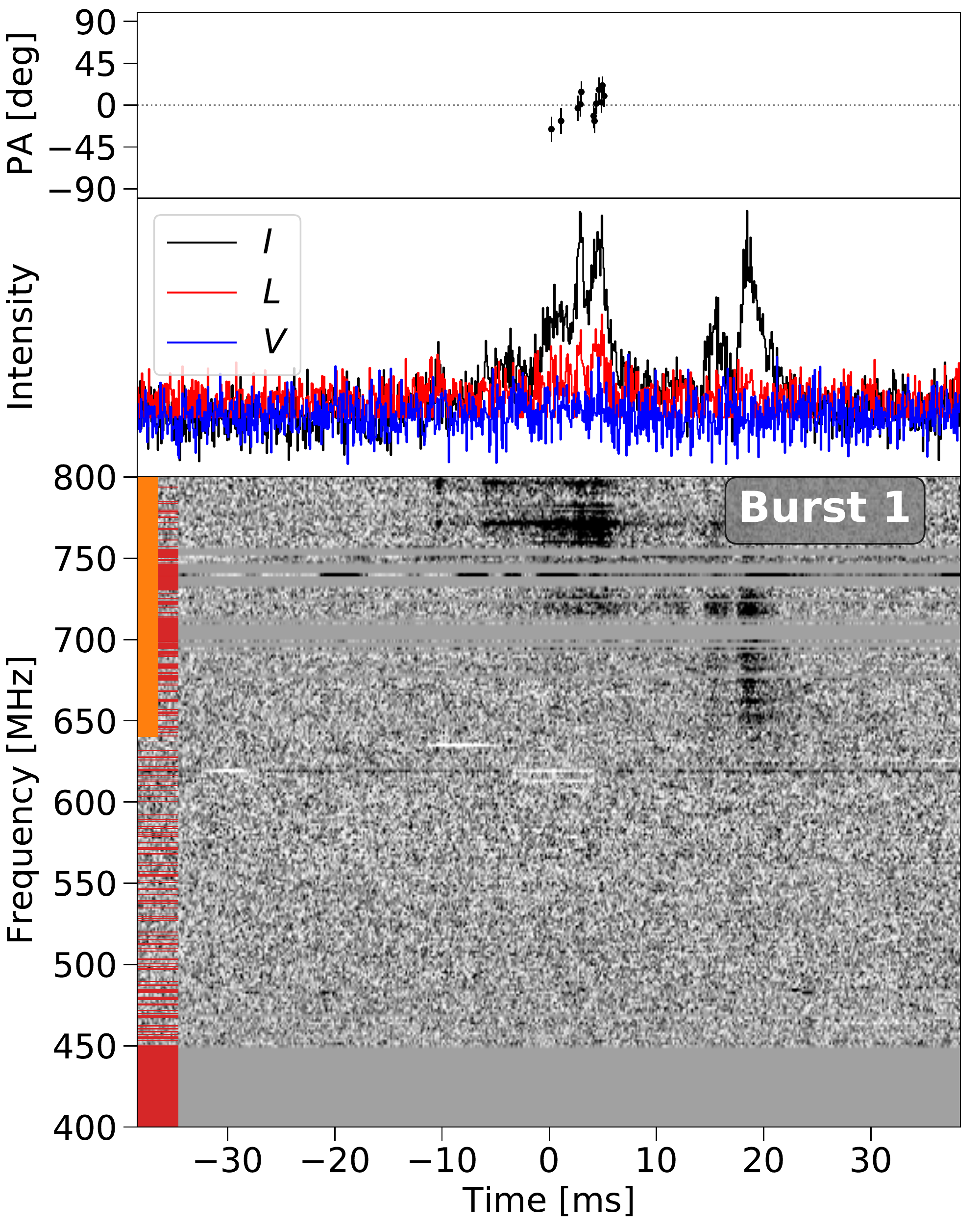}
    \includegraphics[width=0.19\textwidth]{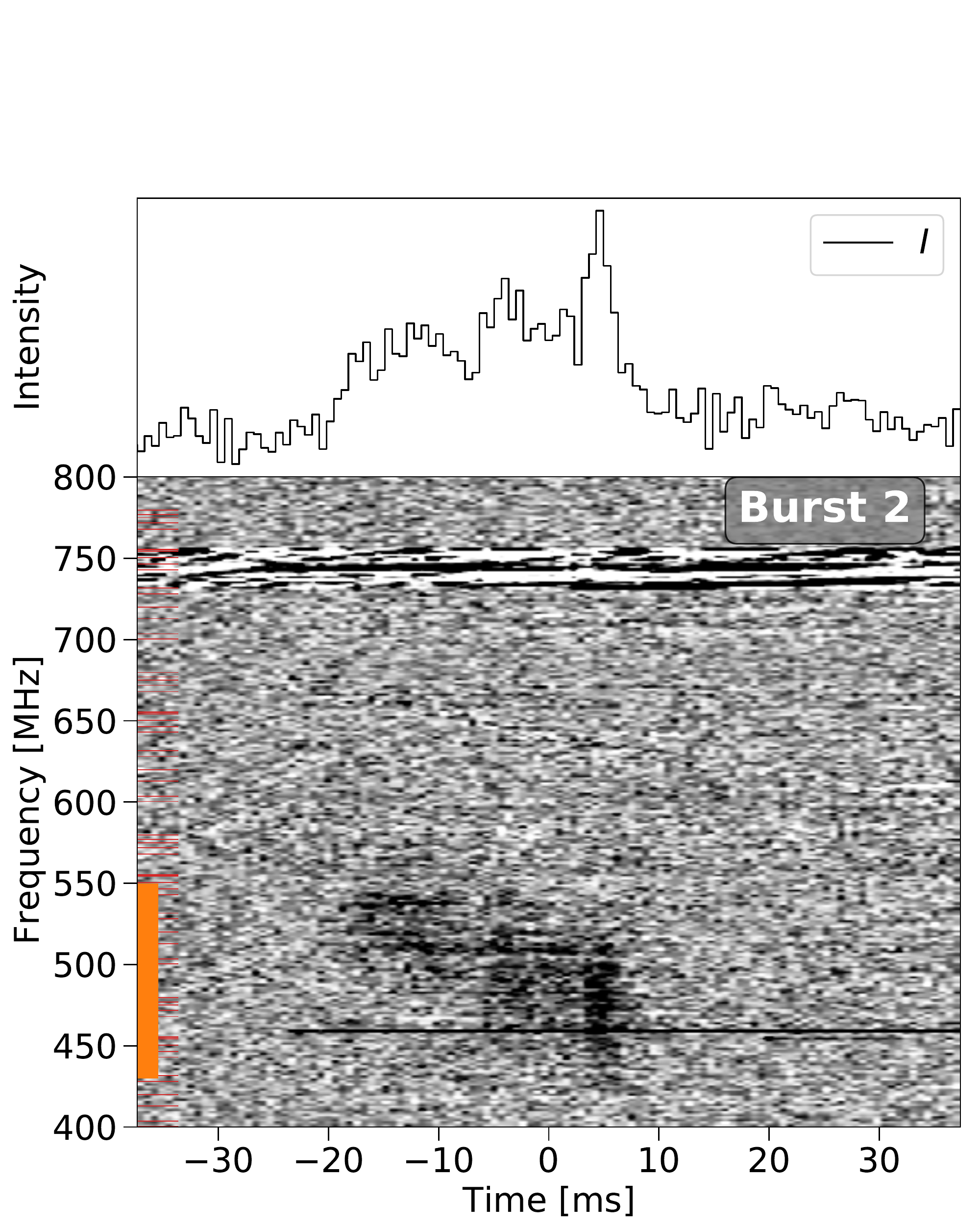}
    \includegraphics[width=0.19\textwidth]{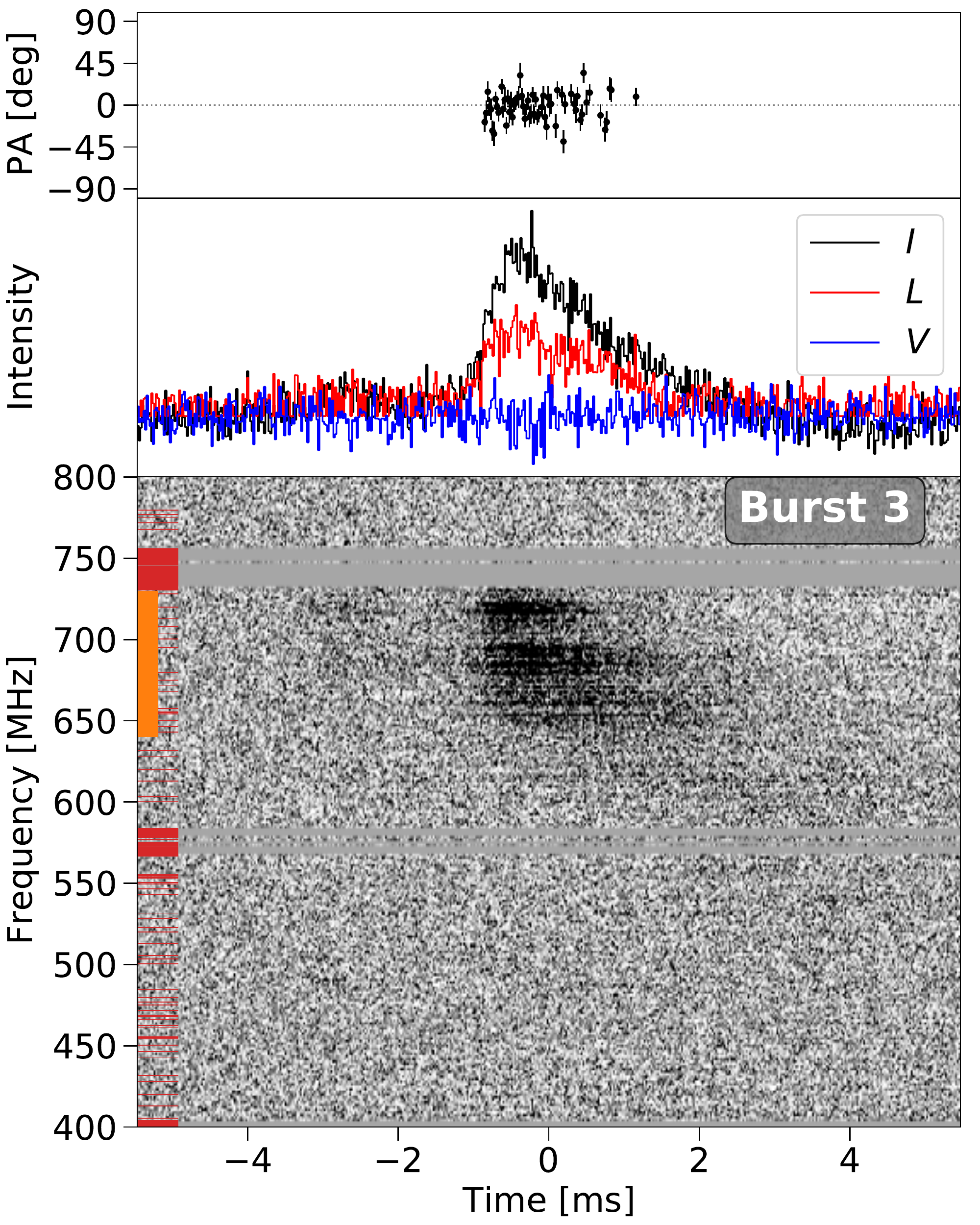} 
    \includegraphics[width=0.19\textwidth]{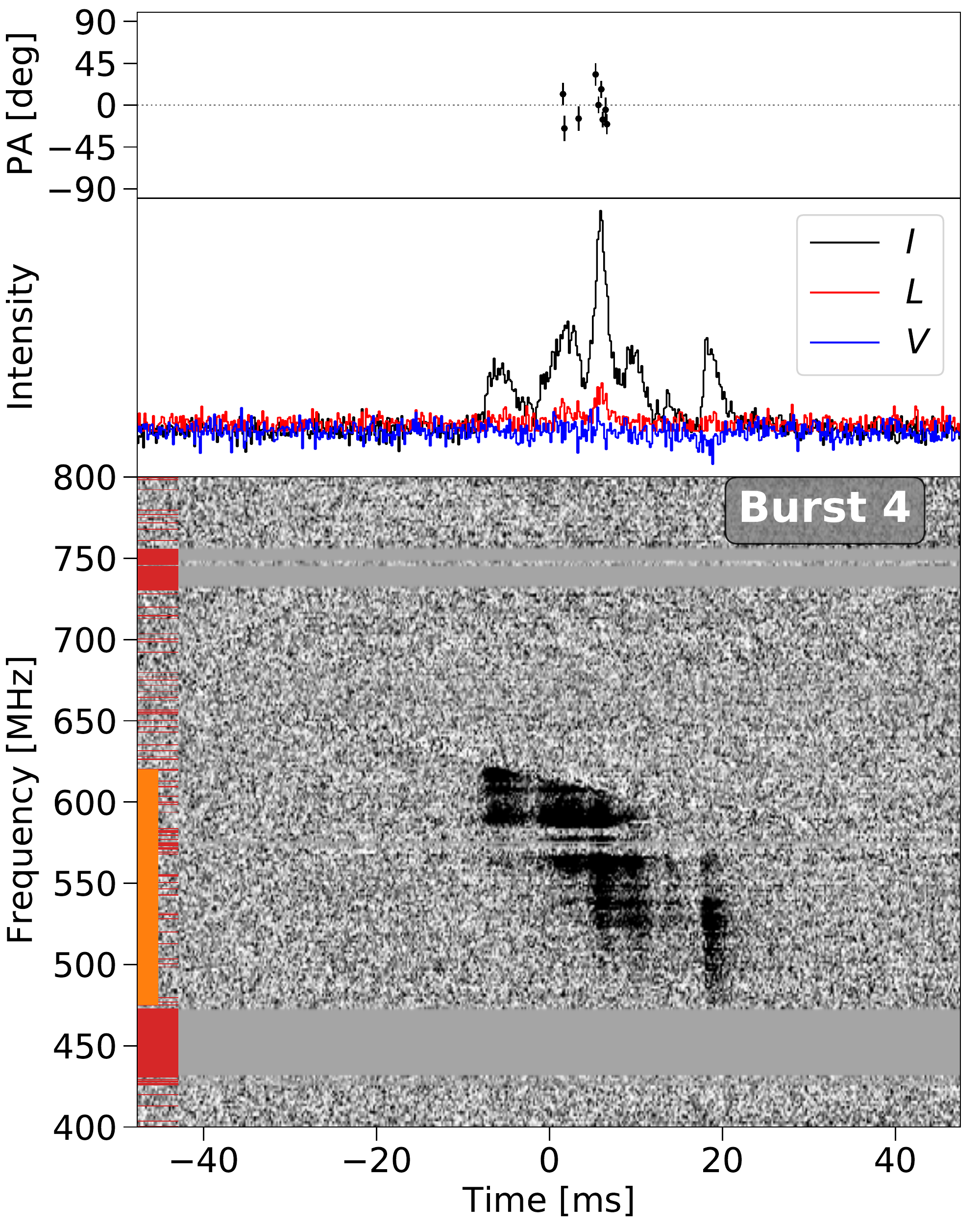}
    \includegraphics[width=0.19\textwidth]{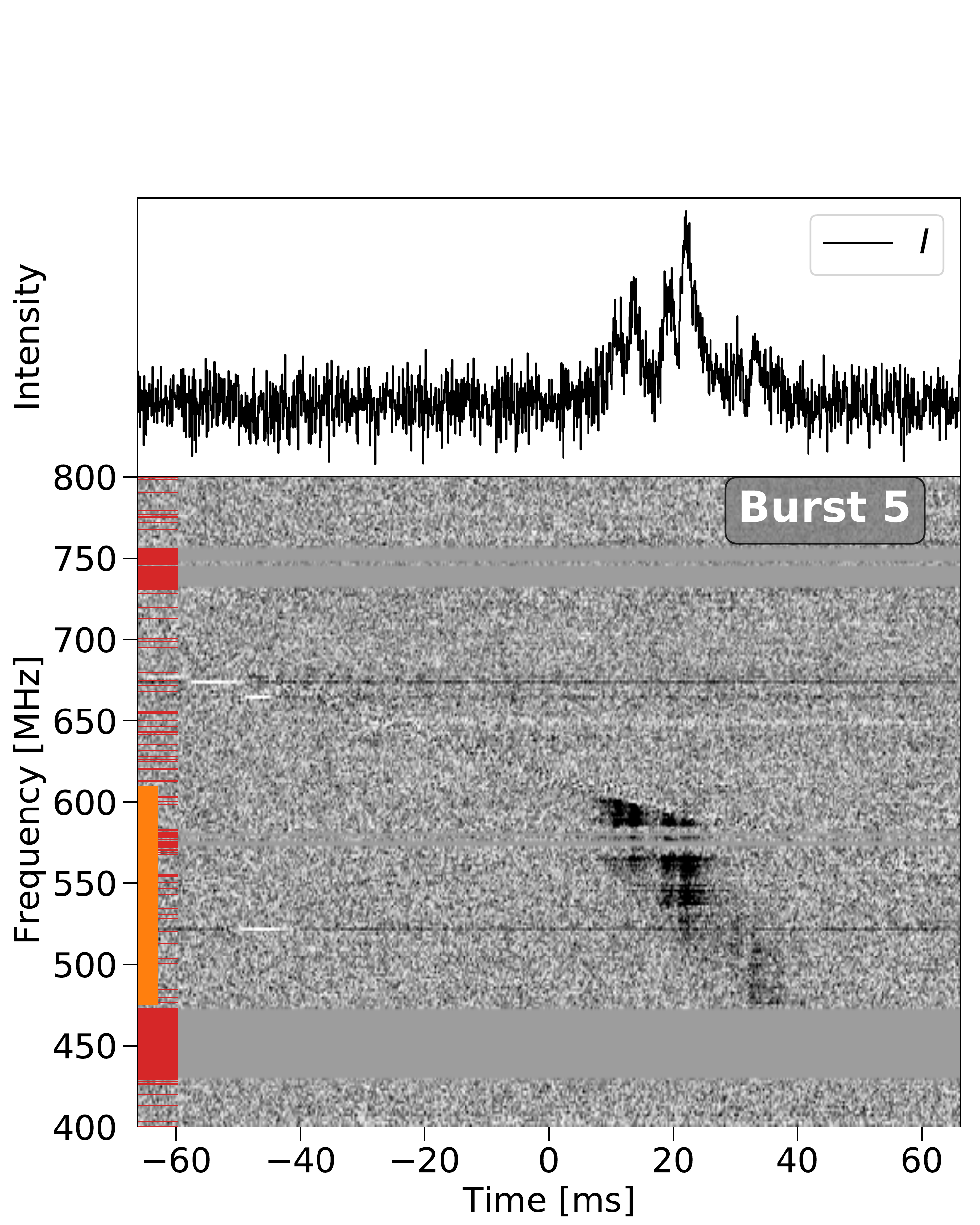}
    \includegraphics[width=0.19\textwidth]{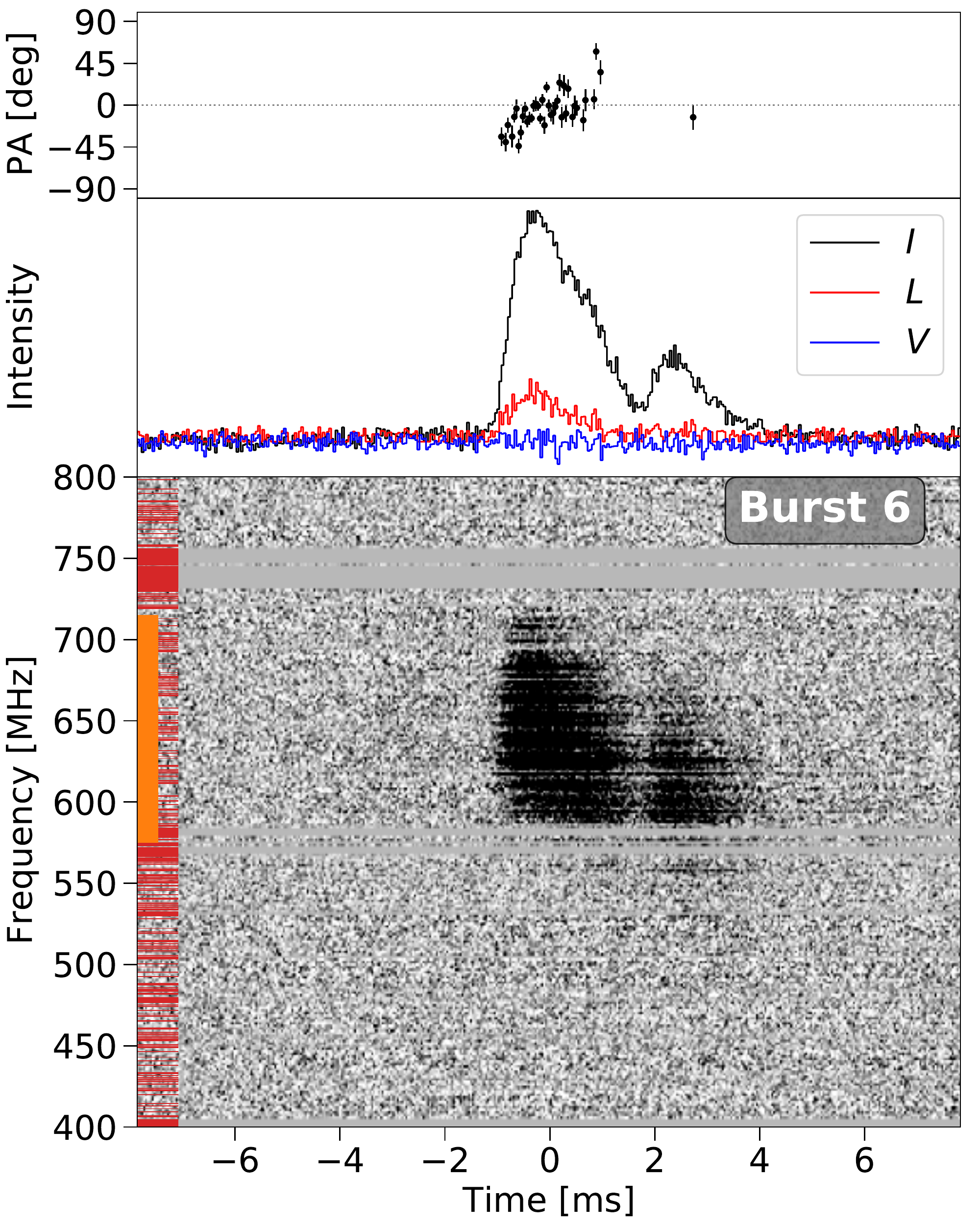}
    \includegraphics[width=0.19\textwidth]{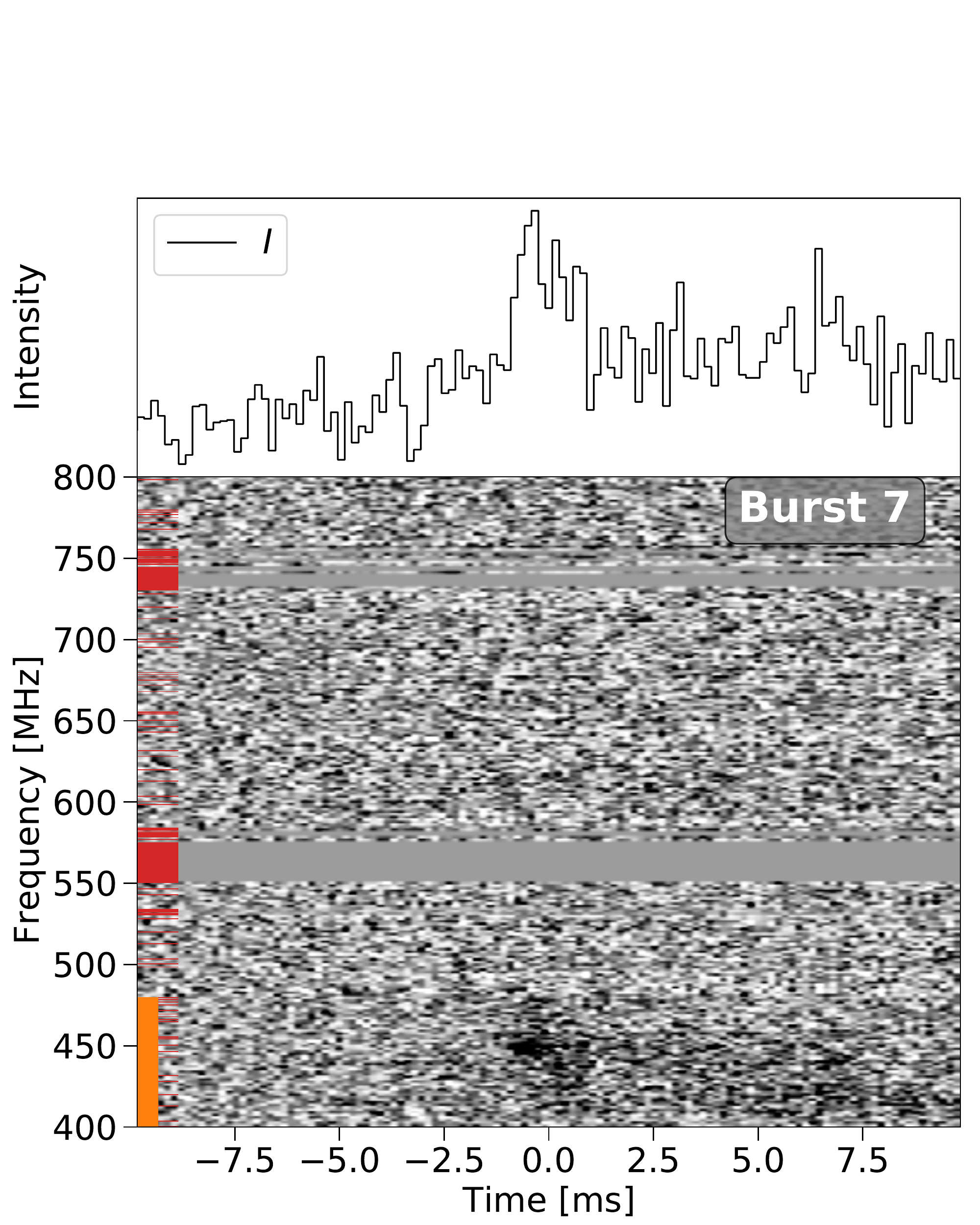}
    \includegraphics[width=0.19\textwidth]{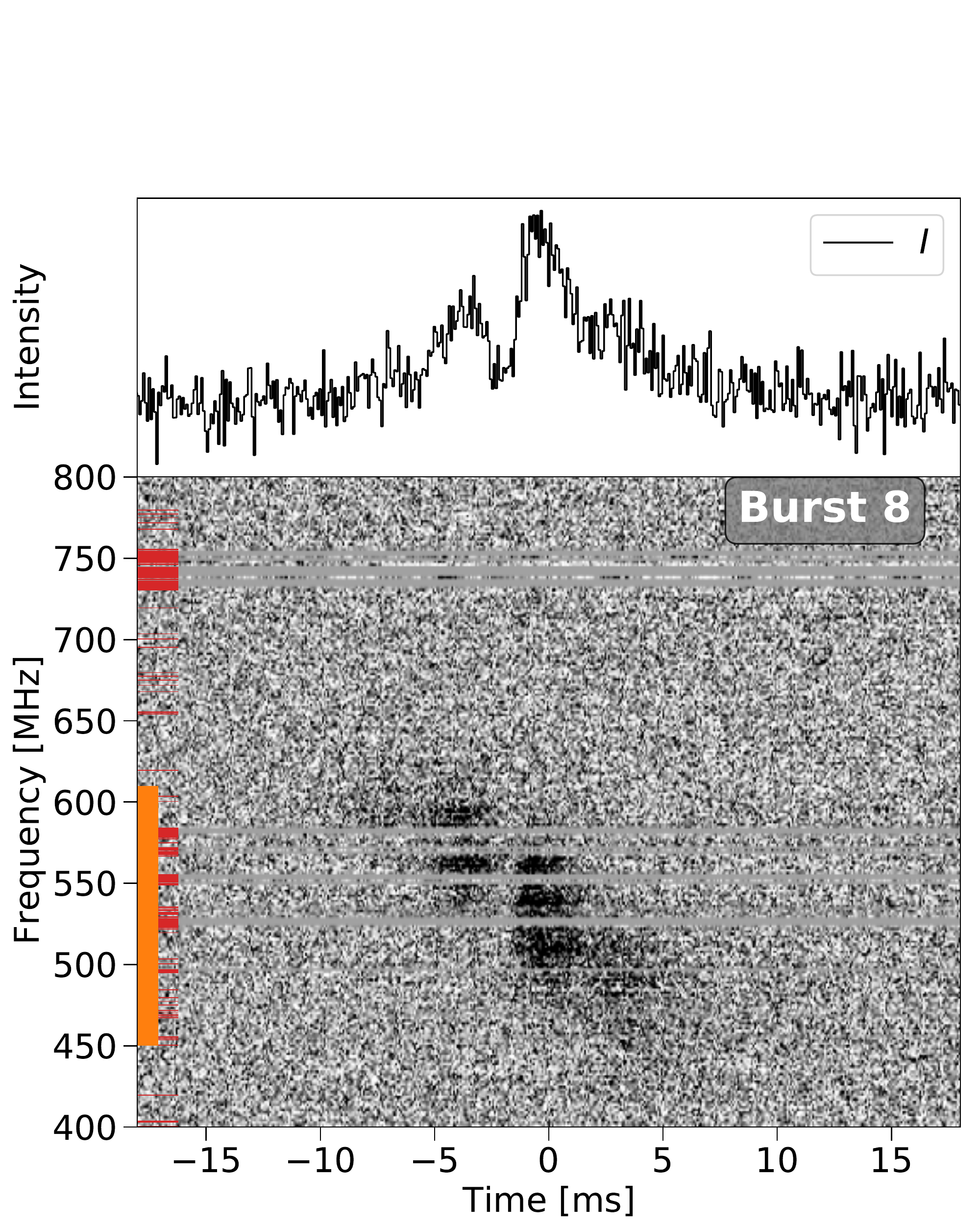}
    \includegraphics[width=0.19\textwidth]{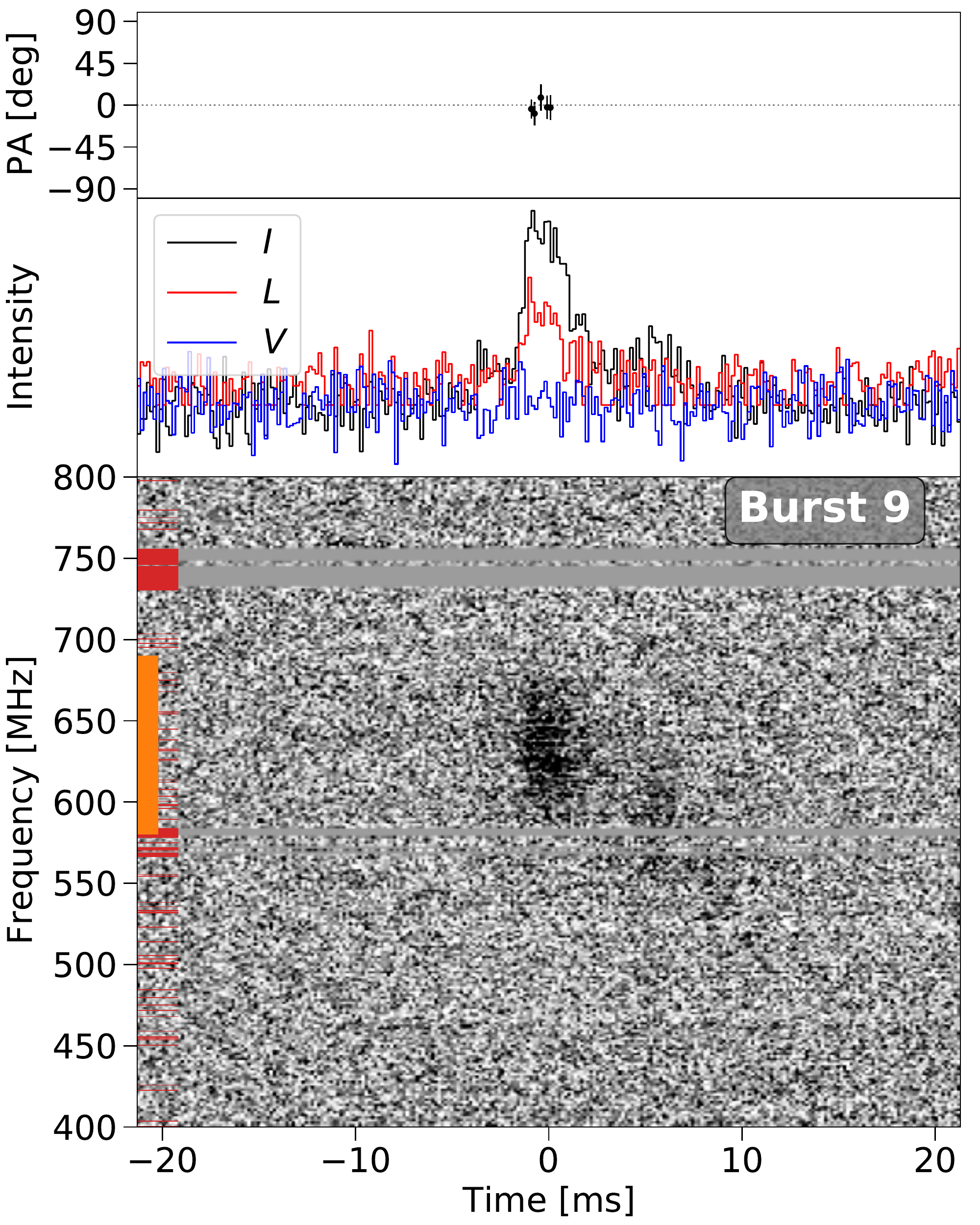}
    \includegraphics[width=0.19\textwidth]{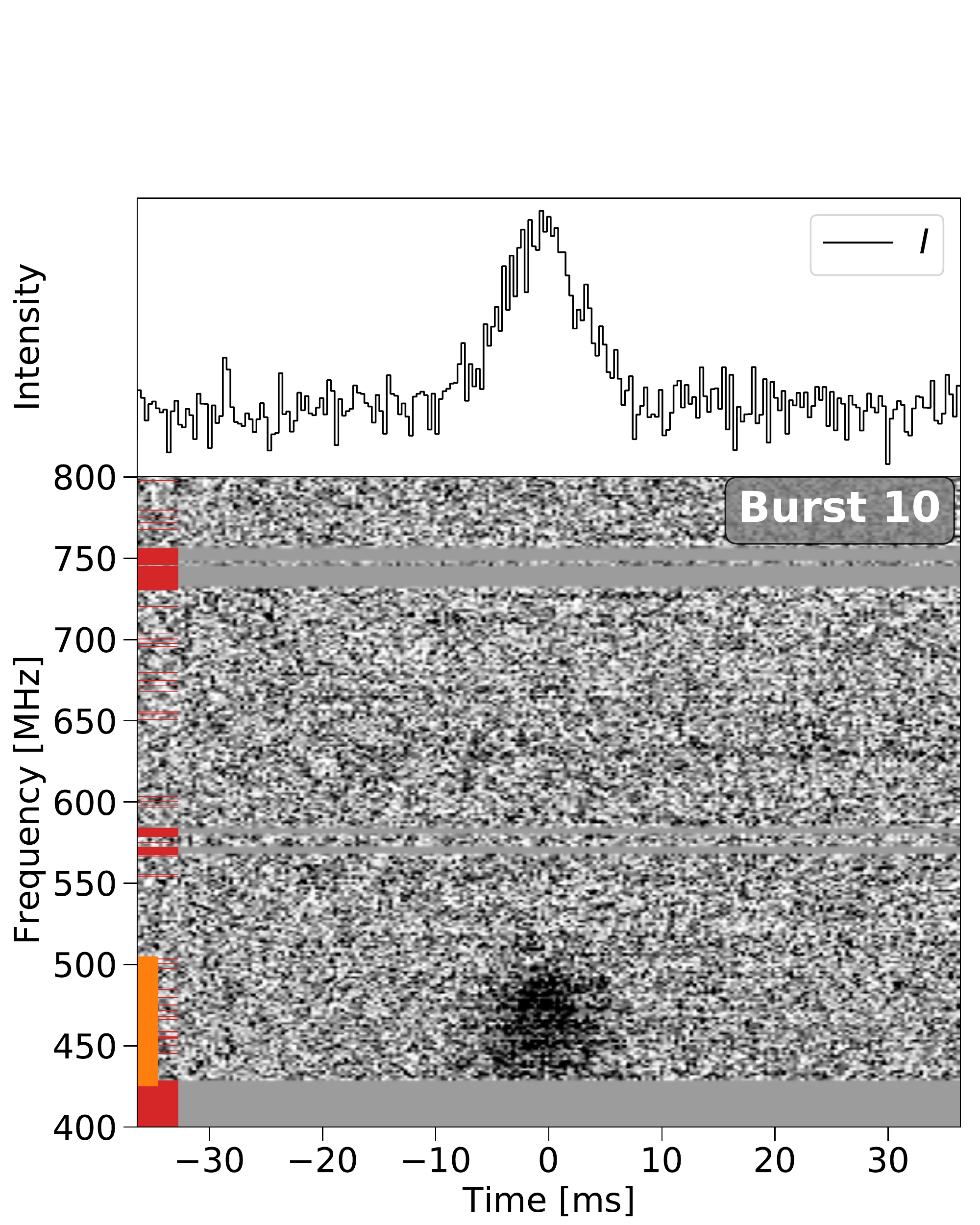}
    \includegraphics[width=0.19\textwidth]{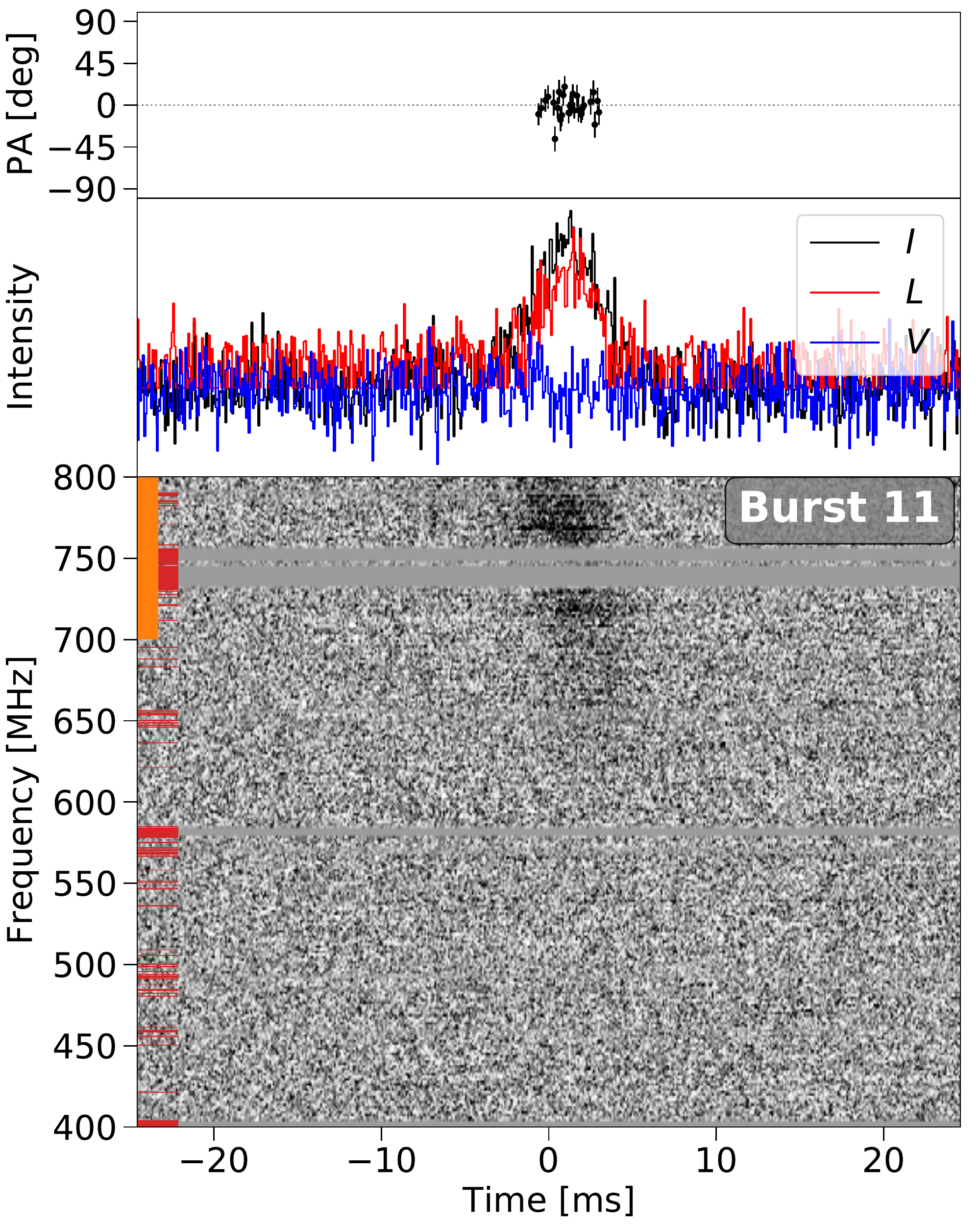}
    \includegraphics[width=0.19\textwidth]{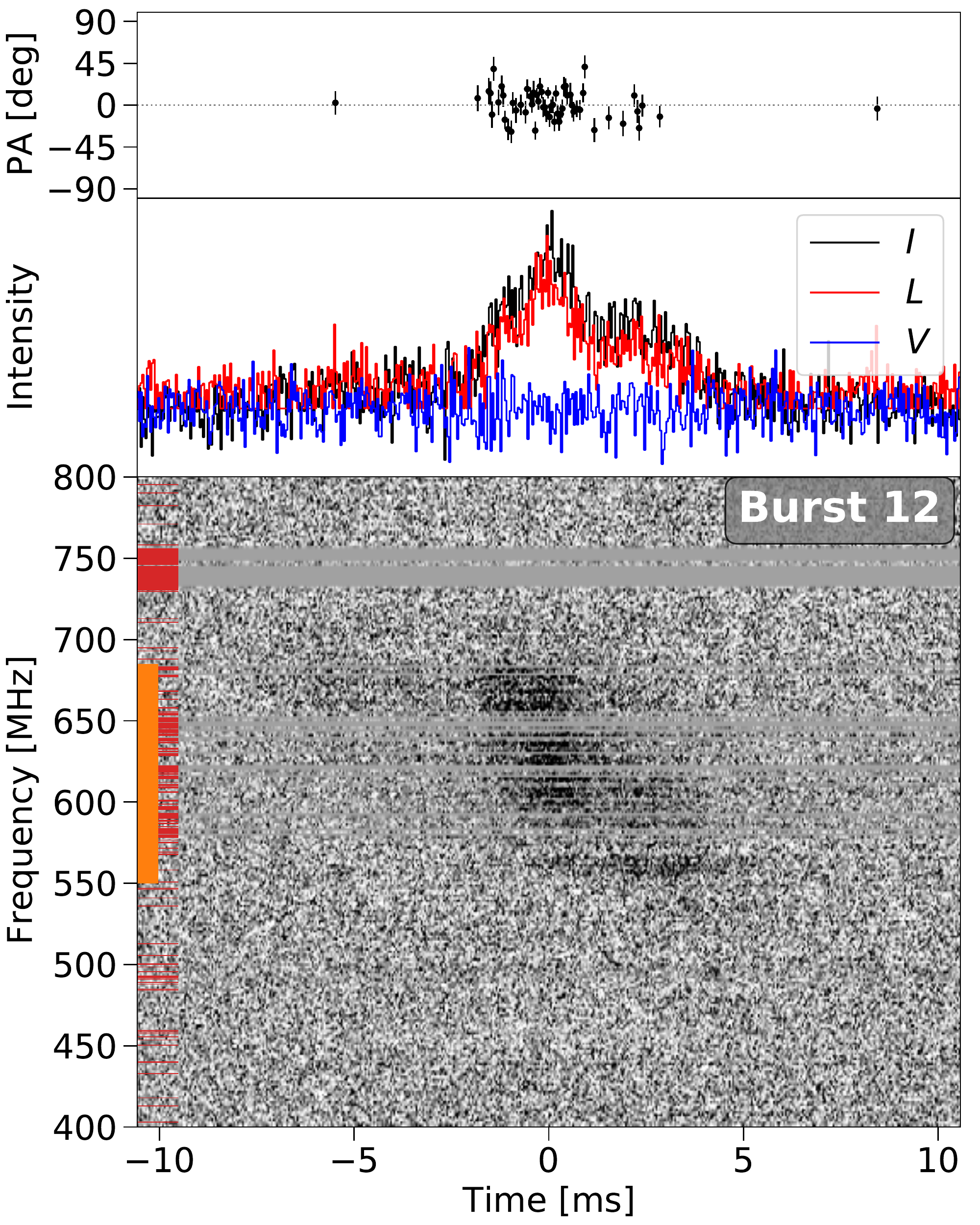}
    \includegraphics[width=0.19\textwidth]{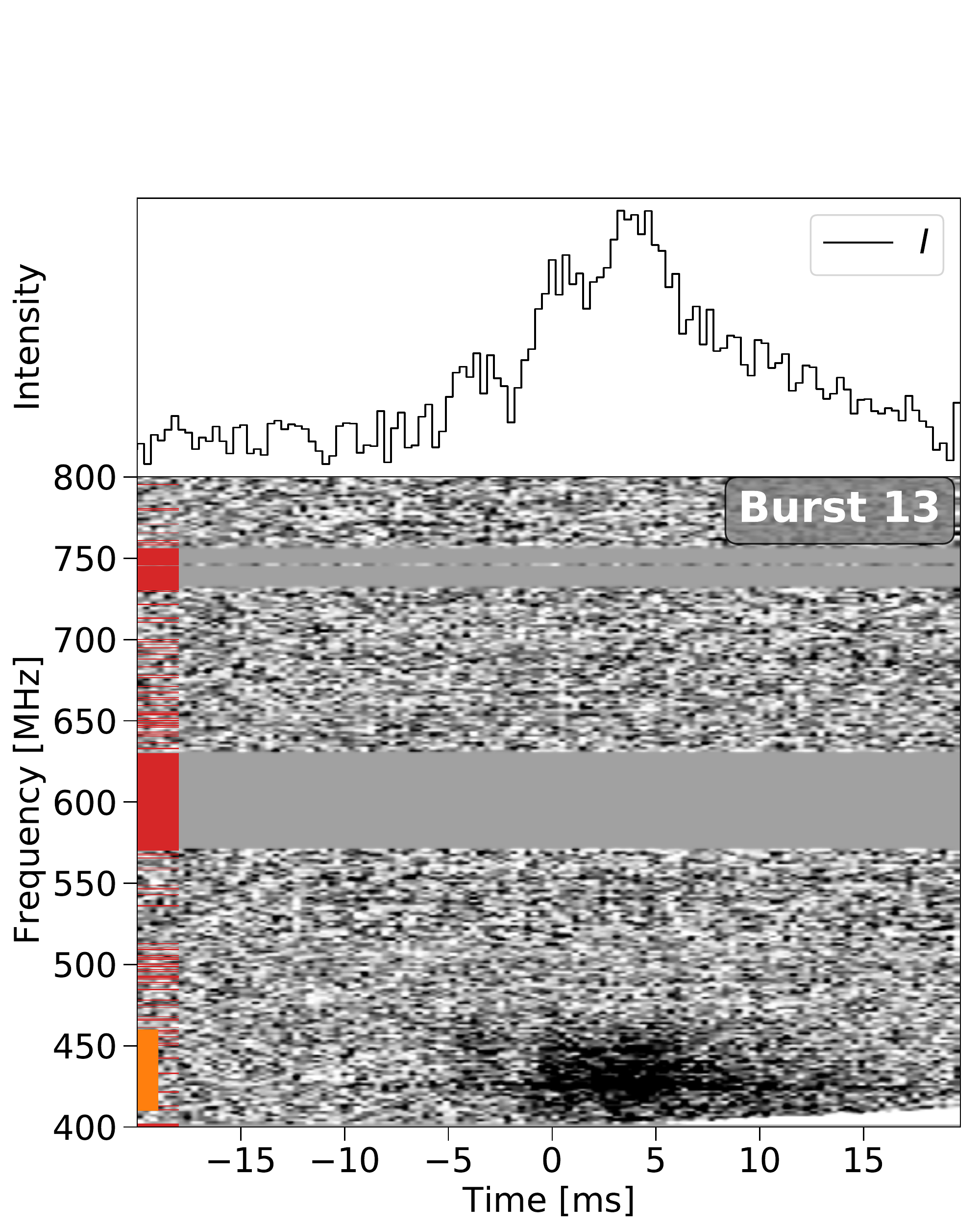}
    \includegraphics[width=0.19\textwidth]{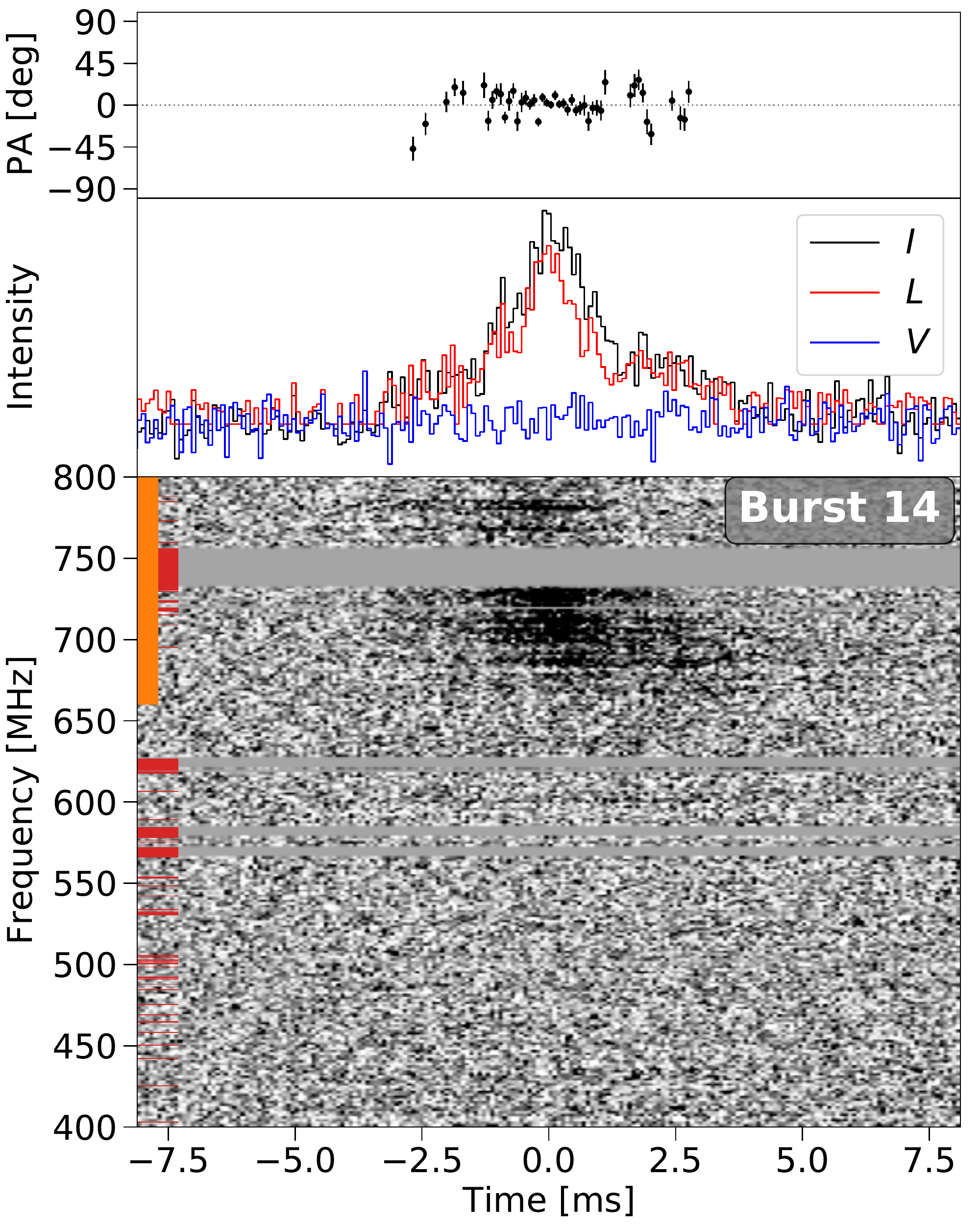}
    \includegraphics[width=0.19\textwidth]{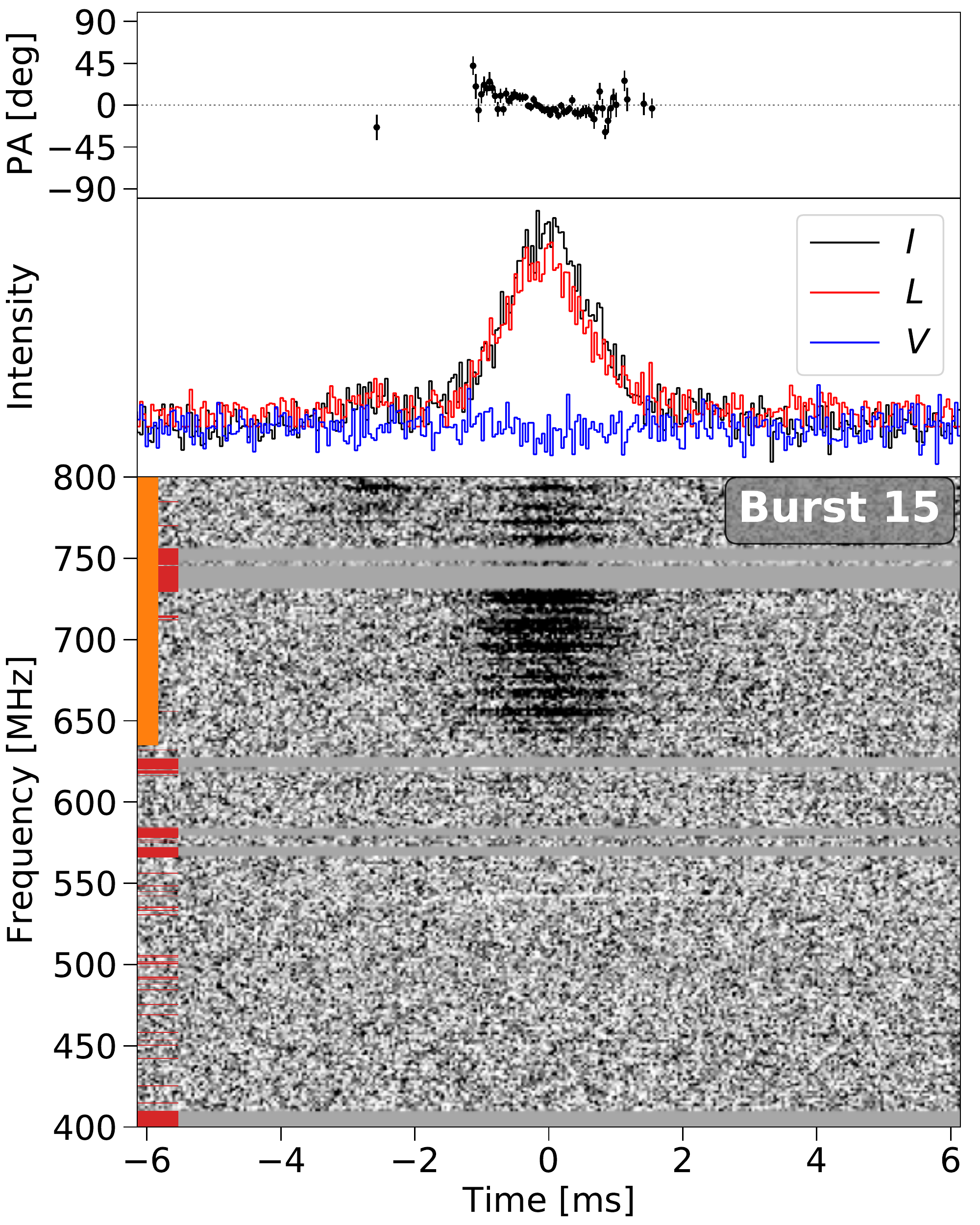} 
    \includegraphics[width=0.19\textwidth]{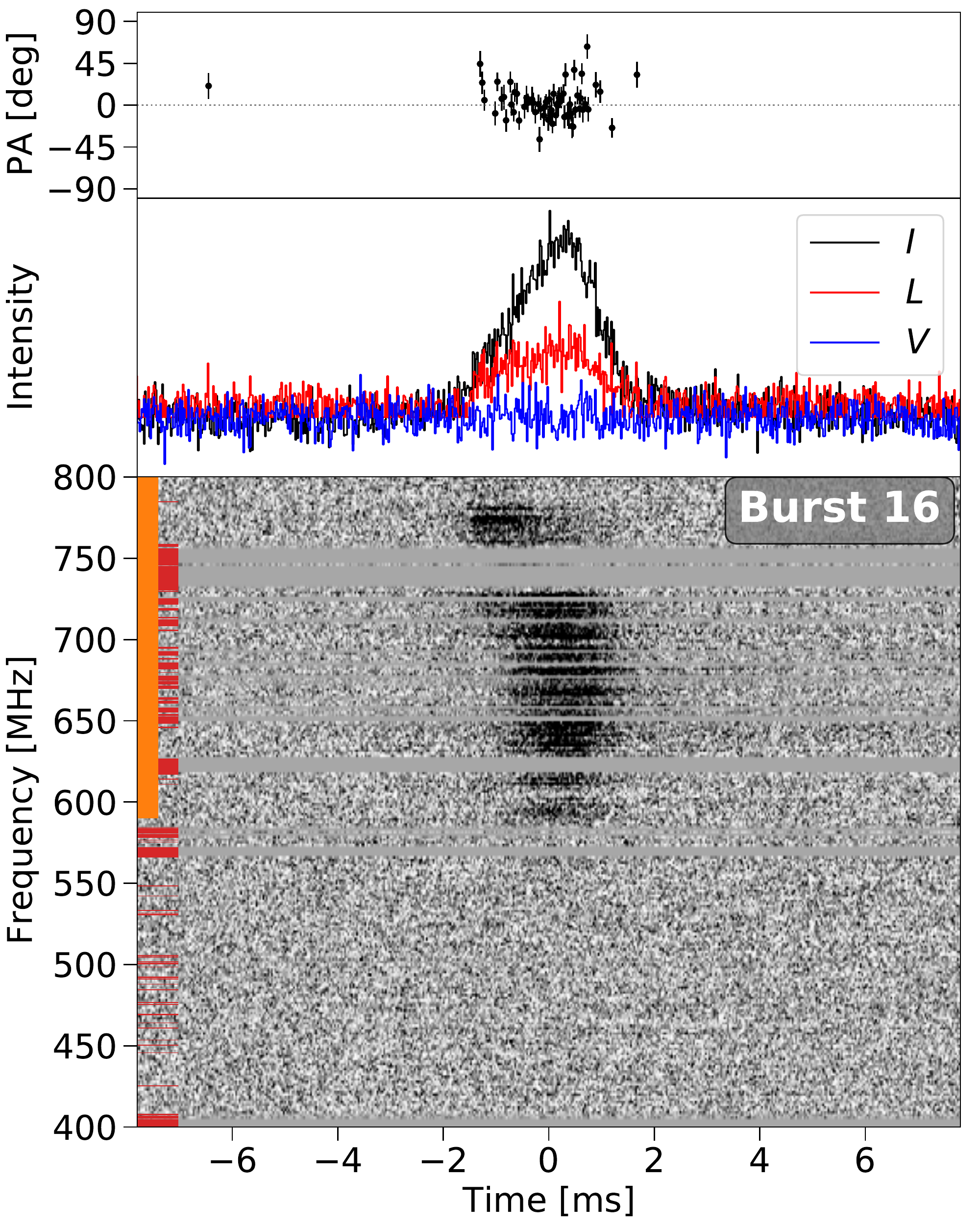}
    \includegraphics[width=0.19\textwidth]{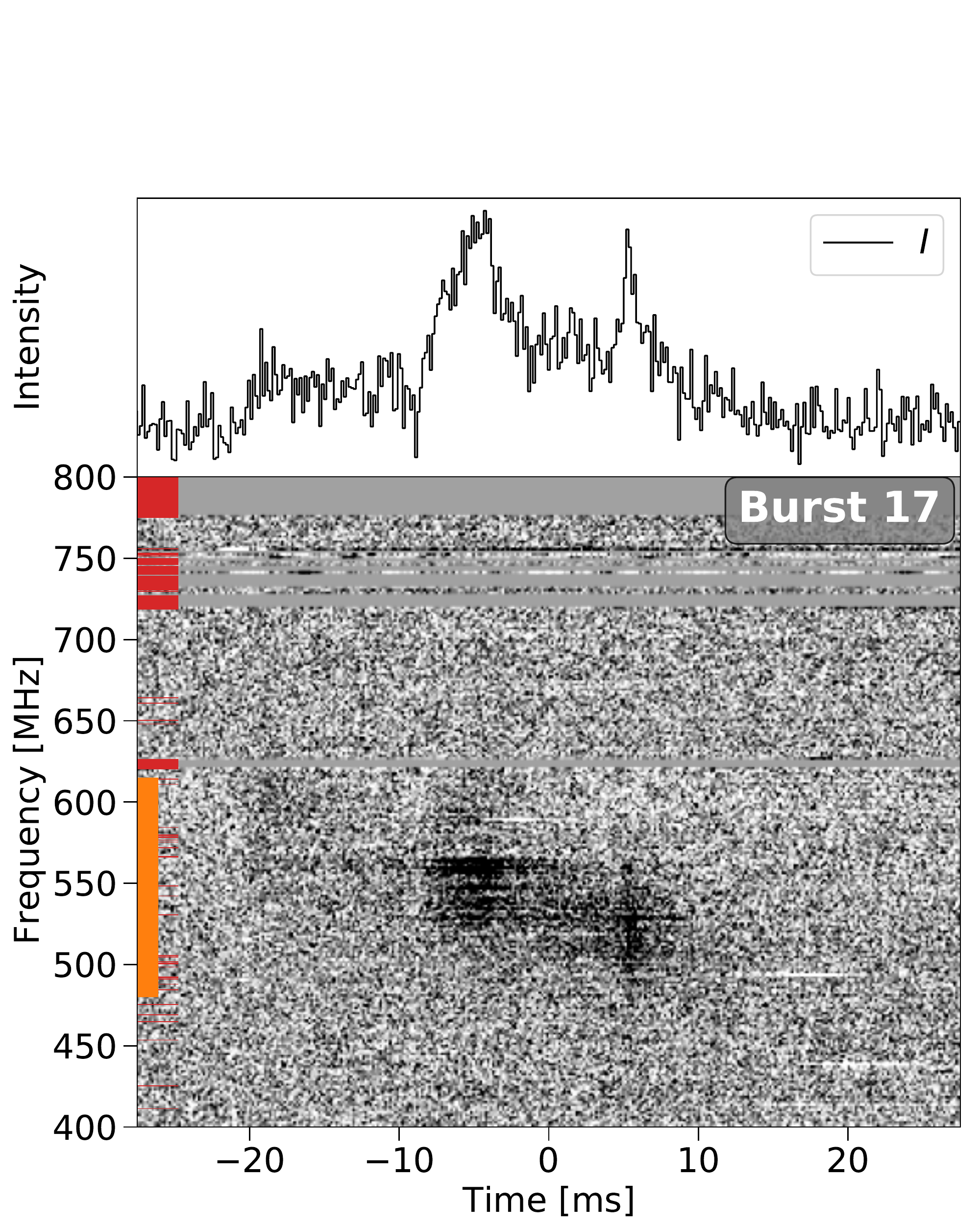}
    \includegraphics[width=0.19\textwidth]{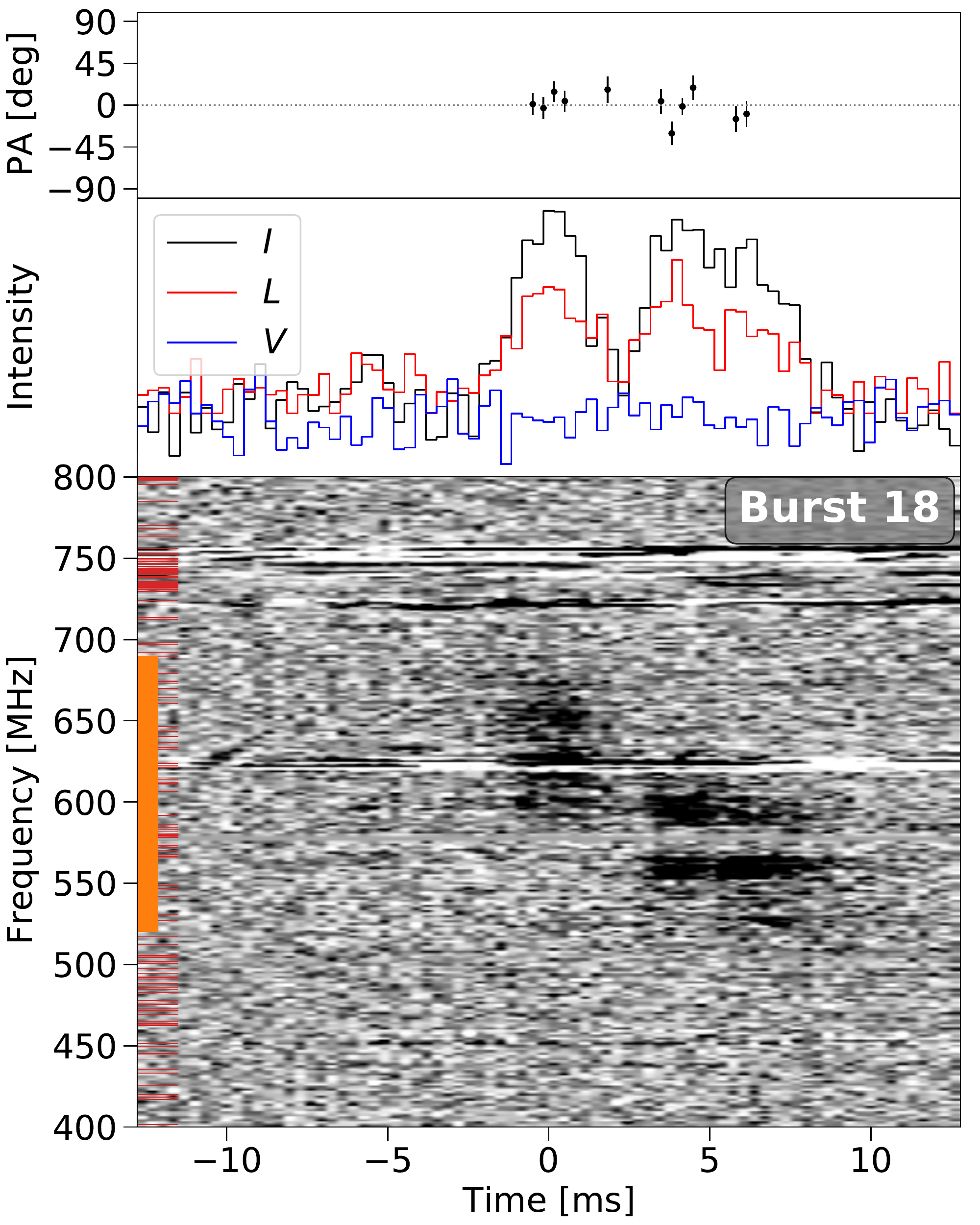}
    \includegraphics[width=0.19\textwidth]{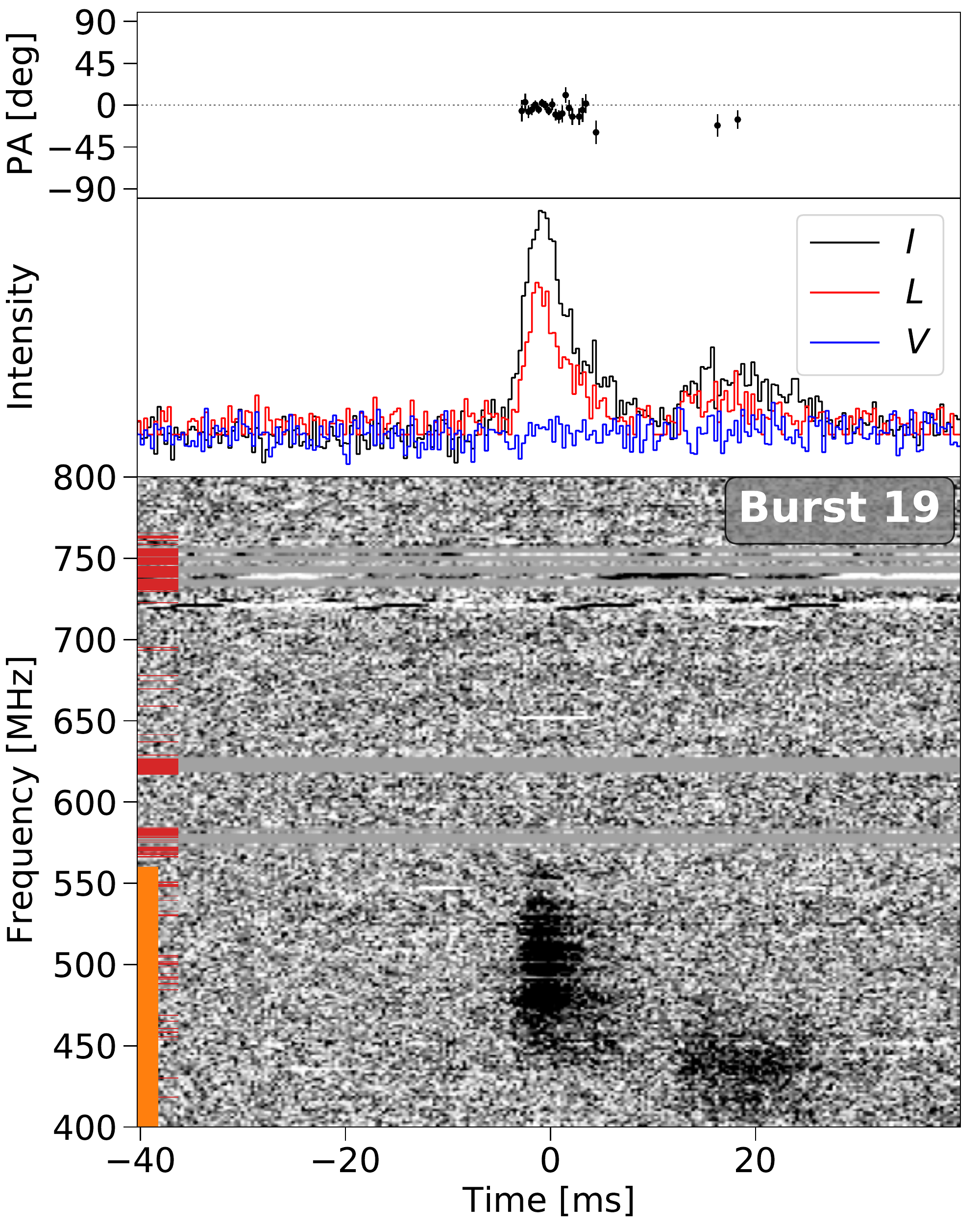}
    \includegraphics[width=0.19\textwidth]{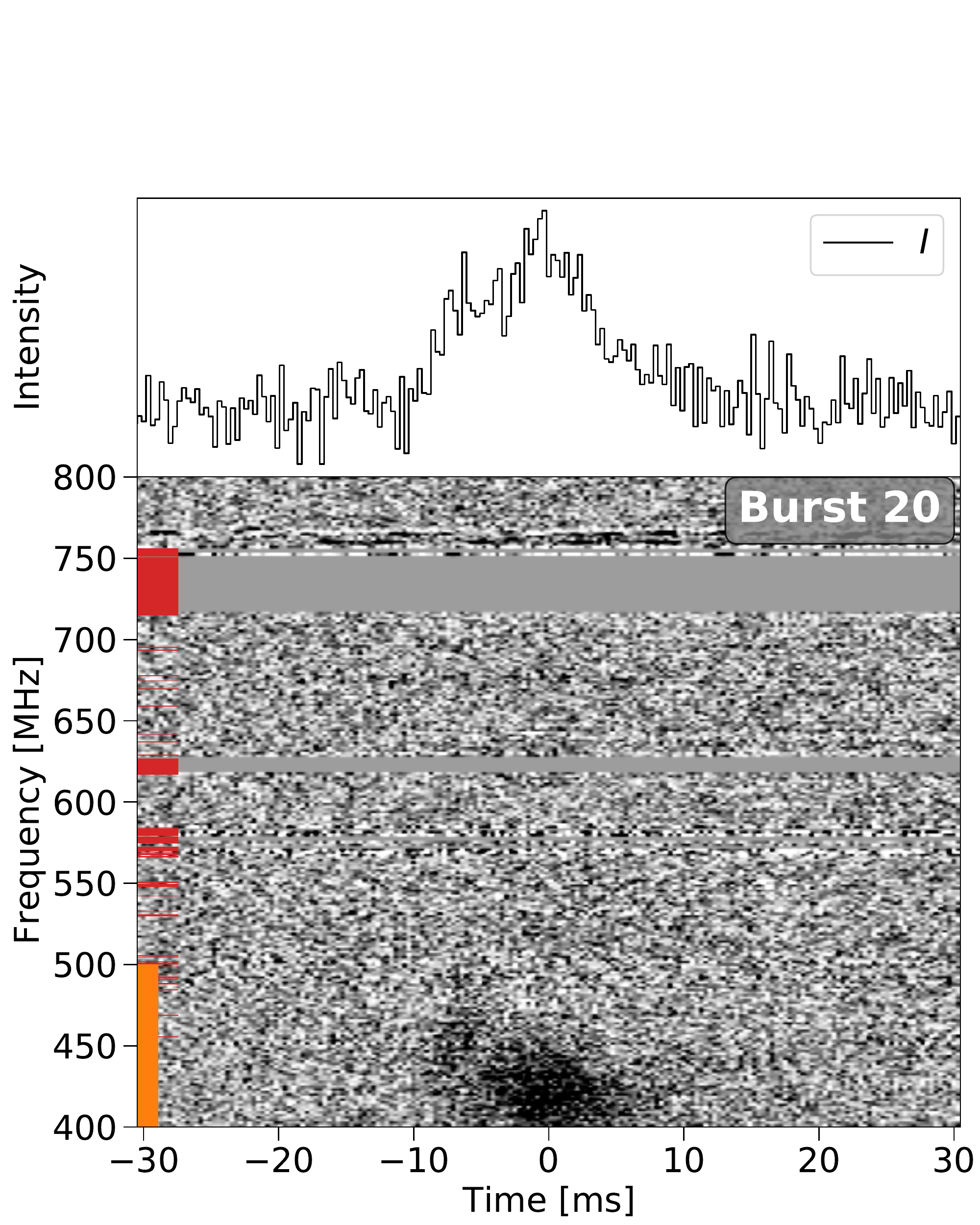}
    \caption{Waterfall plots in Stokes I of bursts from repeating source FRB 20190303A in chronological order and dedispersed to their structure optimized DMs ($\rm{DM_{struct}}$; listed in Table~\ref{ta:bursts}). Data have been rebinned to a fixed frequency resolution of $\nu_{\rm{res}} = 1.5625$ MHz. Time has been rebinned according to $n_{down}$ of each burst, such that $t_{\rm{res}} = n_{down}\times 2.56\; \rm{\mu s}$. Panels above the spectra display burst profiles of total intensity (black). Bursts with significant RM detections are also accompanied by linear polarized (red), and circularly polarized (blue) intensity profiles (peak normalized), as well as polarization angle (PA) curves.
    Burst profiles are obtained by adding signal over the spectral limits of the burst, indicated by orange lines along the frequency axis. Masked frequency channels are indicated by red lines along the vertical axis. Each panel is labeled with the corresponding burst number from Table~\ref{ta:bursts}. Waterfall plots of bursts from other repeating sources in our sample are available in the online journal.}
 \label{fig:waterfalls_R17}
\end{center}
\end{figure*}

 The CHIME/FRB baseband system captured $20$ bursts from FRB 20190303A between 2019 July and 2021 November. These bursts, displayed in Figure~\ref{fig:waterfalls_R17}, show a wide range of linear polarization fractions and regularly exhibit complex burst morphology, with several displaying clear downward drifting (``sad trombone") structure commonly seen in other repeating sources. Seven bursts from this sample are seemingly unpolarized while the remaining 10 have polarization fractions between $20\%$ and $100\%$. 

 Bandwidth depolarization only becomes significant for CHIME/FRB at $\rm{|RM|} \gtrsim 2000 \; \rm{rad\, m^{-2}}$ ($\sim 50\%$ loss in linear polarization at 400 MHz) and therefore cannot explain the range of $L/I$ seen from the polarized sample. Meanwhile, for the unpolarized sample we apply a semi-coherent RM search routine \citep{Mckinven2021} to overcome bandwidth depolarization. We find no significant RM detection ($S/N>5$) by searching over a range $-10^6\leq \rm{RM} \leq 10^6 \; rad \, m^{-2}$. While this does not rule out polarized events with extreme RMs, we consider this an unlikely scenario given the $L/I$ variability and RM range seen in the polarized subsample. The most likely scenario is that the bursts observed as unpolarized are not extreme RM events but simply unpolarized to the limits of our sensitivity. As such, for each unpolarized burst we use the the total intensity burst S/N to calculate the $L/I$ that would be required to produce a polarization detection at a level of $\rm{S/N=5}$. This value is used as a conservative upper-limit on $L/I$ for unpolarized bursts in Table~\ref{ta:bursts}. A low level of circular polarization can be seen from inspecting the polarized burst profiles of this source. This seems to disfavour the observed $L/I$ variability being produced by Faraday conversion, where linear polarization is converted to circular under special conditions of a magneto-ionic environment \citep{Gruzinov2019, Vedantham2019}. An alternate scenario for producing variable $L/I$ via transmission through a magnetized scattering screen \citep{Melrose1998,Beniamini2021} is discussed in Section~\ref{sec:R17_polL}. Here, we conservatively estimate an upper limit of $<10\%$ on the astrophysical circular polarization and note the strong possibility that most of the observed Stokes $V$ is contamination of cross-polarization leakage from imperfect feed alignment. 
 
 PA behavior displayed from this source is more diverse than that seen from FRB 20180916B. Several bursts (e.g., bursts 6, 14, 15) display PA evolution over the burst duration, similar to the diverse PA behavior seen from FRB 20180301A \citep{Luo2020}. Subtle PA variations of $\lesssim 20^{\circ}$ on short ($\ll 1\; \rm{ms}$) timescales also appear in many bursts. For some bursts (e.g., bursts 4, 6, 11), these variations are in excess of 40$^\circ$, suggesting a possible relation between variability in PA and the changing linear polarized fraction. Given the limited sample size, it is unclear whether this PA behavior may be related to the burst morphology.

\subsubsection{Burst Properties vs. Time}

\begin{figure}
	\centering
\begin{center}
    \includegraphics[width=0.5\textwidth]{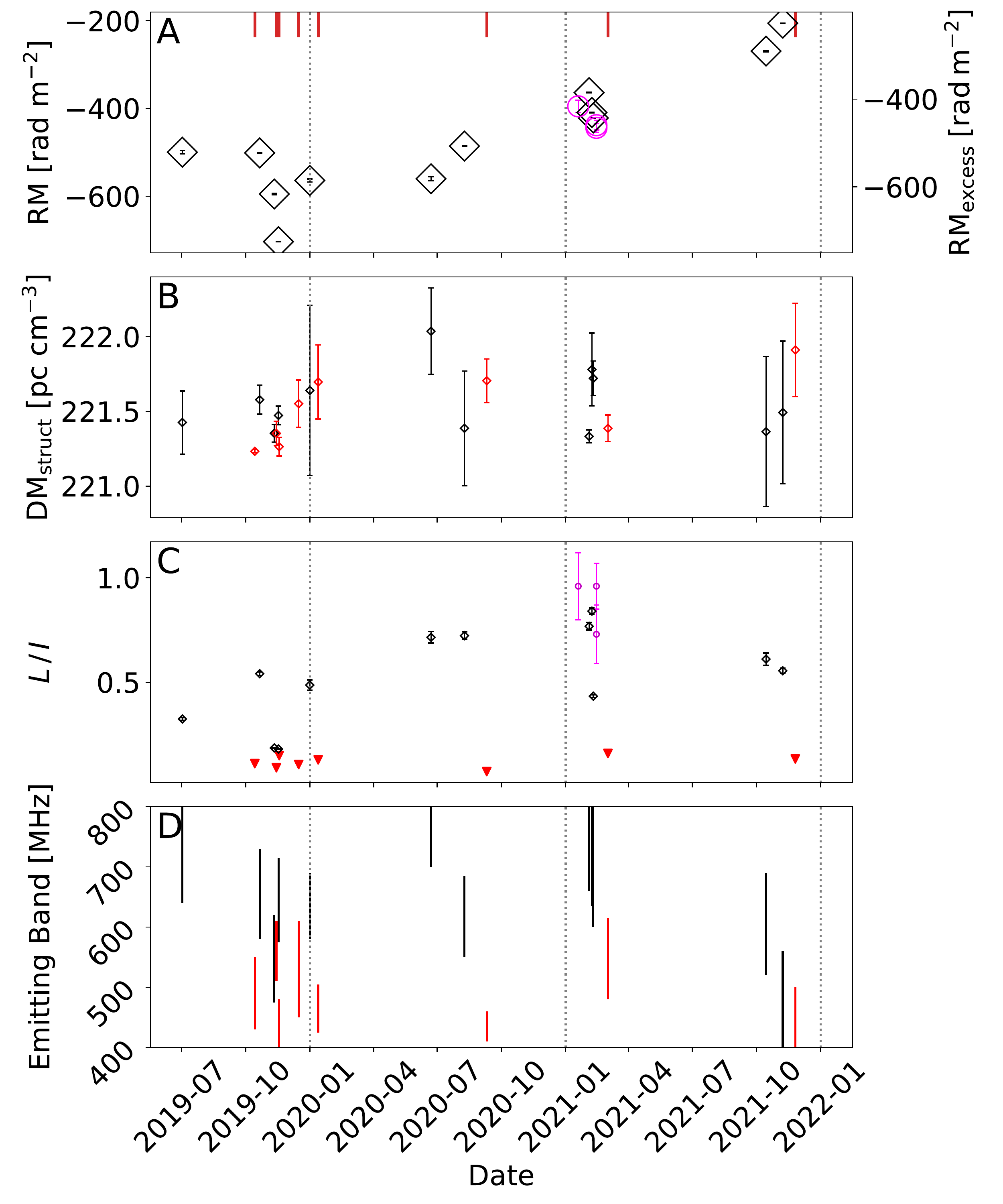}
    \caption{
    Burst properties as a function of time for baseband data recorded from FRB 20190303A, displaying the ionospheric corrected RM (panel A; black diamonds), $\rm{DM_{struct}}$ (panel B), linear polarized fraction ($L/I$; panel C) and emitting band (panel D; uncorrected for the non-uniform bandpass of CHIME). Times are in Coordinated Universal Time (UTC) format with vertical dotted lines indicating the start of each calendar year. The RMs of Panel A (black points) are determined from QU-fitting ($\mathrm{RM_{QU}}$) applied to the $20$-burst sample and are displayed alongside previously published RMs obtained from FAST observations \citep[magenta points;][]{Feng2022}. The extent of the error bars in Panel A appear invisible for some data points due to the the large RM variations displayed from this source. Red data points in panels B, C and D correspond to the unpolarized sample where the inverted red triangles of panel C indicate upper limits on $L/I$. The times of arrival (TOA) of the unpolarized bursts are indicated as lines along the top horizontal axis (panel A).} 
    % A line connecting measurements in panel A emphasizes the structured RM evolution.}
 \label{fig:RMvstime_R17}
\end{center}
\end{figure}

Figure~\ref{fig:RMvstime_R17} displays the individual burst properties of our $20$-burst sample from FRB 20190303A. Panel A summarizes the RM evolution, where individual RM values range over $\sim -700 \lesssim \rm{RM} \lesssim -200 \; \rm{rad\, m^{-2}}$. 
% a fractional change of $100\%$ relative to the mean RM of the sample. 
The arrival times of unpolarized bursts are indicated by red lines on the top horizontal axis. There is no indication of a greater number of unpolarized bursts (relative to polarized) at any point in the RM evolution. RM variations exhibit intervals of secular evolution that are interspersed with non-monotonic behavior, with substantial RM changes occurring over short timescales. Figure~\ref{fig:FDFvstime} provides a summary of the RM measurements, showing the Faraday Dispersion Functions (FDFs) of both the polarized and unpolarized bursts and highlighting the general agreement between the RMs obatined from the FDFs, $\rm{RM_{FDF}}$, and those obtained from QU-fitting, $\rm{RM_{QU}}$ (green lines).  

Over a six-day period from 2019 November 10 to 2019 November 16, the $|\rm{RM}|$ increased by more than $\sim 100 \; \rm{rad \,m^{-2}}$, implying a temporal RM gradient of $|\rm{\partial RM/ \partial t}|\gtrsim 17\; \rm{rad \, m^{-2} \, day^{-1}}$. This period is followed by an interval of over a year where $|\rm{RM}|$ steadily decreased to its minimum near $\rm{RM\sim-200\; rad \, m^{-2}}$, implying an average gradient of $|\rm{\partial RM/ \partial t}|\gtrsim 0.7\; \rm{rad \, m^{-2} \, day^{-1}}$. Subtracting the Galactic $\rm{RM_{MW}}$ estimate obtained from Table~\ref{ta:dmrmgal} implies an $\rm{RM_{excess}}=-227 \pm 5 \;\rm{rad\, m^{-2}}$ for the most recent observation.

Panel B of Figure~\ref{fig:RMvstime_R17} displays $\rm{DM_{struct}}$ for the $20$-burst sample. We do not find strong evidence for significant DM variations relative to the formal measurements errors. Weighting by the inverse square of these errors, we determine the average DM to be $\rm{DM_{struct} = 221.27 \pm 0.01}\; \rm{pc\, cm^{-3}}$, which is slightly lower than the value ($\rm{222.4 \pm 0.7\; pc\, cm^{-3}}$) previously determined from CHIME/FRB intensity data \citep{Fonseca2020}. Such a discrepancy is consistent with well-known biases introduced from unresolved complex structure in FRBs.

Limiting our analysis to the polarized subsample, we do not find significant correlation between $\rm{DM_{struct}}$ and $\rm{RM}$. This result is similar to the equivalent analysis recently applied to CHIME/FRB observations of FRB 20180916B \citep{Mckinven2022}. The lack of correlated variability between $\rm{RM}$ and $\rm{DM_{struct}}$ suggests that the observed RM variations may be produced by local variations in $B_{\parallel}$. However, we cannot exclude the alternate scenario in which $B_{\parallel}$ remains fixed and the observed $\rm{RM}$ variations are produced from subtle fluctuations in $n_e$ or path length, which are not detectable with our DM measurement precision. In either scenario, a substantial magnetic field is required to produce the observed variability in RM, with the estimated strength directly dependent on the assumed DM contribution of the local Faraday-active medium. We estimate the average strength of this magnetic field by applying the equation,
\begin{equation}
\begin{aligned}
    \langle B^{local}_{\parallel} \rangle &= 1.23 \frac{\rm{RM_{local}}}{\rm{DM_{local}}}(1+z) \\
    &\sim 1.23\frac{\Delta \rm{RM}}{\rm{DM_{\Delta RM}}}(1+z) \\ &\gtrsim 3.6 \; \mu\rm{G}.
\end{aligned}
\label{eqn:Bdyn}
\end{equation}
Here, $\Delta \rm{RM}$ is the maximum variation in the observed RM, which we take to be $\Delta \rm{RM} \sim 500 \; \rm{rad \, m^{-2}}$ and is a reasonable proxy for $\rm{RM_{local}}$ given the timescale of the RM evolution. $\rm{DM_{\Delta RM}}$ is the unknown DM contribution of the local Faraday-active medium, which is unconstrained by our observations since observed variations in DM and RM do not appear to be correlated. We arrive at a conservative lower limit for $\langle B^{local}_{\parallel} \rangle$ by setting $\rm{DM_{\Delta RM}} = \rm{DM_{excess}}  =  \rm{DM} - \rm{DM_{MW}} - \rm{DM_{halo}} \lesssim 170\; \rm{pc\, cm^{-3}}$ (see Table~\ref{ta:dmrmgal}). The $(1+z)$ term that accounts for the differential dilution of cosmological expansion on DM and RM is safely ignored in our estimate given the low redshift of the source\footnote{This source has yet to be localized to a host galaxy.} as inferred from its modest $\rm{DM_{excess}}$.

We emphasize that the $\rm{DM_{\Delta RM}}$ value assumed here very likely overestimates the DM contribution of the Faraday-active medium, such that $\rm{DM_{\Delta RM}} << \rm{DM_{excess}}$. For instance, $\langle B^{local}_{\parallel} \rangle \sim 3000\; \mu\rm{G}$ is required if we assume that only the variable component of the DM inferred from the root-mean-square of our $\rm{DM_{struct}
}$ measurements (i.e., $\Delta \rm{DM} = 0.2\; pc\, cm^{-3}$), contributes to the observed RM evolution of this source. 
% Furthermore, this $\Delta \rm{DM}$ measurement may possibly be overestimated, which would lead us to underestimate $B^{local}_{\parallel}$. 
This revised $\langle B^{local}_{\parallel}\rangle$ value is larger than the equivalent estimate of $\sim 100\; \rm{\mu G}$ obtained from recent observations of FRB 20180916B \citep{Mckinven2022} and approaches the $3-17\; \rm{mG}$ constraint arrived at by \citet{Katz2021} from polarimetric observations of FRB 20121102A \citep{Hilmarsson2021}. Further discussion on the implication of these measurements can be found in Section~\ref{sec:B_parallel}. Panel C and D display the linear polarized fraction and emitting band, which again appear to be uncorrelated with the RM evolution of this source.

% However, we do see evidence for a correlation between $L/I$ and the emitting band of each burst, further discussed in the section below.

\subsubsection{A Frequency Dependence in $L/I$}
\label{sec:R17_polL}

\begin{figure*}
\centering     %%% not \center
\subfigure[FRB 20190303A]{\label{fig:depola}\includegraphics[width=0.45\textwidth]{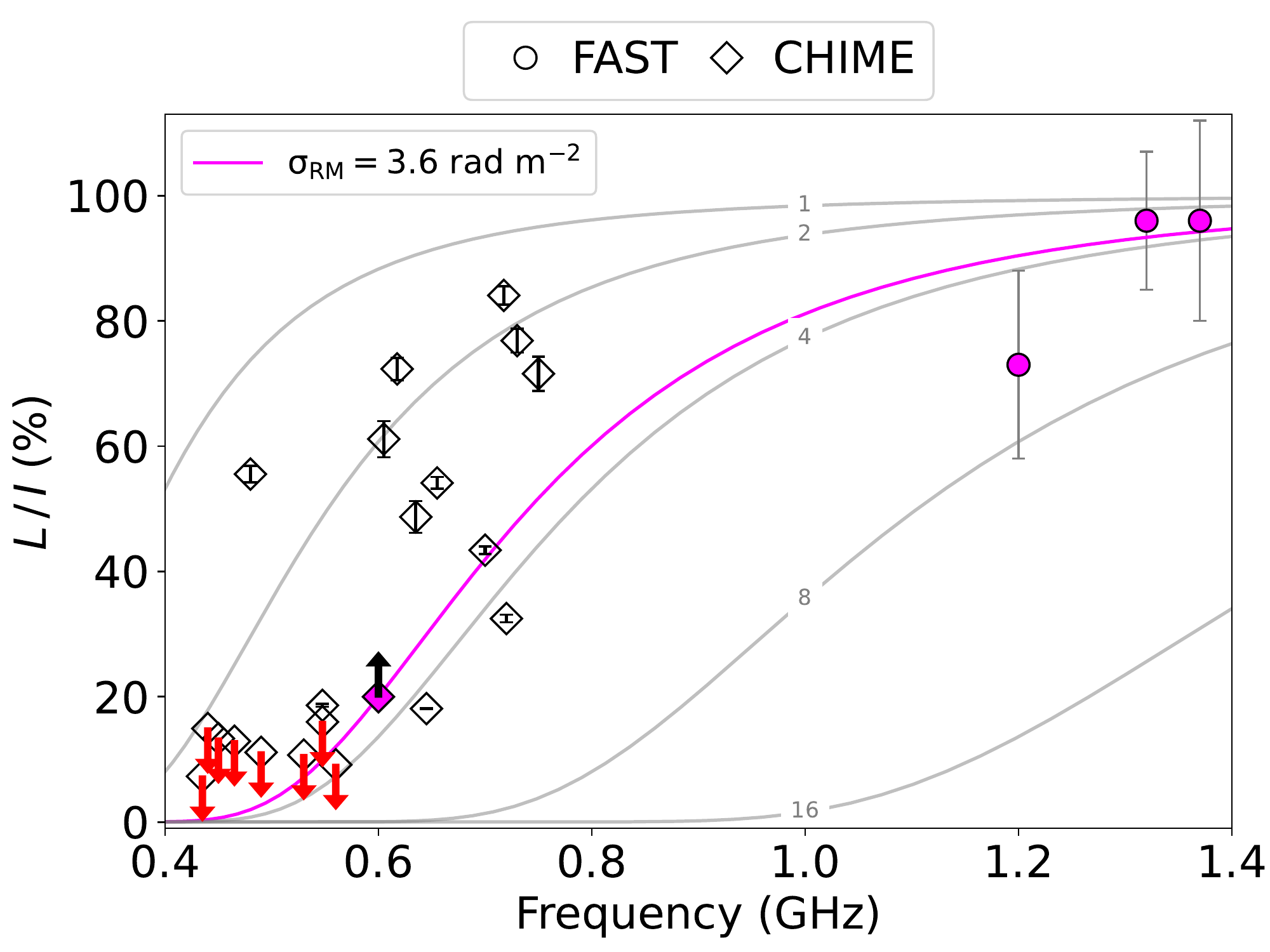}}
\hspace{0.005\textwidth}
\subfigure[FRB 20190417A]{\label{fig:depolb}\includegraphics[width=0.45\textwidth]{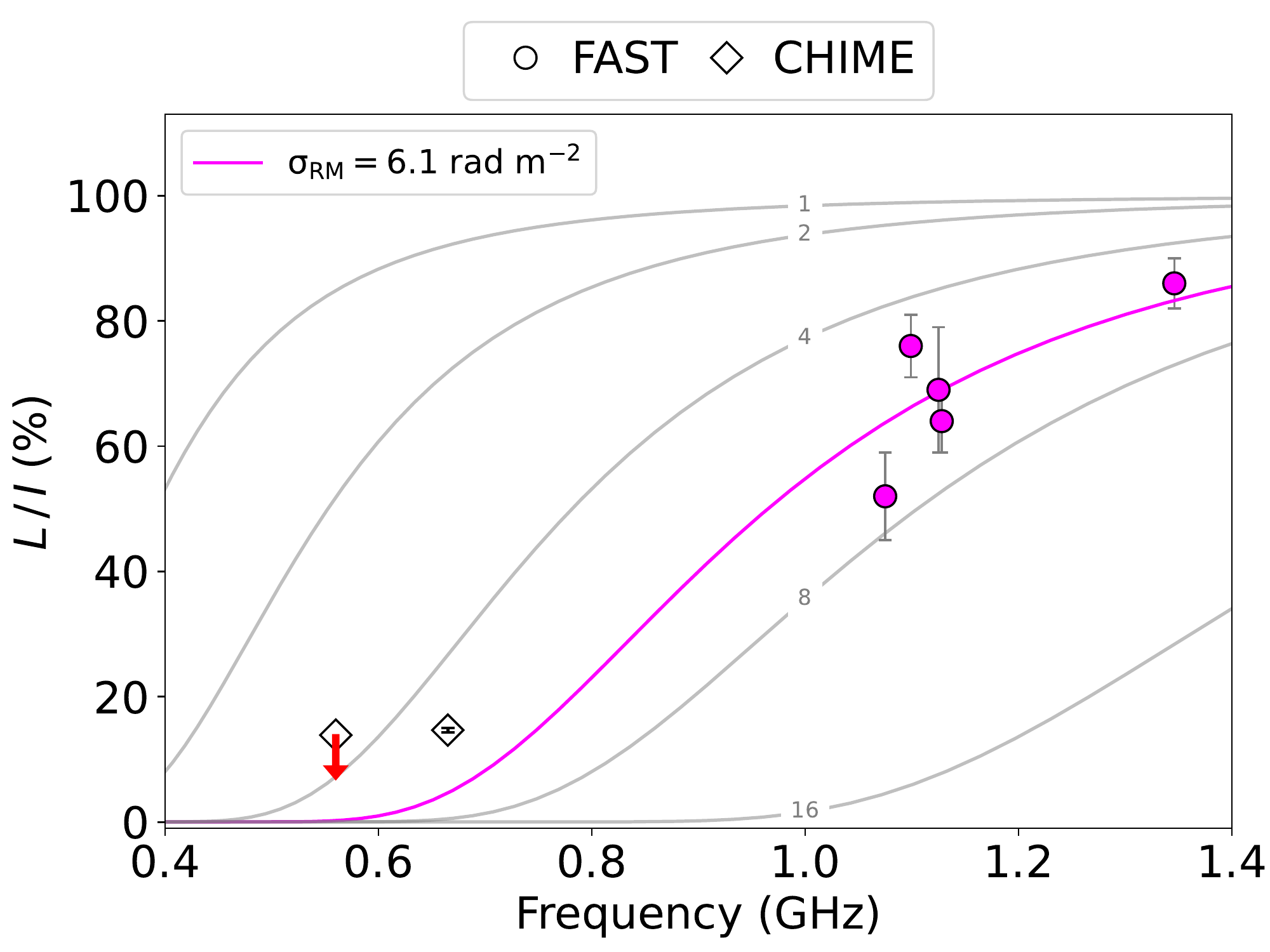}} 
\caption{Degree of linear polarization as a function of frequency for multiband observations of FRB 20190303A (panel a) and FRB 20190417A (panel b). The best-fit depolarization models determined by \citet{Feng2022} are indicated by magenta lines and were calculated from previous measurements (magenta data points) of these sources \citep[see][for details]{Feng2022}. These measurements include a single lower limit measurement for FRB 20190303A (magenta point with upward arrow) obtained from a previously published CHIME/FRB observation \citep{Fonseca2020}. The linear polarization measurements of our sample are represented as black diamonds with the upper limits of our unpolarized subsample indicated by red downward arrows. Depolarization curves for different values of $\sigma_{\rm{RM}}$ (1, 2, 4, 8 $\&$ 16 rad m$^{-2}$) are indicated by grey lines. The excursions of $L/I$ measurements from values predicted by best-fit models is further discussed in Section~\ref{sec:depol}.}
\label{fig:FracPol_R17}
\end{figure*}

Our $20$-burst sample from FRB 20190303A shows $L/I$ to be clearly correlated with frequency, with all the unpolarized events located in the bottom half ($400-600$ MHz) of the CHIME band (see Figure~\ref{fig:RMvstime_R17}). Figure~\ref{fig:depola} summarizes this result by displaying $L/I$ measurements as a function of the central frequency of the emitting band of each burst from both CHIME ($400-800$ MHz; diamonds) and FAST ($1.0-1.5$ GHz; circles) observations. From this combined sample we calculate a significant Spearman correlation coefficient of 0.80 (p-value $< 0.001$) between $L/I$ and frequency. Such a dependence is consistent with expectations for scattering screen models \citep[e.g.][]{Melrose1998,Beniamini2021,Yang2022} whereby multi-path propagation through a turbulent magneto-ionic medium produces stochastic Faraday rotation and partial depolarization that is more significant at lower frequencies and has several compelling pulsar analogues \citep{Xue2019,Sobey2019,Sobey2021}. Within this framework, the amount of depolarization is determined by a single parameter, $\sigma_{\rm{RM}}$, such that $f_{depol} = 1 - \exp(-\rm{2\lambda^4\sigma_{RM}^2})$. Recently, \citet{Feng2022} fit this model to a small subset of $L/I$ measurements shown in Figure~\ref{fig:FracPol_R17} (magenta data points) and arrived at a best-fit $\sigma_{\rm{RM}}=3.6\; \rm{rad\, m^{-2}}$ for this source.  

The additional $L/I$ measurements reported here (black diamonds), although generally consistent with a depolarization imprint, exhibit significant excursions from the predicted trend of the best-fit depolarization model. This suggests that either the intrinsic variability in $L/I$ is significant or that the observed excursions are produced by variations in $\sigma_{RM}$. Given the magnitude of the RM variations observed from this source, it would not be surprising if $\sigma_{RM}$ displayed a similar level of instability in time. Indeed, only a modest range of $1.5\; \rm{rad\, m^{-2}} \lesssim \sigma_{\rm{RM}}\lesssim 4.5\; \rm{rad\, m^{-2}}$ value is required to explain the observed excursions in $L/I$ from the benchmark model of \citet{Feng2022}. This range of $\sigma_{\rm{RM}}$ values is well within the day-to-day RM variations observed from this source and suggests the possibility that the medium producing the quasi-secular RM evolution may also generate sufficient scattering to partially depolarize emission through stochastic Faraday rotation. This scenario assumes that the observed day-to-day RM variations are due to the relative motion between a depolarizing/inhomogeneous plasma screen and the FRB source. Another possibility is that these variations are produced by a plasma screen with a uniform but evolving magnetic field and/or electron density in which case the screen would still produce RM variations but not depolarize the signal. In this latter scenario, any fluctuations in $L/I$ would have to be intrinsic to the source, reminiscent of the unstable polarized fractions (and PAs) reported from radio emitting magnetars \citep[e.g.][]{Camilo2007,Levin2012,Dai2018,Dai2019}.

% Indeed, the feasibility of this depolarization model to describe the frequency dependence of $L/I$ observed from this source and others has several compelling pulsar analogues \citep{Xue2019,Sobey2019,Sobey2021} and is further discussed in Section~\ref{sec:depol}.  

Studying a larger sample of bursts from this source over a wider range of frequencies and incorporating information from other burst properties will illuminate the origins of variability in $L/I$. In particular, correlating $L/I$ or $\sigma_{\rm{RM}}$ with temporal scattering measurements of bursts will test the extent to which variations in $L/I$ are an imprint of a foreground structure or intrinsic to the source. 
% Scintillation analysis may also be applied in the future to characterize the location and relative motion of this structure.

%%%%%%%%%%%%%%%%% Other repeaters %%%%%%%%%%%%%%%

\subsection{Other Repeating FRB Sources}
\label{sec:other_rep}

%%%%%%%%%%%%%%%%%%%% RM vs. time of other repeaters %%%%%%%%%%%%%

\begin{figure*}
\centering     %%% not \center
\subfigure[FRB 20180814A]{\label{fig:a}\includegraphics[width=0.24\textwidth]{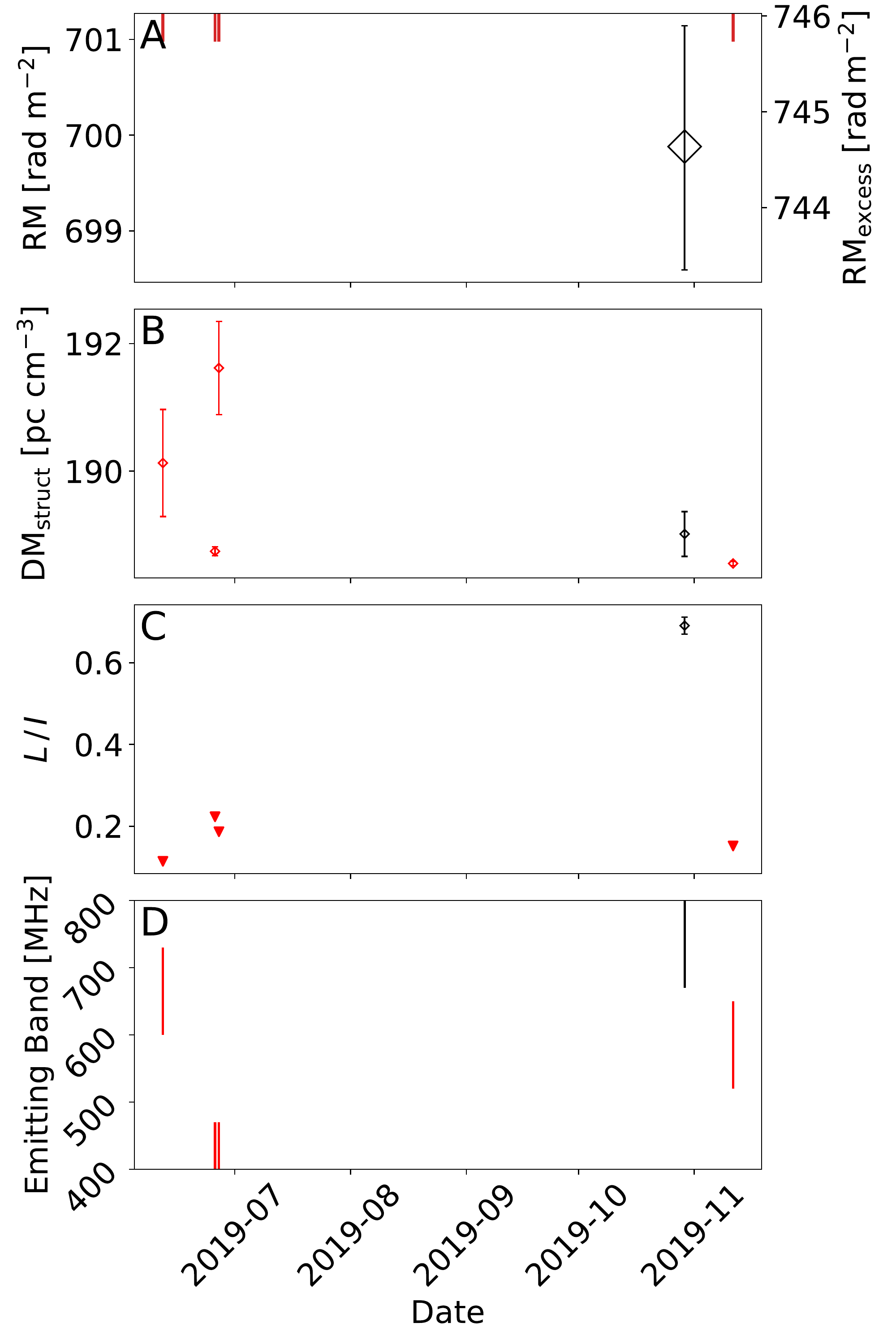}}
\hspace{0.005\textwidth}
\subfigure[FRB 20181030A]{\label{fig:b}\includegraphics[width=0.24\textwidth]{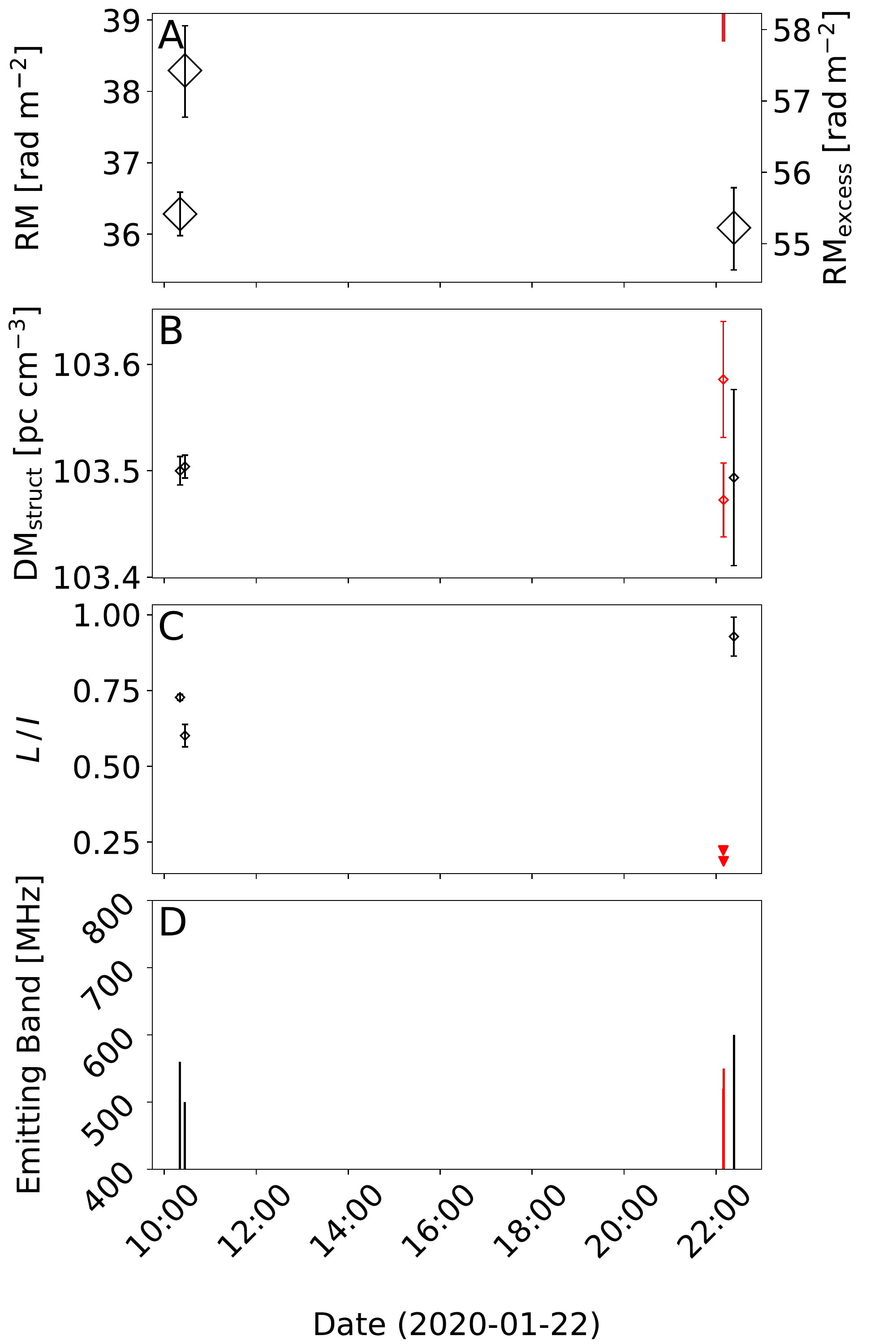}} 
\hspace{0.005\textwidth}
\subfigure[FRB 20181119A]{\label{fig:c}\includegraphics[width=0.24\textwidth]{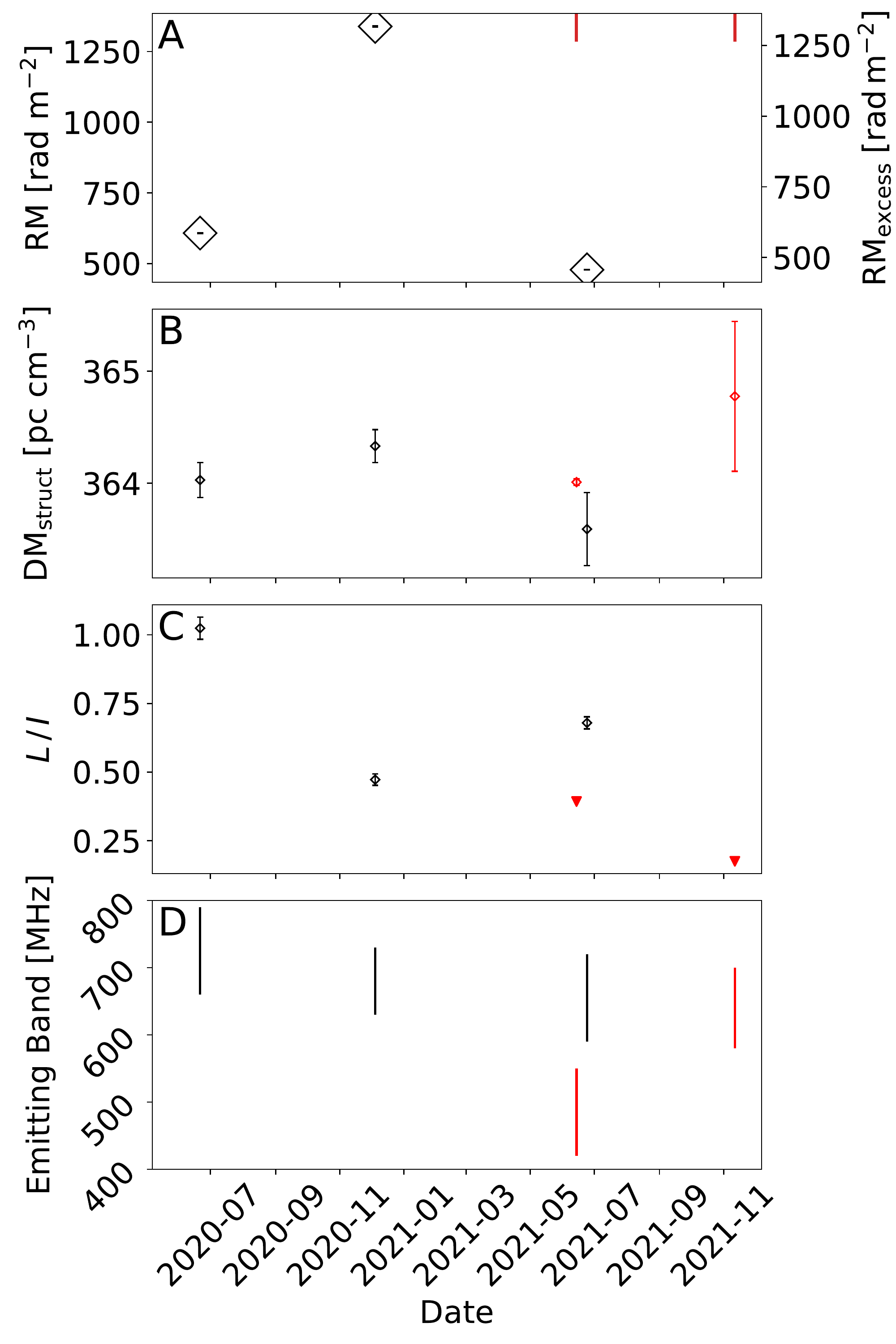}}
% \hspace{0.005\textwidth}
\subfigure[FRB 20190208A]{\label{fig:d}\includegraphics[width=0.24\textwidth]{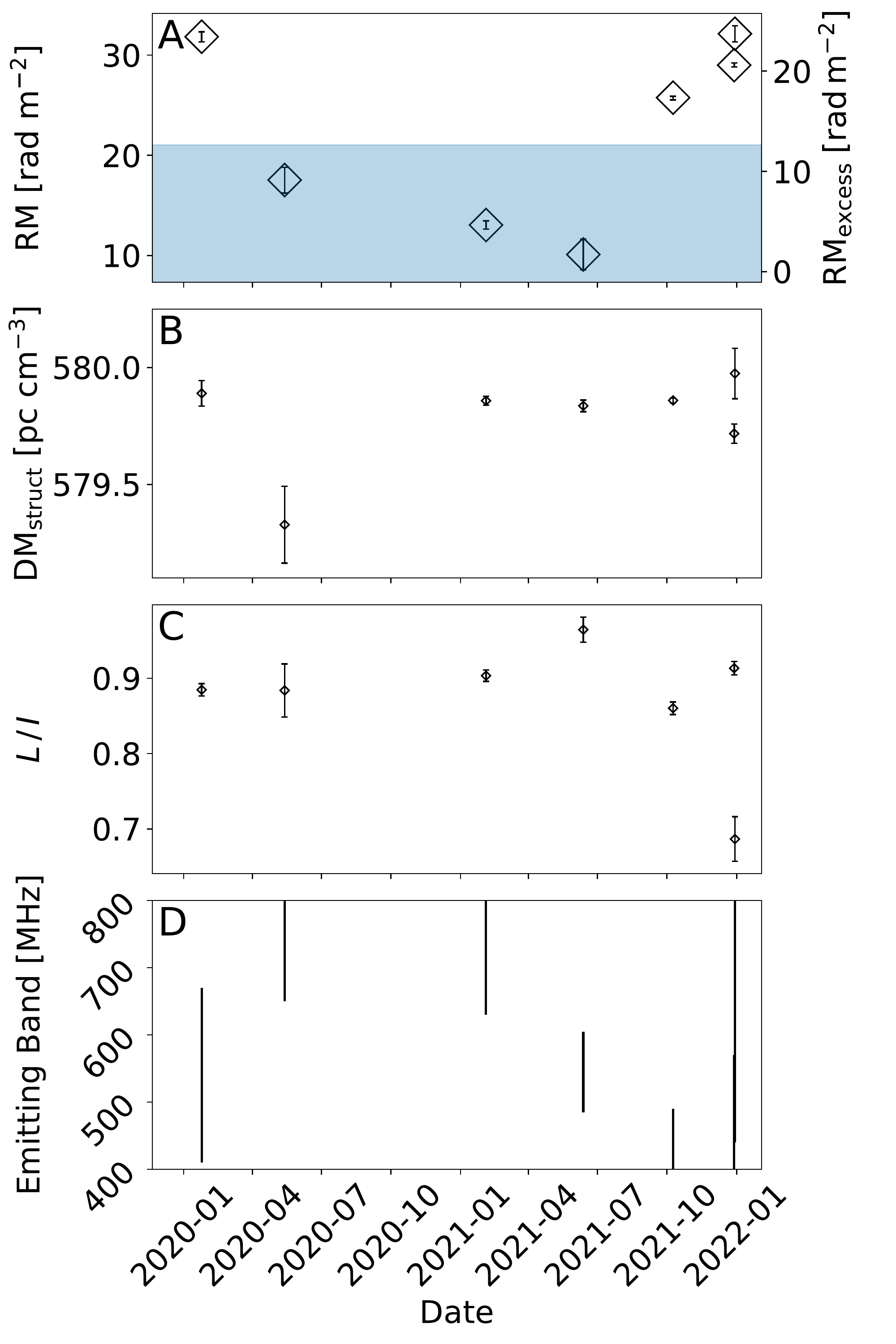}} 
\hspace{0.005\textwidth} \\
\subfigure[FRB 20190213B]{\label{fig:e}\includegraphics[width=0.24\textwidth]{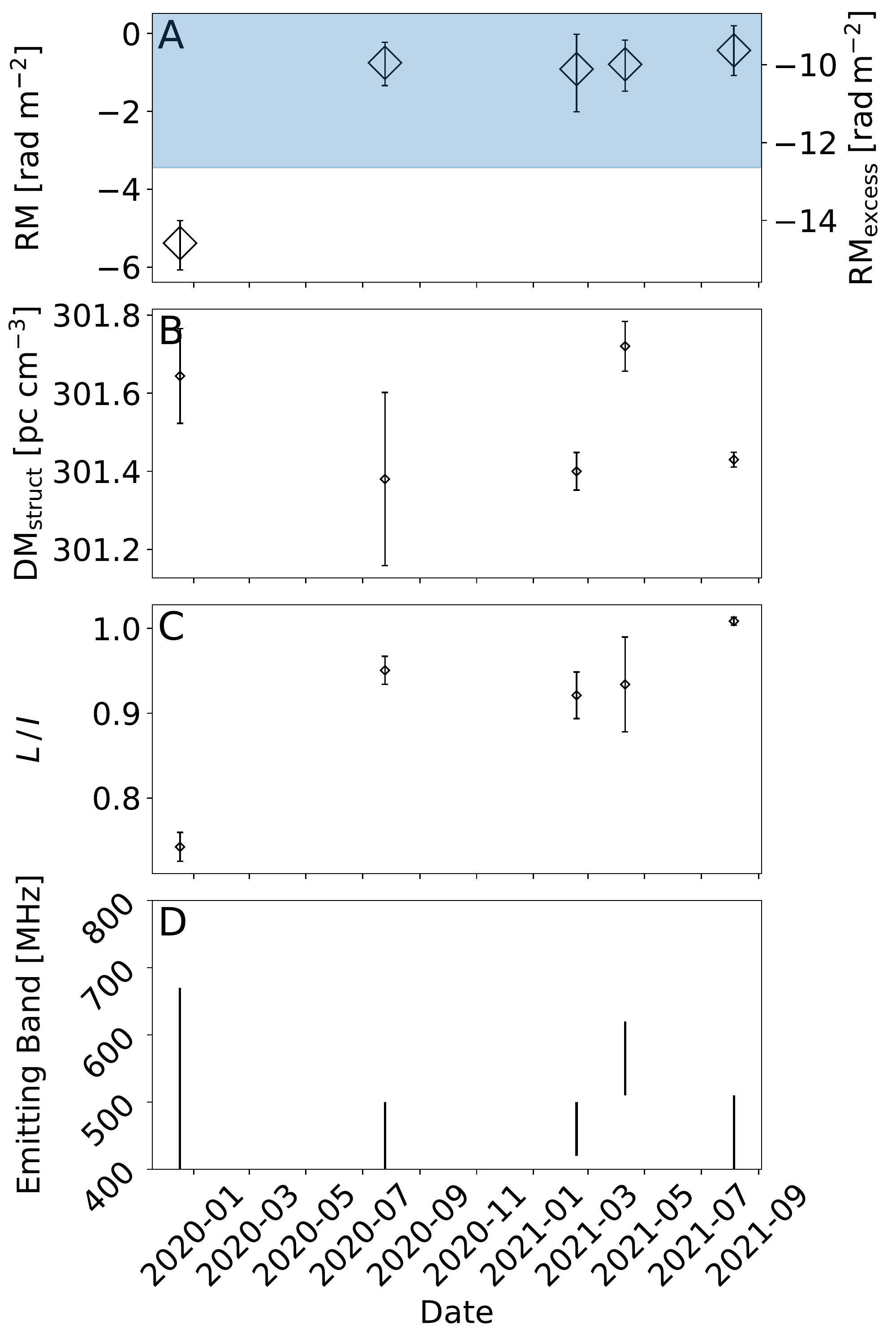}}
\hspace{0.005\textwidth}
\subfigure[FRB 20190117A]{\label{fig:f}\includegraphics[width=0.24\textwidth]{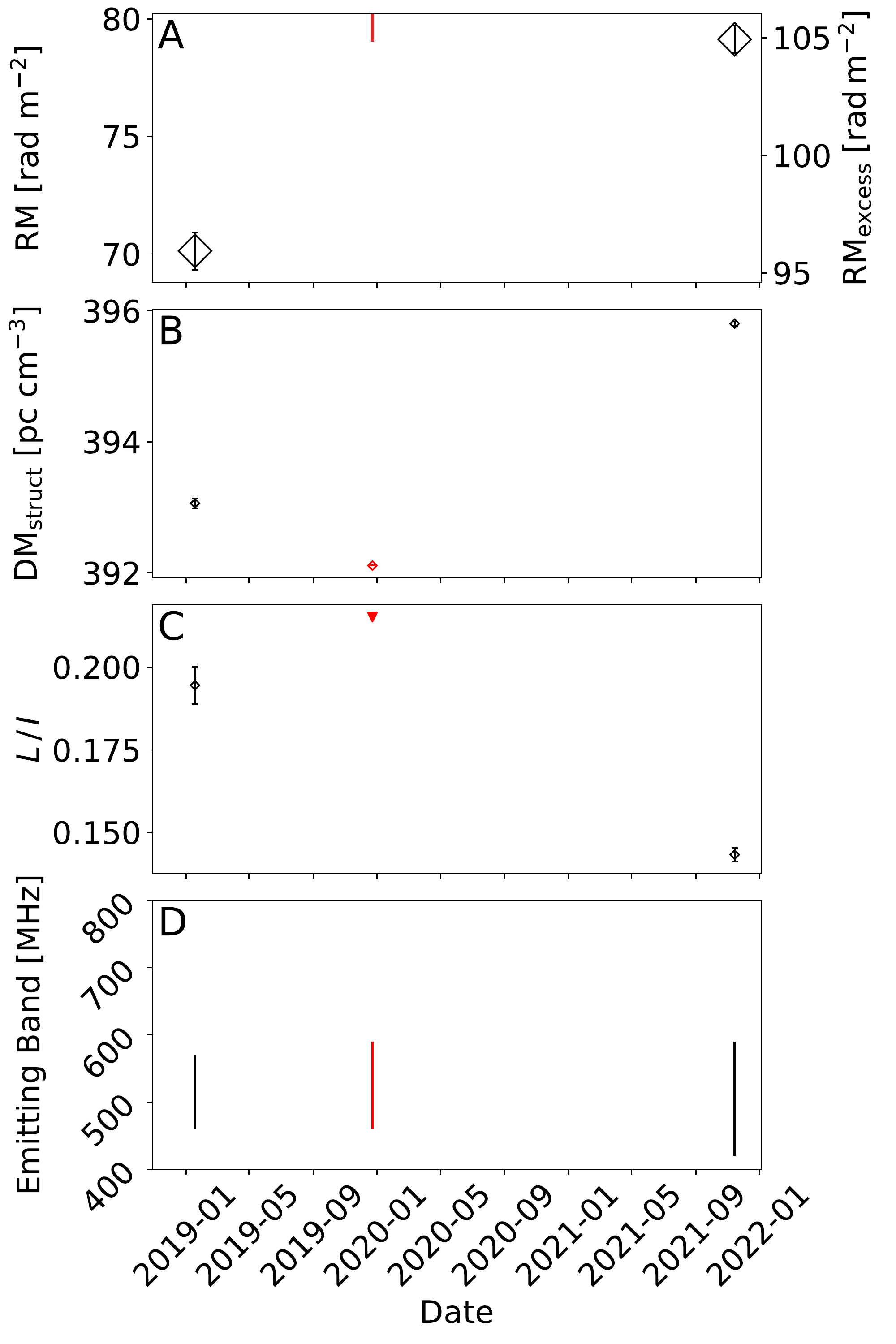}} 
\hspace{0.005\textwidth}
\subfigure[FRB 20190417A]{\label{fig:g}\includegraphics[width=0.24\textwidth]{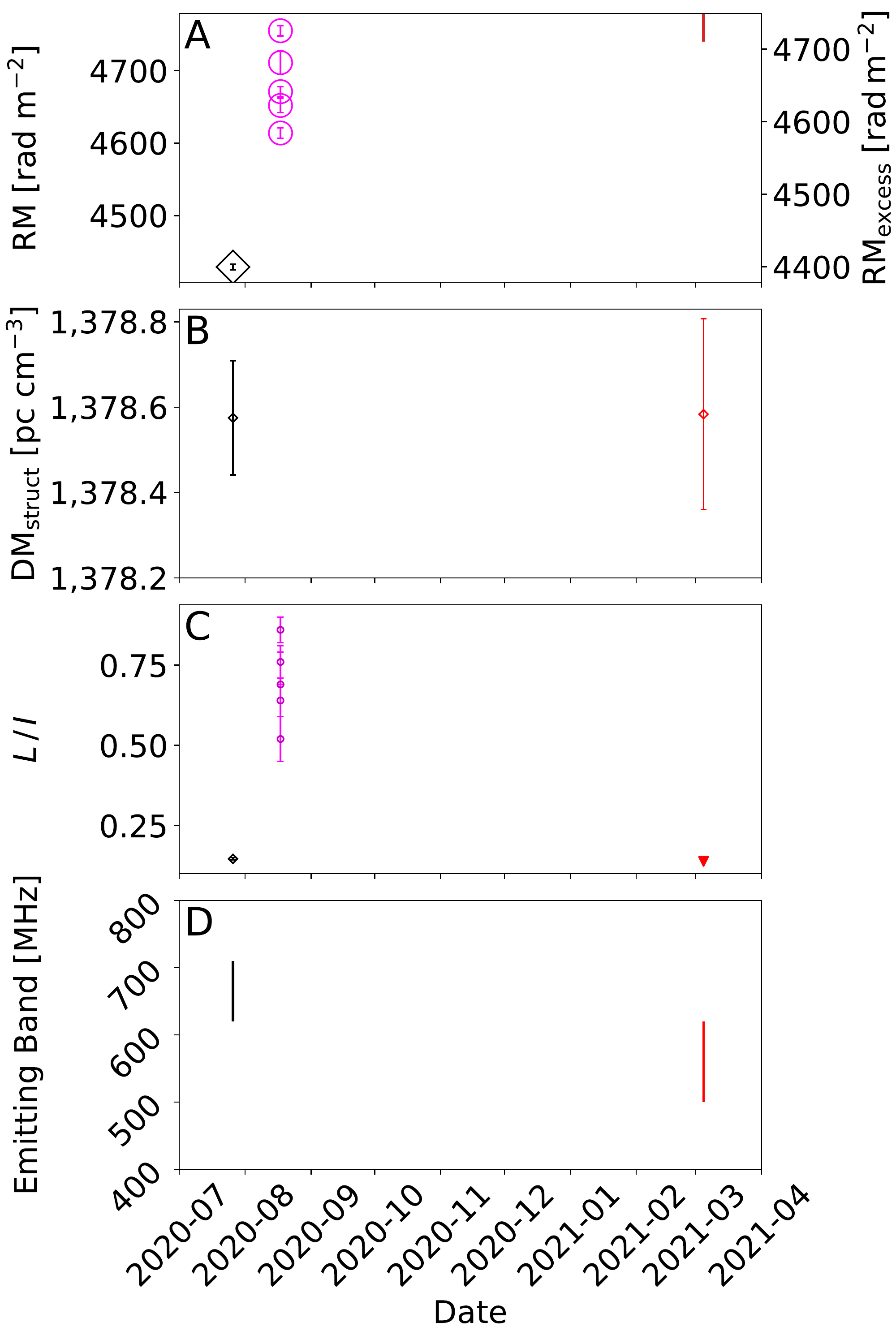}} 
\caption{Similar to Figure~\ref{fig:RMvstime_R17}, summary plots displaying burst properties as a function of time for other repeating sources of our sample with two or more baseband events. Red data points in panels B, C and D correspond to the unpolarized sample where the inverted red triangles of panel C indicate upper limits on $L/I$. The times of arrival (TOA) of the unpolarized bursts are indicated as lines along the top horizontal axis (panel A). The magenta points of panel (g) indicate additional measurements of FRB 20190417A, recently reported by FAST \citep{Feng2022}. For sources where $\rm{RM_{excess}}\sim 0\; \rm{rad\, m^{-2}}$ (panels (c) \& (e)), blue shaded regions are displayed to indicate the uncertainty in these measurements as determined from $\rm{RM_{MW}}$ estimates (see Table~\ref{ta:dmrmgal}).}
\label{fig:RMvstime_other}
\end{figure*}

This section summarizes observations of the remaining bursts of our repeating CHIME/FRB sample. Waterfall plots for each burst can be found in Appendix~\ref{appendix:waterfall}. Similar to Figure~\ref{fig:RMvstime_R17} for FRB 20190303A, Figure~\ref{fig:RMvstime_other} summarizes burst properties ($\rm{RM}$, $\rm{DM}$, $L/I$, emitting band) as a function of time for the subsample of repeaters with two or more baseband events. In all cases a semi-coherent RM search was applied to bursts that were observed to be unpolarized, ruling out $\rm{|RM|}$ values as large as $10^6\; \rm{rad\, m^{-2}}$ at our sensitivity.\footnote{This does not rule out the possibility of a marginally polarized signal (S/N$<5$) at large $\rm{|RM|}$ values.}   

\bigbreak

\noindent \textbf{FRB 20180814A:} Five baseband events were recorded from this source over a 5 month period between 2019 June 11 and 2019 November 11. 
% Burst envelopes from this source are generally wide, spanning several tens of ms, and contain multiple subbursts. 
 The complex burst morphology previously reported for this source \citep{chime/frb2019,chime/frb2019b} is replicated here with baseband data (see Figure set 1). Bursts are detected over the entire $400-800$ MHz band and several bursts display downward drifting subbursts. A systematic study of this time-frequency structure is beyond the scope of this paper and is deferred to a future analysis of repeater morphology at high time-resolution. We calculate an average $\mathrm{DM_{struct}}$ value of $188.60 \pm 0.03\; \rm{pc \, cm^{-3}}$ that is consistent with previous constraints determined from CHIME/FRB intensity data and not too dissimilar from a representative value ($188.8\; \mathrm{pc\, cm^{-3}}$) recently determined from analysis of individual subbursts of this source \citep{Chamma2021}. Only a single burst (burst 3) appears significantly linearly polarized and produces an RM detection at $\mathrm{RM} \sim +700\; \mathrm{rad\, m^{-2}}$, implying $\mathrm{RM_{excess}} \sim +745 \pm 18 \; \mathrm{rad\, m^{-2}}$. 
The remaining bursts appear to be unpolarized. Figure~\ref{fig:a} summarizes the bursts' properties as a function of time for this source.

\noindent \textbf{FRB 20181030A:} Five baseband events were captured from this source in a single day. This source is located at a sufficiently high declination ($>+70^{\circ}$) such that it enters CHIME's field-of-view (FoV) twice in a sidereal day, at both upper and lower transit \citep{chime/frb2019}. As such, two bursts were detected in the lower transit and three were detected approximately 12 hours later in upper transit. Bursts preferentially occupy the bottom half of the CHIME bandpass. Significant RM detections are made in three bursts (bursts 1,2 \& 5), all of which are observed to be $\sim 100\%$ polarized. The remaining bursts appear unpolarized. Several events (bursts 1, 2 \& 4; see Figure set 1) display multiple subbursts with substantial ($\gtrsim 30\; \rm{ms}$) separation in time. Interestingly, bursts 1 and 2 show evidence of PA evolution between these separated subbursts. There is no evidence for substantial RM variations relative to an average $\rm{RM}$ of $+36.6 \pm 0.2\; \rm{rad\, m^{-2}}$. Figure~\ref{fig:b} summarizes the bursts' properties as a function of time for this source.

\noindent \textbf{FRB 20181119A:} Five baseband events were detected from this source over 2020 June - 2021 December. Three of these bursts are significantly polarized with the remaining two observed as unpolarized. The most recent unpolarized burst is sufficiently bright to put an upper limit of $L/I<0.18$.  The remaining unpolarized burst is much less bright and, therefore, the corresponding upper limit on $L/I$ is significantly higher. The RM from this source appears to be highly variable, increasing by over $700\; \rm{rad\, m^{-2}}$ to a maximum value of $\rm{RM} = +1339.3 \pm 2.7\; \rm{rad\, m^{-2}}$ over six months and decreasing by an approximately equal amount over a similar time interval. Such large variations imply temporal RM gradients of at least $|\rm{\partial RM/ \partial t}| \gtrsim 4\; \rm{rad\, m^{-2}\, day^{-1}}$ and are not accompanied by any significant DM changes. Figure~\ref{fig:c} summarizes this source's burst properties as a function of time.  

\noindent \textbf{FRB 20190222A:} A single baseband event was detected from this source on 2019 March 1. The burst was observed to be partially linearly polarized ($\gtrsim 30\%$); however, the significant circular polarization seen from this burst (see Figure set 1) suggests that the linear polarization could be higher if instrumental leakage is present. An RM of $+567.7 \pm 2.3\; \rm{rad\, m^{-2}}$ was detected for this source.  

\noindent \textbf{FRB 20190208A:} Seven baseband events were captured from this source between 2020 January and 2021 December. Bursts occured across the CHIME band (see Figure set 1) and are highly linearly polarized ($L/I\lesssim0.7$). There is slight evidence for small variations in the PA of several bursts; however, the PA excursions are undersampled and tend to occur at the edges of the burst where the S/N is lower. Figure~\ref{fig:d} summarizes this source's burst properties as a function of time. The RM of this source appears as a ``U" shaped evolution over a two year interval where the evolution ``bottoms-out" near $RM_{\rm{excess}}\sim 0\; \rm{rad\, m^{-2}}$ and is followed by a recovery to previous RM levels. This RM evolution implies a modest temporal RM gradient of $|\rm{\partial RM/ \partial t}| \sim  0.1-0.2\; \rm{rad\, m^{-2}\, day^{-1}}$ and there is no evidence for correlated changes in $\rm{DM_{struct}}$ over this period.

% The evolution of this source is recent RM values now indicate a change in sign of $\rm{RM_{excess}}$, implying a flip in the orientation of $B_{\parallel}$ similar to FRB 20190520B \citep{Reshma2022, Dai2022}. 

\noindent \textbf{FRB 20190604A:} A single baseband event was detected from this source on 2019 June 6. The polarized properties of this burst were originally reported by \citet[their source 2;][]{Fonseca2020}. Refined analysis applied here demonstrates this burst to be $100\%$ linearly polarized with a flat PA curve (see Figure set 1). We detect an RM of $-17.8 \pm 1.2\; \rm{rad \, m^{-2}}$ that is not significantly different from the previous reported value.  

\noindent \textbf{FRB 20190213B:} Five baseband events were detected from this source between 2019 December and 2021 August. This source (see Figure set 1) exhibits fairly complex morphology, with several bursts displaying strong frequency drifting in the emission (e.g. Burst 1). Bursts with more complex morphology appear to be accompanied by evolution of their PA, however; this may also be a coincidence given the small size of our sample. Significant circular polarization is also observed in several sources (bursts 1, 2 \& 4). This may be a real signal from the source but is at odds with our simultaneous $L/I \sim 100\%$ measurement and suggests the presence of cross polarization. Figure~\ref{fig:e} summarizes this source's burst properties as a function of time. RMs from this source are fairly modest ($\sim 0\; \rm{rad\, m^{-2}}$) and would possibly be regarded as false detections from instrumental polarization were it not for the high linear polarized fractions observed. Furthermore, the RM of the first observation is significantly ($>6\sigma$) different from the remaining observations, which is further evidence that these RM detections are astrophysical. For the most recent observation, we determine $\rm{RM_{excess}} = -10 \pm 13\; \rm{rad\, m^{-2}}$.

\noindent \textbf{FRB 20180908B:} A single baseband event was captured from this source on 2019 June 21. The burst occupies the bottom 50 MHz of the CHIME band and is observed to be unpolarized. The $\rm{DM_{struct}} = 195.3 \pm 0.4\; \rm{pc\, cm^{-3}}$ for this source is consistent with previous measurements determined with CHIME/FRB intensity data \citep{Fonseca2020}.

\noindent \textbf{FRB 20190117A:} We captured three baseband events from this source between 2019 January and 2021 November. Bursts appear to be marginally polarized ($L/I\lesssim 30\%$), leading to significant RM detection in only two events. The remaining event is significantly fainter and a similar level of polarization as seen from the other two bursts would not be detectable. PA evolution within the two polarized bursts is unclear given the low S/N of the linear polarized signal. Figure~\ref{fig:f} summarizes this source's burst properties as a function of time. The RM is observed to increase by $\sim 9\; \rm{rad\, m^{-2}}$ and is accompanied by a significant increase in the DM of $\sim 2.75\; \rm{pc\, cm^{-3}}$. After correcting for DM and RM contributions from the Milky Way this amounts to an approximately $9\%$ change in the RM and $0.7\%$ in the DM.  

\noindent \textbf{FRB 20190417A:} We captured two baseband events from this source between 2020 July and 2021 March with one of these bursts providing a significant RM detection near $\rm{RM} \sim +4430\; \rm{rad\, m^{-2}}$. This RM is similar to recent measurements obtained at higher frequencies by FAST \citep{Feng2022}. However, bursts from this source appear significantly less polarized when observed at the lower frequencies of CHIME. This result is summarized in Figure~\ref{fig:g}, which shows burst properties as a function of time and compares $\rm{RM}$ and $L/I$ of bursts observed by CHIME and FAST. Figure~\ref{fig:depolb} displays the $L/I$ measurements as a function of frequency. The lower degree of linear polarization of this source when observed in the CHIME band appears to be generally consistent with predicted behavior from depolarization \citep{Feng2022}. From this combined sample of FAST and CHIME observations, we calculate a significant Spearman correlation coefficient of 0.86 (p-value $= 0.01$) between $L/I$ and frequency.
If depolarization is indeed responsible for the observed frequency dependence of $L/I$, future observations of this source with CHIME will be helpful for constraining the nature of the depolarizing medium, including the precise value of $\sigma_{\rm{RM}}$ and its stability over time. 

\noindent \textbf{FRB 20190907A:} A single baseband event was detected from this source on 2020 July 29. Significant RFI occupies the top half of the CHIME band and is masked in the corresponding waterfall plot (see Figure set 1). The burst is observed to be unpolarized. The $\rm{DM_{struct}}$ determined here is found to be in agreement with previous measurements from CHIME/FRB intensity data \citep{Fonseca2020}.

\section{Discussion}
\label{sec:discussion}

We summarize the main results of our analysis below.
\begin{enumerate}
    \item Multi-year polarimetric monitoring of CHIME/FRB repeating sources \citep{chime/frb2019b, chime/frb2019,Fonseca2020} has provided a sample of 100 bursts from 13 repeaters. 
    A significant fraction of these bursts originate from the periodic source FRB 20180916B (44 bursts), which is described in a companion paper \citep{Mckinven2022}. The remaining sample studied here include many repeaters for which these are the first reported polarized measurements (FRB 20180814A, FRB 20181030A, FRB 20181119A, FRB 20190222A, FRB 20190208A, FRB 20190213B, FRB 20180908B, FRB 20190117A \& FRB 20190907A) in addition to a handful with previously reported results \citep[FRB 20190604A, FRB 20190303A \& FRB 20190417A;][]{Fonseca2020, Feng2022}.
    \item The repeating FRBs studied here exhibit a high degree of variability in their linear polarized fractions which appears to be intrinsic to the sources. However, some repeaters also display clear frequency dependence in $L/I$ (see Figure~\ref{fig:FracPol_R17}) that is consistent with the general behavior expected from propagation through a depolarizing medium. Meanwhile, the circular polarization displayed in some bursts is very likely to be, at least in part, instrumental.
    \item Significant burst-to-burst RM variations are observed in the majority of our repeating sources. These variations often appear to be secular on week-month timescales rather than stochastic and are, in general, not accompanied by correlated change in $\rm{DM}$, $L/I$ or emitting band.
    \item Some repeating sources display extreme RM variations of several hundred $\rm{rad\, m^{-2}}$. The $20$-burst sample from FRB 20190303A exhibits an RM that has changed by nearly $500 \; \rm{rad \, m^{-2}}$ over a two year period (Figure~\ref{fig:RMvstime_R17}). FRB 20181119A displays significant RM variations, changing by as much as $\sim 860\; \rm{rad\, m^{-2}}$ over $\sim 6$ month intervals (Figure~\ref{fig:c}). FRB 20190208A, meanwhile, displays a ``U" shaped RM evolution over the course of two years, bottoming out near $\rm{RM_{excess}}\sim 0\; \rm{rad\, m^{-2}}$ before recovering to previously measured values ($\rm{RM_{excess}}\sim +30\; \rm{rad\, m^{-2}}$; see Figure~\ref{fig:d}).
    % . with most recent observations indicating a flip in orientation of the perturbing medium's magnetic field, as in recent observations of FRB 20190520B \citep{Reshma2022, Dai2022}.
\end{enumerate}

\subsection{Implications of RM Variations}
\label{sec:RM_variation}

Burst-to-burst RM variations appear to be a ubiquitous feature of repeating FRB sources. In many cases these RM changes can be extreme, changing by several hundred $\rm{rad\, m^{-2}}$ on weeks$-$months timescales and implying a dynamic local environment, which may or may not be related to the central engine. Examples of such sources in our sample include FRB 20190303A, FRB 201811119A and FRB 20190417A. These sources join other repeaters FRB 20121102A \citep{Michilli2018, Hilmarsson2021, Plavin2022} and FRB 20190520B \citep{Reshma2022, Dai2022} in establishing a significant subsample of the repeater population that reside in dynamic magneto-ionic environments. For these sources, negligible (or very little) change is observed in the DM. This suggests that either the observed RM evolution is produced mainly from changes in $B_{\parallel}$ or that the small but significant changes in the electron number density, $n_e$, or path length through the perturbing medium
exist but are too small to be detected with our DM measurement precision. In reality, the observed RM variations maybe some combination of these two extreme scenarios with changes in $B_{\parallel}$, $n_e$ and path length each imparting individual contributions to the observed RM variation. If, however, a change in the sign of the $\rm{RM_{excess}}$ is observed then this necessarily implies a change in the orientation of $B_{\parallel}$ (field reversal). Indeed, such a scenario has been observed from FRB 20190520B \citep{Reshma2022, Dai2022} and several sources in our sample come very close to displaying this behavior (e.g., FRB 20190208A, FRB 20190213B).\footnote{See also recent RM evolution from FRB 20180916B \citep{Mckinven2022}} 

Our best sampled sources commonly display intervals of quasi-secular RM evolution (e.g., FRB 20190208A, 20190303A) that are similar to the remarkably linear RM evolution recently observed from FRB 20180916B \citep{Mckinven2022}. The month$-$year timescales of these secular intervals seem to suggest the presence of ionized structures permeated by a reasonably ordered, but changing, magnetic field. Such behavior contrasts the seemingly stochastic RM variability displayed by repeating sources FRB 20180301A \citep{Luo2020} and FRB 20201124A \citep{Hilmarsson2021, Kumar2021}, albeit on much shorter ($<1$ day) timescales. Whether FRB 20190208A and/or FRB 20190303A display similar stochasticity on these timescales remains unknown from the sparse sampling of these sources in time. Subsequent observations by FAST of FRB 20201124A \citep{Lee2021} have shown secular intervals of RM evolution similar to that displayed here for FRB 20190303A, changing by hundreds of $\rm{rad\, m^{-2}}$ on week$-$month timescales. There are several interesting similarities between FRB 20201124A and FRB 20190303A that may suggest similar local environments and/or progenitors. Both sources display large RM variations on relatively short timescales, variable polarization fractions, and significant PA evolution across burst durations. There are even morphological similarities with both sources regularly exhibiting the frequency drift in subbursts commonly associated with repeaters. However, unlike FRB 20201124A, FRB 20190303A displays evidence for a secular decrease in the RM contribution from its local environment. This behavior is reminiscent of that seen from FRB 20121102A, which displays dramatic RM evolution that generally follows a decreasing trend but is interspersed by erratic periods where the RM can change by $\rm{\sim 1000 \; \rm{rad\, m^{-2}\, week^{-1}}}$ \citep{Hilmarsson2021}. 
Curiously, the temporal RM gradients, $\rm{\partial RM/ \partial t}$, displayed by FRB 20190303A are of comparable amplitude to the erratic RM excursions reported from FRB 20121102A. The extent to which this is a coincidence or an actual characteristic feature reflecting spatial scales of magnetized structures in a similar local environment remains to be seen. 

The origins of such strong and varying magnetic fields required to explain the RM varibility observed here and elsewhere are topics of on-going study. Popular scenarios commonly involve an FRB-emitting neutron star in which RM variability is supplied by inhomogeneities in the surrounding circum-burst medium that may be associated with a supernova remnant, a pulsar wind nebula or coronal wind from a binary stellar companion \citep{Reshma2022}. Alternatively, the RM variations may be an imprint of an intermediate-mass black hole with strong outflows, as for PSR J1745$-$2900 whose significant RM variations are believed to be an imprint of magnetized outflows from the nearby supermassive black hole, Sagittarius A$^{\star}$, at the center of the Galaxy \citep{Pang2011, Eatough2013}. 
One important detail for distinguishing among these competing scenarios is the presence of compact persistent radio source (PRS) associated with repeaters that display the most extreme RM behavior\footnote{Whether that be average RM, burst-to-burst RM variability and/or RM-scatter inferred from frequency-dependent depolarization.}\citep{Feng2022}. Such a result indicates a possible relation between the luminosity of the PRS and the magnetization of the source's local environment. This could be leveraged in the future to prioritize high sensitivity follow-up observations that seek to identify more potential FRB-PRS associations \citep{Yang2020c,Yang2022}. In Section~\ref{sec:B_parallel} this topic is further discussed in the context of $B^{local}_{\parallel}$ estimates and equivalent estimates gleaned from a limited sample of young pulsars associated with SNRs.

\subsection{Depolarization via Stochastic Faraday Rotation}
\label{sec:depol}

\begin{figure*}
	\centering
\begin{center}
    \includegraphics[width=0.7\textwidth]{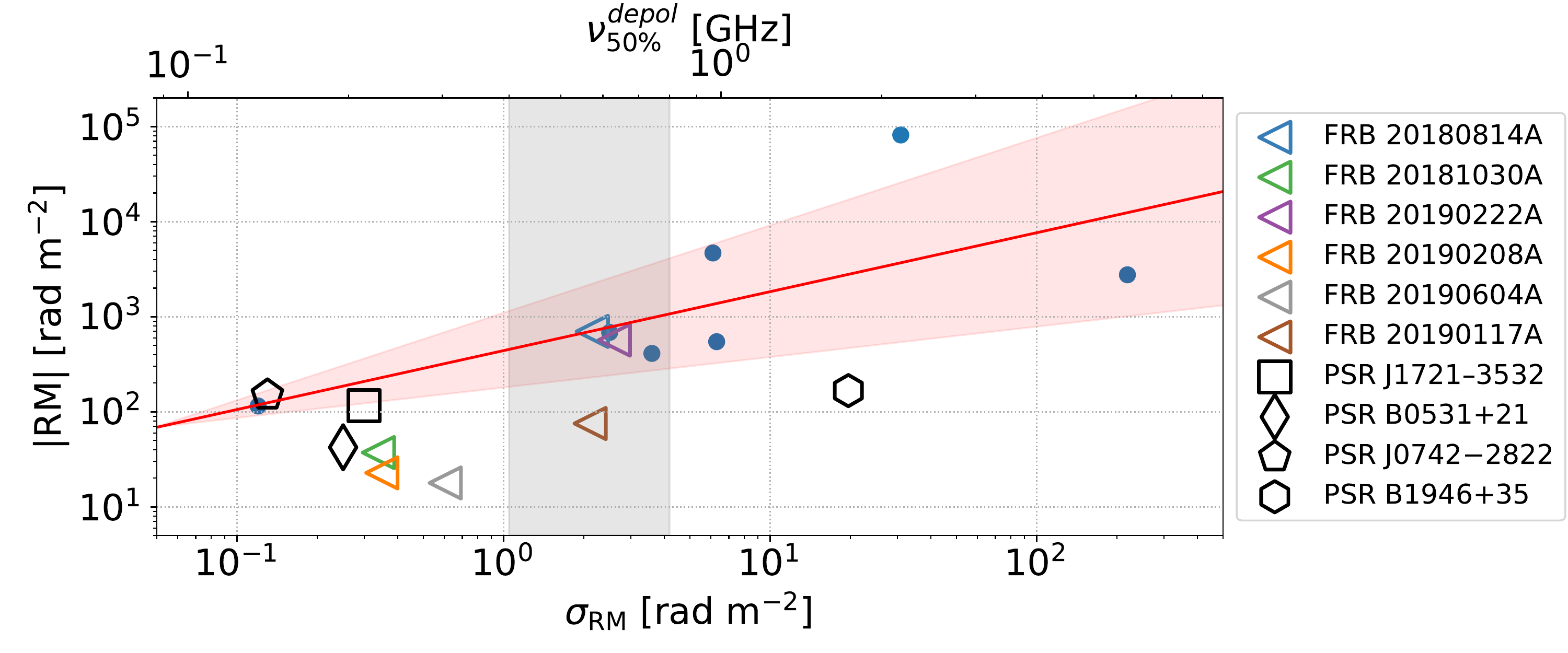}
    \caption{The $\sigma_{\rm{RM}}$, $|\rm{RM}|$ relation of \citet{Feng2022}, updated to include $\sigma_{\rm{RM}}$ upper limits from additional sources studied here. Measurements of \citet{Feng2022} are indicated by blue circles with the red line indicating the best-fit linear model (slope$=0.62\pm 0.30$). Upper limits on $\sigma_{\rm{RM}}$ for six additional sources studied here are indicated by leftward triangles and are seen to be consistent with the predicted trend of the linear model. Equivalent measurements for a small sample of pulsars (PSRs J1721$-$3532, B0531+21, J0742$-$2822, B1946+35) are indicated by open, black markers. The top horizontal axis indicates the frequency at which $50\%$ depolarization occurs and is highlighted as a gray region over the 400-800 MHz bandpass of CHIME. Figure updated and adapted with permission from \citet{Feng2022}.}
 \label{fig:sigmarm_rm}
\end{center}
\end{figure*}

In Figure~\ref{sec:depol}, we showed that $L/I$ measurements of two repeaters (FRB 20190303A, FRB 20190417A) are systematically lower than equivalent observations at higher frequencies. This feature appears to be replicated in recent $L/I$ measurements of FRB 20180916B with CHIME/FRB \citep{Mckinven2022}. These observations support depolarization via multi-path propagation as a major factor affecting $L/I$ measurements. However, we also find significant excursions in $L/I$ measurements relative to the best-fit depolarization trends recently determined for each of these sources \citep{Feng2022}. These excursions reflect some combination of intrinsic variability and evolving properties of the depolarizing medium. A changing magneto-ionic environment was recently proposed by \citet{Laha2022} to explain the marginal linear polarization ($<10\%$) observed from FRB 20180301A \citep{Laha2022}, a source that has previously displayed significant linear polarization at similar frequencies \citep{Luo2020}. Indeed, many of the $L/I$ measurements reported here cover multi-year timescales. It would not be surprising therefore if the significant burst-to-burst RM variations observed from FRB 20190303A, FRB 20190417A and FRB 20180916B translated into similar variability in the depolarization parameter, $\sigma_{\rm{RM}}$. In the future, this scenario can be tested by fitting a depolarization model to discrete epochs of observation and looking for significant changes in $\sigma_{\rm{RM}}$ and any correlated behavior with the RM evolution of the source.

Apart from FRB 20190303A, FRB 20190417A and FRB 20180916B, no other repeating source in our sample contains a sufficient number of observations and/or bandwidth to sensibly fit a depolarization model. However, we can estimate upper limits on $\sigma_{\rm{RM}}$ for six additional sources by assuming $100\%$ intrinsic linear polarization and calculating the $\sigma_{\rm{RM}}$ that would be required to remain consistent with the $L/I$ measurements of each repeater. The results of this analysis are summarized in Figure~\ref{fig:sigmarm_rm} where upper limits on $\sigma_{\rm{RM}}$ for each repeater are indicated as leftward triangles. These upper limits are all positioned below the best-fit $\sigma_{\rm{RM}}$, $|\rm{RM}|$ trendline recently established from a small sample of repeaters with $\sigma_{\rm{RM}}$ measurements \citep{Feng2022}. Equivalent measurements obtained from a small sample of pulsars \citep[PSRs J0742-2822, B0531+21, B1946+35, $\&$ J1721-3532][]{Xue2019,Sobey2019,Sobey2021} are also shown. The large scatter of our upper limits relative to this trendline does not disprove the existence of a correlation between $\sigma_{\rm{RM}}$ and $|\rm{RM}|$ but instead likely reflects the limitations of determining $\rm{\sigma_{RM}}$ for observations with insufficient bandwidth. In fact, the deficit of CHIME/FRB repeaters with RMs greater than several thousand $\rm{rad\; m^{-2}}$ supports the notion that $\sigma_{\rm{RM}}$ is correlated with $|\rm{RM}|$. The gray region in Figure~\ref{fig:sigmarm_rm} corresponds to the range of $\sigma_{\rm{RM}}$ values at which $50\%$ depolarization occurs over the CHIME band. Sources with $\sigma_{RM}$ to the right of this region are therefore significantly depolarized at CHIME frequencies. This implies that unpolarized repeaters in our sample (e.g., FRB 20180908B \& FRB 20190907A) may in fact have significant $\sigma_{\rm{RM}}$, making them good candidates for followup at higher frequencies where the deleterious effects of depolarization are less substantial. Conversely, sources with $\sigma_{\rm{RM}}$ to the left of the gray region will likely need to be observed at frequencies below 400 MHz for a robust measurement of $\sigma_{\rm{RM}}$. Meanwhile, the absence of correlation between $\sigma_{\rm{RM}}$ and $|\rm{RM}|$ in the pulsar sample is perhaps not surprising given their preferential sampling of the Galactic plane where the RM is dominated by the large scale magnetic field permeating the ISM.

\subsection{Repeating vs. Non-repeating FRB Samples}

\begin{figure}
	\centering
\begin{center}
    \includegraphics[width=0.45\textwidth]{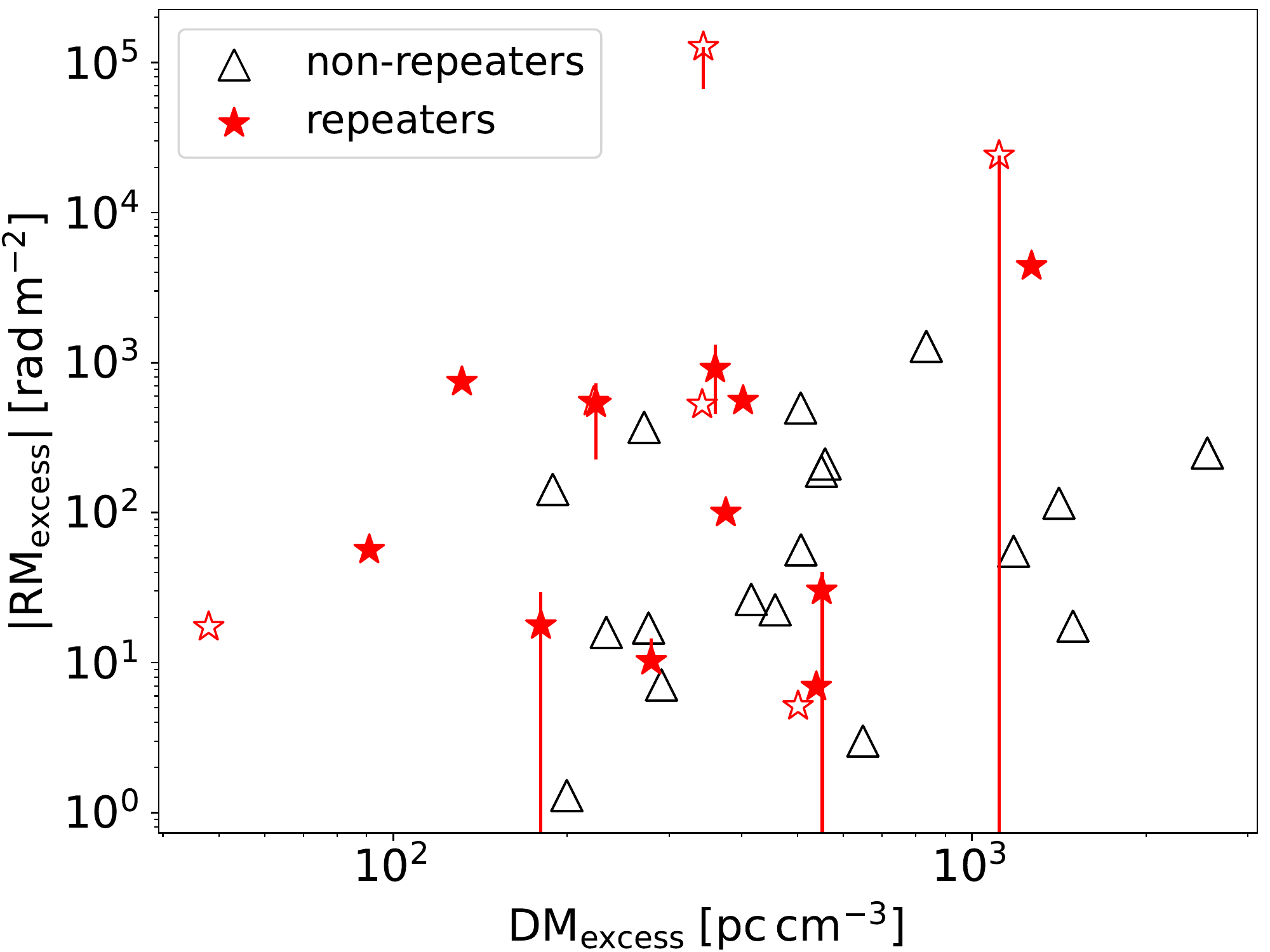}
    \caption{$\rm{DM_{excess}}$, $|\rm{RM_{excess}}|$ distribution (observer's frame) of the current sample of published FRBs showing the non-repeating (black triangles) and repeating (red stars) subsamples. Empty stars correspond to repeating sources other than the sample studied here (see text). Red vertical lines indicate the range of values observed from repeaters displaying substantial RM variations. $\rm{DM_{MW}}$ estimates are obtained from Table~\ref{ta:dmrmgal} using the NE2001 model of \citet{Cordes2002,Cordes2003} and assuming a fixed $30 \; \rm{pc\, cm^{-3}}$ contribution from the Galactic halo. $\rm{RM_{MW}}$ estimates are obtained from the Galactic RM map of \citet{Hutschenreuter2022}.}
 \label{fig:RM_DM_dist}
\end{center}
\end{figure}

Combining the polarized observations from the repeating sources studied here with previously reported measurements from additional repeaters; FRB 20121102A \citep{Michilli2018, Hilmarsson2021, Plavin2022}, FRB 20180301A \citep{Luo2020}, FRB 20190520B \citep{Reshma2022, Dai2022}, FRB 20190711A \citep{Day2020}, FRB 20200120E \citep{Bhardwaj2021, Nimmo2022} and FRB 20201124A \citep{Hilmarsson2021b, Kumar2021, Lee2021, Kumar2022}, gives us a sample of 16 sources with unambiguous RM detections that can be compared to a non-repeating counterpart sample. Such analysis is motivated by the possibility of different progenitors for the two populations, which could appear as an imprint in the RM distribution. We investigate this possibility in Figure~\ref{fig:RM_DM_dist} by displaying the joint $\rm{DM_{excess}}$, $\rm{RM_{excess}}$ distributions of the repeating and non-repeating samples. $\rm{DM_{excess}}$ estimates are obtained using the NE2001 model \citep{Cordes2002, Cordes2003} with an additional $\rm{DM_{halo}}$ contribution of $30\; \rm{pc\, cm^{-3}}$ while $\rm{RM_{excess}}$ estimates are obtained from the Galactic Faraday foreground model of \citet{Hutschenreuter2022} (see Table~\ref{ta:dmrmgal}). 
The resulting joint $\rm{DM_{excess}}, \rm{RM_{excess}}$ distribution of the repeating sample is compared to a sample of 18 non-repeating FRBs that is obtained by combining results from \citet{Wang2020} with observations of two additional one-off events, FRB 20191219F and FRB 20200917A \citep{Mckinven2021}. 

Based on the modest sample shown here, non-repeating FRBs appear to be more highly represented at larger $\rm{DM_{excess}}$ values. We performed an Anderson-Darling (AD) test to determine the significance of differences in the $\rm{DM_{excess}}$ distributions of the repeating and non-repeating samples and found a marginal p-value of $\sim 0.03$. This simple comparison is misleading however since it omits differences in the observing telescopes/surveys of the two samples. For instance, the absence of repeaters at very large $\rm{DM_{excess}}$ values can easily be explained by the disproportionate representation (relative to the non-repeating sample) of CHIME/FRB observations in our repeating sample, which is known to suffer significant sensitivity loss for higher DM events \citep{chime/frb2021}. Similarly, we find no obvious difference between the $|\rm{RM_{excess}}|$ distributions of the repeating and non-repeating samples ($\rm{p-value} > 0.13$), with both samples well represented at fairly modest $|\rm{RM_{excess}}|$ values. RM variations in the repeating sample are denoted by vertical lines covering the range of observed values. The absence of any substantial differences between the $|\rm{RM_{excess}}|$ distributions of the repeating and non-repeating samples suggests that the underlying sources reside in similar host galaxies and local environments. This result is consistent with the current localized FRB subsample, which has yet to display significant differences in the host properties of repeating and apparently nonrepeating FRBs \citep[e.g.,][]{Shivani2022}.    

We calculate the Spearman correlation coefficient between $\rm{RM_{excess}}$ and $\rm{DM_{excess}}$ and find no significant relation in either the repeating ($\rm{p-value<0.86}$), non-repeating ($\rm{p-value<0.22}$) or combined samples ($\rm{p-value<0.71}$). This result suggests that, on average, the media producing $\rm{DM_{excess}}$ and $\rm{RM_{excess}}$ are, at most, only weakly associated. This agrees nicely with similar conclusions that can be independently inferred from the relative variability of $\rm{DM}$ and $\rm{RM}$ of the repeating sample (Section~\ref{sec:B_parallel}). This conclusion is perhaps unsurprising given that in most cases $\rm{DM_{excess}}$ is dominated by the signal's transit through the IGM, where accumulated RM contributions should be well below $10\; \rm{rad\, m^{-2}}$ \citep{Akahori2016}. With a much larger sample, one might expect to see anti-correlation between $\rm{DM_{excess}}$ and $\rm{RM_{excess}}$ due to the the diluting effect of cosmological expansion on the RM contribution of the host galaxy / local environment of FRBs at increasing distances (i.e., $\sim \frac{1}{(1+z)^{2}}$)\footnote{The DM is less affected by this since: 1) the scaling (i.e., $\sim \frac{1}{(1+z)}$) is less extreme and 2) the contribution of the host galaxy / local environment to the total DM is likely substantially smaller than the equivalent contribution to the total RM.}.
The strength of such an anti-correlation will depend on how reliable $\rm{DM_{excess}}$ is as a proxy for distance and the degree to which $\rm{RM_{excess}}$ dominates the host galaxy / local environment. Alternatively, such a feature may be investigated with a smaller sample of localized FRBs that circumvent the need of using $\rm{DM_{excess}}$ as a proxy for distance.  

\subsection{Implications of $B^{local}_{\parallel}$ Estimates and RM Variability}
\label{sec:B_parallel}

\begin{table*}[t]
\begin{center}
\caption{Comparison of $B^{local}_{\parallel}$ estimates for a selection of repeating FRB sources and pulsars$^a$}
\hspace{-1.in}
\resizebox{1.1\textwidth}{!}{ 
\begin{tabular}{lcccccccc} \hline\hline
    Source & $B^{local}_{\parallel}$ & \multicolumn{2}{c}{$|\Delta \rm{RM}|$} & Timescale & RM evolution & Associated Structure & Reference \\
    & [$\rm{\mu G}$] & $\rm{[rad \, m^{-2}]}$ & $\%$ & & & &  \\\hline
        \multicolumn{8}{c}{} \\ %\hline
\bf{FRB 20200120E}$^b$ & $-$ & $\sim 30$ & $\sim 100$ & weeks$-$months & $-$ & globular cluster & \citet{Bhardwaj2021, Nimmo2022} \\
\color{gray}{PSR J1825$-$1446} & $\sim 0.2$ & $\sim 20$ & $\sim9$ & $\sim2$ years & Secular & SNR & \citet{Johnston2021} \\
\bf{FRB 20190117A} & $\gtrsim 4$ & $\sim9$ & $\sim 9$ & 1031 days & $-$ & $-$ & this work\\
\bf{FRB 20190213B} & $\gtrsim 16$ & $\sim 4.6$ & $\sim 10$ & 220 days & $-$ & $-$ & this work\\
\color{gray}{PSR J0908$-$4913 (B0906$-$49)} & $\sim 20$ & $\sim 4$ & $\sim 40$ & $\sim 3000$ days & Secular & SNR & \citet{Johnston2021b} \\
\color{gray}{PSR B0833$-$45$^c$} & $\gtrsim 22$ & $\sim 10$ & $\sim26$ & $\sim$15 years & Secular & SNR & \citet{Hamilton1985} \\
FRB 20180301A & $\gtrsim 50$ & $\sim43$ & $\sim 8$ & $<1$ day & Stochastic & -- & \citet{Luo2020} + this work \\
\bf{FRB 20180916B} & $\gtrsim 55$ & $\sim 50$ & $\sim40$ & $\sim9$ months & Secular \& Stochastic & -- & \citep{Mckinven2022} \\
\bf{FRB 20190208A} & $\gtrsim80$ & $\sim 35$ & $\sim100$ & $\sim 200$ days & Secular (non-monotonic) & $-$ & this work\\
FRB 20201124A$^d$ & $\gtrsim 100$ & $\sim 500$ & $\gtrsim$ 100 & $\lesssim 0.5$ months & Secular & SF region / PRS$^e$ & \citet{Hilmarsson2021, Kumar2021, Lee2021} \\
\color{gray}{PSR J0540$-$6919 (B0540$-$69)} & $115 \pm 15$ & $\sim 15$ & $\sim 6$ & $\sim5$ months & Secular & SNR & \citet{Geyer2021} \\
\color{gray}{PSR B0531+21$^f$} & $150-200$ & 6.6 & $\sim 14$ & 20 months & Secular & SNR & \citet{Rankin1988} \\
\bf{FRB 20181119A} & $\gtrsim 2200$ & $\sim 860$ & $\sim 100$ & $\sim200$ days & $-$ & $-$ & this work\\
\bf{FRB 20190303A} & $\gtrsim 3000$ & $\sim 500$ & $\sim 100$ & $\sim 2$ years & Secular (non-monotonic) & -- & this work \\
\color{gray}{PSR B1259$-$63} & $\sim 500-10,000$ & $\lesssim 15,000$ & $-$ & $\sim0.5$ months & $-$ & pulsar-Be star binary & \citet{Johnston2005} \\
FRB 20190520B & $\gtrsim 4000$ & $\sim 26,000$ & $-$ & $\sim 7$ months & Secular & PRS & \citet{Reshma2022,Dai2022} \\ 
\color{gray}{PSR J1745$-$2900$^g$} & $\gtrsim 10,000$ & $\sim3500$ & $\sim 5$ & $\sim16.5$ months & Secular & Sgr A$^{\star}$ & \citet{Desvignes2018,Katz2021} \\
FRB 20121102A  & $3000-17,000$ & 15,000 / 4000 & 20 / 5 & 160 days / 450 days & Secular & SF region / PRS$^{h}$ &\citet{Hilmarsson2021, Katz2021} \\
% \bf{FRB 20181030A (R4)} & ??  & ?? & ?? & ?? & ?? & ?? & this work\\
\hline
\end{tabular}}
\label{ta:B_parallel}
\end{center}
$^a$ Ordered by increasing $B^{local}_{\parallel}$. Pulsars are indicated by gray font while CHIME/FRB repeaters are highlighted with boldface font. \\  
$^b$ $B^{local}_{\parallel}$ \& RM evolution columns are undetermined for FRB 20200120E due to lack of sufficient observations and $\rm{DM_{struct}}$ measurements. \\
$^c$ The `Vela pulsar'. \\ 
$^d$ \citet{Hilmarsson2021b} quote a representative DM for their entire sample rather than determine a DM for each burst. We therefore estimate the DM variability as the uncertainty on their DM measurement, $0.6 \;\rm{pc\, cm^{-3}}$. \\
$^e$ \citet{Ravi2021} claim that the spatial extent and luminosity of the persistent radio source (PRS) associated with FRB 20201124A is consistent with star-formation activity. \\
$^f$ The `Crab pulsar'. \\
$^g$ The Galactic center magnetar. \\
$^h$ FRB 20121102A is associated with a star-forming region within
its host galaxy as well as a persistent radio source (PRS) suggestive of
a pulsar wind nebula or supernova remnant \citep{Bassa2017, Chatterjee2017}.
\end{table*}

The analysis pursued in Section~\ref{sec:res_R17} leverages $\Delta \rm{DM}$ and $\Delta \rm{RM}$ measurements of FRB20190303A to constrain the LoS magnetic field of the medium producing the observed variations. In principle, this magnetic field could exist at any point along the LoS. Indeed, multi-epoch observations of Galactic pulsars regularly reveal stochastic RM variations that are attributed to scattering and some combination of fluctuations in $B_{\parallel}$ and $n_e$ in the Galactic ISM \citep[e.g.][]{Yang2011,Wahl2022}. However, these RM variations rarely exceed $\sim \rm{few}\; \rm{rad \, m^{-2}}$, significantly less than the RM variability observed from many repeaters reported here. Therefore, given the amplitude of RM variations and the relatively short timescales seen here for FRBs, it is highly likely that the magneto-ionic fluctuations producing the RM variability occur in the immediate environment of the FRB source. $B^{local}_{\parallel}$ estimates reported here are therefore likely a direct probe of the magnetization of the FRB source environment and can be compared to equivalent estimates made from observations of other repeating FRB sources and certain Galactic pulsars. 

Table~\ref{ta:B_parallel} summarizes this analysis, where we have calculated $B^{local}_{\parallel}$ estimates for all repeating sources with multiple RM measurements and compared them to equivalent measurements for a sample of pulsars. Importantly, Equation~\ref{eqn:Bdyn} is undefined when $\rm{DM_{\Delta RM}}$ is not a priori known. An exception exists when DM and RM variations are correlated such that $B^{local}_{\parallel}$ can be estimated by substituting $\rm{DM_{\Delta RM}}=\Delta \rm{DM}$ \citep[e.g.,][]{Hamilton1985, Katz2021}.\footnote{Such estimates assume that the observed DM, RM variations are an imprint of a Faraday active medium with a changing $n_e$ and a static magnetic field.} However, when DM and RM variations are uncorrelated an appropriate value for $\rm{DM_{\Delta RM}}$ remains unknown. In such cases, we make the simplifying assumption that the same medium is responsible for the observed DM and RM variability and estimate $\rm{DM_{\Delta RM}}$ from the root-mean-square of the $\rm{DM}$ measurements. Table~\ref{ta:B_parallel} displays the resulting $B^{local}_{\parallel}$ estimates in increasing order, and lists additional quantities such as the RM variation ($\Delta \rm{RM}$), its characteristic timescale, the nature of the RM evolution (stochastic, secular) and whether an associated structure has been identified that could possibly produce the observed variability.

Filamentary structure in SNRs has been invoked to explain the RM variations observed in most of the pulsars in this sample. These SNRs may either be associated with the source, as in the case of the Crab \citep{Rankin1988} or Vela \citep{Hamilton1985} pulsars, or be unassociated, as in the case of the fortuitous alignment of PSR J1825$-$1446 with a foreground SNR, G16.8$-$1.1 \citep{Moldon2012}.\footnote{See also the putative association of PSR J0908$-$4913 (B0906$-$49) with the previously unknown SNR G270.4$-$1.0 \citep{Johnston2021b}.} For such systems, $B^{local}_{\parallel}$ estimates up to several hundred $\mu G$ have been inferred. Such a range is comparable to equivalent estimates inferred from some repeaters in Table~\ref{ta:B_parallel}. While it remains plausible that an SNR-like structure may be associated with FRB sources, a significant subsample of repeaters (FRBs 20121102A, 20190520B, 20190303A, 20181119A) display RM variations that are far more extreme than what has hitherto been observed from relatively young pulsar-SNR systems. 

RM variations from repeaters generally appear to be more significant and prevalent than that seen from pulsars where stability in RMs from epoch to epoch is a necessary condition for their application as probes of the large scale Galactic magnetic field \citep[e.g.,][]{Han2018}. 
% Indeed, our $B^{local}_{\parallel}$ estimates of Table~\ref{ta:B_parallel} are several orders of magnitude larger than the $\rm{\mathcal O(\mu G)}$ field strengths occupying a Milky Way-like ISM \citep{Han2017}. 
Furthermore, pulsars displaying substantial RM variations remain quite rare. Recent multi-epoch monitoring of several hundred young, non-recycled pulsars using the Parkes telescope found only one pulsar (PSR J1825$-$1446) exhibiting significant RM variability \citep[see Table~\ref{ta:B_parallel};][]{Johnston2021}. This is difficult to reconcile with the ubiquity of RM variability displayed from the repeating FRB sample and suggests a magneto-ionic medium for repeating FRBs distinct from the environment of young pulsars. These observations may either indicate an earlier evolutionary stage of FRB--SNR systems, where magneto-ionic fluctuations are substantially larger, or an entirely different structure responsible for the strong RM evolution observed in many repeating sources. This novelty in FRB environments has been independently noted based on a joint analysis of FRB dispersion and scattering. Most recently, a population synthesis study of the CHIME/FRB catalog \citep{chime/frb2021b} found that the ISM of FRB host galaxies alone could not reproduce the observed scattering timescales of FRBs but instead required a substantial scattering contribution from the circumgalactic medium (CGM) or an extreme circum-burst environment dissimilar to that associated with Galactic pulsars \citep{Chawla2022}.

\begin{figure}
	\centering
\begin{center}
    \includegraphics[width=0.45\textwidth]{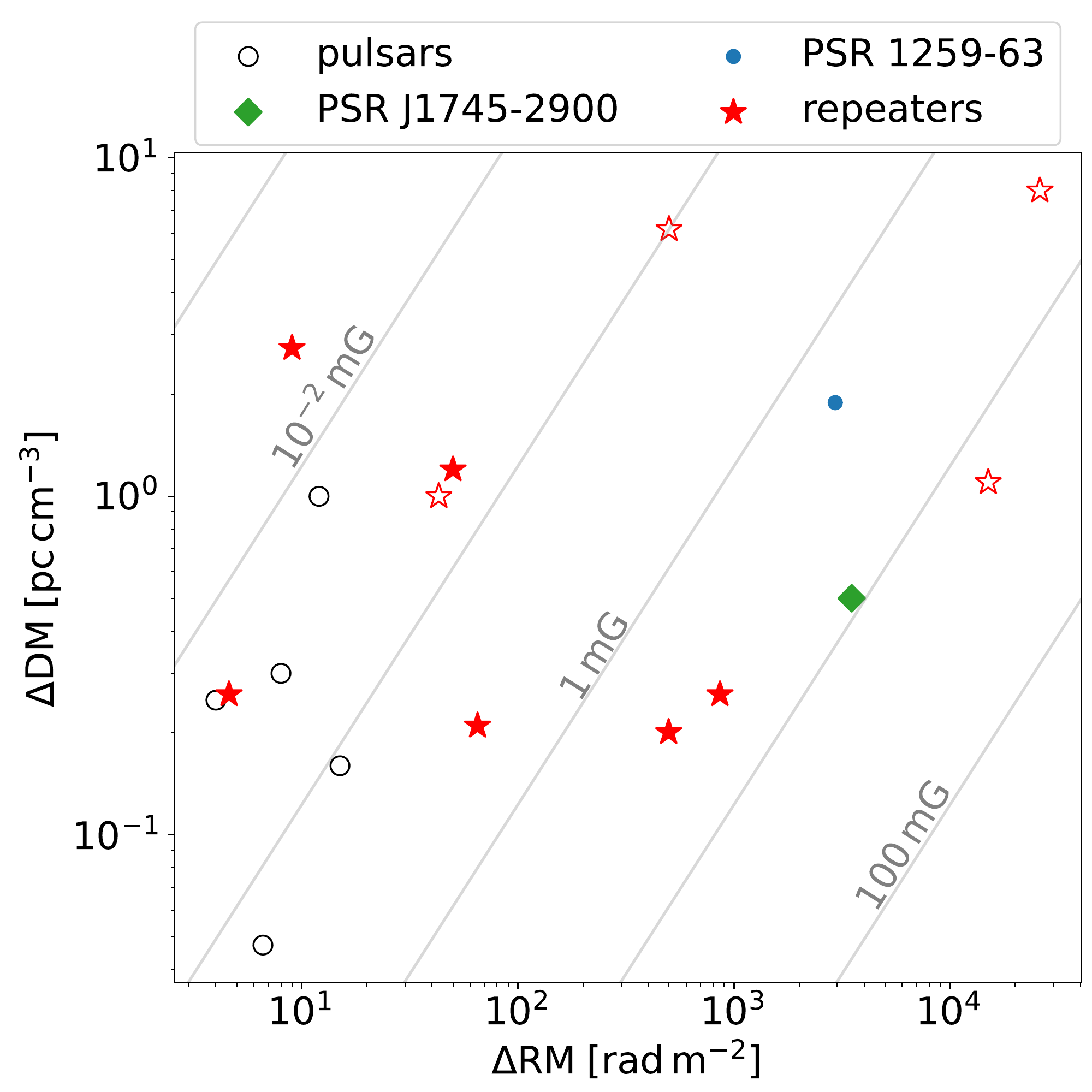}
    \caption{$\rm{\Delta RM}$, $\rm{\Delta DM}$ distribution of repeating FRB sources (red stars) compared to a sample of pulsars from which significant RM variations have been observed (black circles). Empty stars correspond to repeating sources other than the sample studied here. Special pulsars include the Galactic center magnetar (PSR J1745$-$2900; green diamond), and measurements from PSR B1259$-$63 (blue points), a pulsar in a close binary orbit with a Be star companion. Gray diagonal lines indicate different strength levels of the parallel component of the magnetic field ($B^{local}_{\parallel}$) required to explain $\rm{\Delta RM}$, $\rm{\Delta DM}$ measurements.}
 \label{fig:delrm_deldm_dist}
\end{center}
\end{figure}

The discrepancy in $\rm{RM}$ variability between repeating FRBs and pulsars is illustrated in Figure~\ref{fig:delrm_deldm_dist}, which compares the DM and RM variability of the two samples. $\Delta \rm{RM}$ of the repeater sample regularly exceeds several tens of $\rm{rad\, m^{-2}}$ while pulsar-SNR systems rarely exceed 10 $\rm{rad\, m^{-2}}$. Two exceptions are the Galactic center magnetar, PSR J1745$-$2900, and PSR B1259$-$63. Both these sources display significant variability in their RMs that can be explained by peculiarities of their immediate environments. In particular, PSR J1745$-$2900 is located at a projected distance of 0.12 pc from Sgr A$^{\star}$ \citep{Eatough2013}. The substantial RM variability recently observed from this source \citep{Desvignes2018} is expected to be an imprint of inhomogeneities produced by the accretion flow onto Sgr A$^{\star}$ \citep{Pang2011}. PSR B1259$-$63, meanwhile, displays significant changes in its DM and RM that are produced from its orbital passage through the dense circumstellar disc of its Be-star companion. Interestingly, multiple epochs of observations have established differences in the observed properties from one periastron to the next, indicating significant changes in disc density and magnetic field structure on its $\sim 1237$ day orbital timescale \citep{Connors2002, Johnston2005}. This source is particularly interesting given its relevance to ``cosmic comb" models that have been put forth to explain both the FRB emission mechanism and periodic bursting cycles observed from some repeaters \citep{Ioka2020, Lyutikov2020, Wada2021}. Indeed, these models predict that the RM modulation of FRBs will correlate with orbital phase, since the former is caused by the companion wind. However, as observations of PSR B1259$-$63 have shown, RM variations can be erratic due to substantial inhomogeneities in the magnetic field strength and free electron density in the ambient medium of the companion. It may therefore be naive to expect smooth RM evolution tied to orbital phase as would be assumed for the above FRB models.

Without additional information, several key parameters characterizing the local environment of repeating FRBs remain degenerate vis-a-vis RM variations (i.e., $n_e$, $B_{\parallel}$, path length / relative velocity). 
For prolifically repeating sources, future observations may further constrain these properties by extending our analysis to include variability in scattering timescales, as was done for the Crab pulsar \citep{McKee2018}. This would greatly help our understanding of repeater environments by enabling correlation of the scattering, RM and DM to be simultaneously probed. Any significant correlated behavior would indicate a common structure producing the variability, with scattering/scintillation measurements providing further information on the structure, namely, its physical size and distance from the source.

\subsection{Repeater Burst Morphology}
\label{sec:morph}

The morphology of bursts (the change in flux as a function of time and frequency) from our sample of repeating FRBs in the CHIME band has been discussed before 
\citep{chime/frb2019,chime/frb2020c,Fonseca2020,Pleunis2021b}. However, similar to our recent analysis of FRB 20180916B, the baseband data presented here have several properties that are amenable to burst morphology analysis (see \citet{Mckinven2022} for details). Inspection of Figure~\ref{fig:waterfalls_R17} and the additional waterfall plots for the remaining sources (see Figure set 1) reveals that bursts from these repeaters show similar characteristics in the CHIME band as were identified at coarser time resolution. The bursts typically show wide burst emission envelopes, of up to $\sim$40\,ms (e.g., burst 1 of FRB 20190213B). Some more narrow subbursts with widths down to $\lesssim80\,\mu$s are present, e.g., in bursts 1, 2 and 4 of FRB 20181030A -- as have been detected before in bursts from FRBs 20180916B and 20200120E \citep{Nimmo2021,Nimmo2022}. We note that these are three of the closest repeating sources of FRBs, and that the CHIME/FRB detection pipeline with its $\sim1-$\,ms time resolution may not be sensitive to similarly narrow emission from more distant sources \citep[see also][]{Connor2019}. The repeater bursts show narrowband emission ($\sim$50--200\,MHz) with varying central frequency. Downward-drifting subbursts (the ``sad trombone effect'') are ubiquitous, with burst 7 of FRB 20190208A being a prominent example. Drift rates are consistent with previous measurement in the CHIME band, with linear drift rates of a few to a few tens of MHz ms$^{-1}$. We have detected pre- and post-cursors that are significantly fainter than the main component of the burst, e.g., burst 1 of FRB20180814A, burst 1 of FRB 20190222A, bursts 3 and 5 of FRB 20190208A and burst 1 of FRB 20190117A, as has also been detected for other FRB sources by, e.g., \citet{Caleb2020} and \citet{Chawla2020}. As recently noted for FRB 20180916B \citep{Mckinven2022}, the distinction between multiple subbursts under one burst envelope vs. multiple separate burst envelopes without a ``bridge'' in emission remains ambiguous \citep[e.g., bursts 1, 2 and 4 of FRB 20181030A, all seem to be comprised of at least two burst envelopes, whereas this is less obvious in the case of bursts 1 and 3 of FRB 20190208A, and burst 1 of FRB 20190303A; see also discussion in][]{Pleunis2021b}. In general, the interpretation of burst structure remains dependent on the detection S/N, with fainter bursts requiring more substantial downsampling that possibly `washes-out' fine scale but faint substructure. This likely affects our $\rm{DM_{struct}}$ measurements, with downward-drifting subbursts remaining unresolved and biasing the DM measurement high. Burst-to-burst $\rm{DM_{struct}}$ variations are therefore very likely over-estimated here due to confusion with unresolved structure. Regardless, temporal DM variations from the FRB sources studied here appear to be significantly lower than those observed elsewhere \citep[e.g., $\Delta \rm{DM} \sim$ few pc cm$^{-3}$;][]{Oostrum2020}, a somewhat surprising result considering the multi-year timescales of these observations and the ubiquity of substantial RM variations.

\section{Conclusion}
\label{sec:conclusion}

Multi-year polarimetric monitoring of a sample of CHIME/FRB repeaters has provided a sample of 100 bursts from 13 previously published sources \citep{chime/frb2019b, chime/frb2019, Fonseca2020}. The morphology of the bursts from these sources in the CHIME band show qualitatively similar characteristics at the higher time resolution presented here to what has been described before by \citet{Pleunis2021b} based on data with coarser time resolution. In general, bursts from our sample are significantly linearly polarized, with only two sources observed as (apparently) unpolarized. Significant burst-to-burst variability in $L/I$ is observed from many sources along with marginal circular polarization ($\lesssim 20\%$), whose origin is likely instrumental. A systematic study of $L/I$ as a function of frequency for some of our sources (Section~\ref{sec:depol}; see also Section 4.2 of \citet{Mckinven2022}) has provided further support to the existence of depolarization via multi-path propagation \citep{Feng2022}. However, significant excursions in individual $L/I$ measurements relative to the best-fit depolarization models of each source indicate additional sources of variability.

Significant RM variability appears to be a characteristic feature of repeating FRB sources, with our sample displaying secular intervals of RM variations on week to month timescales that are not accompanied by any equivalent DM evolution. Some repeating sources exhibit extreme RM variations of several hundred $\rm{rad\, m^{-2}}$ over multi--month timescales. From our constraints on $\Delta \rm{DM}$ and $\Delta \rm{RM}$, we estimate the strength of the line-of-sight component of the magnetic field ($B^{local}_{\parallel}$) in the FRB source's environment and compare to equivalent estimates from a sample of pulsars. 
% finding values that significantly exceed $\langle B^{local}_{\parallel}\rangle$ determined from $\rm{RM_{excess}}$ estimates and assumptions on $\rm{DM_{host}}$ \citep{Wang2020}. 
Although our $B^{local}_{\parallel}$ estimates are similar to equivalent estimates for pulsars with associated SNRs, there exist a signifcant subsample of repeating FRBs that display RM variations that are far more extreme than what has been hitherto been observed from pulsar-SNR systems. This suggests that repeaters occupy magneto-ionic environments that are significantly different than those associated with our current sample of pulsar--SNR systems, an interpretation that is reinforced by the scattering distribution of FRBs \citep{Chawla2022}. 
% This may either indicate an earlier evolutionary stage of the SNR, where magneto-ionic fluctuations are substantially larger, or an entirely different structure responsible for the strong RM evolution observed in many repeating sources.
Without additional information, several key parameters characterizing the local environments of reeating FRBs remain degenerate vis-a-vis RM variations. Additional measurements such as those inferred from scattering or frequency-dependent depolarization will be invaluable for constraining this parameter space. Finally, if the non-uniform magneto-ionic environments of repeating FRBs extends to the non-repeating sample then this will be unfavorable to efforts to model the FRB RM distribution, since a significant contribution to the RM distribution will be associated with inhomogeneities in the local magneto-ionic environment that are quasi-random and difficult to model.     

% Observations of FRB 20180916B and FRB 20190303A cover time intervals of several years in which we observe significant RM variations of up to $40 \; \rm{rad \, m^{-2}}$ and $500 \; \rm{rad \, m^{-2}}$, respectively. We observe a secular decrease of $\sim 40 \; \rm{rad \, m^{-2}}$ in the $|\rm{RM}|$ of FRB 20180916B over the period April$-$December 2021, with most recent RMs $\sim -70\; \rm{rad\, m^{-2}}$. 

% For FRB 20190303A, bursts display a high degree of variability in their linear polarized fractions ($L/I$), with 7 bursts being (apparently) unpolarized and the remaining bursts displaying values between $20\%$ and $100\%$. RMs observed from this source appear to follow non-monotonic but quasi-secular evolution, with temporal gradients that indicate magneto-active regions near the source. Given the low DM of this FRB, a localization may be possible even with CHIME/FRB and would greatly illuminate the interpretation of such extreme RM evolution. 

\begin{acknowledgements}
The Dunlap Institute is funded through an endowment established by the David Dunlap family and the University of Toronto. R.M. recognizes support from the Queen Elizabeth II Graduate Scholarship and the Lachlan Gilchrist Fellowship. B.M.G. is supported by an NSERC Discovery Grant (RGPIN-2015-05948), and by the Canada Research Chairs (CRC) program. K.W.M. is supported by an NSF Grant (2008031). V.M.K. holds the Lorne Trottier Chair in Astrophysics $\&$ Cosmology, a Distinguished James McGill Professorship, and receives support from an NSERC Discovery grant (RGPIN 228738-13), from an R. Howard Webster Foundation Fellowship from CIFAR, and from the FRQNT CRAQ.
A.B.P. is a McGill Space Institute (MSI) Fellow and a Fonds de Recherche du Quebec -- Nature et Technologies (FRQNT) postdoctoral fellow. Z.P. is a Dunlap Fellow. C.L. was supported by the U.S. Department of Defense (DoD) through the National Defense Science $\&$ Engineering Graduate Fellowship (NDSEG) Program. FQD is supported by the 4YF. E.P. acknowledges funding from an NWO Veni Fellowship. K.S. is supported by the NSF Graduate Research Fellowship Program. 
The polarization analysis presented here makes use of the {\tt RMtools} package\footnote{https://github.com/CIRADA-Tools/RM-Tools}\citep{Purcell2020} written by Cormac Purcell, and maintained by Cameron Van Eck.

\end{acknowledgements}

\bibliographystyle{aasjournal}
\bibliography{references}

\appendix

\renewcommand\thefigure{\thesection.\arabic{figure}}

\section{Additional Waterfall Plots}
\label{appendix:waterfall}
\setcounter{figure}{0} 

\figsetstart
\figsetnum{1}
\figsettitle{Waterfall plots of individual bursts from our repeating FRB source sample.}

%%%%%%% R2

\figsetgrpstart
\figsetgrpnum{1.1}
\figsetgrptitle{FRB 20180814A}
\figsetplot{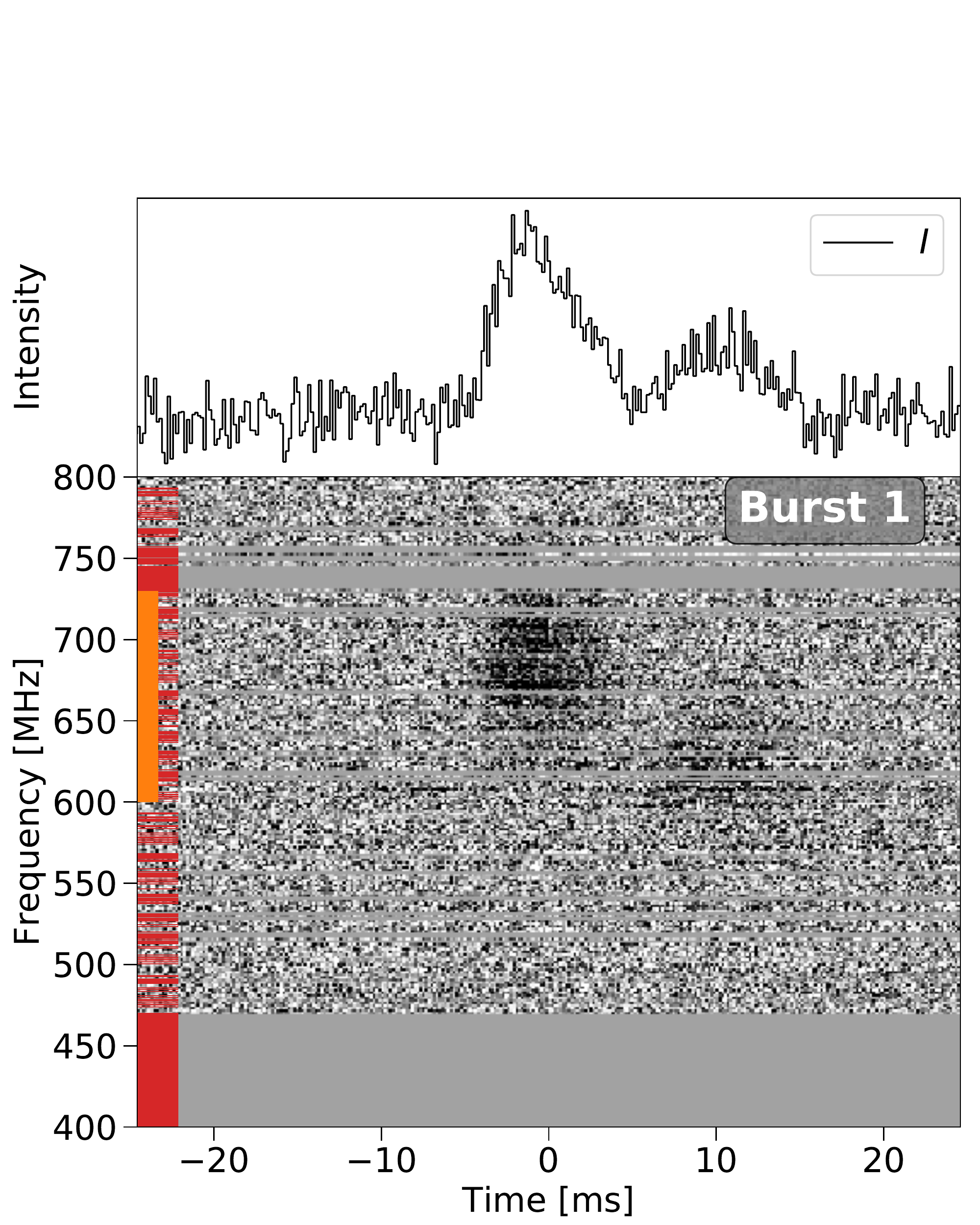}
\figsetgrpnote{}
\figsetgrpend

\figsetgrpstart
\figsetgrpnum{1.2}
\figsetgrptitle{FRB 20180814A}
\figsetplot{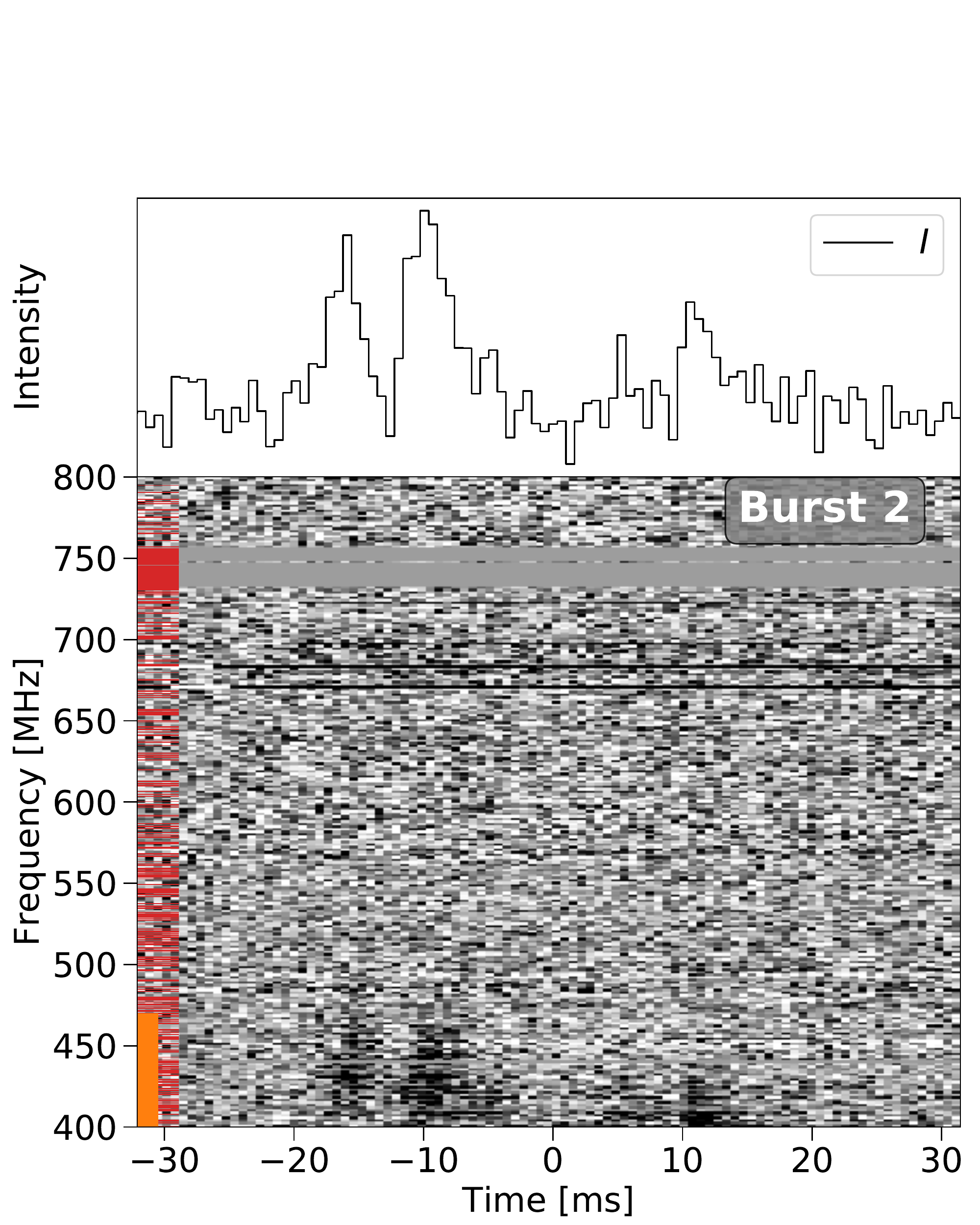}
\figsetgrpnote{}
\figsetgrpend

\figsetgrpstart
\figsetgrpnum{1.3}
\figsetgrptitle{FRB 20180814A}
\figsetplot{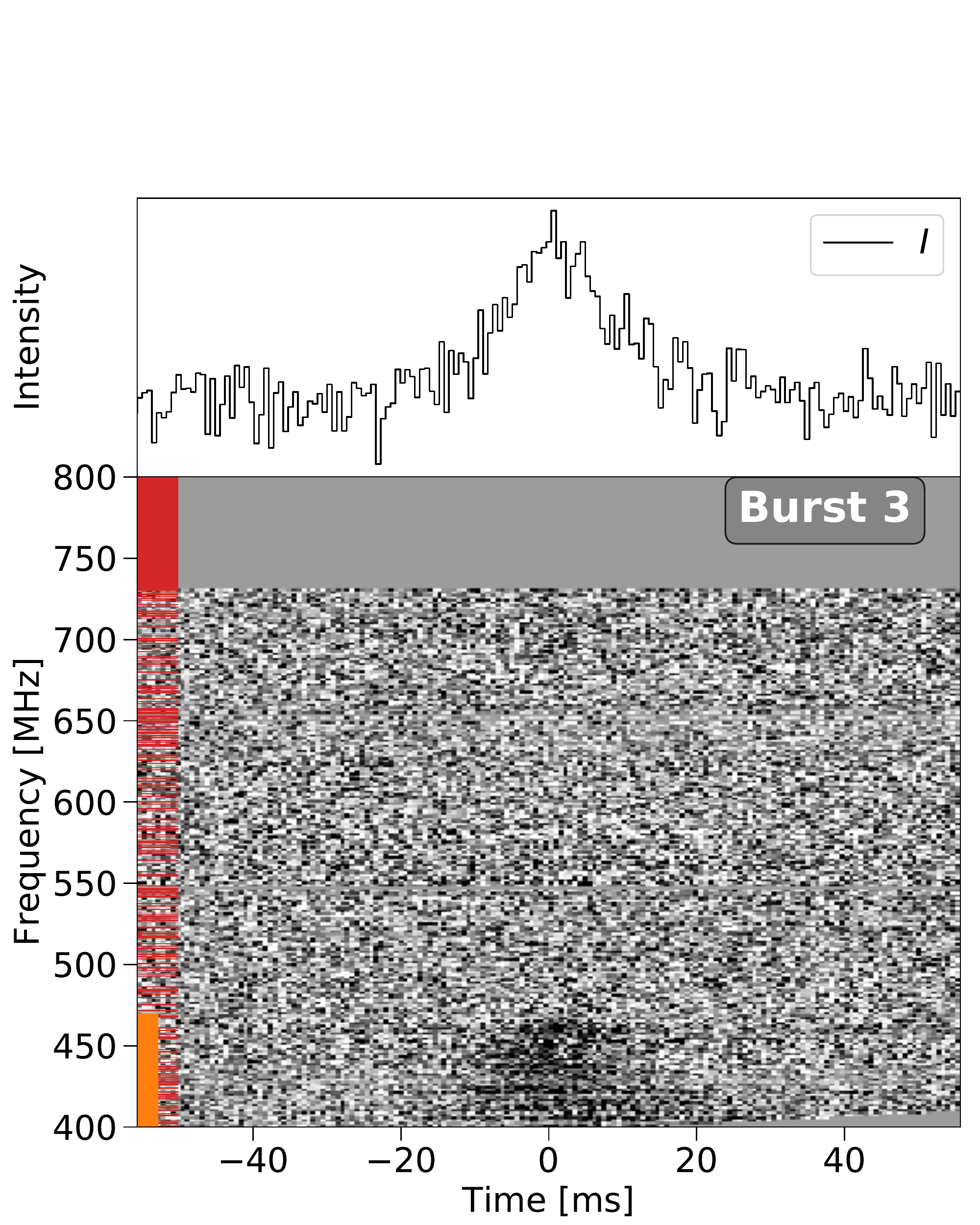}
\figsetgrpnote{}
\figsetgrpend

\figsetgrpstart
\figsetgrpnum{1.4}
\figsetgrptitle{FRB 20180814A}
\figsetplot{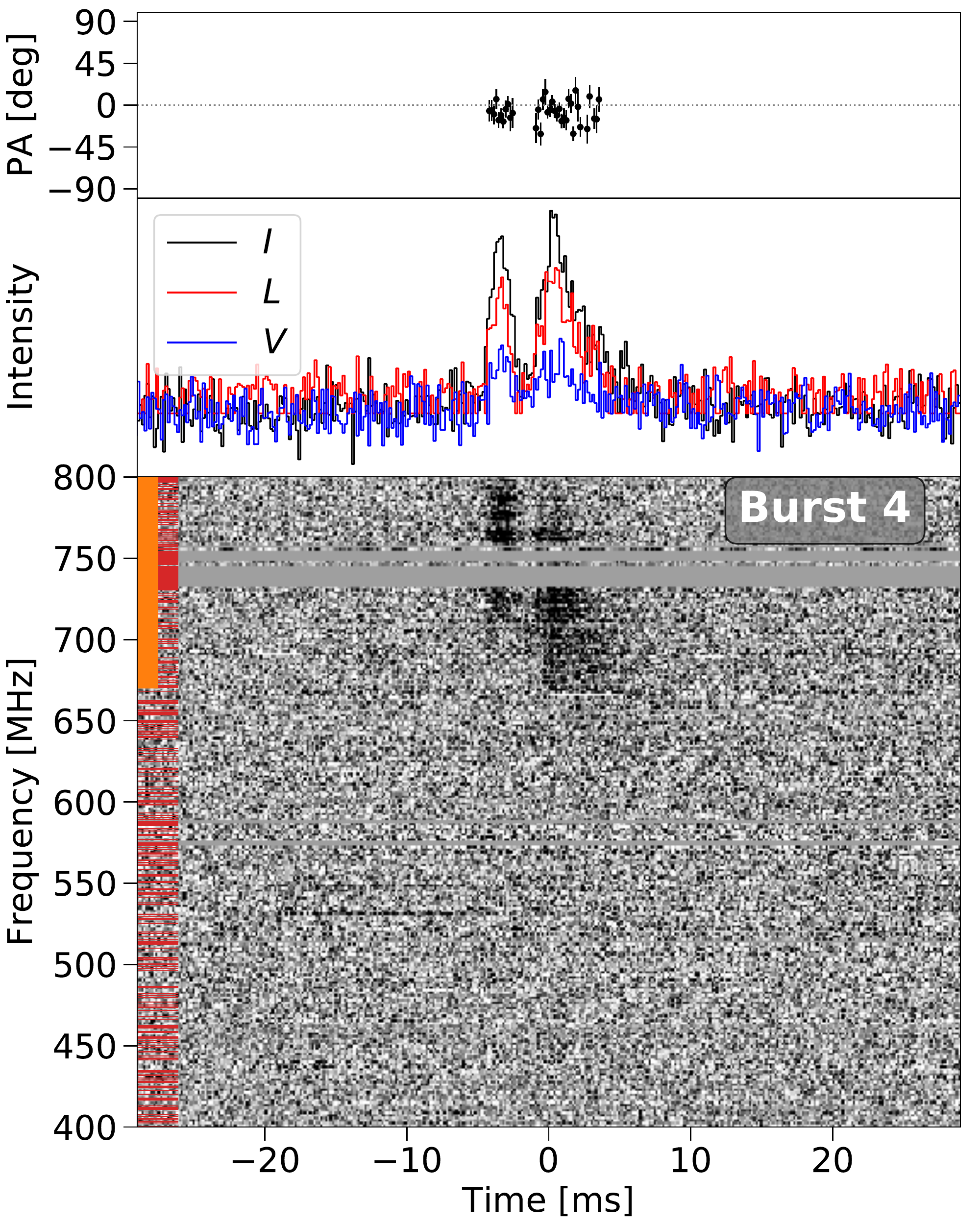}
\figsetgrpnote{}
\figsetgrpend

\figsetgrpstart
\figsetgrpnum{1.5}
\figsetgrptitle{FRB 20180814A}
\figsetplot{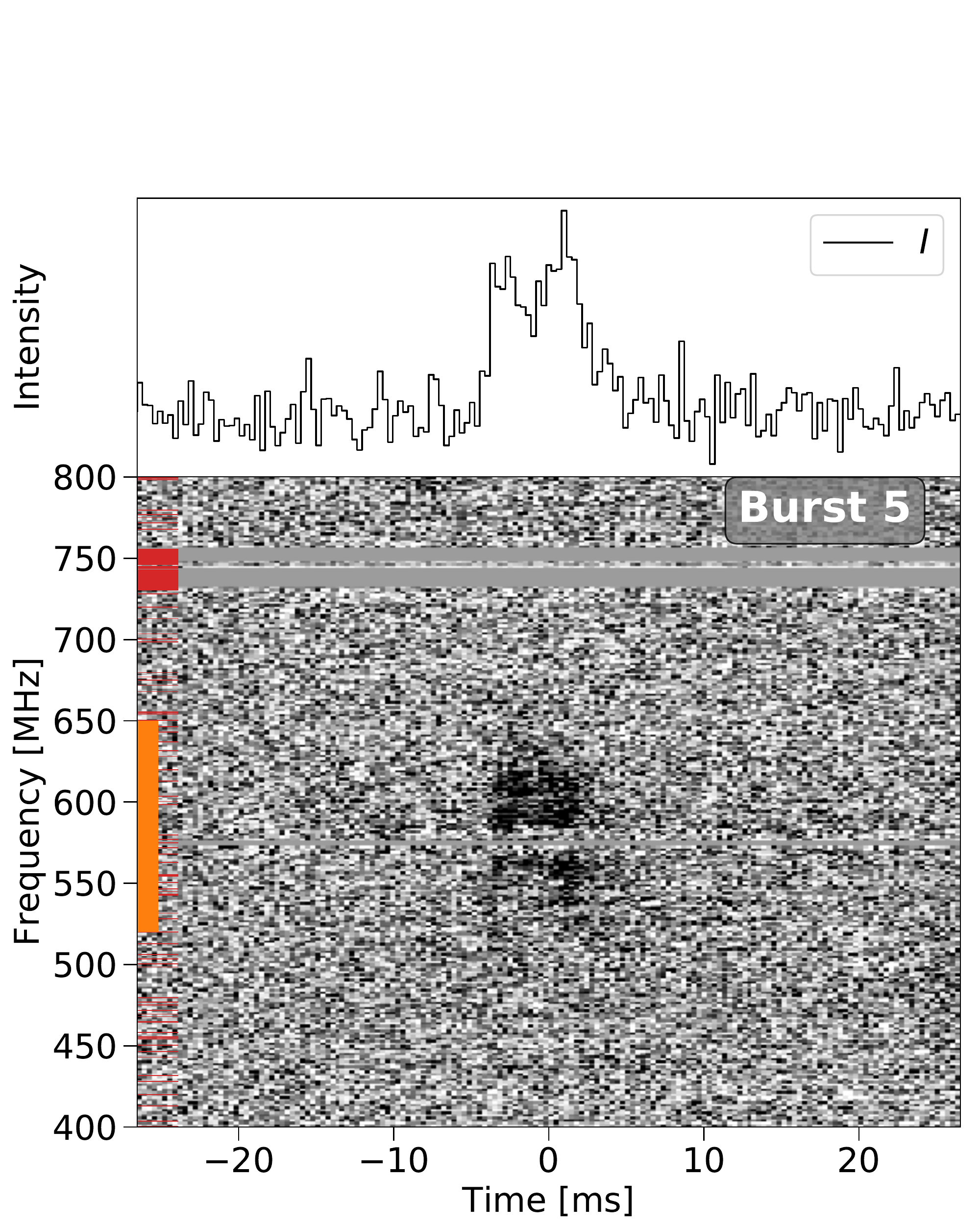}
\figsetgrpnote{}
\figsetgrpend

%%%%%%% R4 

\figsetgrpstart
\figsetgrpnum{1.6}
\figsetgrptitle{FRB 20181030A}
\figsetplot{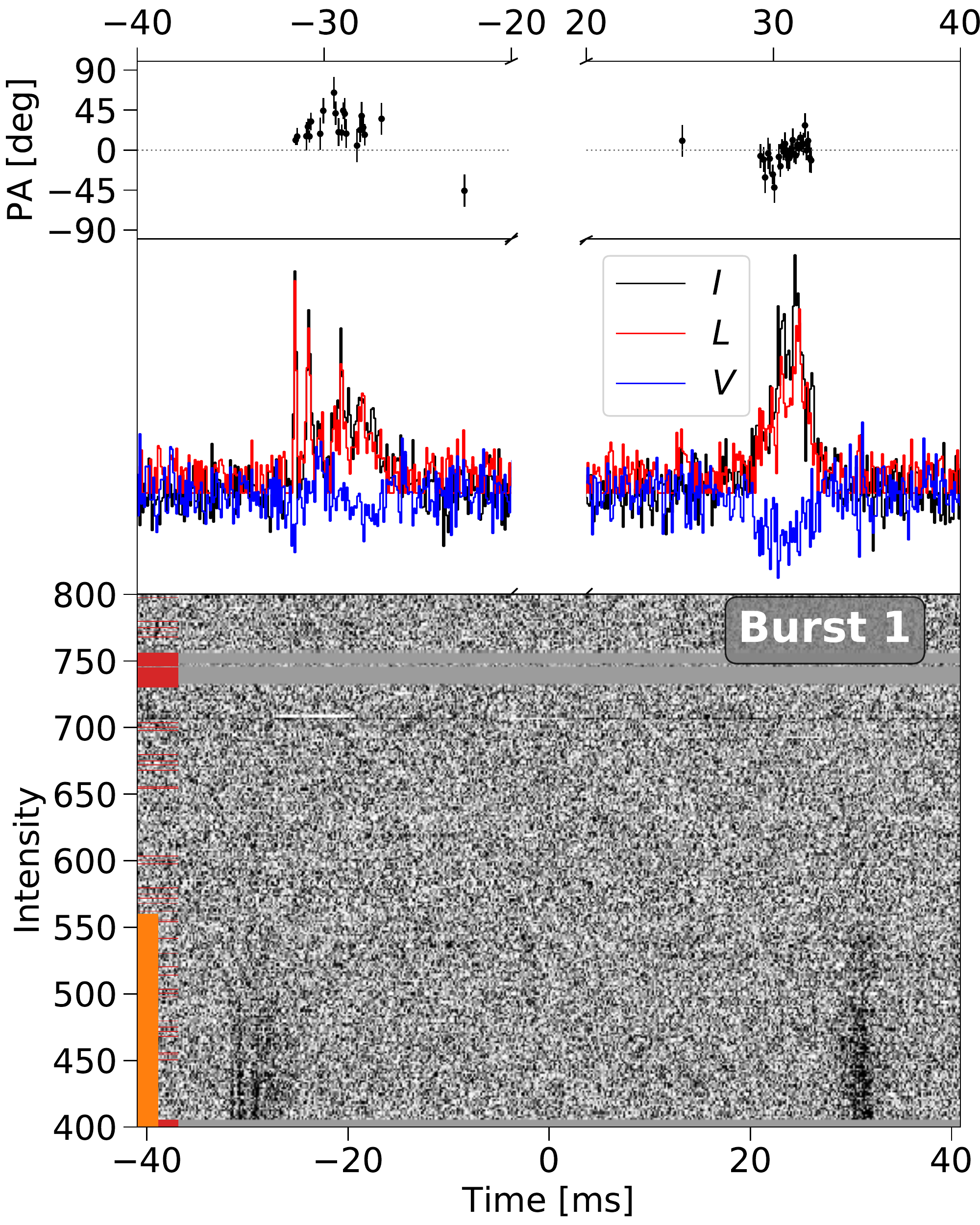}
\figsetgrpnote{}
\figsetgrpend

\figsetgrpstart
\figsetgrpnum{1.7}
\figsetgrptitle{FRB 20181030A}
\figsetplot{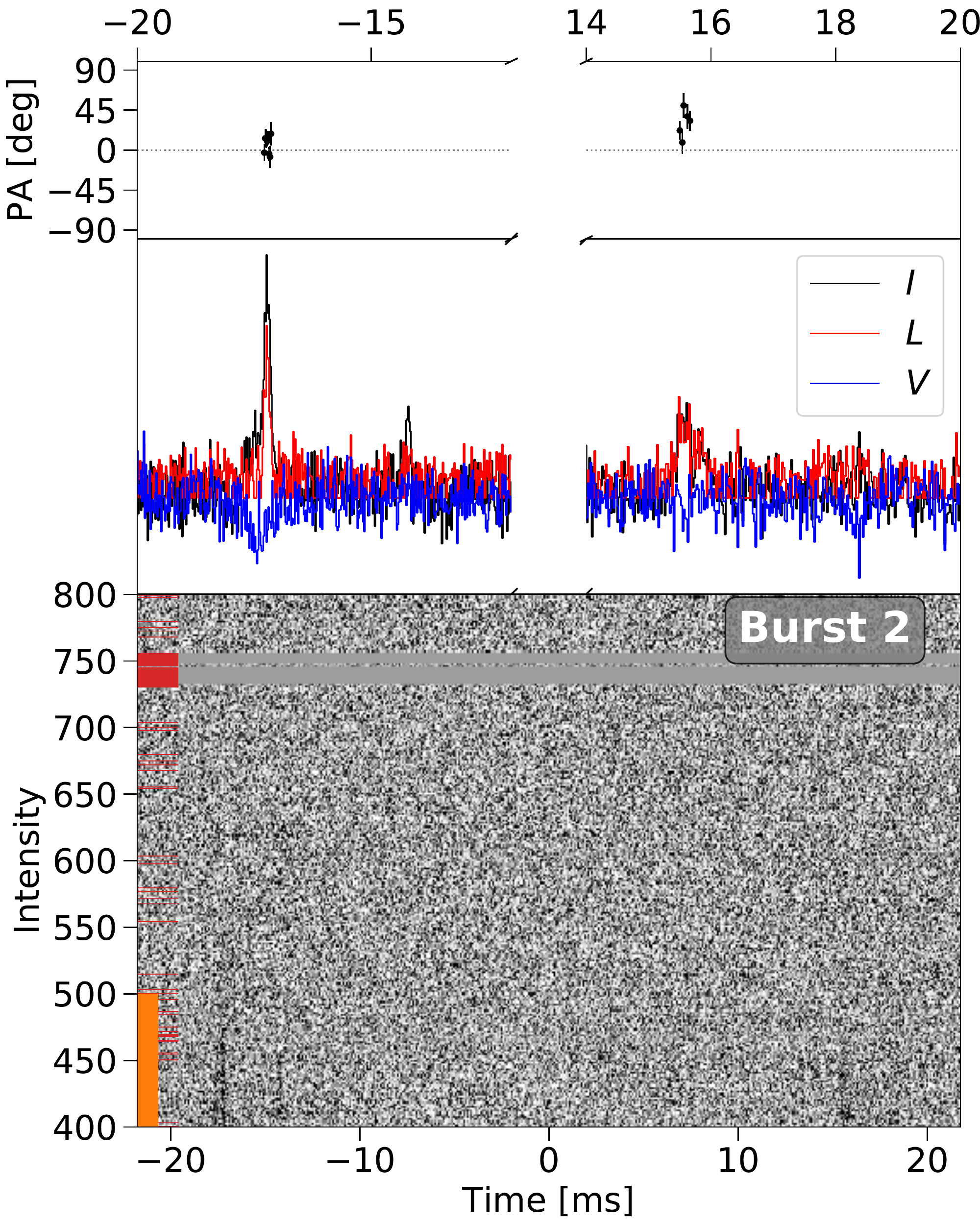}
\figsetgrpnote{}
\figsetgrpend

\figsetgrpstart
\figsetgrpnum{1.8}
\figsetgrptitle{FRB 20181030A}
\figsetplot{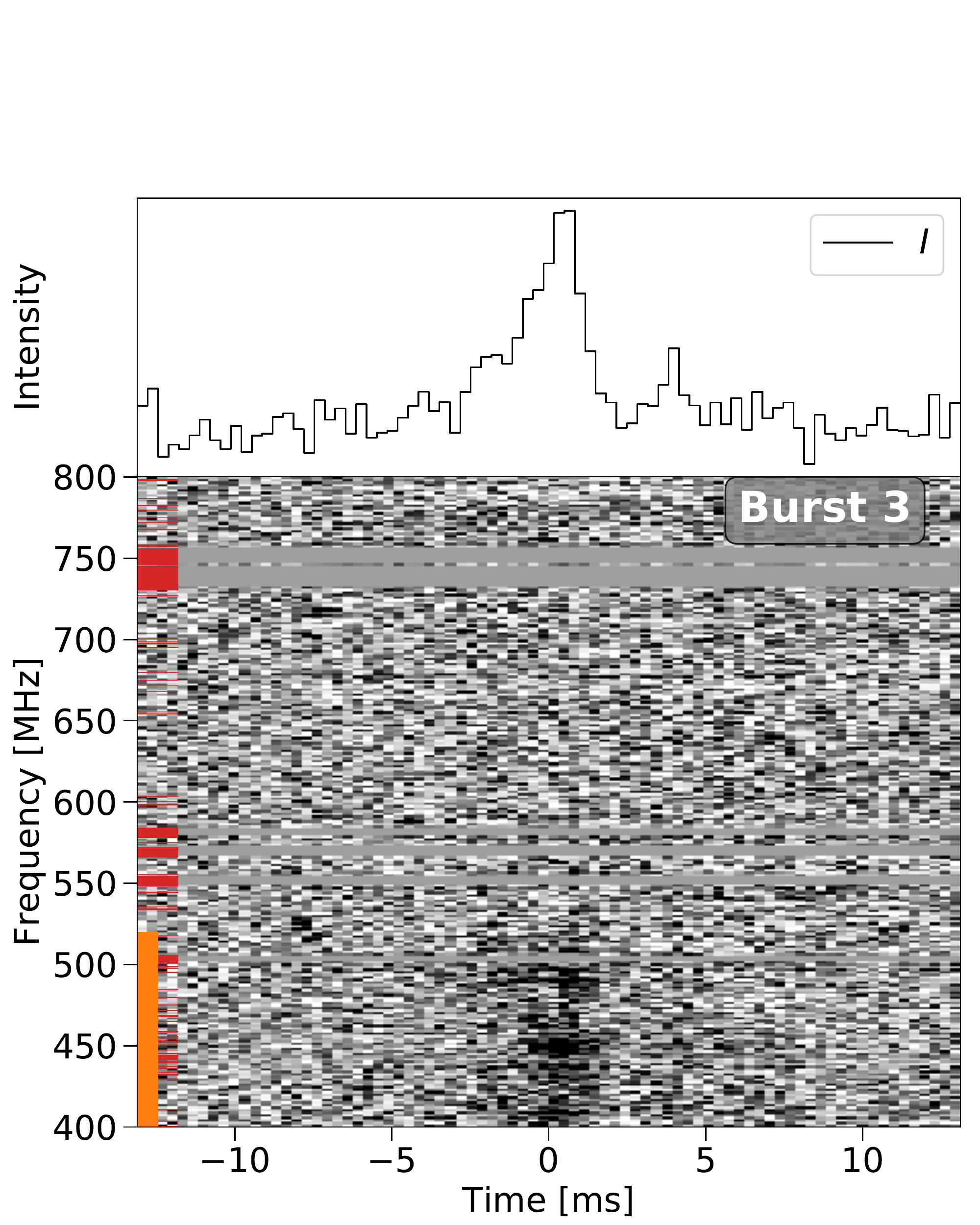} 
\figsetgrpnote{}
\figsetgrpend

\figsetgrpstart
\figsetgrpnum{1.9}
\figsetgrptitle{FRB 20181030A}
\figsetplot{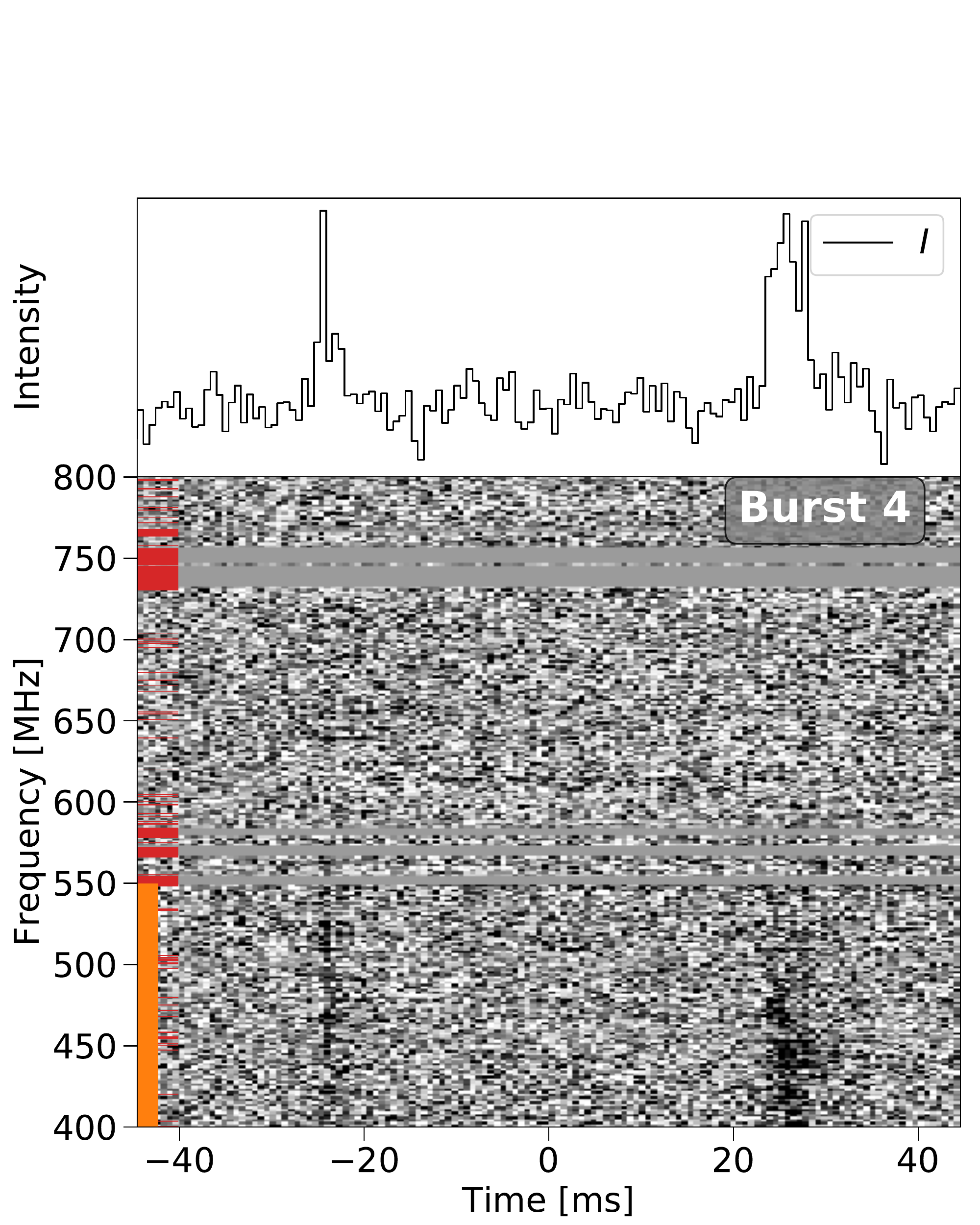}
\figsetgrpnote{}
\figsetgrpend

\figsetgrpstart
\figsetgrpnum{1.10}
\figsetgrptitle{FRB 20181030A}
\figsetplot{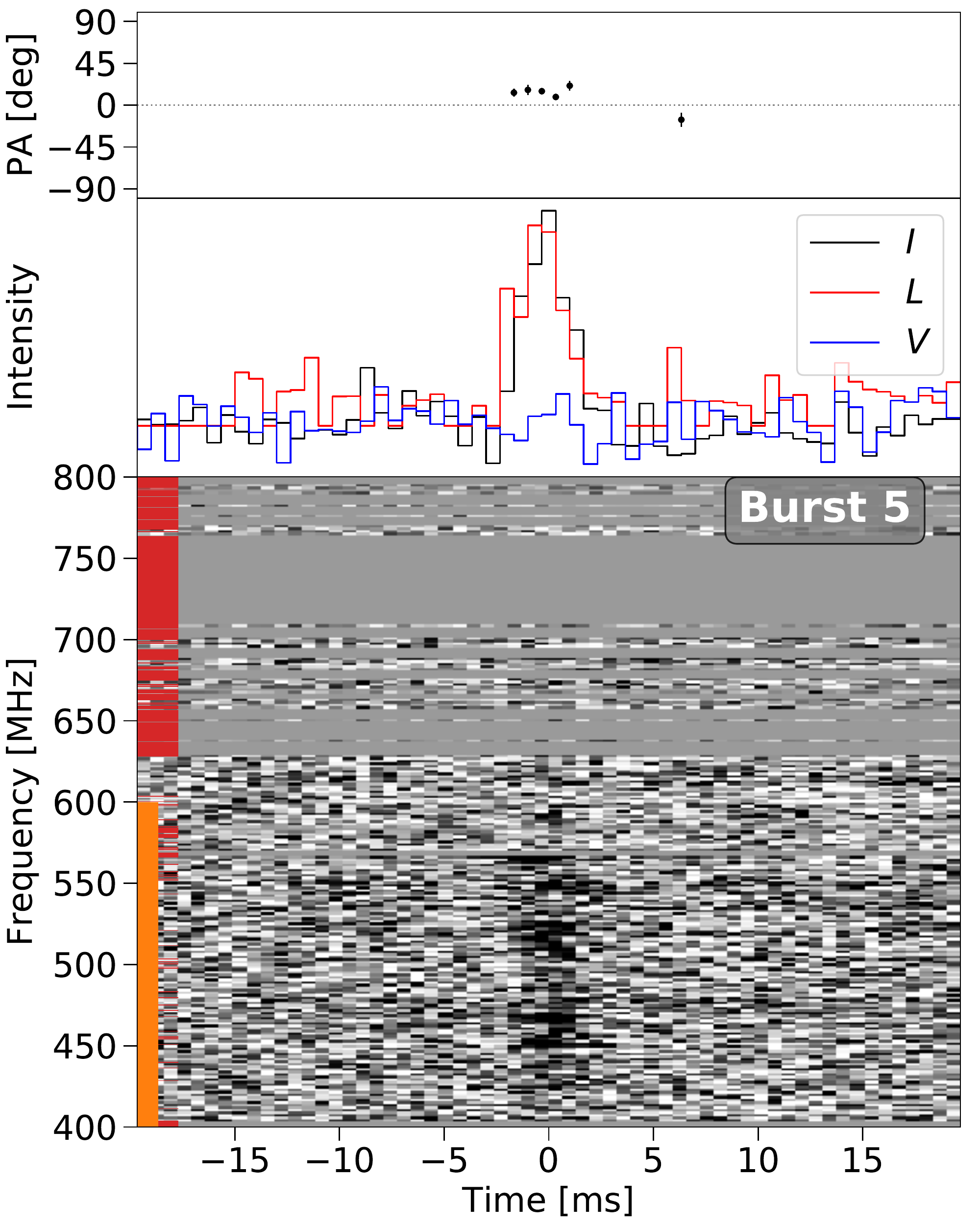}
\figsetgrpnote{}
\figsetgrpend

%%%%%%% R6 

\figsetgrpstart
\figsetgrpnum{1.10}
\figsetgrptitle{FRB 20181119A}
\figsetplot{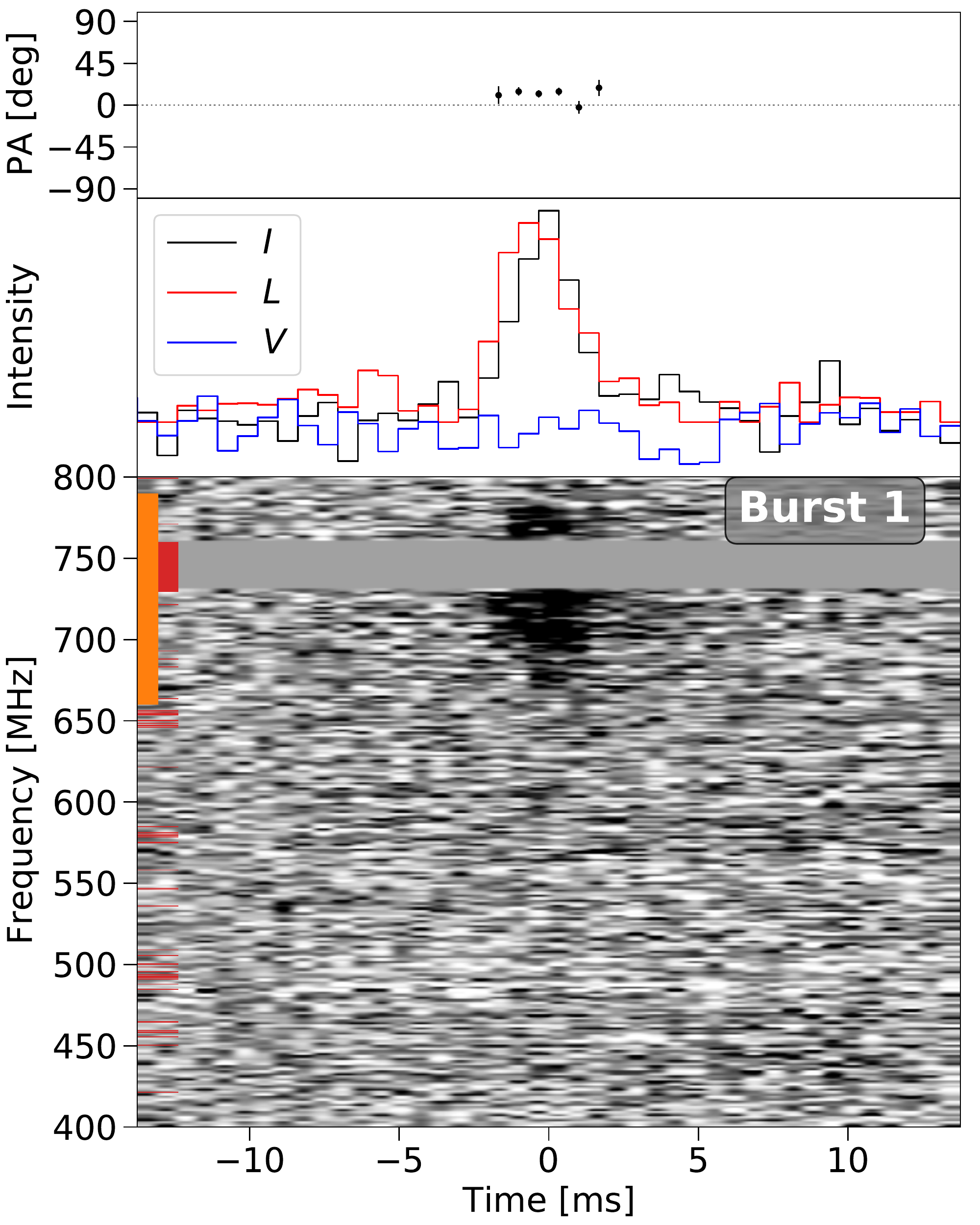}
\figsetgrpnote{}
\figsetgrpend

\figsetgrpstart
\figsetgrpnum{1.11}
\figsetgrptitle{FRB 20181119A}
\figsetplot{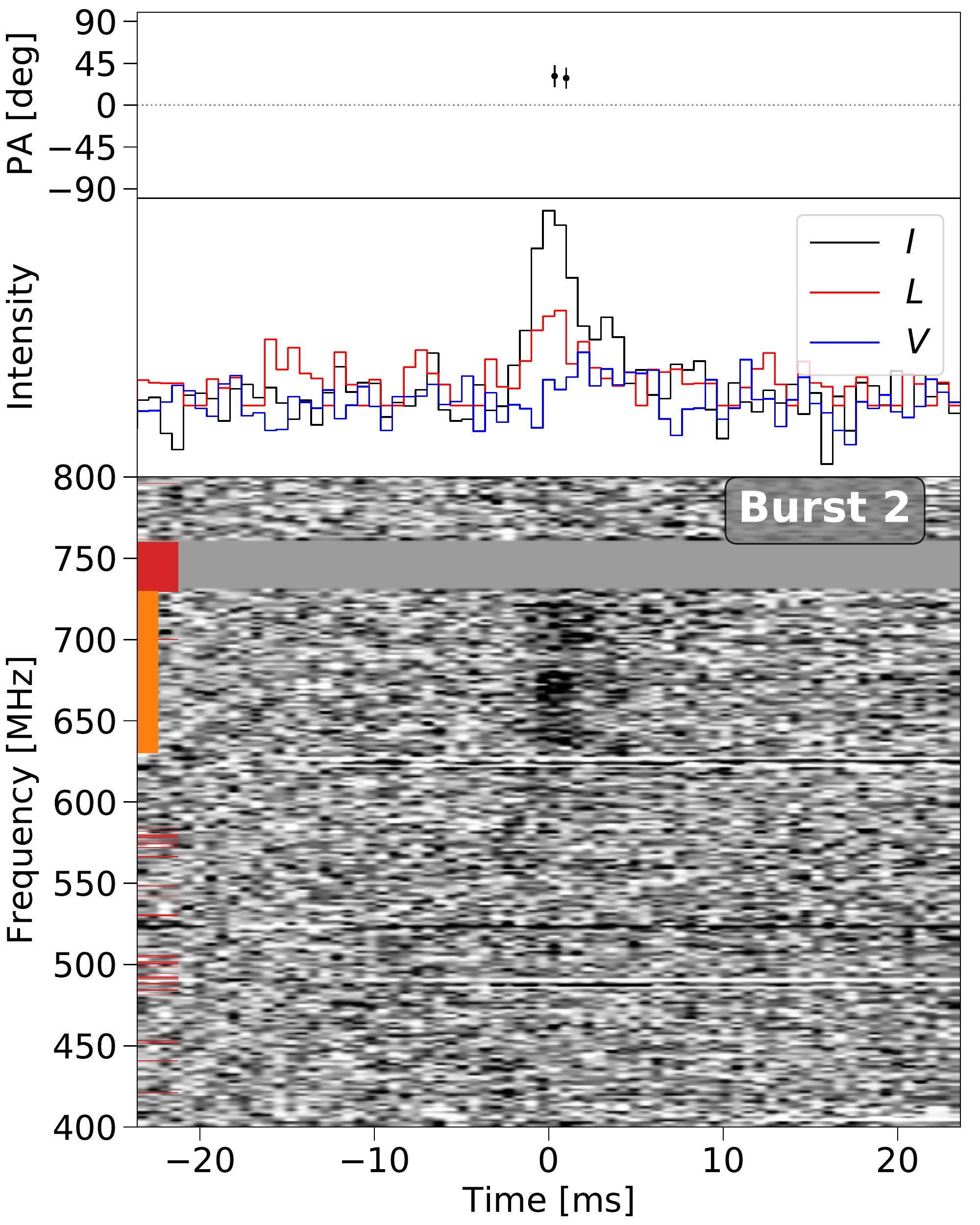}
\figsetgrpnote{}
\figsetgrpend

\figsetgrpstart
\figsetgrpnum{1.12}
\figsetgrptitle{FRB 20181119A}
\figsetplot{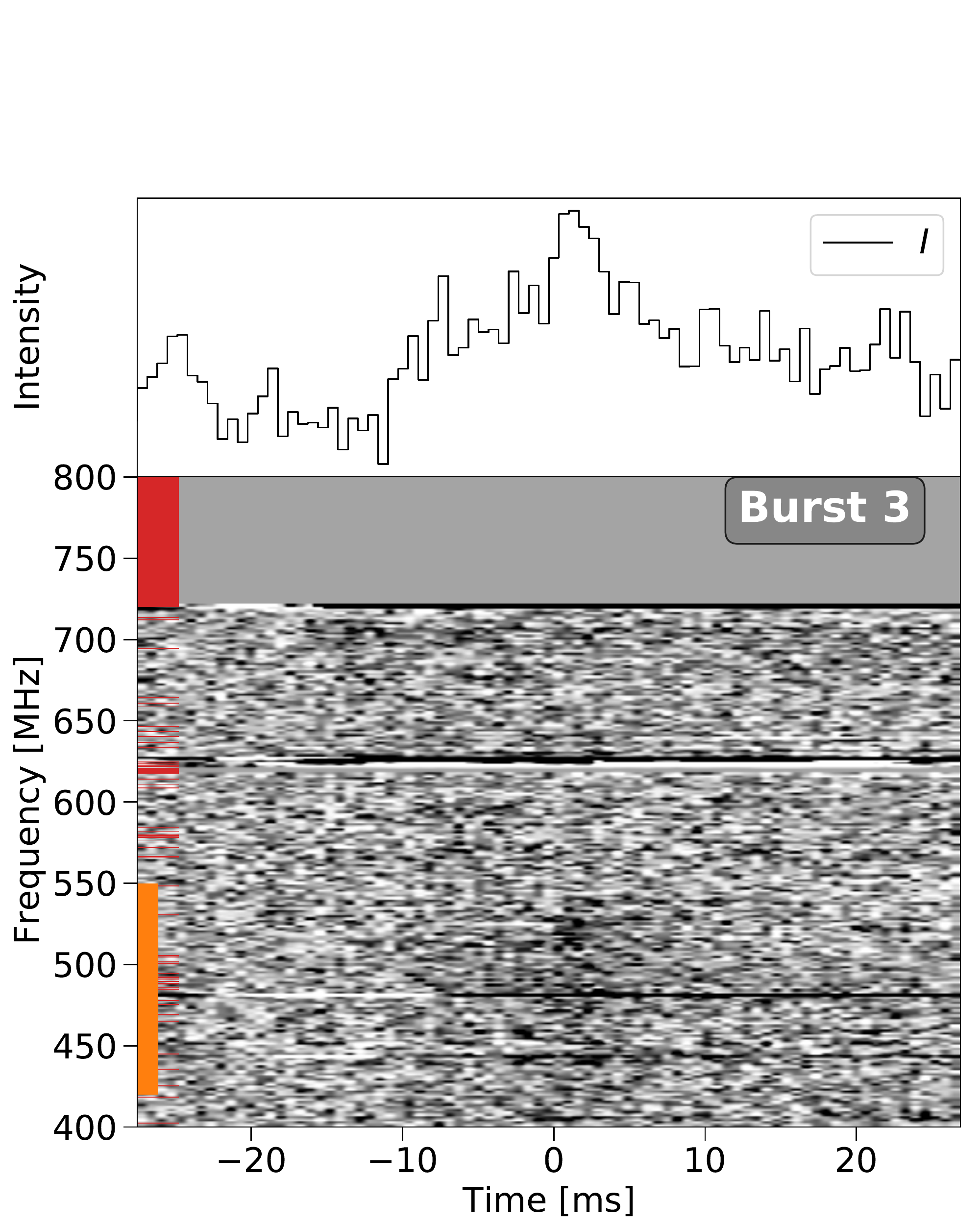} 
\figsetgrpnote{}
\figsetgrpend

\figsetgrpstart
\figsetgrpnum{1.13}
\figsetgrptitle{FRB 20181119A}
\figsetplot{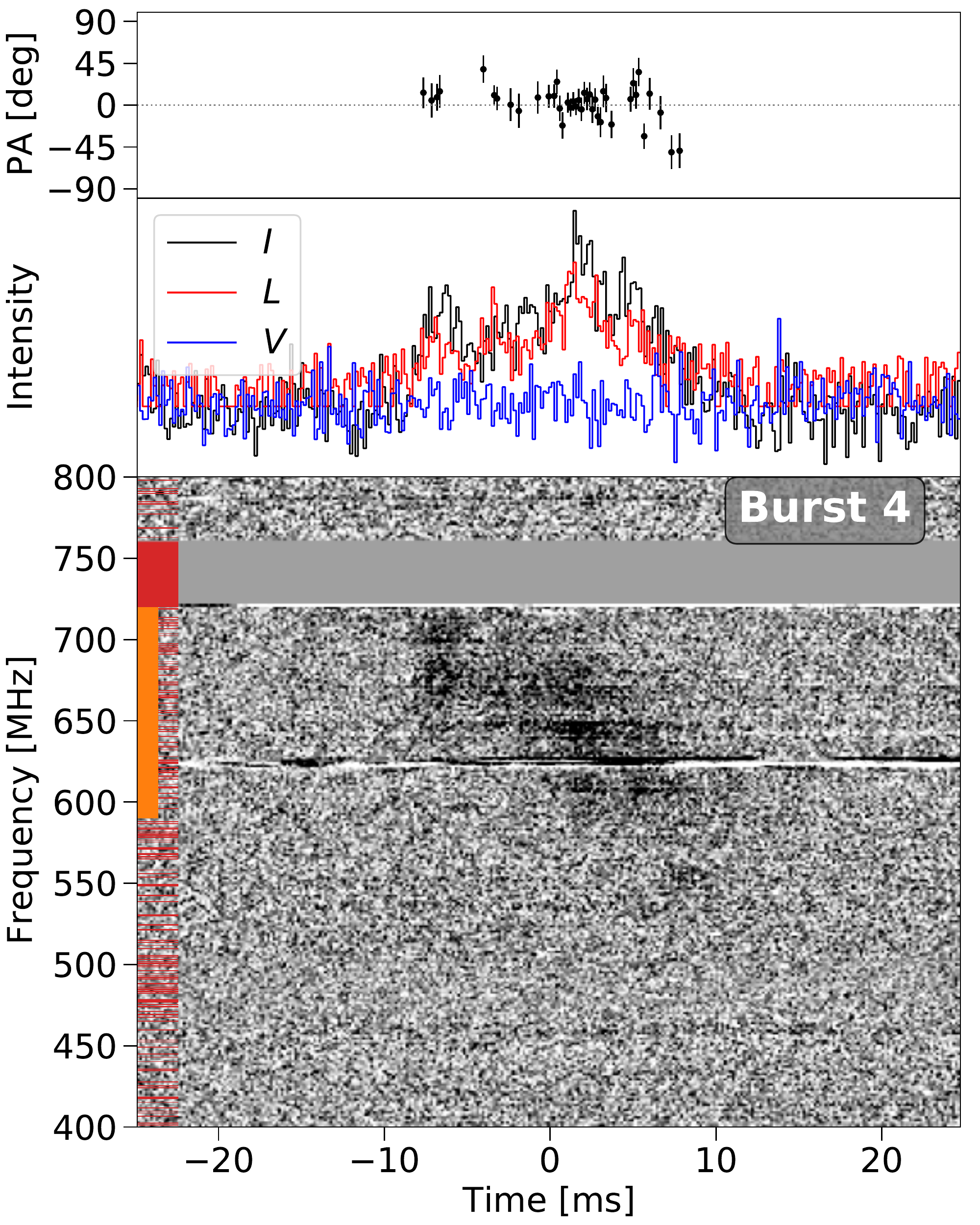}
\figsetgrpnote{}
\figsetgrpend

\figsetgrpstart
\figsetgrpnum{1.14}
\figsetgrptitle{FRB 20181119A}
\figsetplot{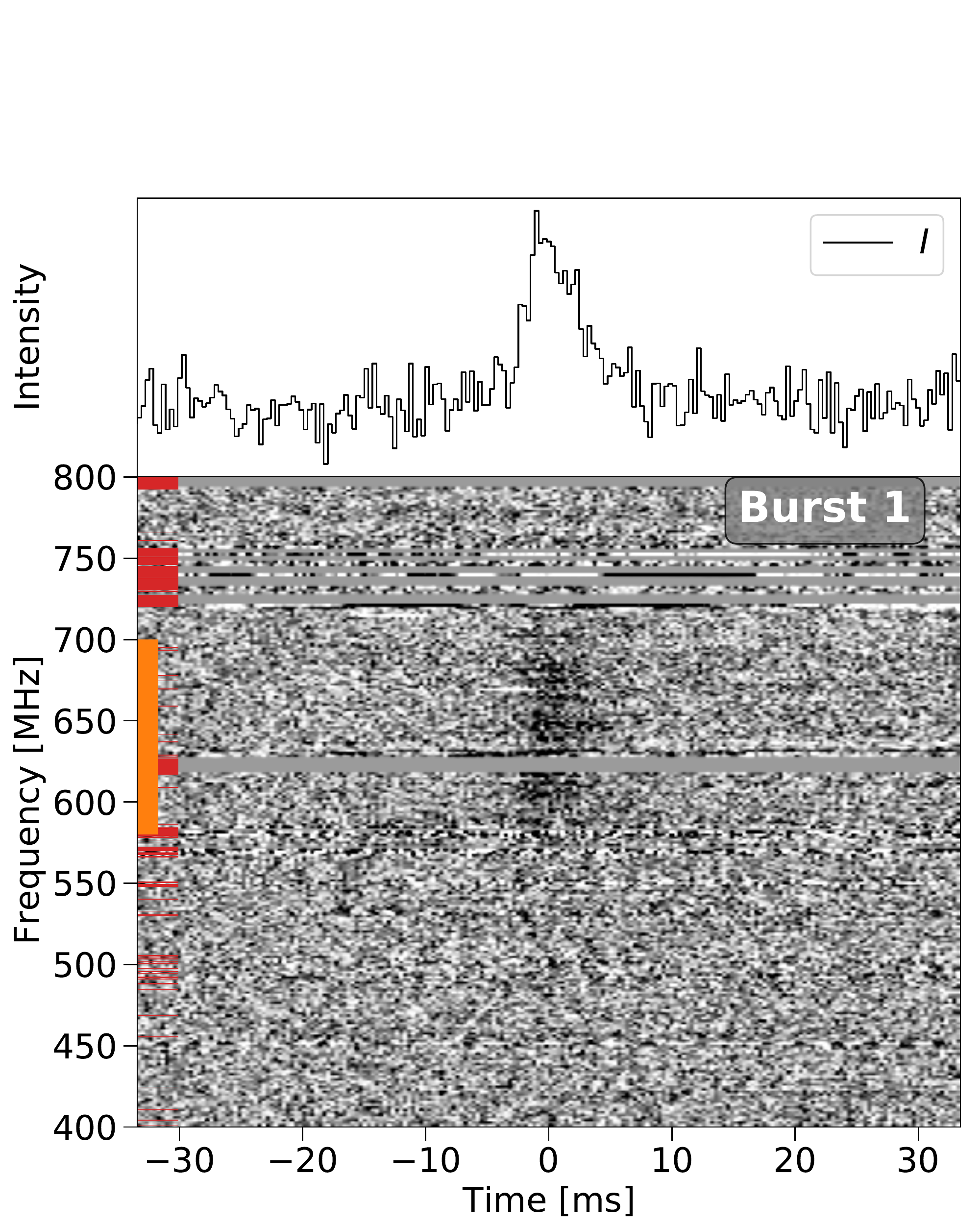}
\figsetgrpnote{}
\figsetgrpend

%%%%%%% R11

\figsetgrpstart
\figsetgrpnum{1.15}
\figsetgrptitle{FRB 20190222A}
\figsetplot{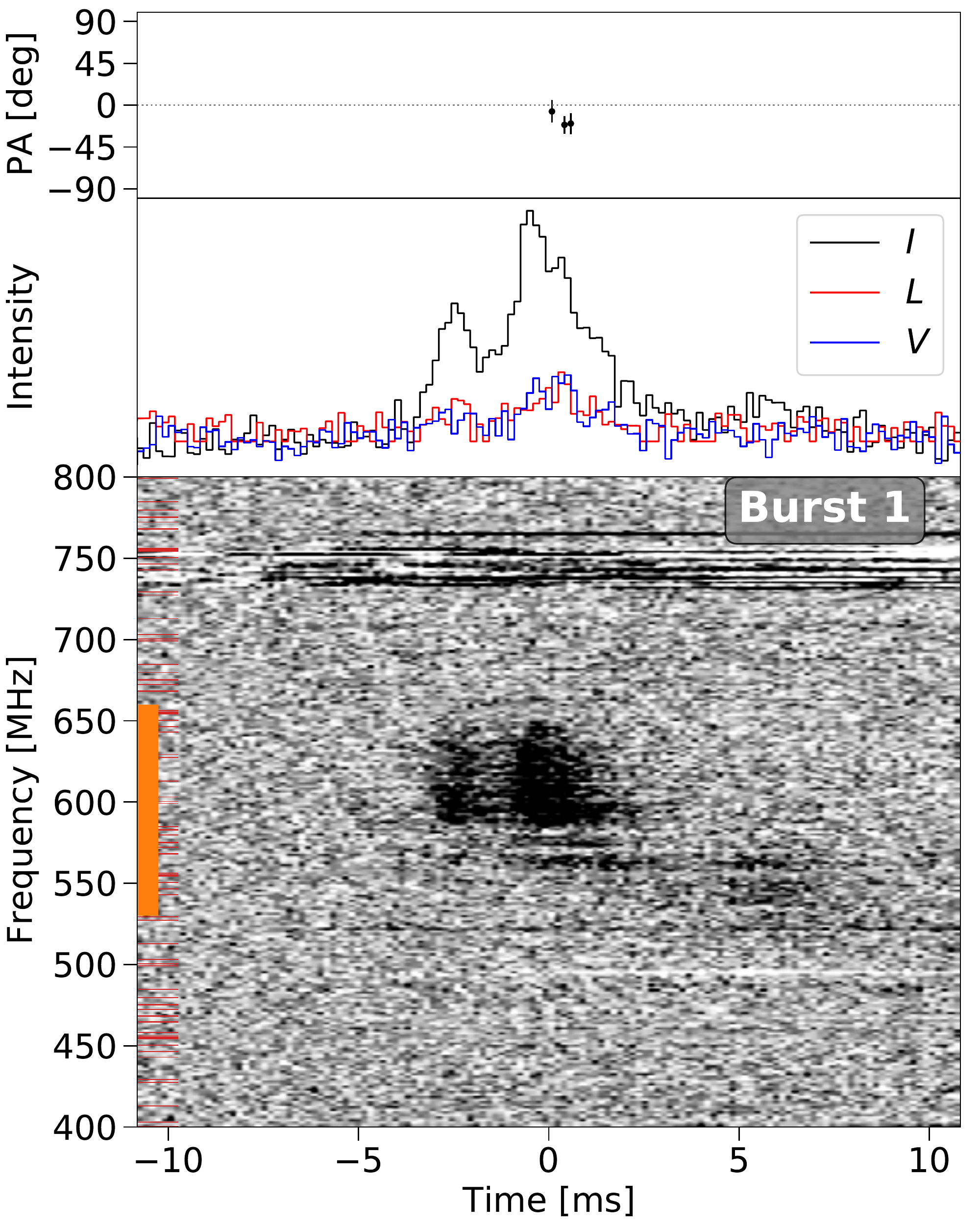}
\figsetgrpnote{}
\figsetgrpend

%%%%%%% R12

\figsetgrpstart
\figsetgrpnum{1.16}
\figsetgrptitle{FRB 20190208A}
\figsetplot{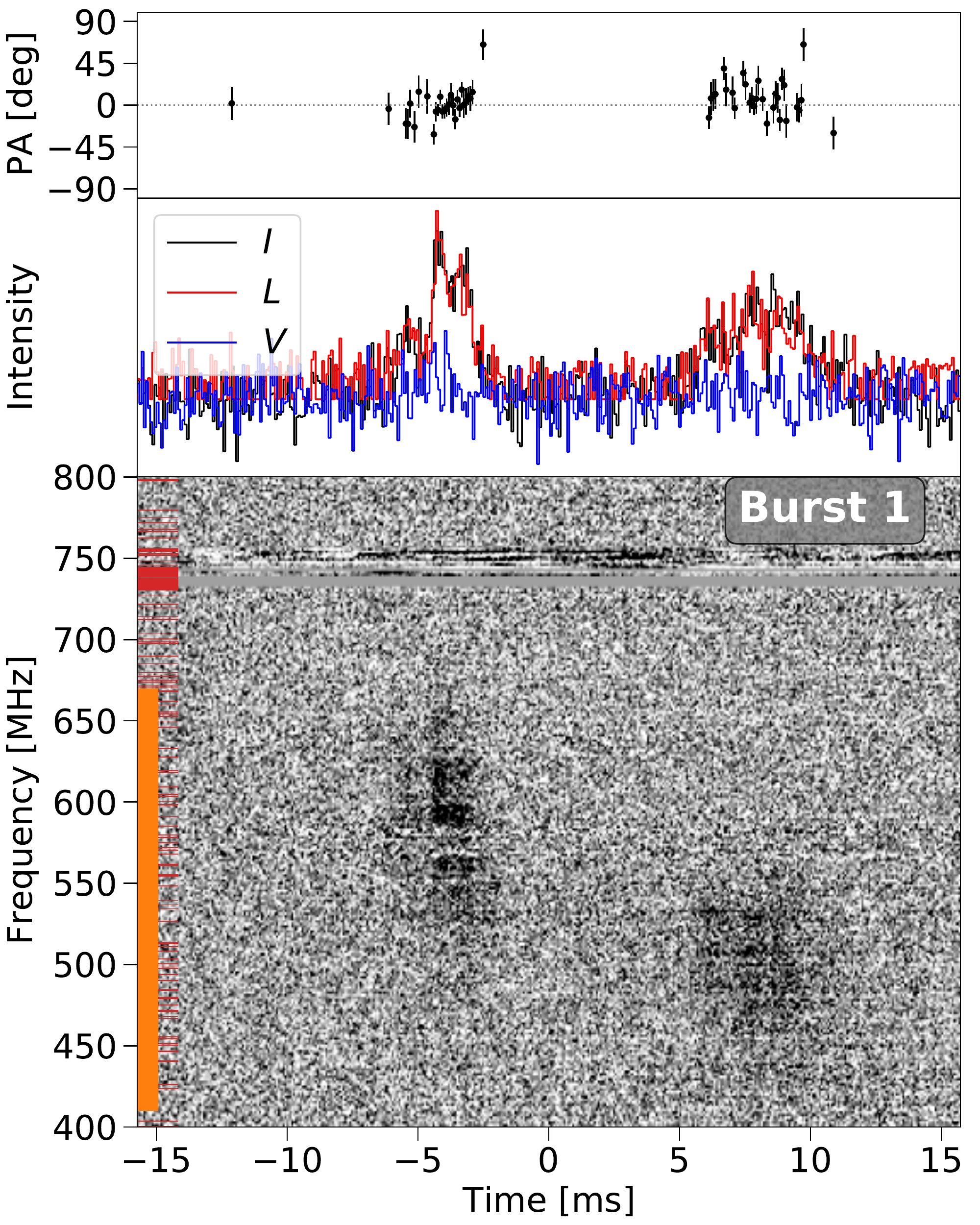}
\figsetgrpnote{}
\figsetgrpend

\figsetgrpstart
\figsetgrpnum{1.17}
\figsetgrptitle{FRB 20190208A}
\figsetplot{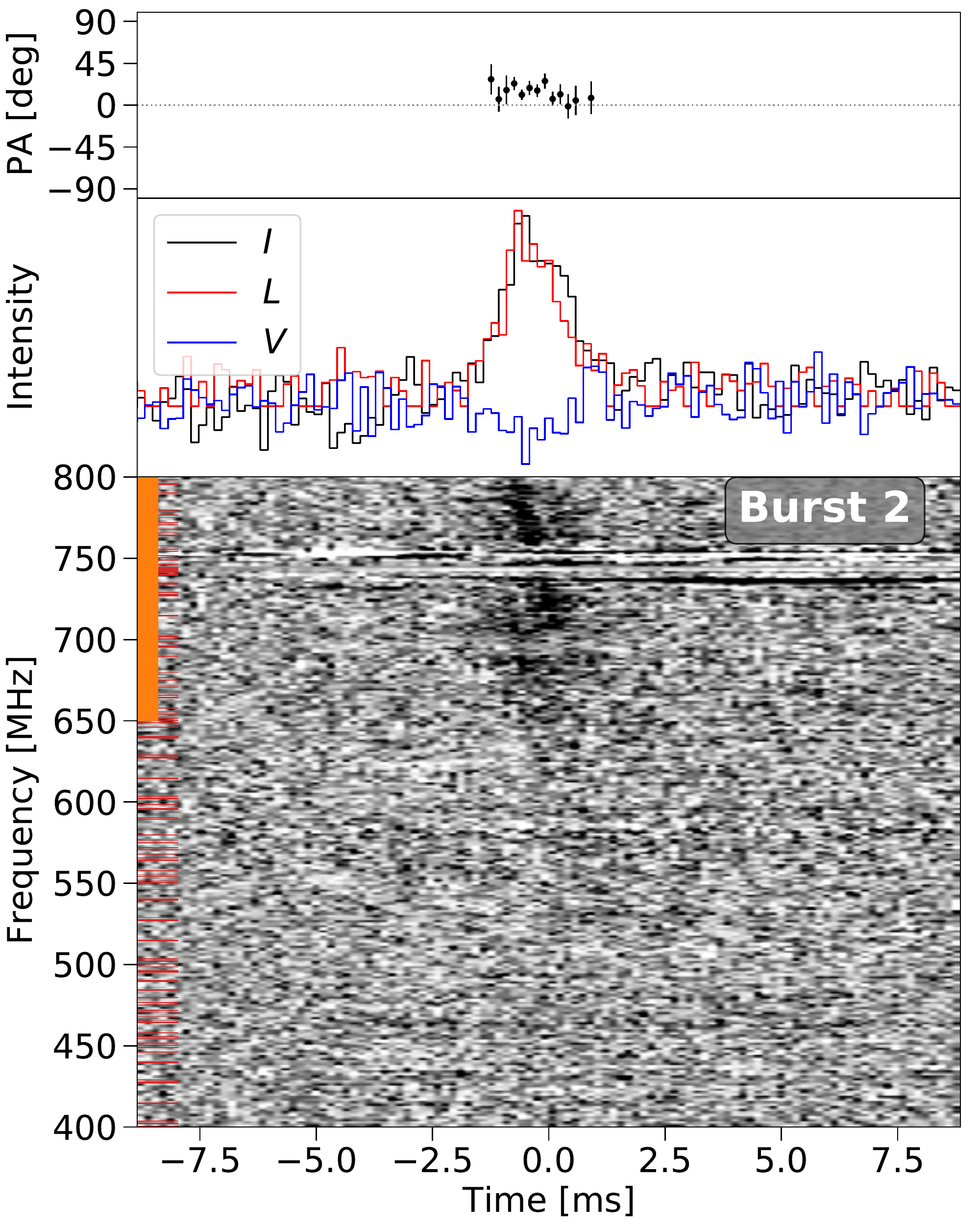}
\figsetgrpnote{}
\figsetgrpend

\figsetgrpstart
\figsetgrpnum{1.18}
\figsetgrptitle{FRB 20190208A}
\figsetplot{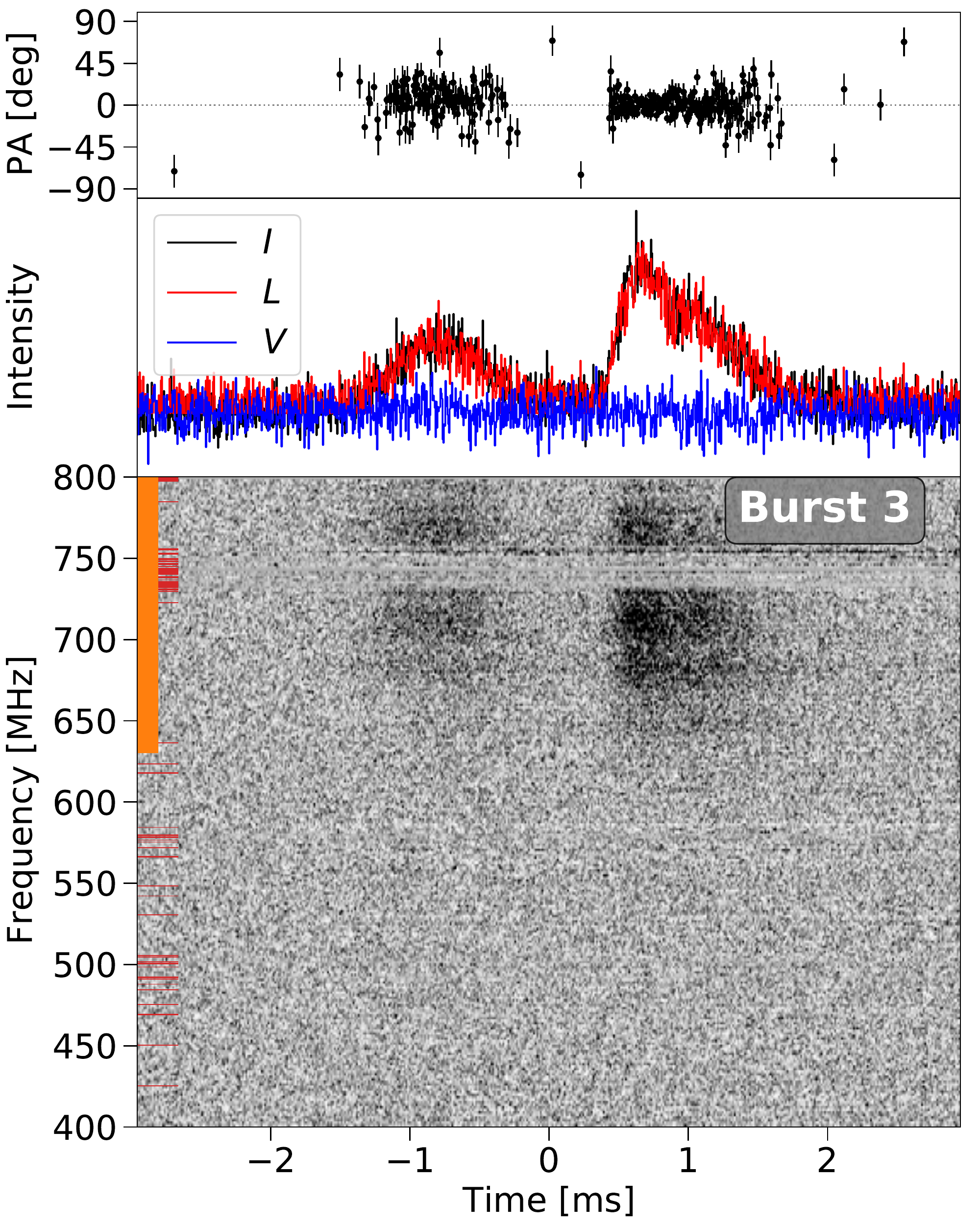} 
\figsetgrpnote{}
\figsetgrpend

\figsetgrpstart
\figsetgrpnum{1.19}
\figsetgrptitle{FRB 20190208A}
\figsetplot{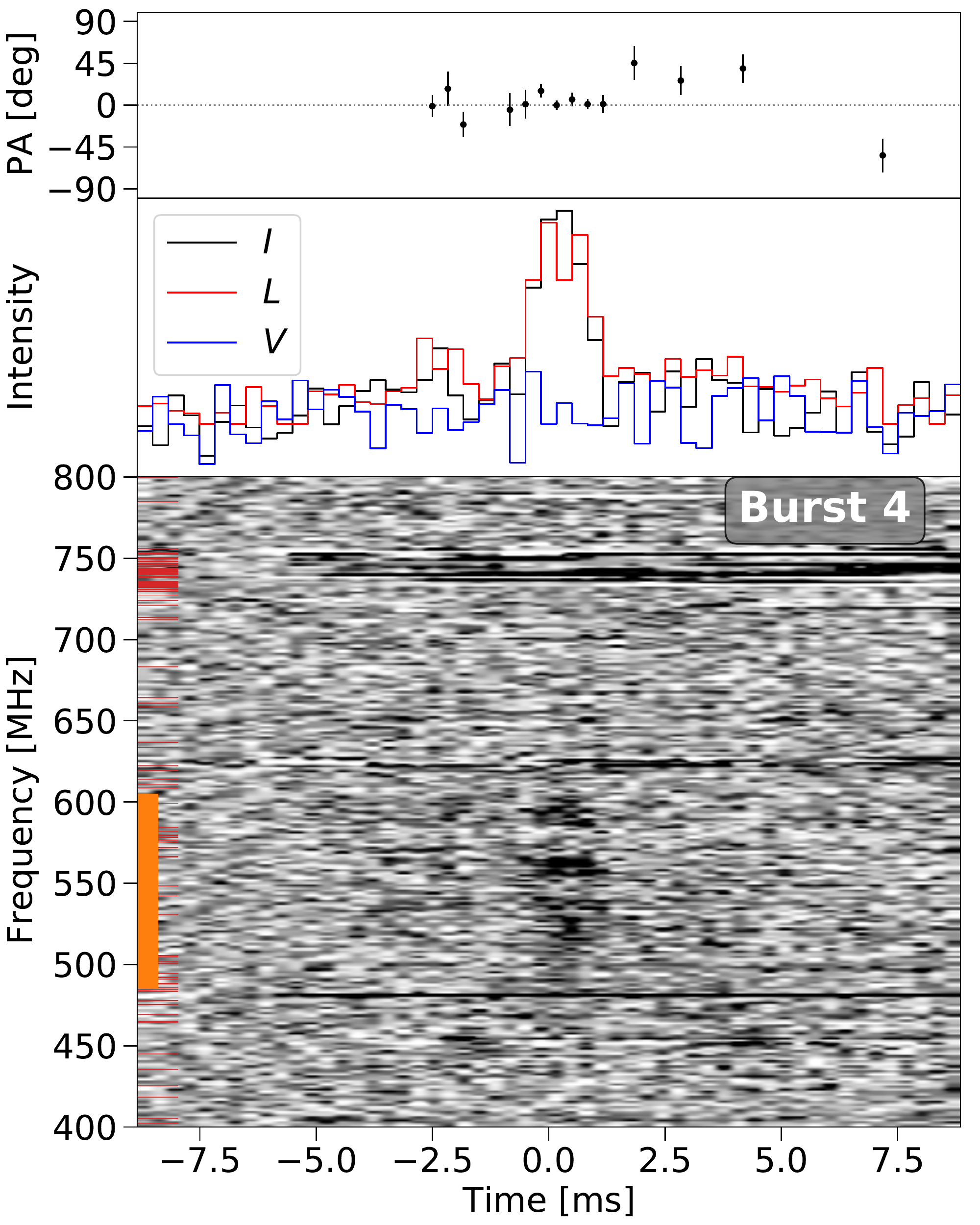}
\figsetgrpnote{}
\figsetgrpend

\figsetgrpstart
\figsetgrpnum{1.20}
\figsetgrptitle{FRB 20190208A}
\figsetplot{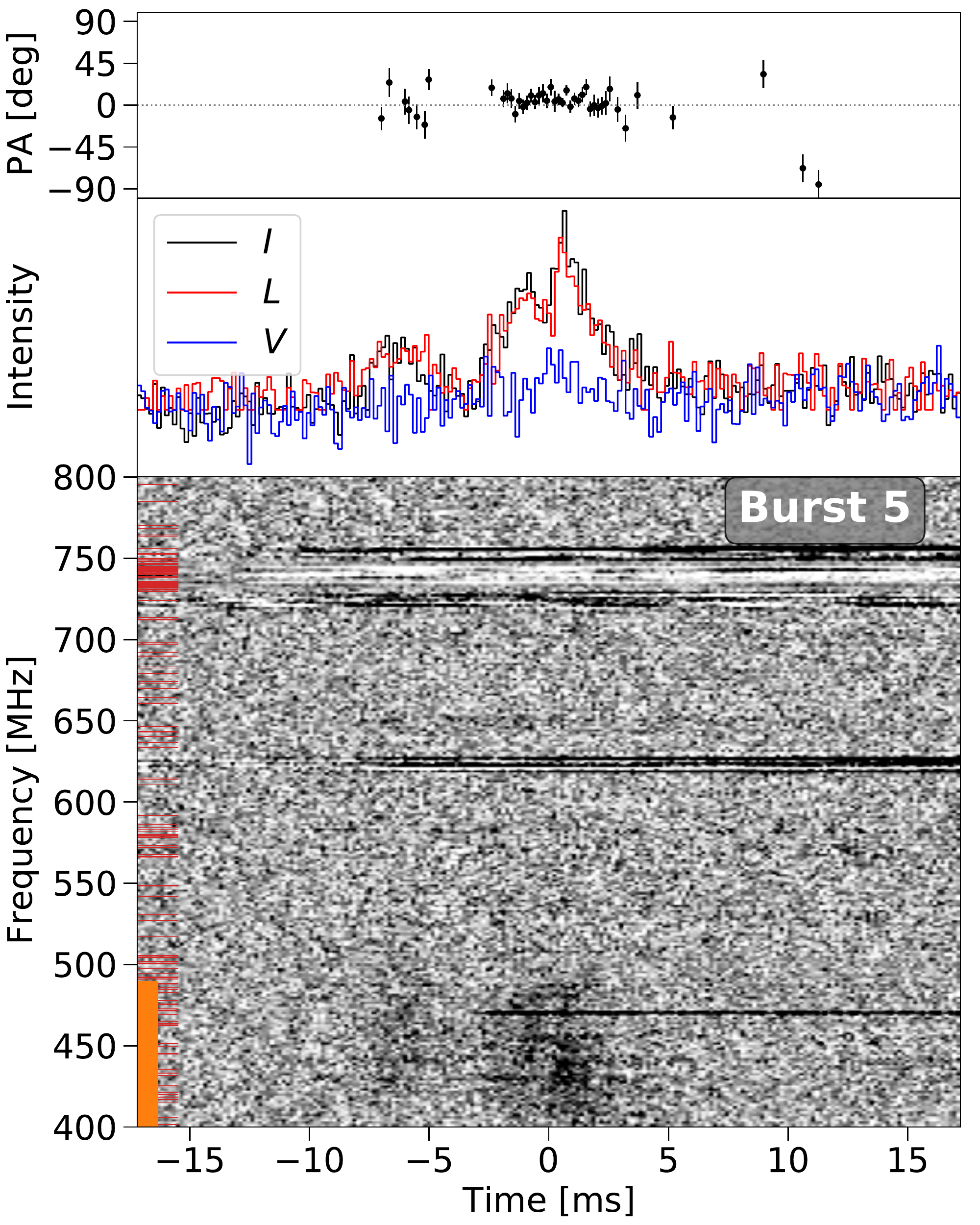}
\figsetgrpnote{}
\figsetgrpend

\figsetgrpstart
\figsetgrpnum{1.21}
\figsetgrptitle{FRB 20190208A}
\figsetplot{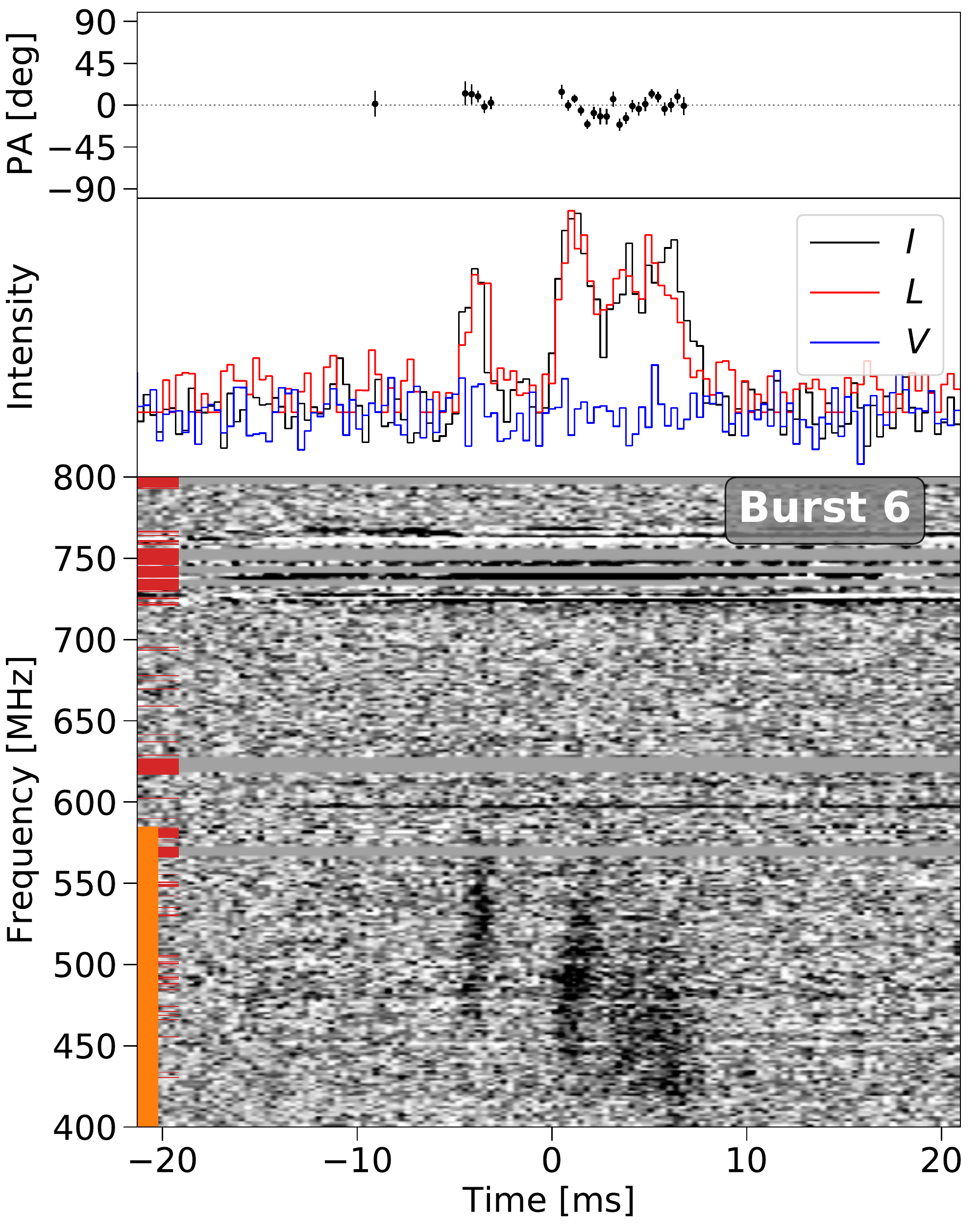}
\figsetgrpnote{}
\figsetgrpend

\figsetgrpstart
\figsetgrpnum{1.22}
\figsetgrptitle{FRB 20190208A}
\figsetplot{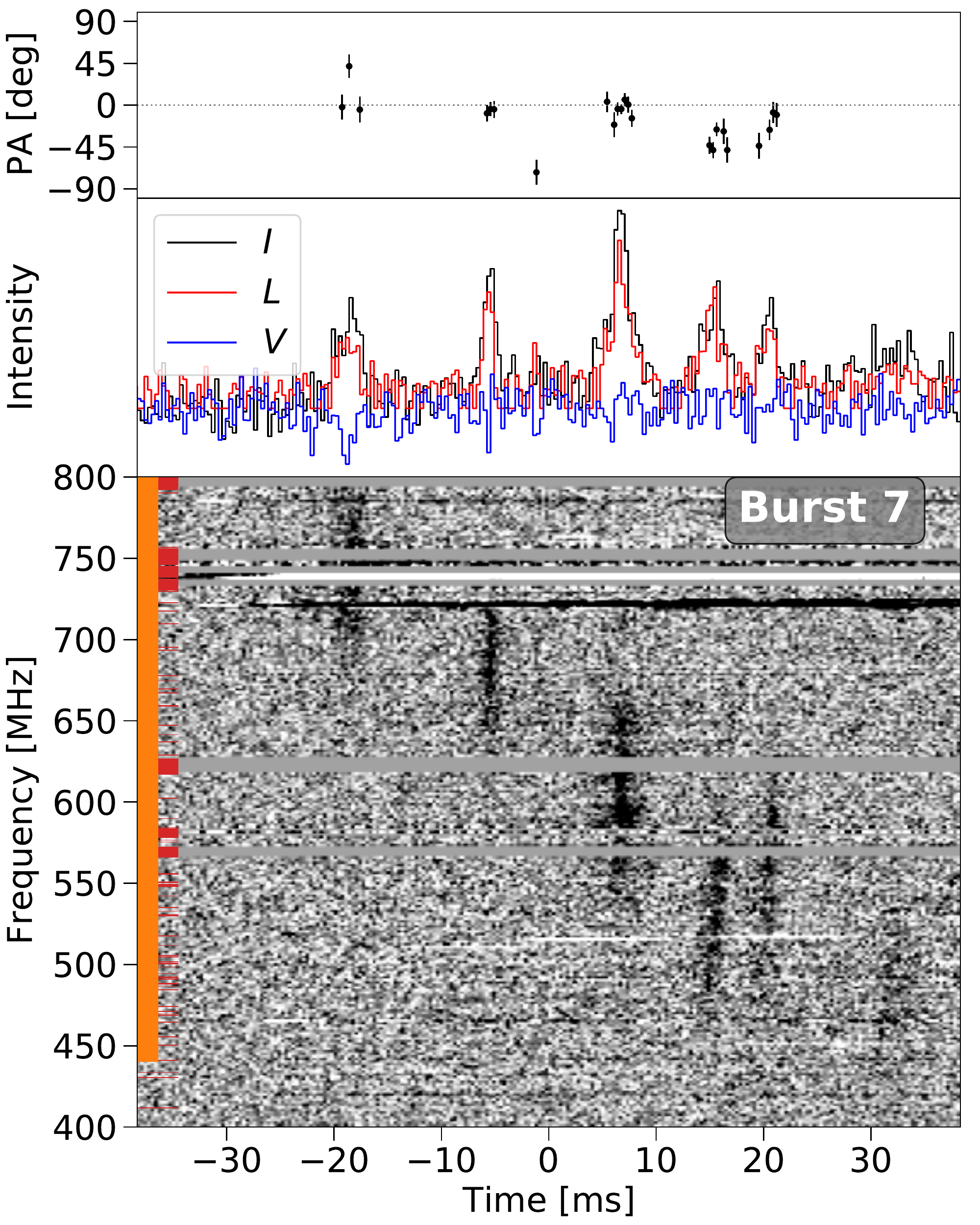}
\figsetgrpnote{}
\figsetgrpend

%%%%%%% R13

\figsetgrpstart
\figsetgrpnum{1.23}
\figsetgrptitle{FRB 20190604A}
\figsetplot{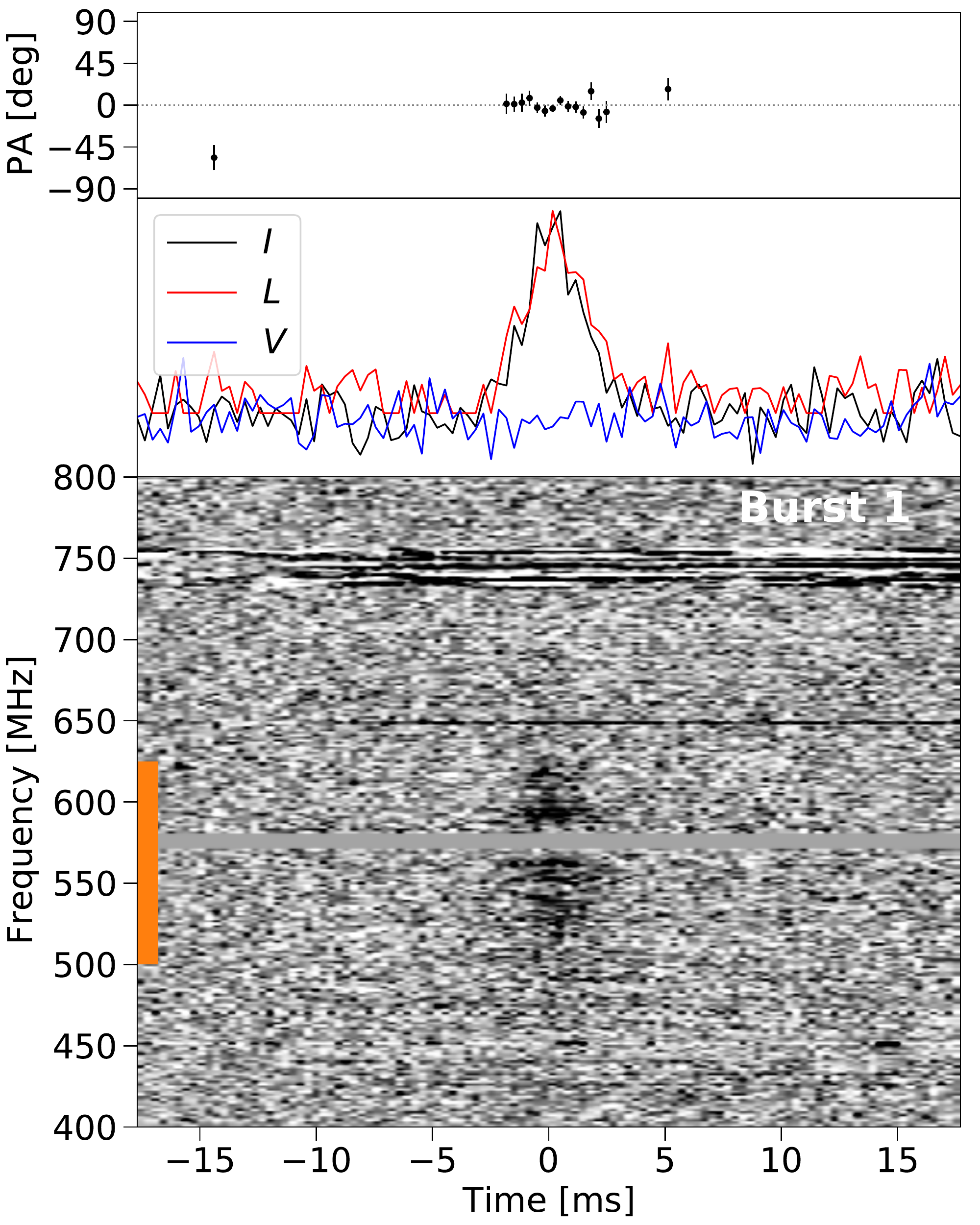}
\figsetgrpnote{}
\figsetgrpend

%%%%%%% R14

\figsetgrpstart
\figsetgrpnum{1.24}
\figsetgrptitle{FRB 20190213B}
\figsetplot{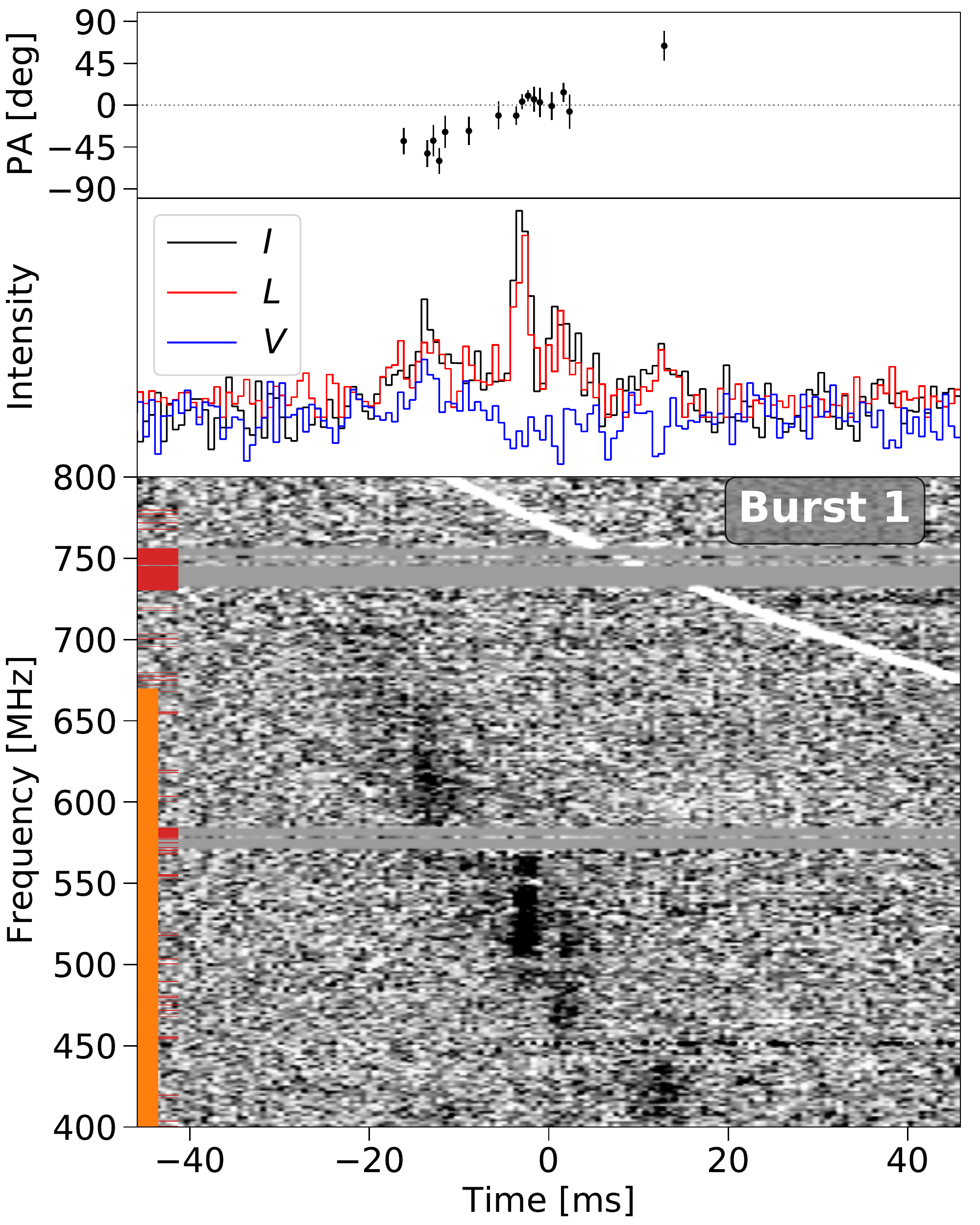}
\figsetgrpnote{}
\figsetgrpend

\figsetgrpstart
\figsetgrpnum{1.25}
\figsetgrptitle{FRB 20190213B}
\figsetplot{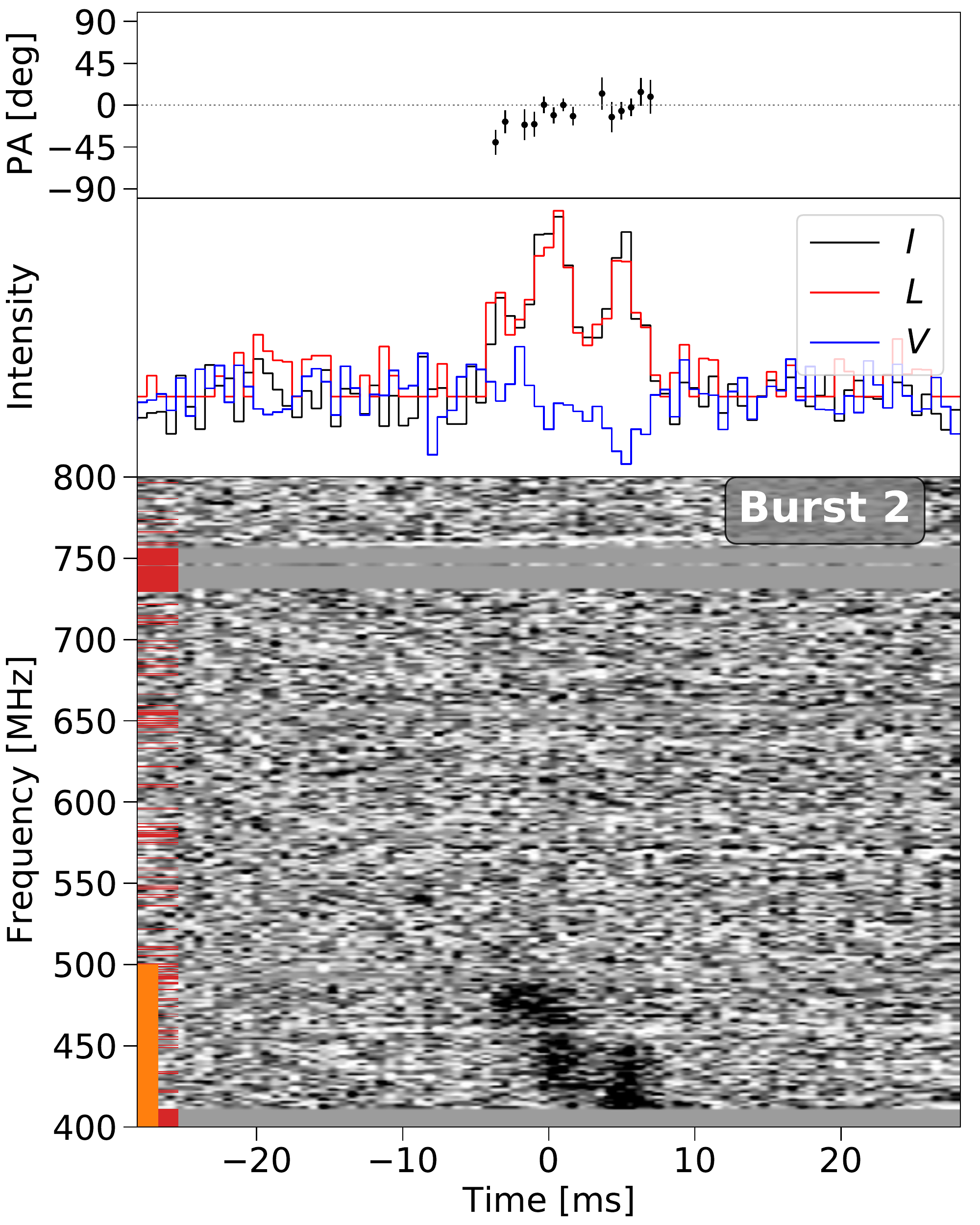}
\figsetgrpnote{}
\figsetgrpend

\figsetgrpstart
\figsetgrpnum{1.26}
\figsetgrptitle{FRB 20190213B}
\figsetplot{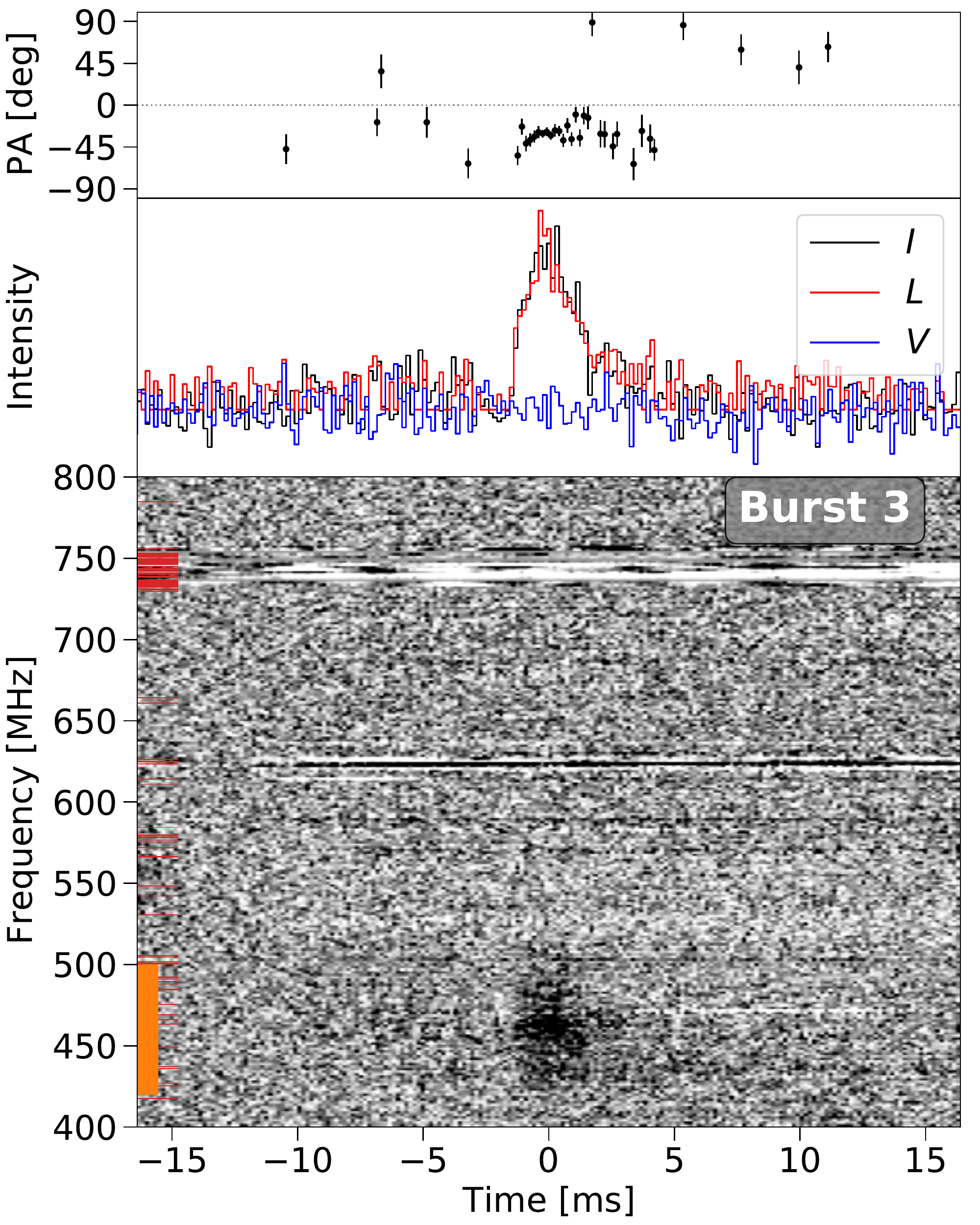} 
\figsetgrpnote{}
\figsetgrpend

\figsetgrpstart
\figsetgrpnum{1.27}
\figsetgrptitle{FRB 20190213B}
\figsetplot{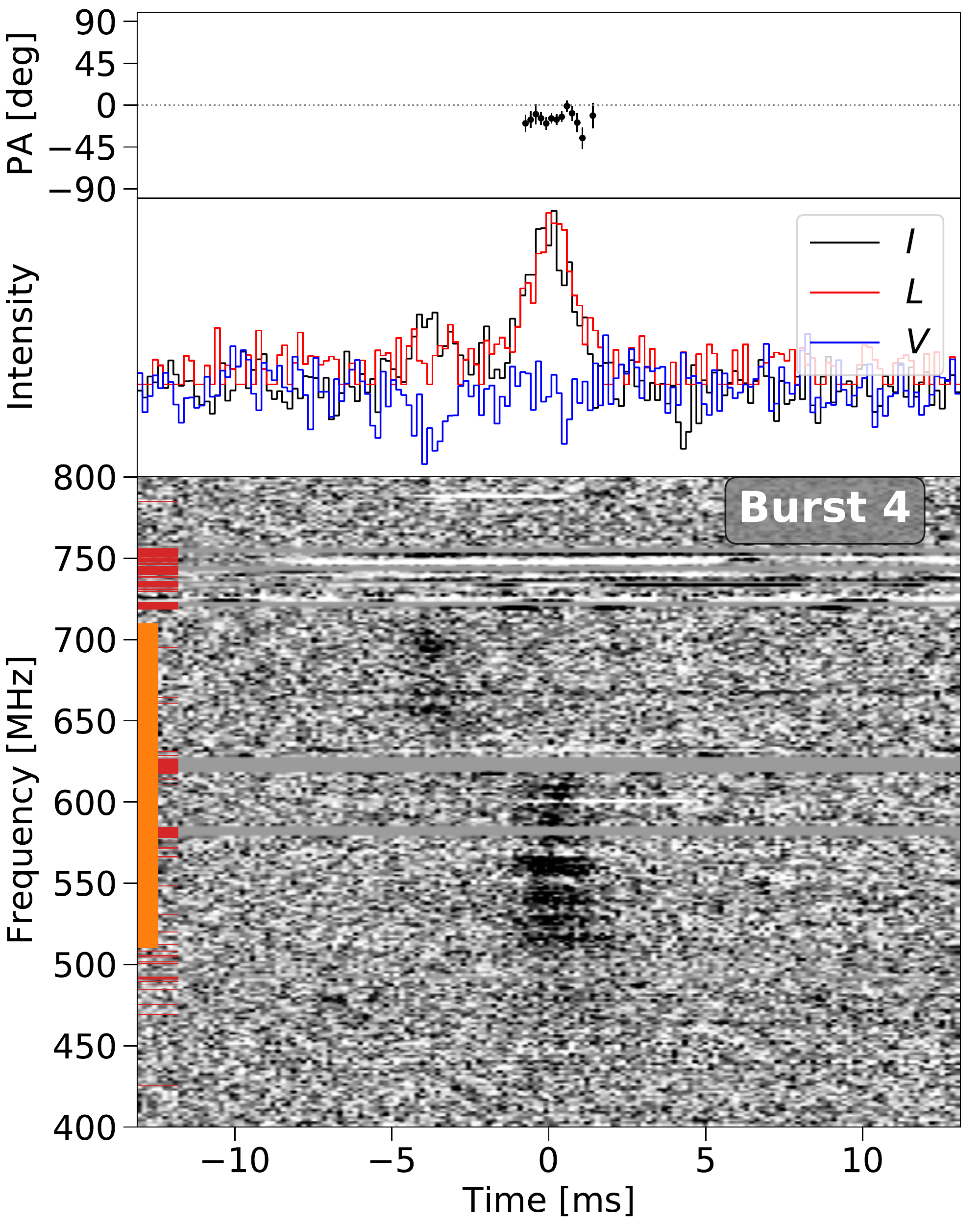}
\figsetgrpnote{}
\figsetgrpend

\figsetgrpstart
\figsetgrpnum{1.28}
\figsetgrptitle{FRB 20190213B}
\figsetplot{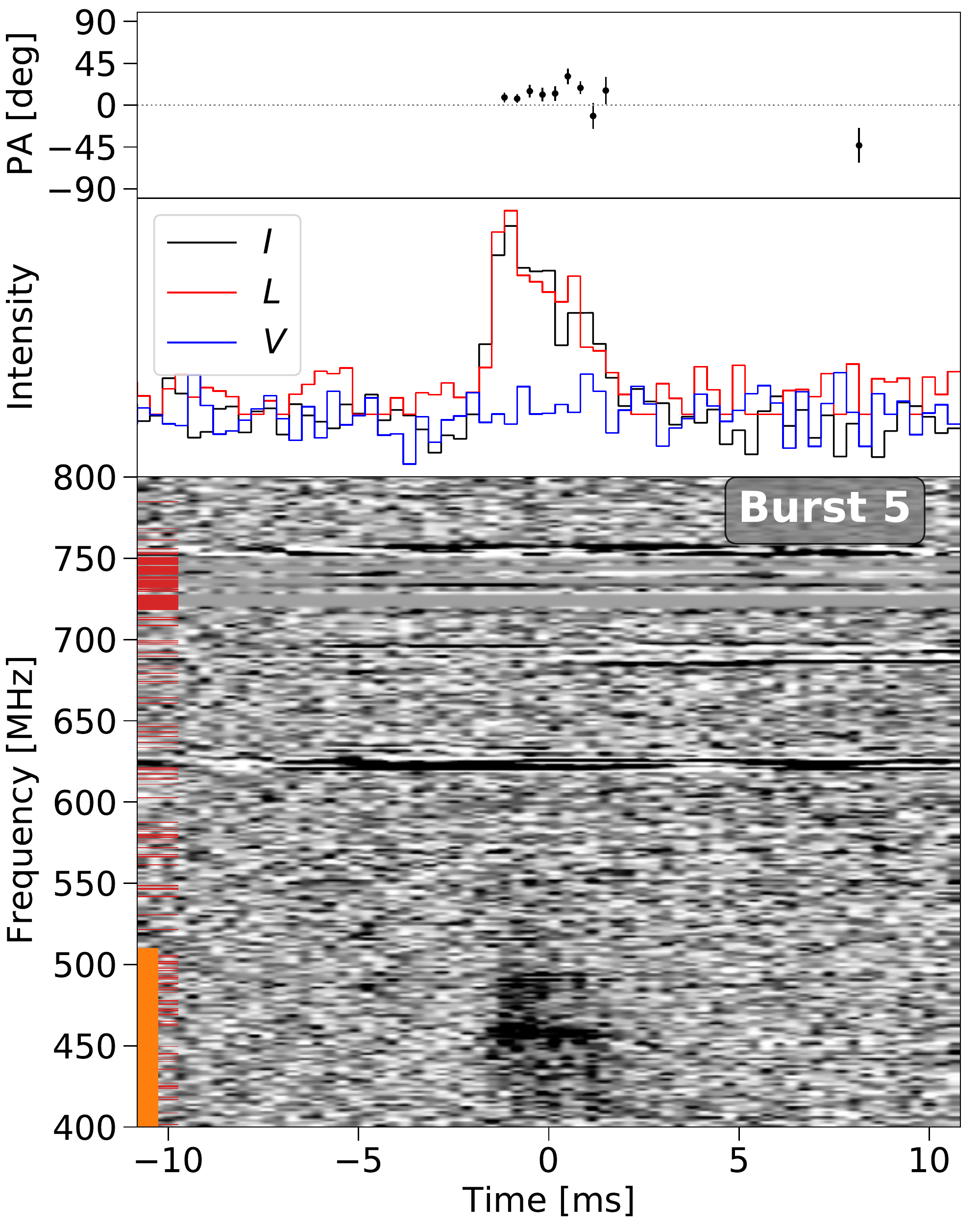}
\figsetgrpnote{}
\figsetgrpend

%%%%%%% R15

\figsetgrpstart
\figsetgrpnum{1.29}
\figsetgrptitle{FRB 20180908B}
\figsetplot{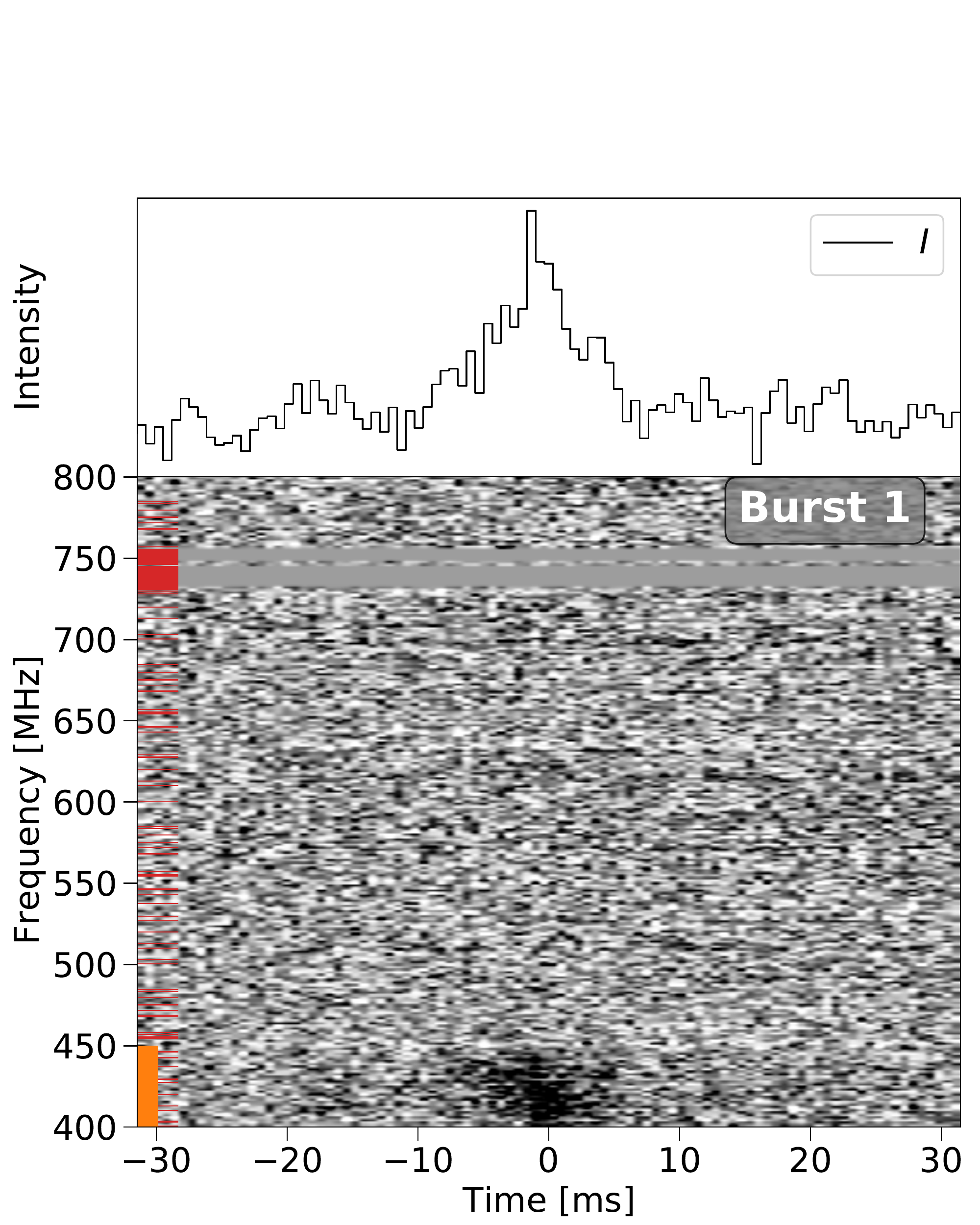}
\figsetgrpnote{}
\figsetgrpend

%%%%%%% R16

\figsetgrpstart
\figsetgrpnum{1.30}
\figsetgrptitle{FRB 20190117A}
\figsetplot{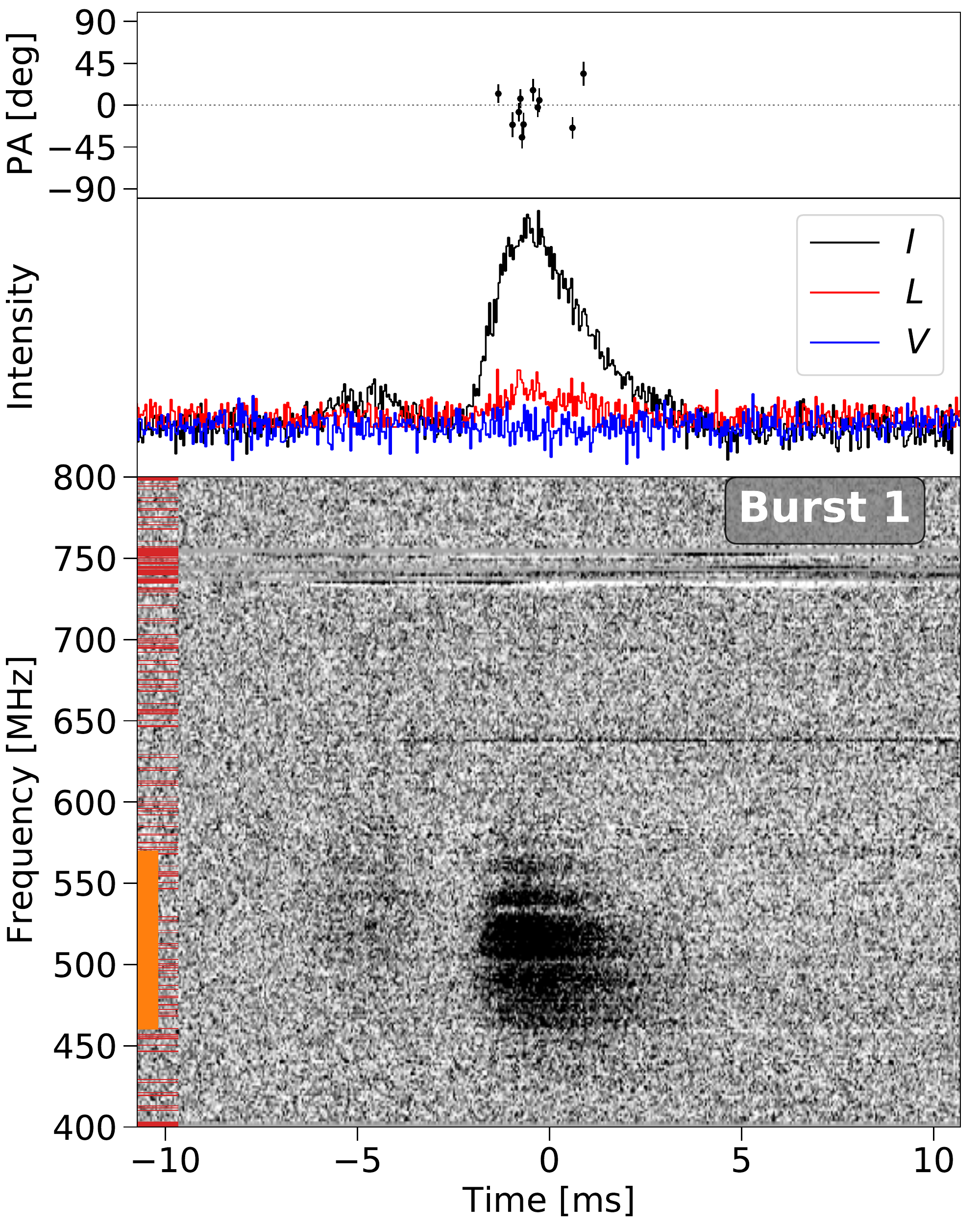}
\figsetgrpnote{}
\figsetgrpend

\figsetgrpstart
\figsetgrpnum{1.31}
\figsetgrptitle{FRB 20190117A}
\figsetplot{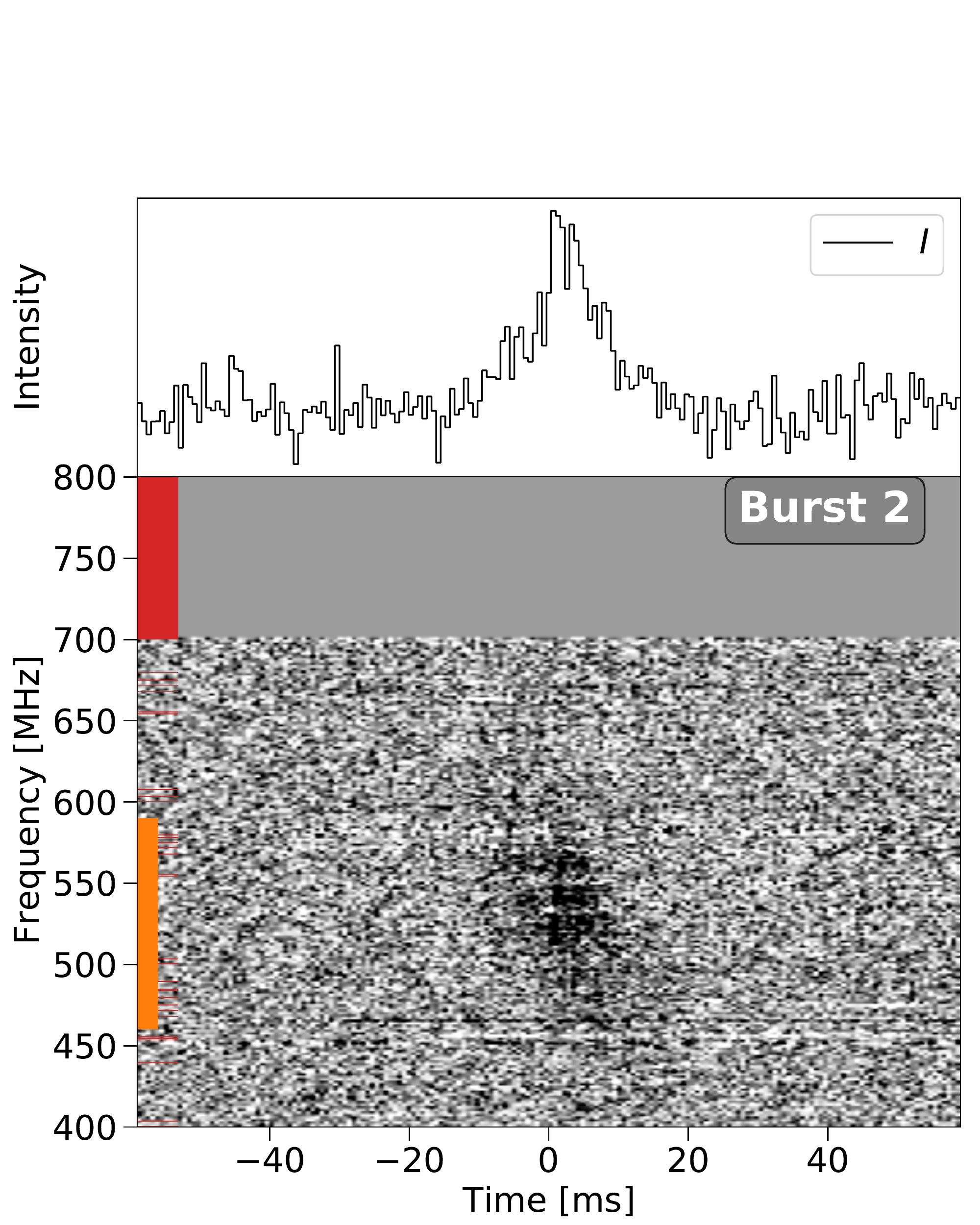}
\figsetgrpnote{}
\figsetgrpend

\figsetgrpstart
\figsetgrpnum{1.32}
\figsetgrptitle{FRB 20190117A}
\figsetplot{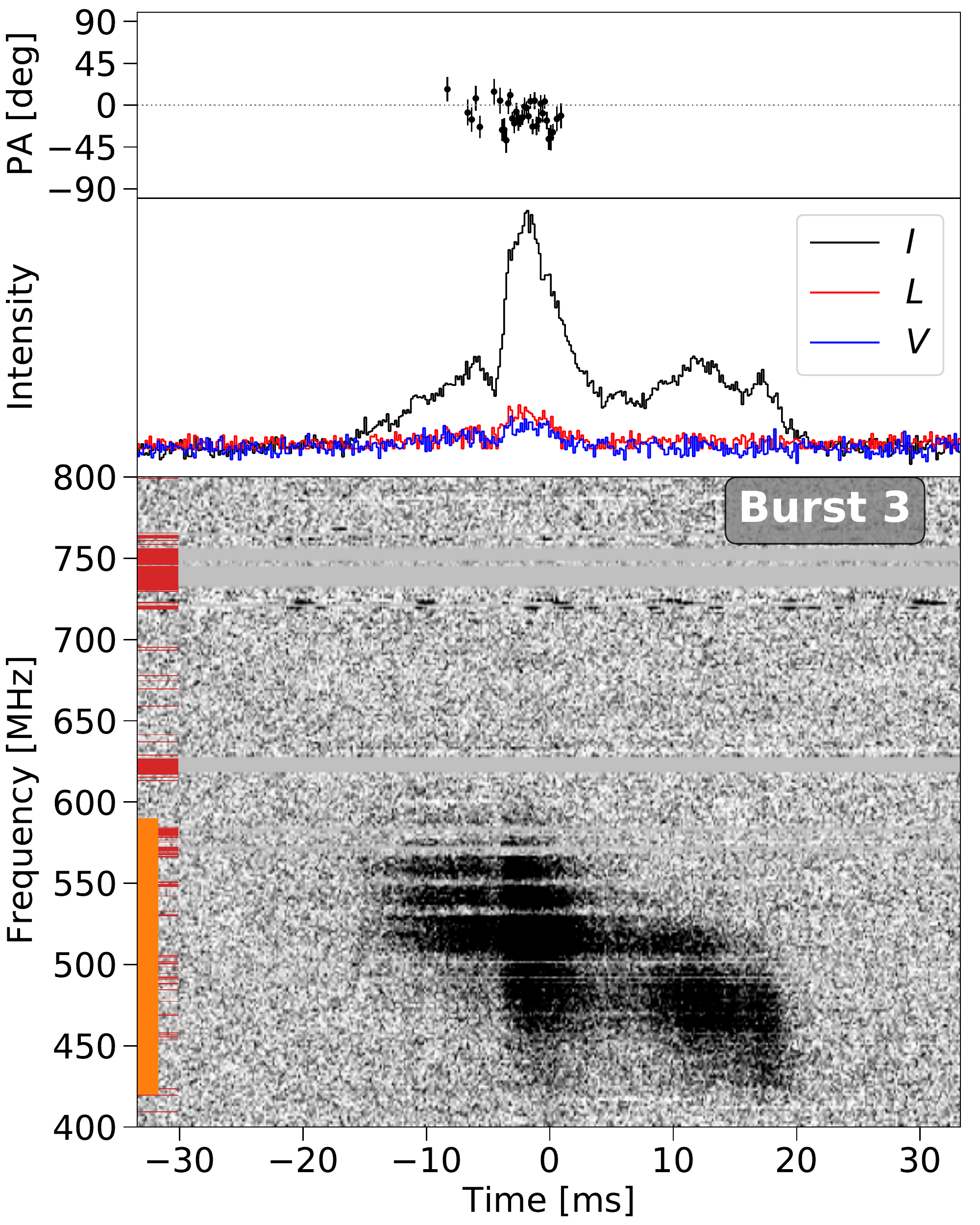} 
\figsetgrpnote{}
\figsetgrpend

%%%%%%% R18

\figsetgrpstart
\figsetgrpnum{1.33}
\figsetgrptitle{FRB 20190417A}
\figsetplot{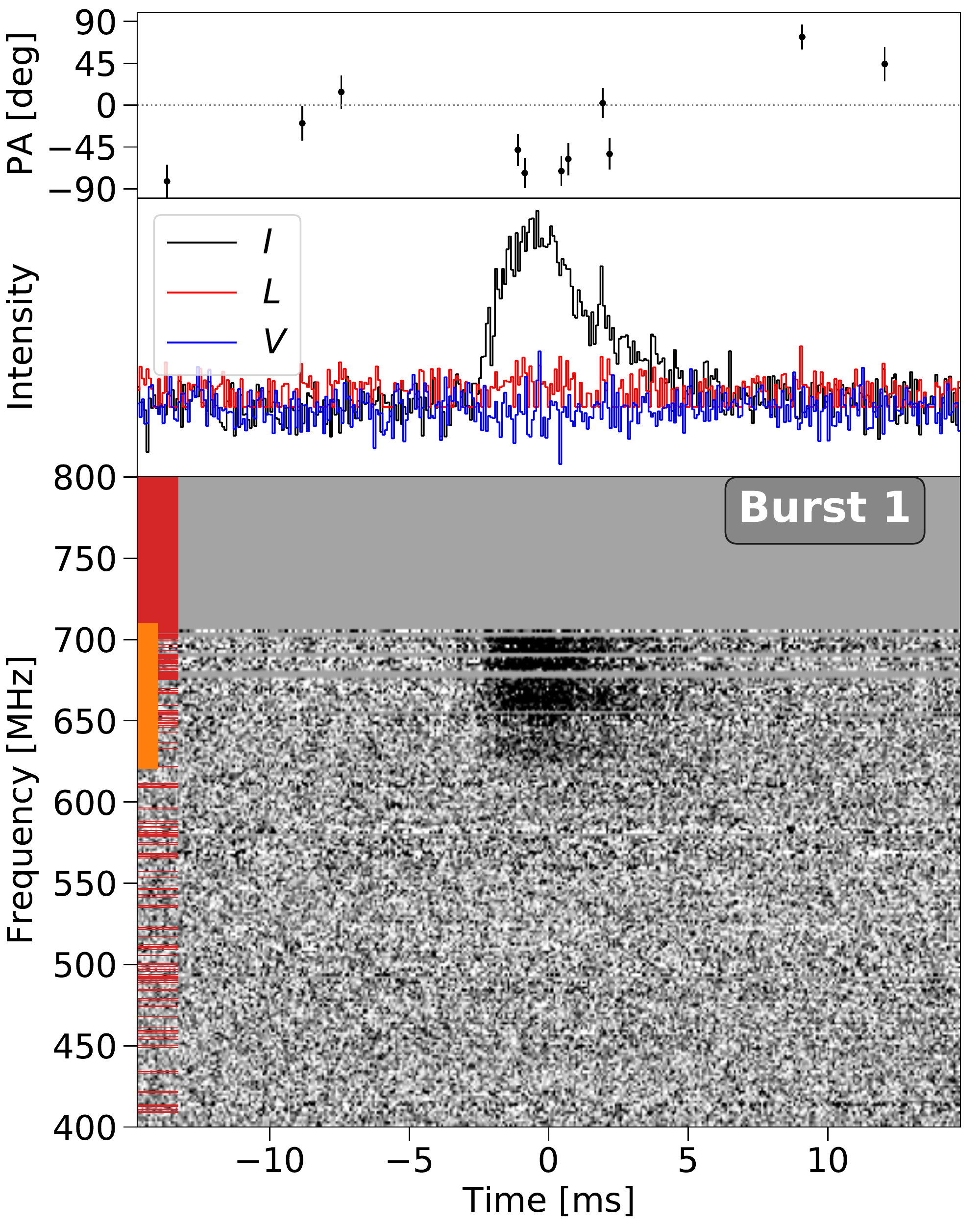}
\figsetgrpnote{}
\figsetgrpend

\figsetgrpstart
\figsetgrpnum{1.34}
\figsetgrptitle{FRB 20190417A}
\figsetplot{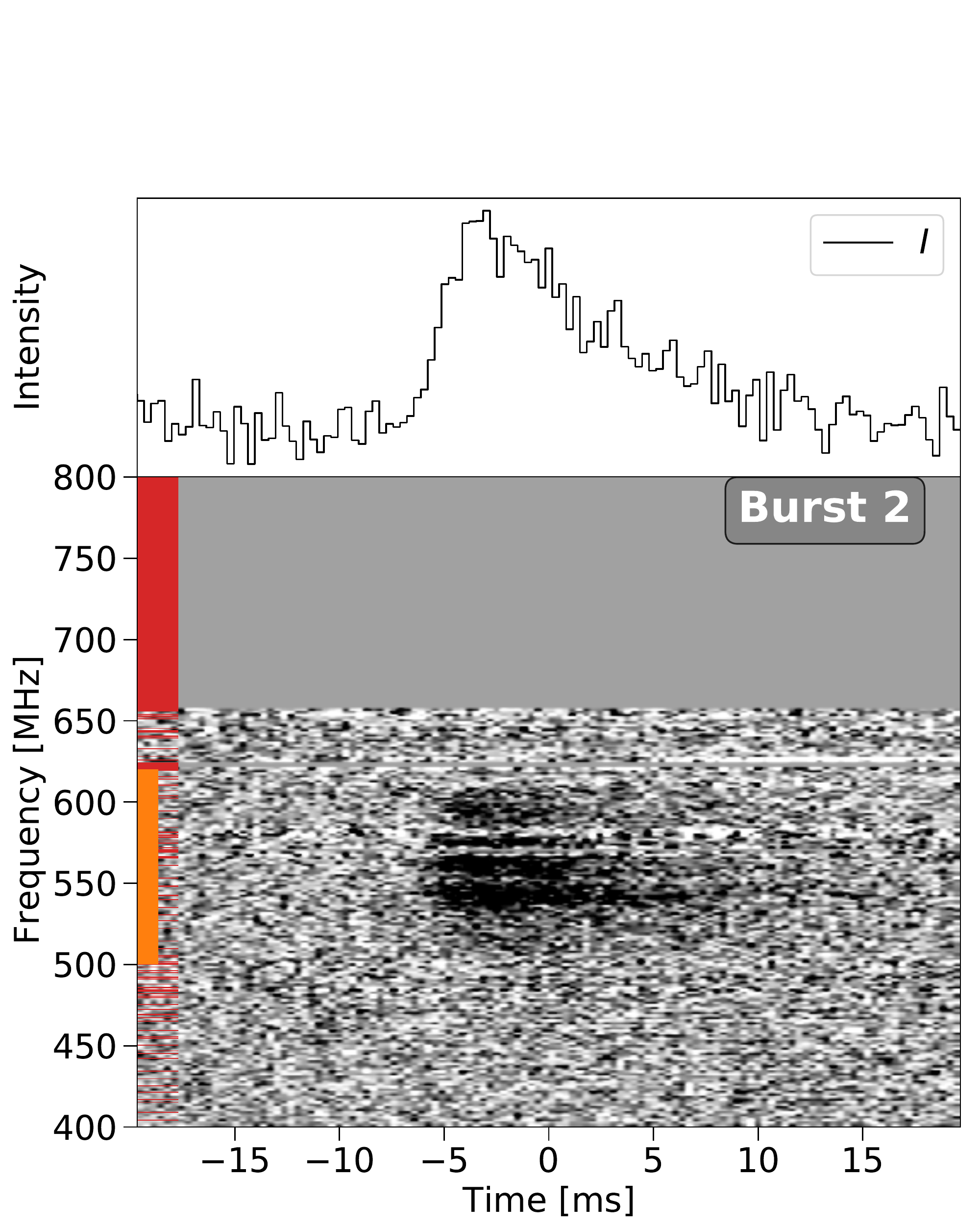}
\figsetgrpnote{}
\figsetgrpend

%%%%%%% R22

\figsetgrpstart
\figsetgrpnum{1.35}
\figsetgrptitle{FRB 20190907A}
\figsetplot{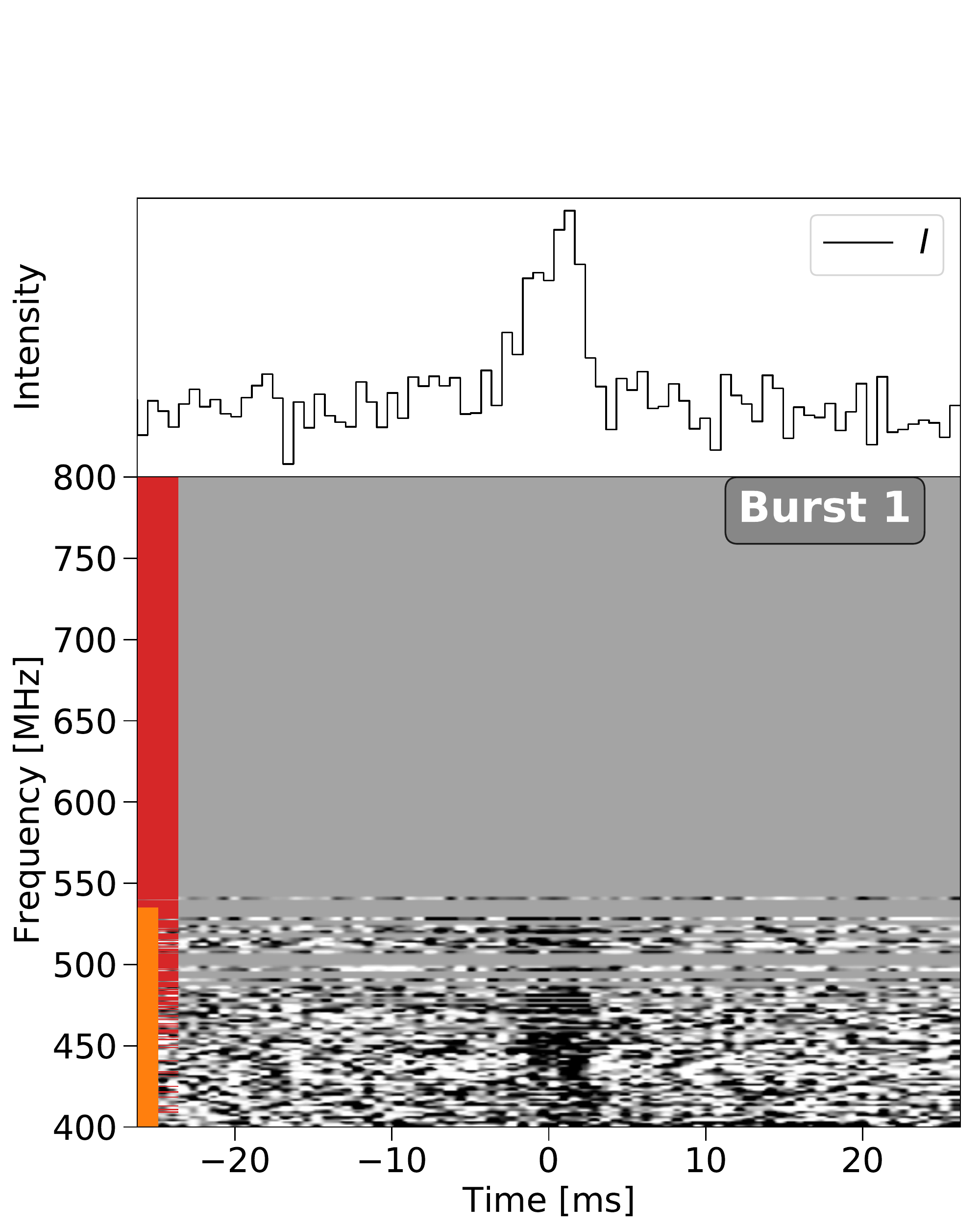}
\figsetgrpnote{}
\figsetgrpend

\renewcommand\thefigure{\thesection.\arabic{figure}} 

\section{Supplemental Figures}
\label{appendix:supplemental}
\setcounter{figure}{0} 

\begin{figure}
	\centering
\begin{center}
    \includegraphics[width=0.49\textwidth]{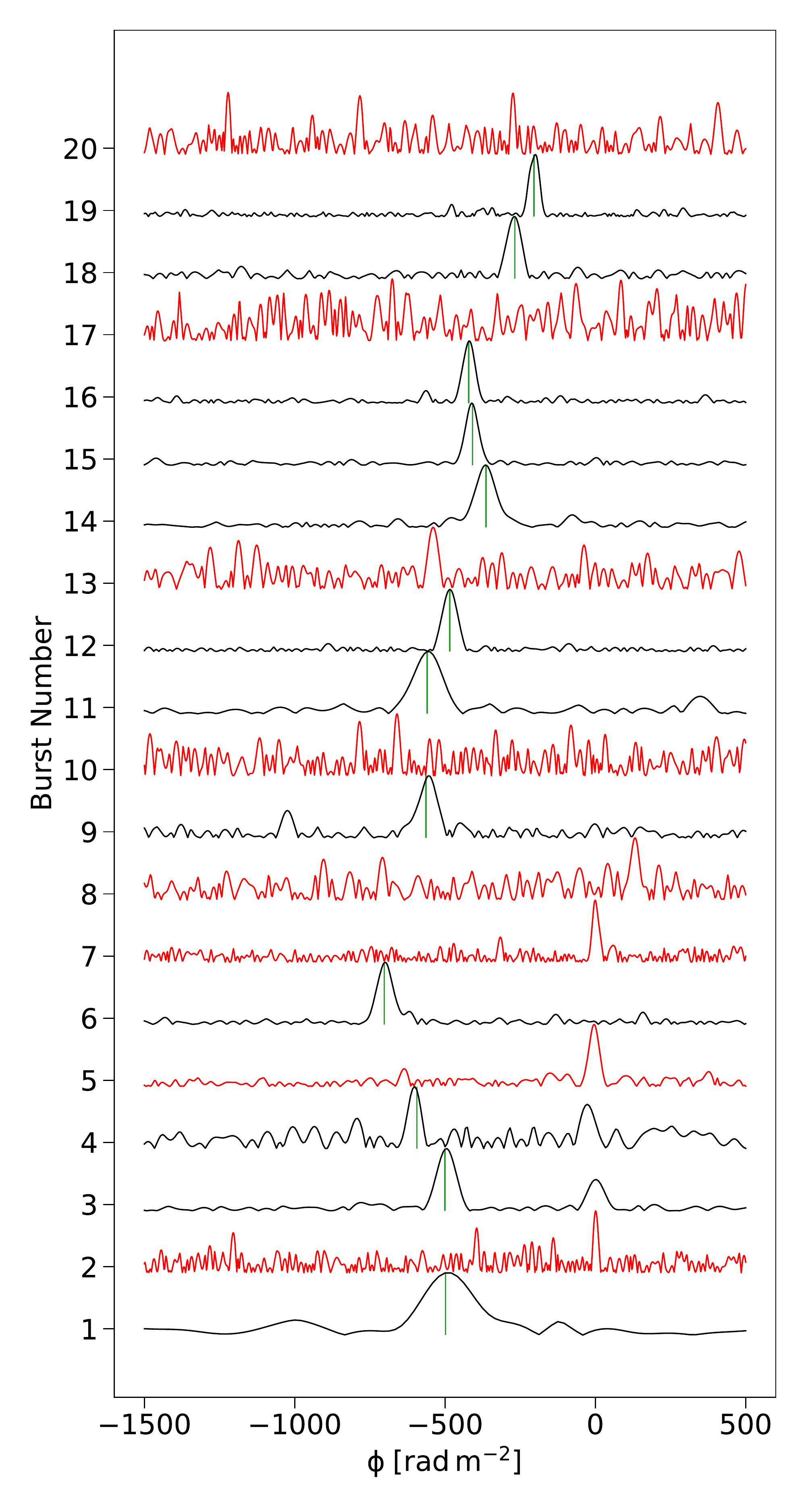}
    \caption{{Cleaned FDFs (see Section 2.2.2 of \citet{Mckinven2022} for bursts from FRB 20190303A. Each curve is normalized to a unitary peak and ordered chronologically by burst number (see Table~\ref{ta:bursts}). $\rm{RM}_{QU}$ measurements are indicated as vertical green lines. Peaks near $\phi\sim 0\; \rm{rad\, m^{-2}}$ correspond to instrumental polarization. Red curves correspond to bursts where no significant RM detection was made. Peaks near $\phi\sim 0\; \rm{rad\, m^{-2}}$ correspond to instrumental polarization.}} 
 \label{fig:FDFvstime}
\end{center}
\end{figure}

%%%%%%%%%%%%%%%%%% Appendix figures %%%%%%%%%%%%%%%%%%

%%%%%%%%%%%%%%%%%% R17 

\begin{figure}[!htb]
	\centering
\begin{center}
    \includegraphics[width=0.9\textwidth]{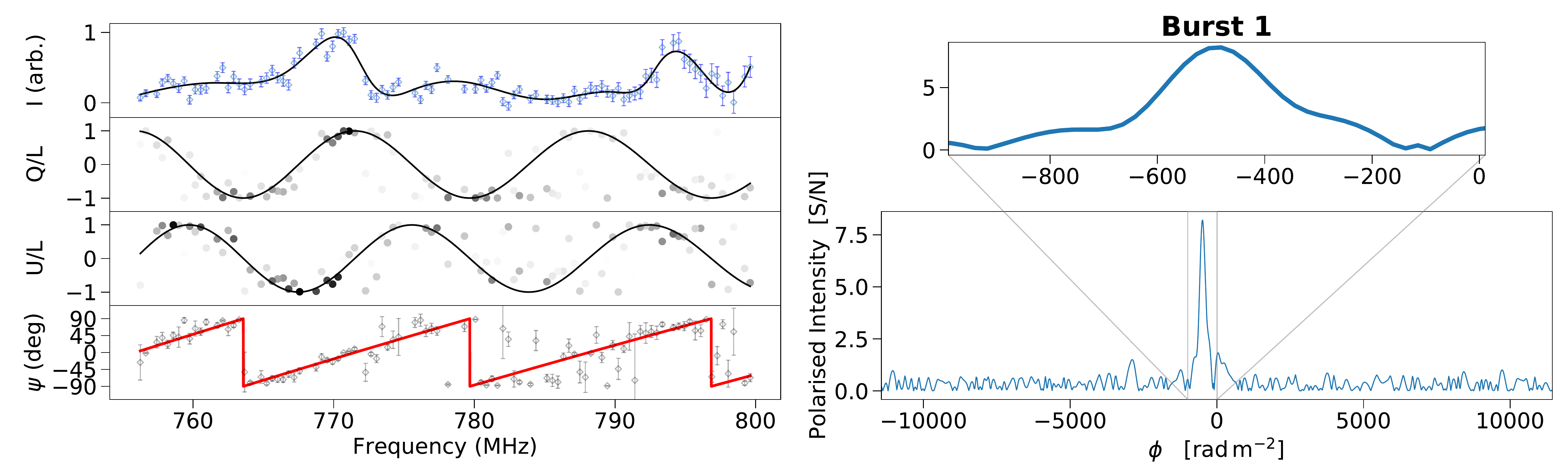}
    \includegraphics[width=0.9\textwidth]{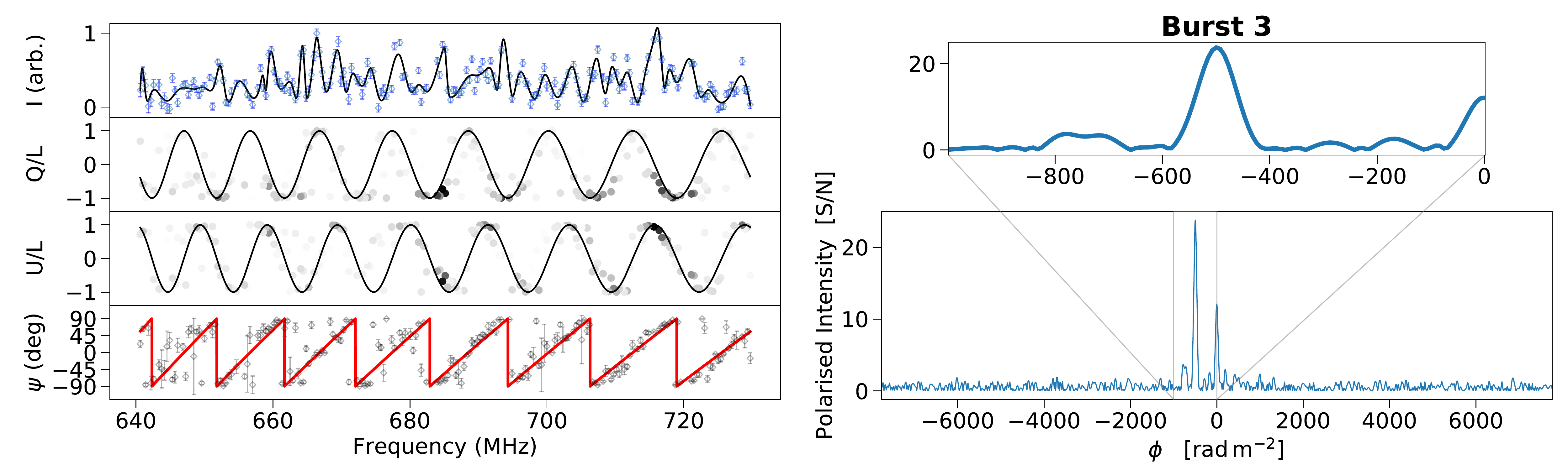}
    \includegraphics[width=0.9\textwidth]{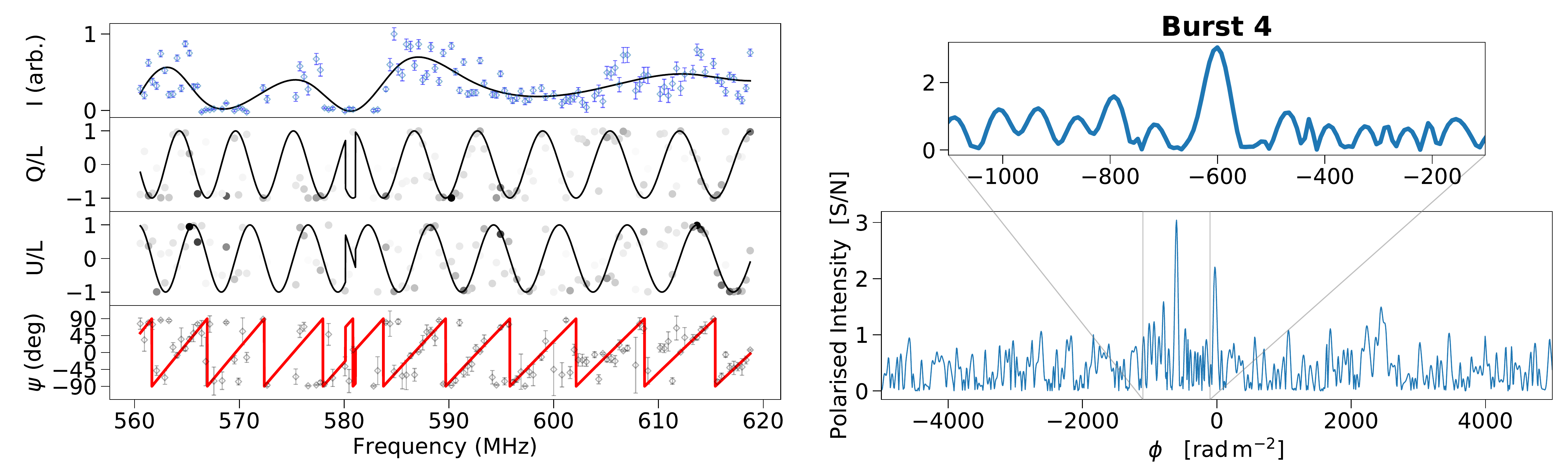}
    \includegraphics[width=0.9\textwidth]{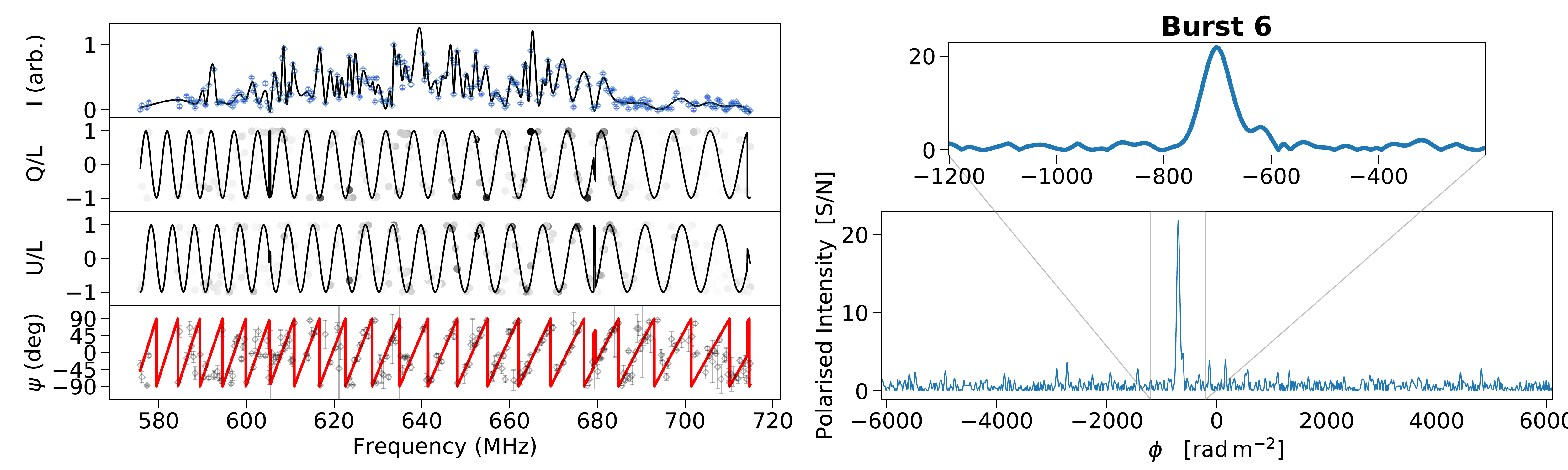}
    \caption{Examples of QU-fitting (left) and RM-synthesis (right) results for individual bursts from repeating source FRB 20190303A after correcting for a non-zero $\tau$ parameter \citep[see][]{Mckinven2022}. Left panel: The polarization spectra displaying, from top to bottom, Stokes $I$ (blue points) and its cubic spline smoothed version (black line), Stokes $Q$ parameter divided by the total linear polarization ($L$), Stokes $U$ divided by the total linear polarization, and the uncalibrated polarization angle ($\psi_0$). Frequency channels with highly polarized signal are indicated with darker points. Best-fit models are indicated as black lines (red lines for $\psi_0$). Right panel: The cleaned FDFs displaying linear polarized intensity as a function of $\phi$ (RM). The upper panel shows the FDFs confined to $\pm 500\; \rm{rad\, m^{-2}}$ of their peak locations. The complete figure set, including those of other repeating sources, is available in the online journal.}
 \label{fig:appendix_R17}
\end{center}
\end{figure} 

\figsetstart
\figsetnum{2}
\figsettitle{QU-fitting and RM-synthesis results for individual bursts from our repeating FRB source sample.}

%%%%%%%%%%%%%%% R17

\figsetgrpstart
\figsetgrpnum{2.1}
\figsetgrptitle{FRB 20190303A}
\figsetplot{plots/supplemental/R17/appendix_43265588.pdf}
\figsetgrpnote{}
\figsetgrpend

\figsetgrpstart
\figsetgrpnum{2.2}
\figsetgrptitle{FRB 20190303A}
\figsetplot{plots/supplemental/R17/appendix_59240332.pdf}
\figsetgrpnote{}
\figsetgrpend

\figsetgrpstart
\figsetgrpnum{2.3}
\figsetgrptitle{FRB 20190303A}
\figsetplot{plots/supplemental/R17/appendix_61303573.pdf}
\figsetgrpnote{}
\figsetgrpend

\figsetgrpstart
\figsetgrpnum{2.4}
\figsetgrptitle{FRB 20190303A}
\figsetplot{plots/supplemental/R17/appendix_62352548.pdf}
\figsetgrpnote{}
\figsetgrpend

\figsetgrpstart
\figsetgrpnum{2.5}
\figsetgrptitle{FRB 20190303A}
\figsetplot{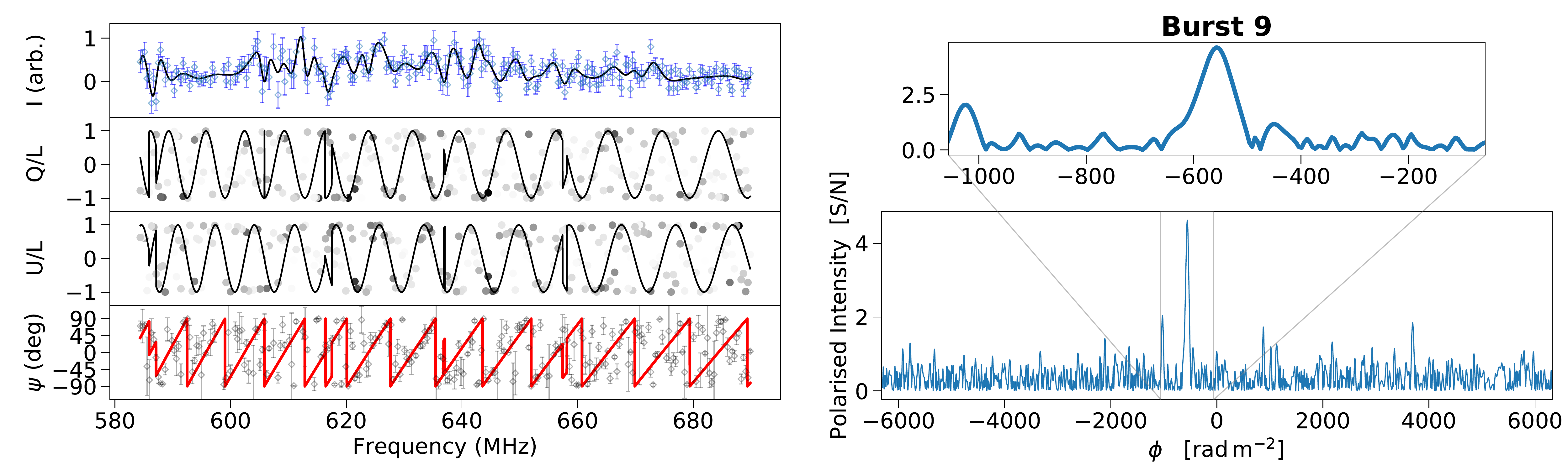}
\figsetgrpnote{}
\figsetgrpend

\figsetgrpstart
\figsetgrpnum{2.6}
\figsetgrptitle{FRB 20190303A}
\figsetplot{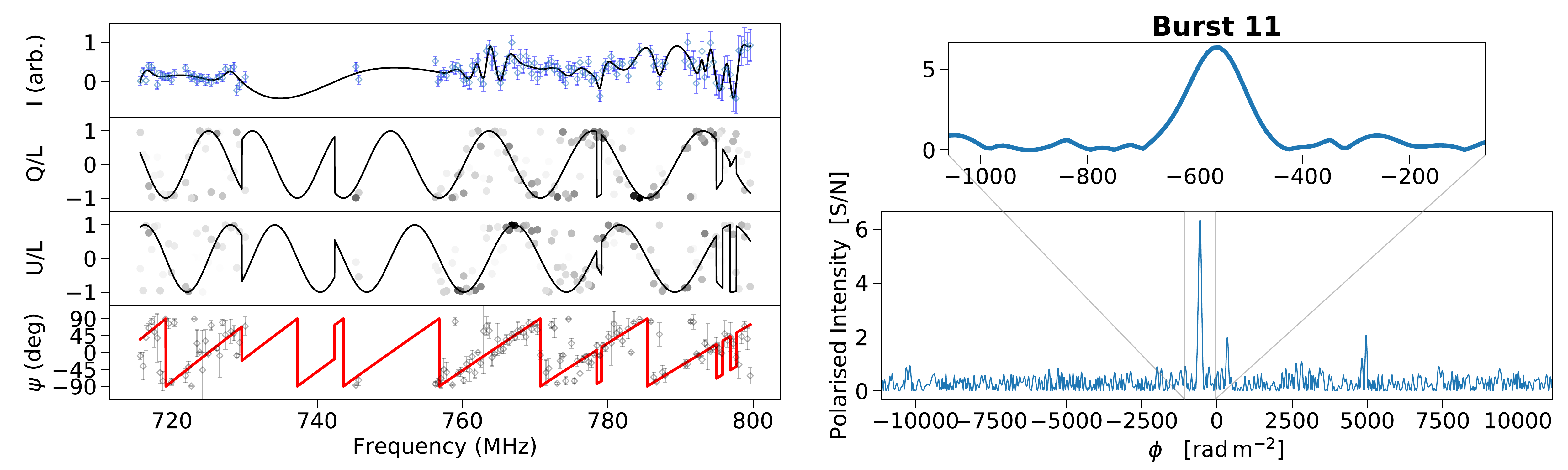}
\figsetgrpnote{}
\figsetgrpend

\figsetgrpstart
\figsetgrpnum{2.7}
\figsetgrptitle{FRB 20190303A}
\figsetplot{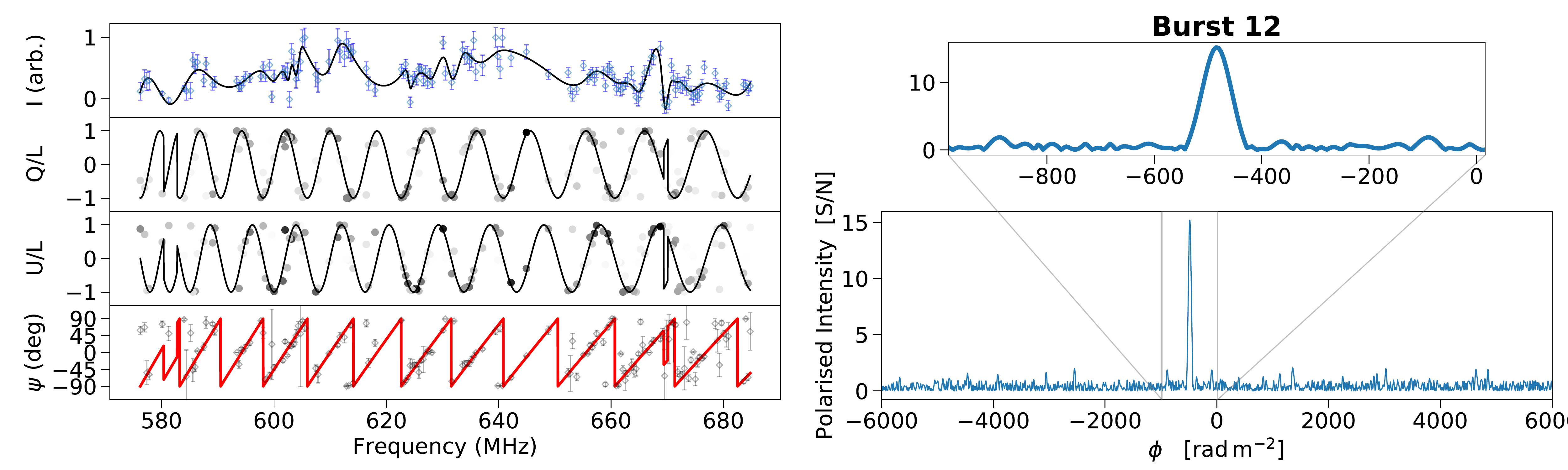}
\figsetgrpnote{}
\figsetgrpend

\figsetgrpstart
\figsetgrpnum{2.8}
\figsetgrptitle{FRB 20190303A}
\figsetplot{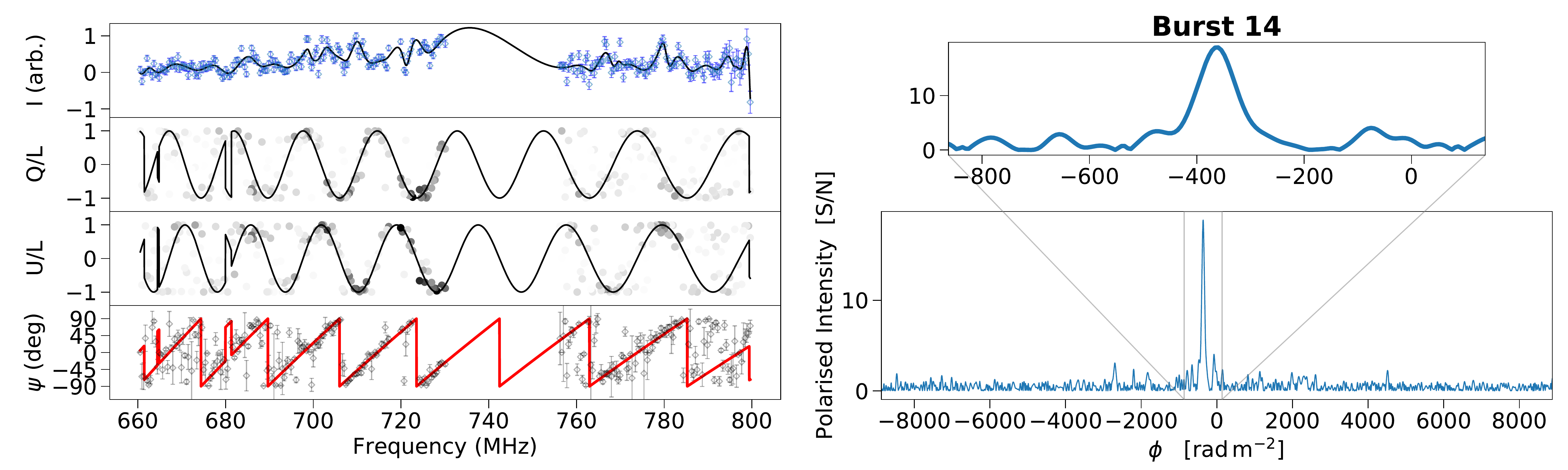}
\figsetgrpnote{}
\figsetgrpend

\figsetgrpstart
\figsetgrpnum{2.9}
\figsetgrptitle{FRB 20190303A}
\figsetplot{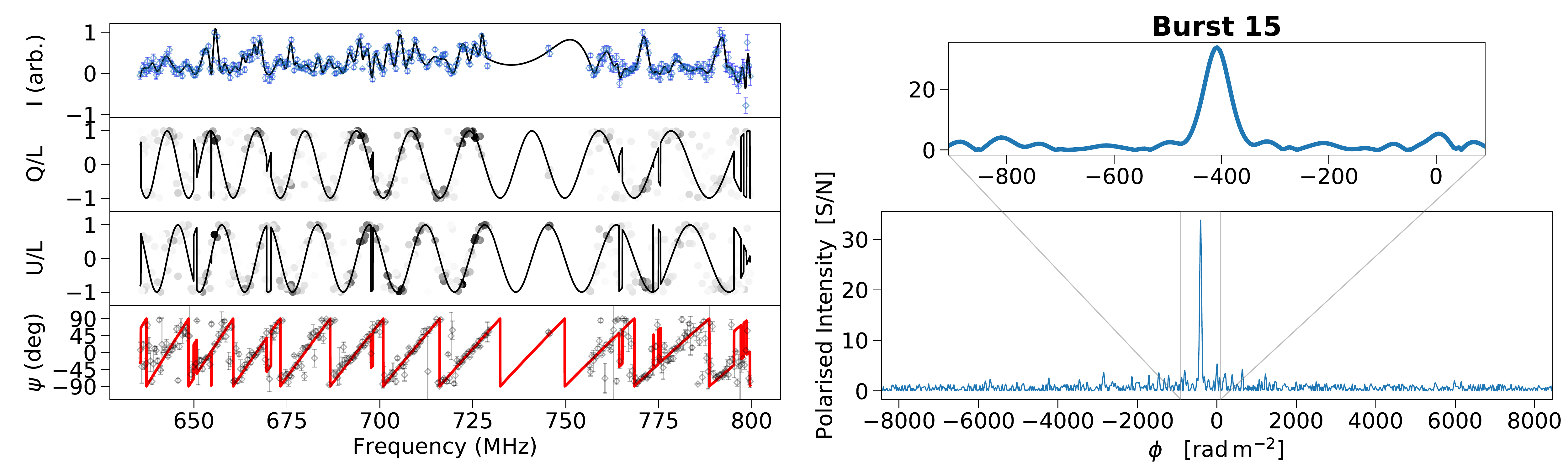}
\figsetgrpnote{}
\figsetgrpend

\figsetgrpstart
\figsetgrpnum{2.10}
\figsetgrptitle{FRB 20190303A}
\figsetplot{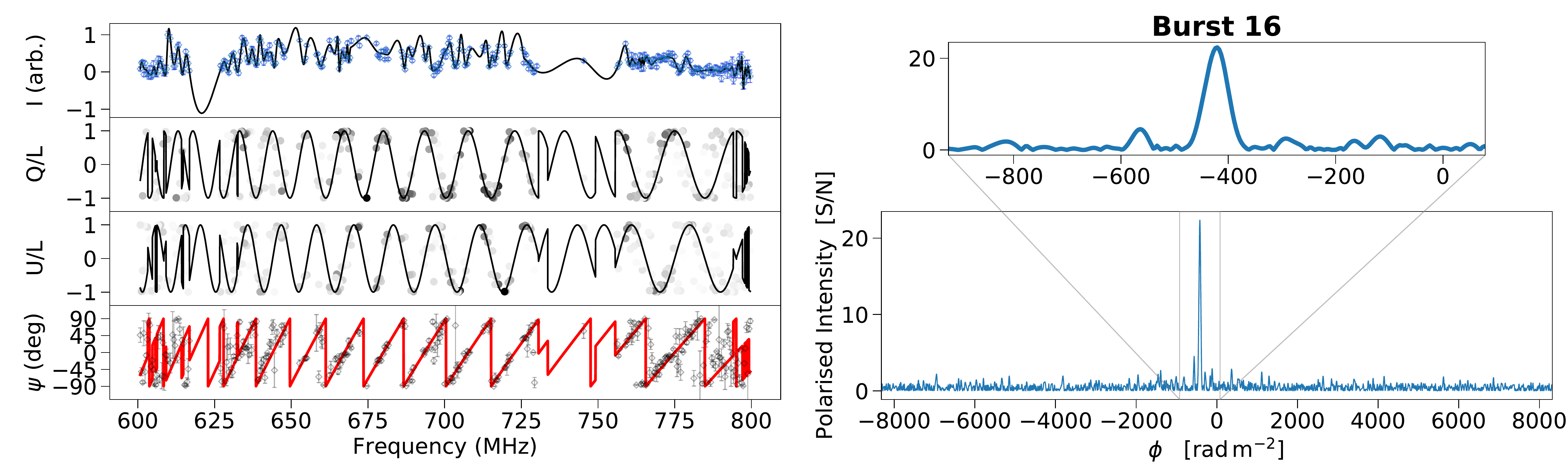}
\figsetgrpnote{}
\figsetgrpend

\figsetgrpstart
\figsetgrpnum{2.11}
\figsetgrptitle{FRB 20190303A}
\figsetplot{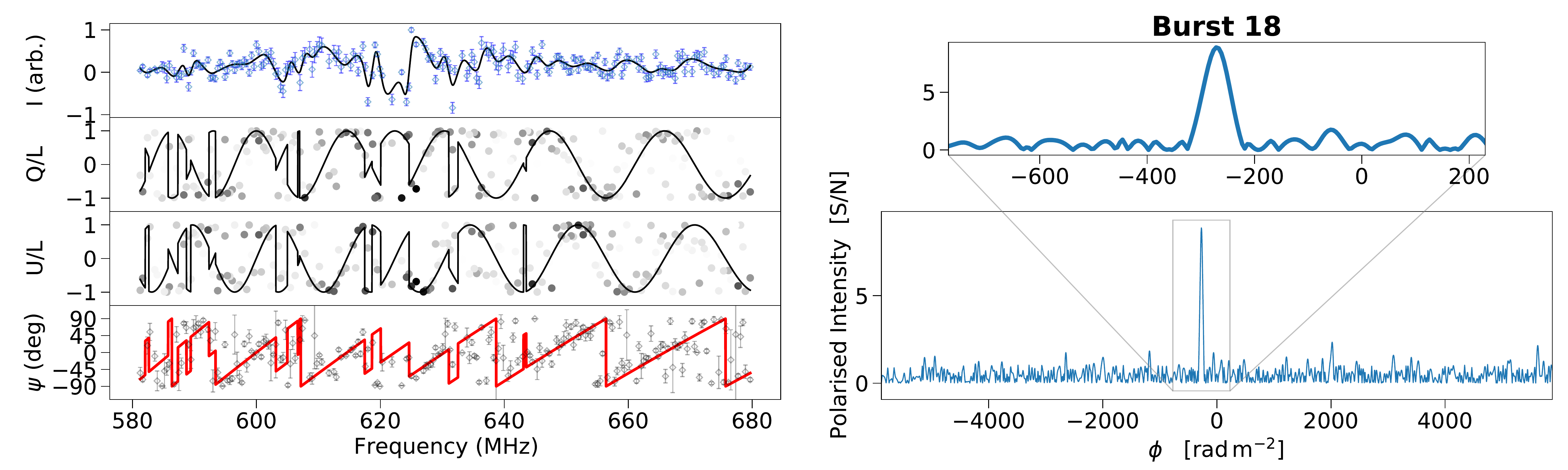}
\figsetgrpnote{}
\figsetgrpend

\figsetgrpstart
\figsetgrpnum{2.12}
\figsetgrptitle{FRB 20190303A}
\figsetplot{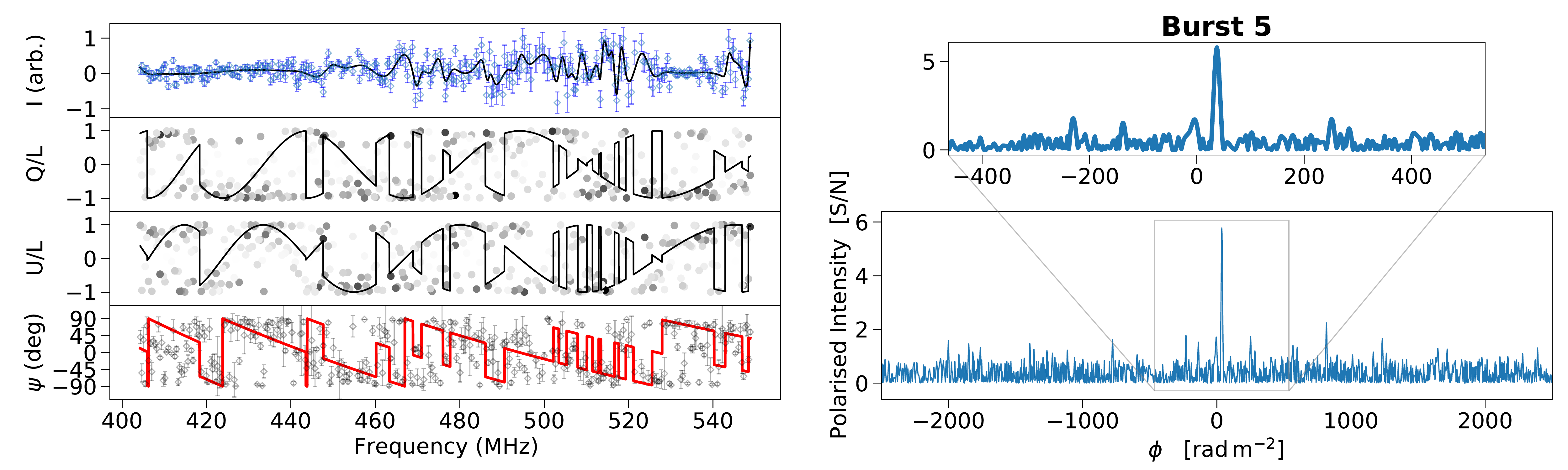}
\figsetgrpnote{}
\figsetgrpend

\figsetgrpstart
\figsetgrpnum{2.13}
\figsetgrptitle{FRB 20190303A}
\figsetplot{plots/supplemental/R4/appendix_69978377.pdf}
\figsetgrpnote{}
\figsetgrpend

%%%%%%%%%%%%%%%% Other repeaters %%%%%%%%%%%%%%%

%%%%%%%%%%%%%%% R2

\figsetgrpstart
\figsetgrpnum{2.14}
\figsetgrptitle{FRB 20180814A}
\figsetplot{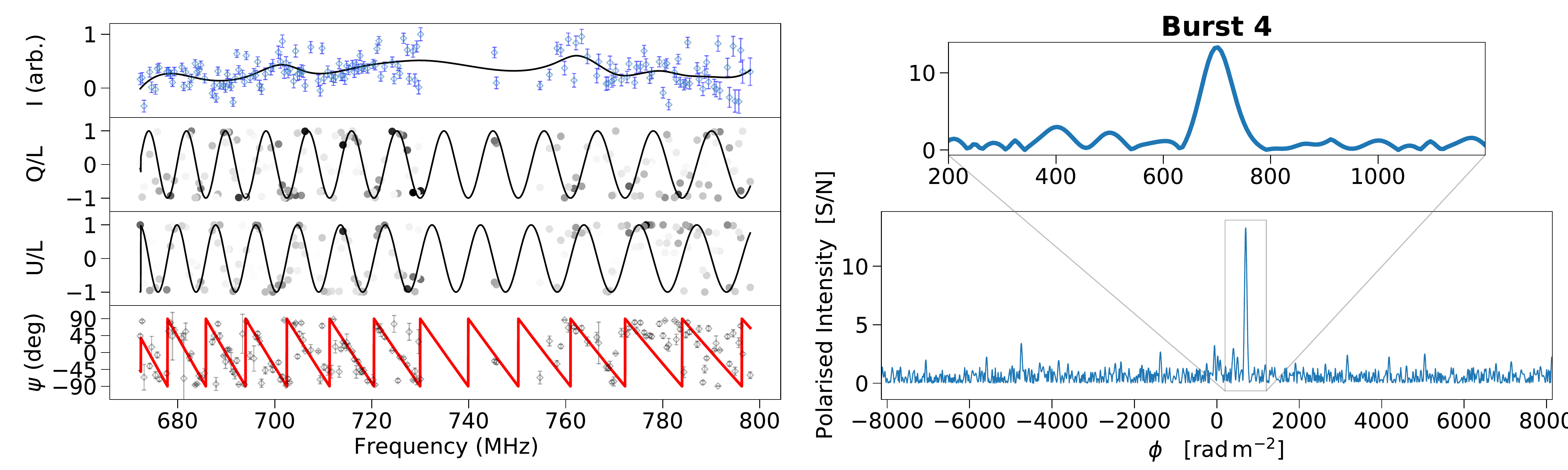}
\figsetgrpnote{}
\figsetgrpend

%%%%%%%%%%%%%%% R4

\figsetgrpstart
\figsetgrpnum{2.15}
\figsetgrptitle{FRB 20181030A}
\figsetplot{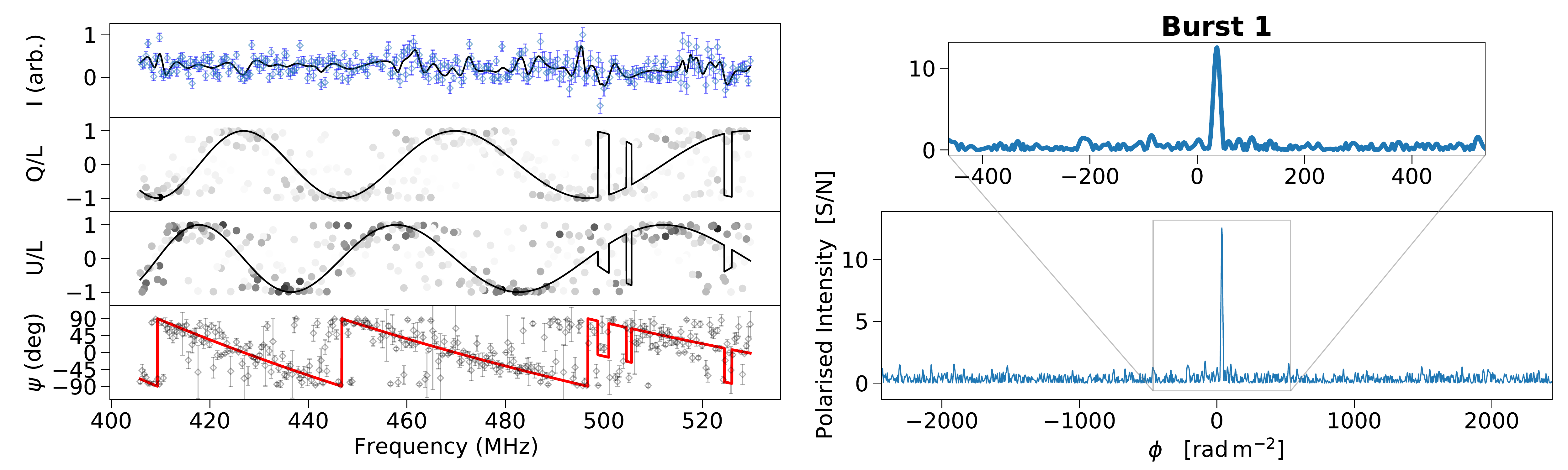}
\figsetgrpnote{}
\figsetgrpend

\figsetgrpstart
\figsetgrpnum{2.16}
\figsetgrptitle{FRB 20181030A}
\figsetplot{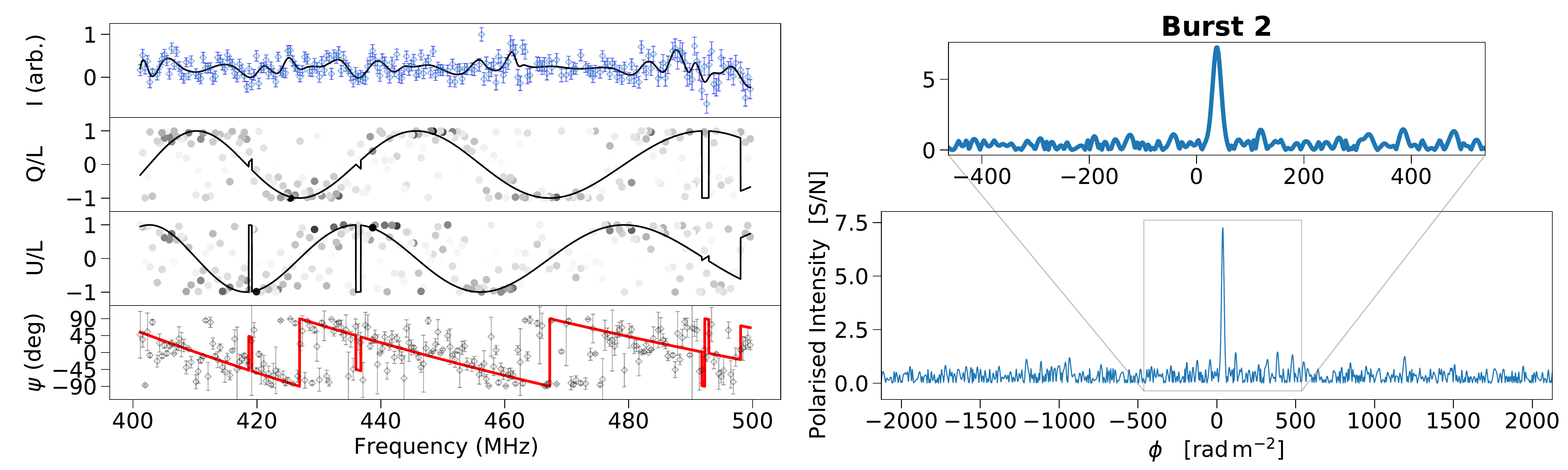}
\figsetgrpnote{}
\figsetgrpend

\figsetgrpstart
\figsetgrpnum{2.17}
\figsetgrptitle{FRB 20181030A}
\figsetplot{plots/supplemental/R4/appendix_69978377.pdf}
\figsetgrpnote{}
\figsetgrpend

%%%%%%%%%%%%%%% R6

\figsetgrpstart
\figsetgrpnum{2.18}
\figsetgrptitle{FRB 20181119A}
\figsetplot{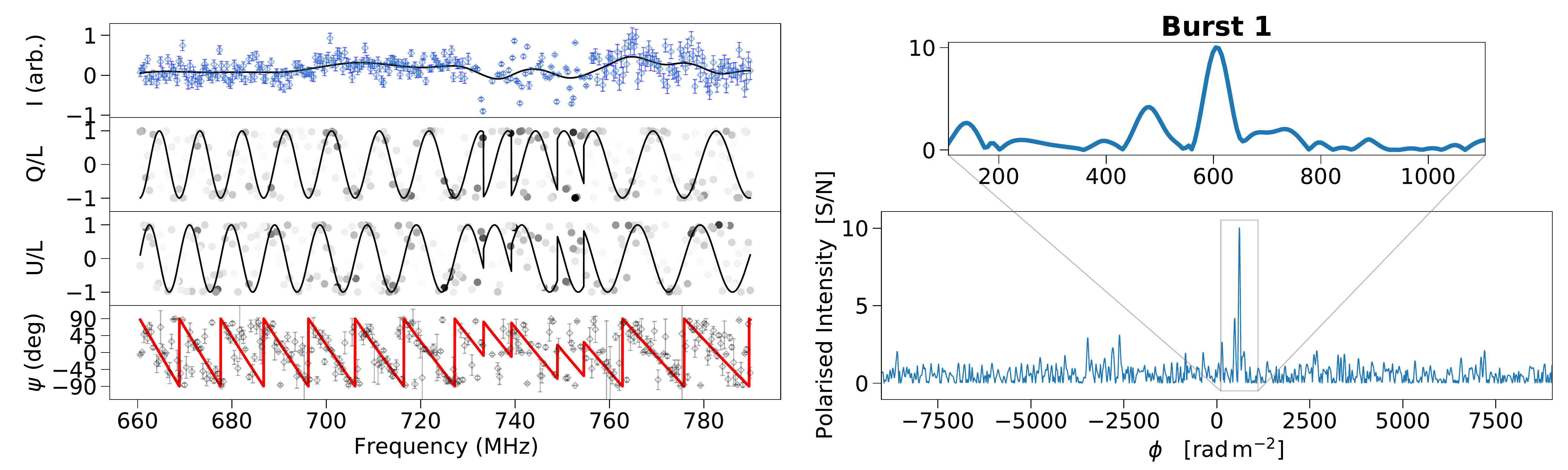}
\figsetgrpnote{}
\figsetgrpend

\figsetgrpstart
\figsetgrpnum{2.19}
\figsetgrptitle{FRB 20181119A}
\figsetplot{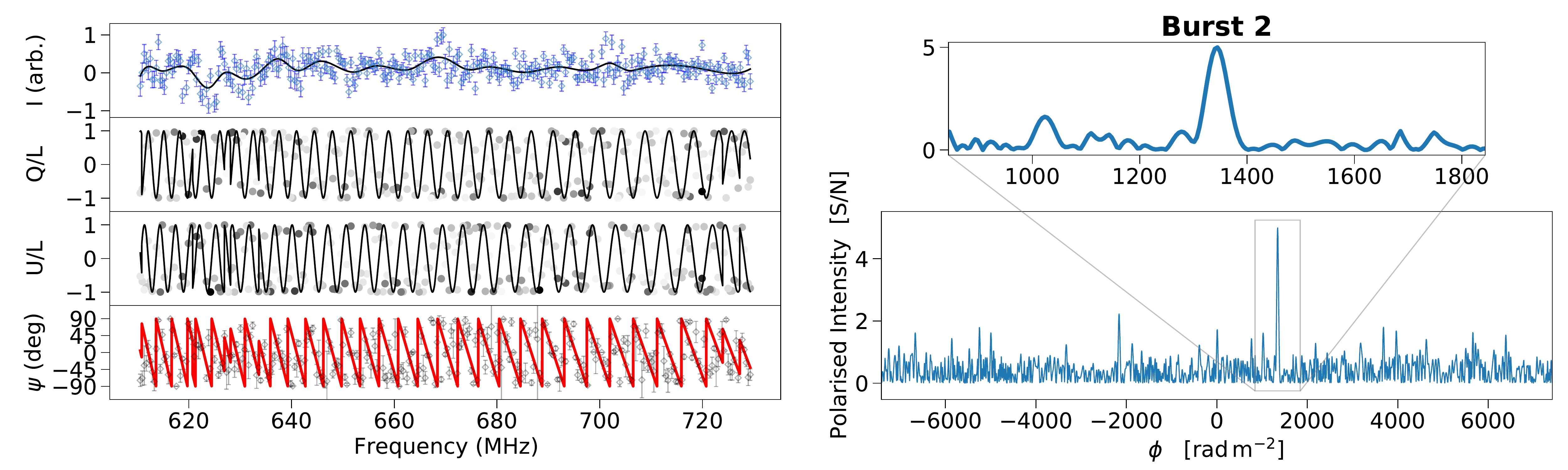}
\figsetgrpnote{}
\figsetgrpend

\figsetgrpstart
\figsetgrpnum{2.20}
\figsetgrptitle{FRB 20181119A}
\figsetplot{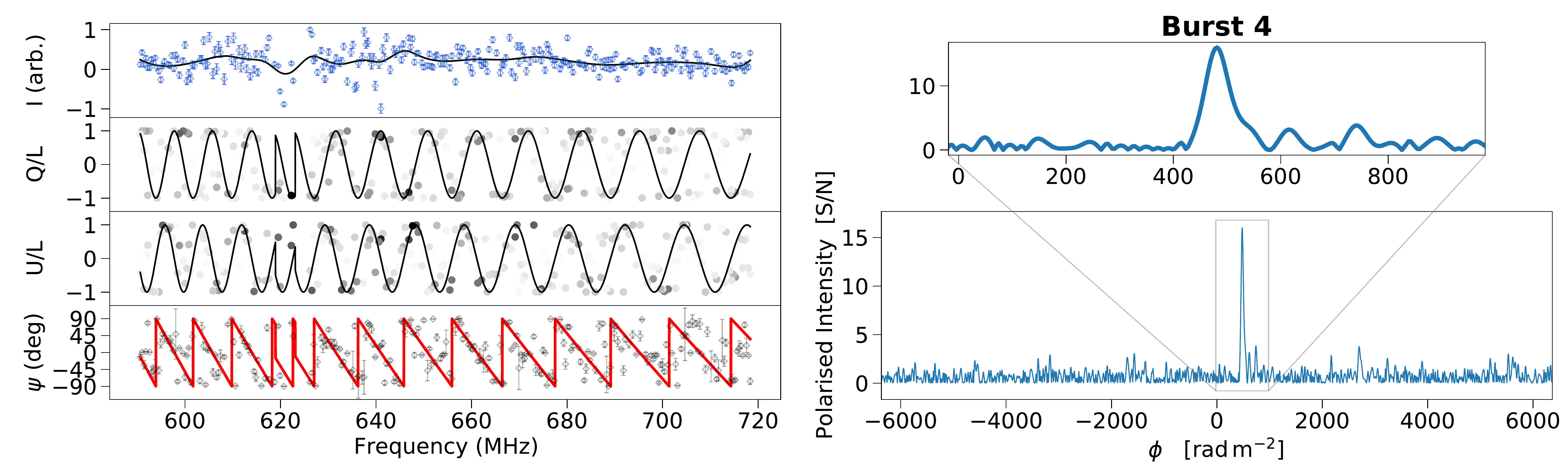}
\figsetgrpnote{}
\figsetgrpend

%%%%%%%%%%%%%%% R11

\figsetgrpstart
\figsetgrpnum{2.21}
\figsetgrptitle{FRB 20190222A}
\figsetplot{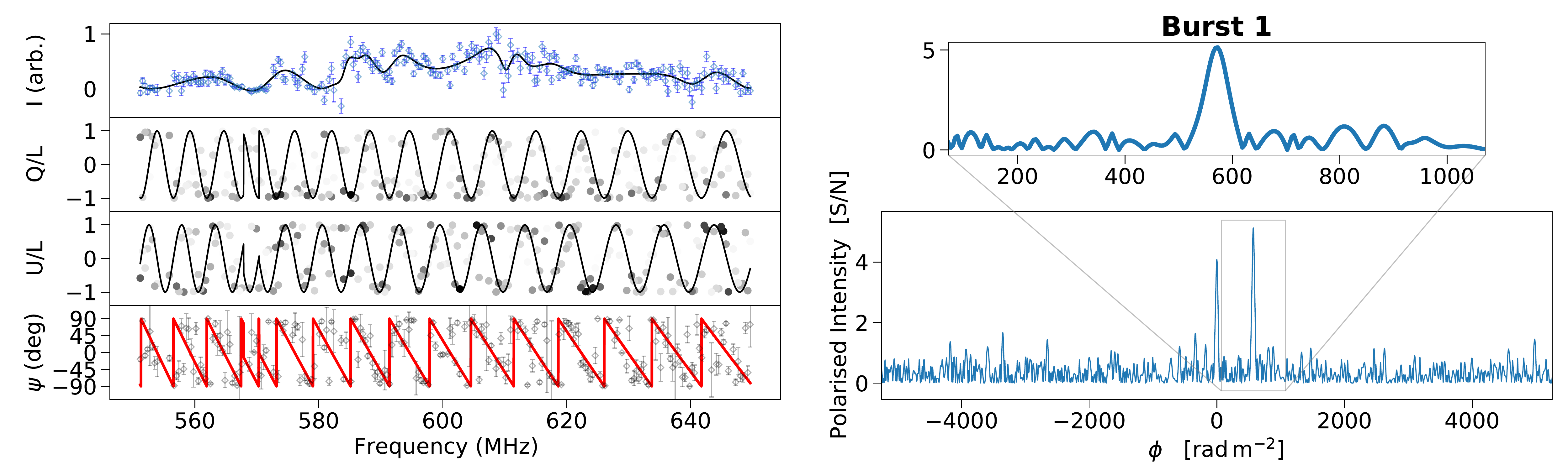}
\figsetgrpnote{}
\figsetgrpend

%%%%%%%%%%%%%%% R12

\figsetgrpstart
\figsetgrpnum{2.22}
\figsetgrptitle{FRB 20190208A}
\figsetplot{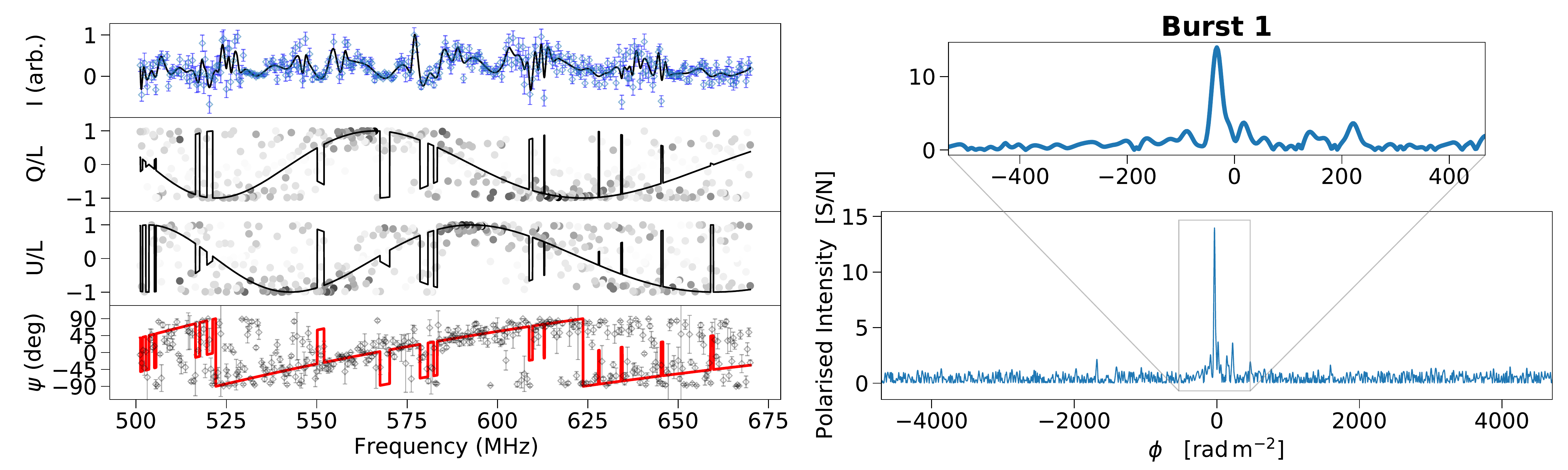}
\figsetgrpnote{}
\figsetgrpend

\figsetgrpstart
\figsetgrpnum{2.23}
\figsetgrptitle{FRB 20190208A}
\figsetplot{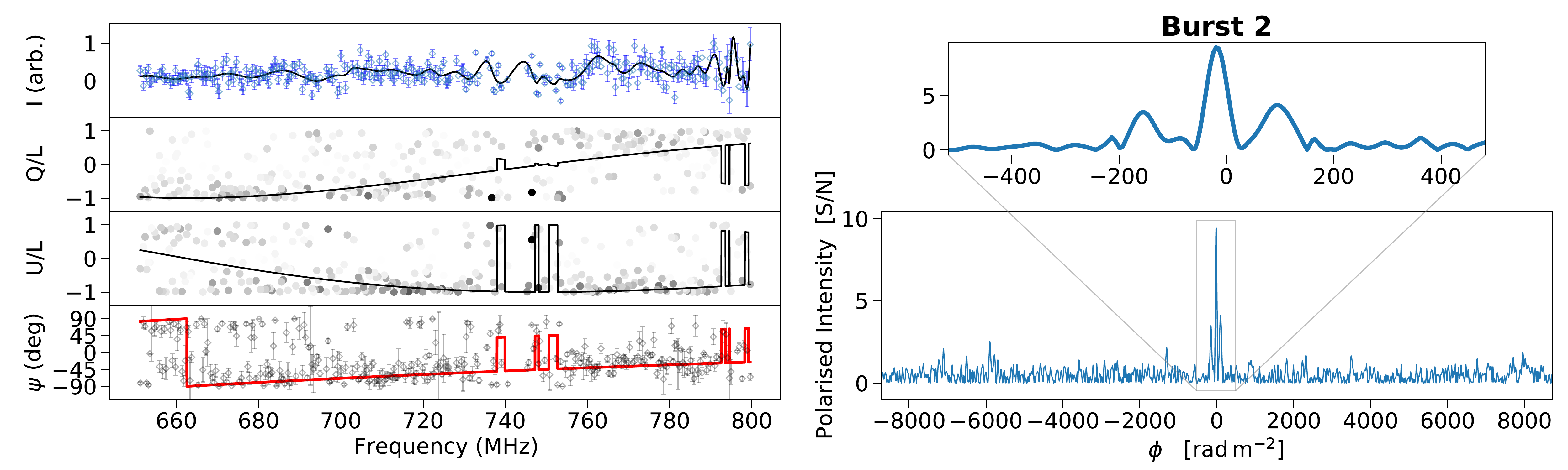}
\figsetgrpnote{}
\figsetgrpend

\figsetgrpstart
\figsetgrpnum{2.24}
\figsetgrptitle{FRB 20190208A}
\figsetplot{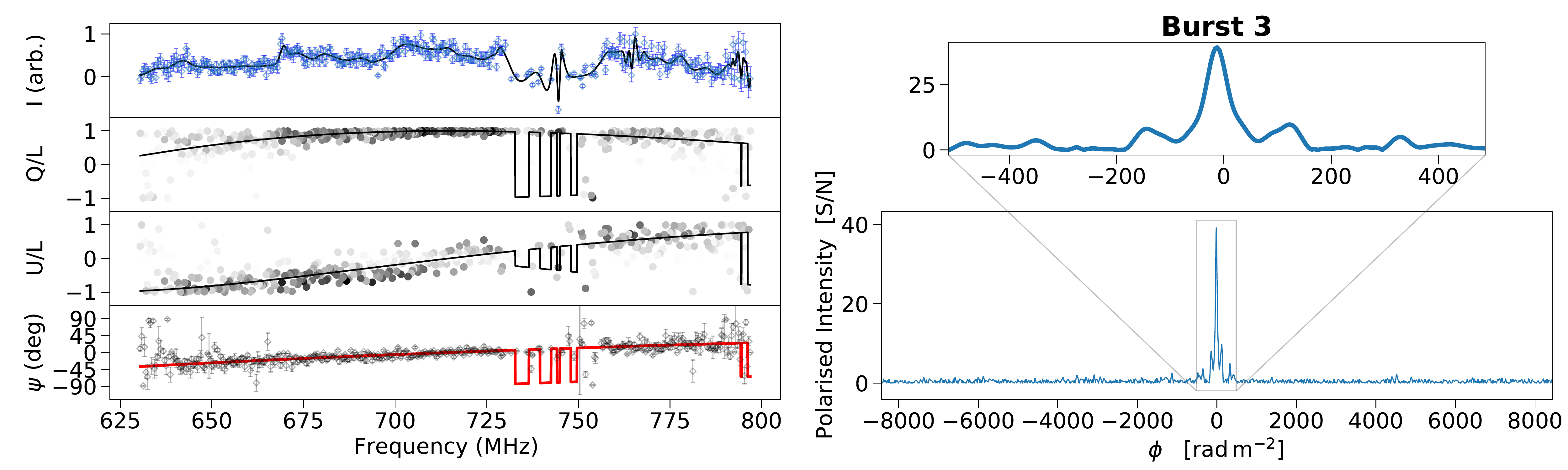}
\figsetgrpnote{}
\figsetgrpend

\figsetgrpstart
\figsetgrpnum{2.25}
\figsetgrptitle{FRB 20190208A}
\figsetplot{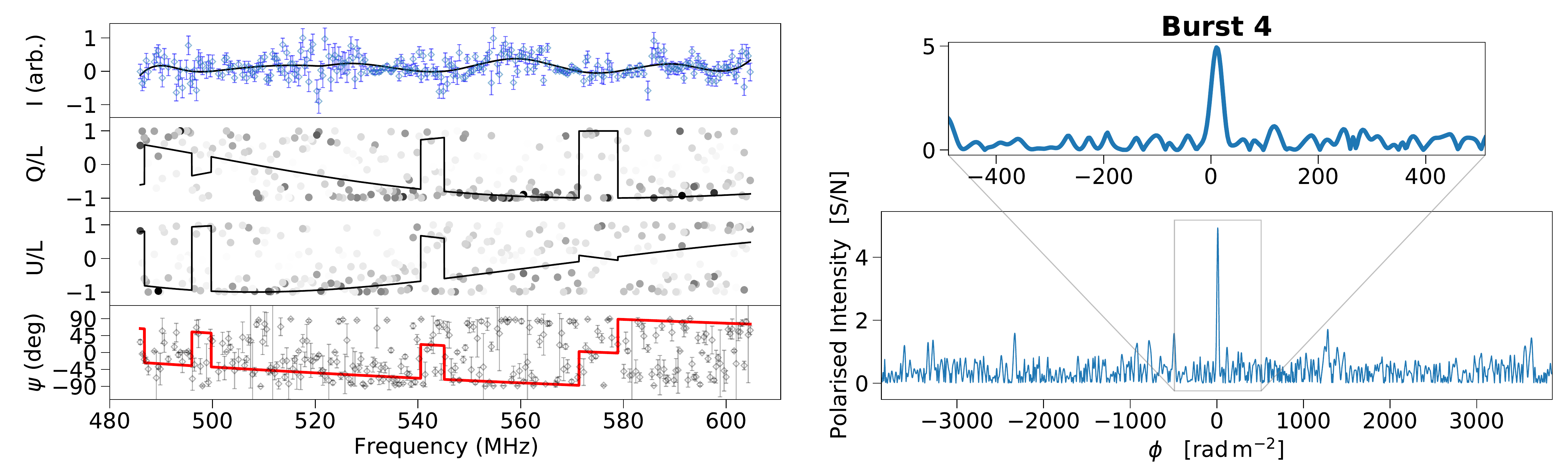}
\figsetgrpnote{}
\figsetgrpend

\figsetgrpstart
\figsetgrpnum{2.26}
\figsetgrptitle{FRB 20190208A}
\figsetplot{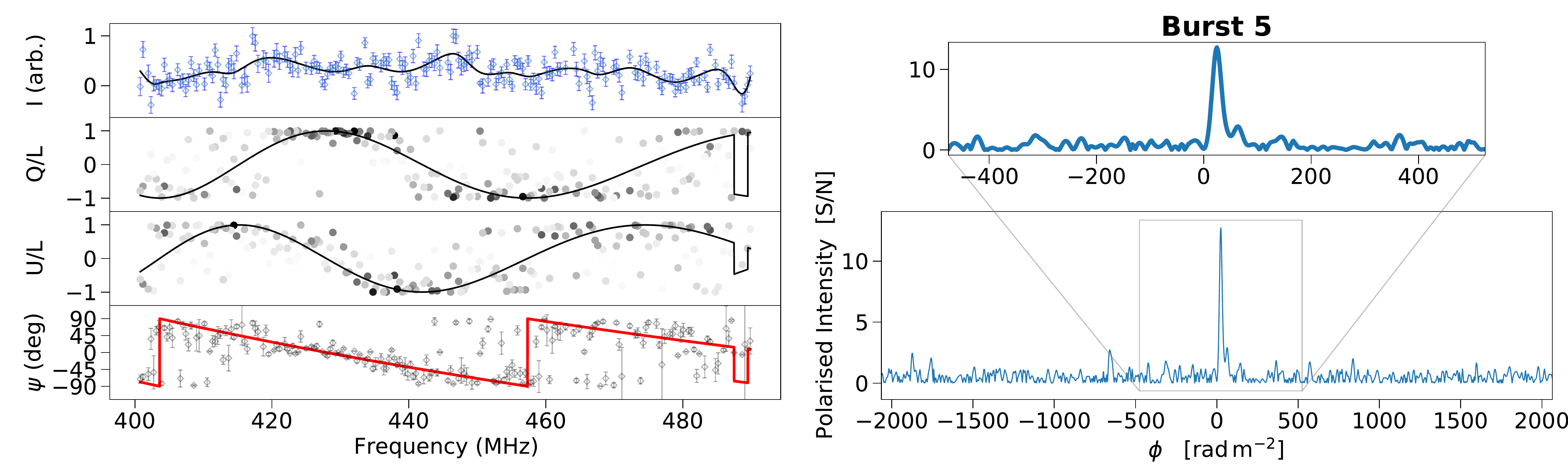}
\figsetgrpnote{}
\figsetgrpend

\figsetgrpstart
\figsetgrpnum{2.27}
\figsetgrptitle{FRB 20190208A}
\figsetplot{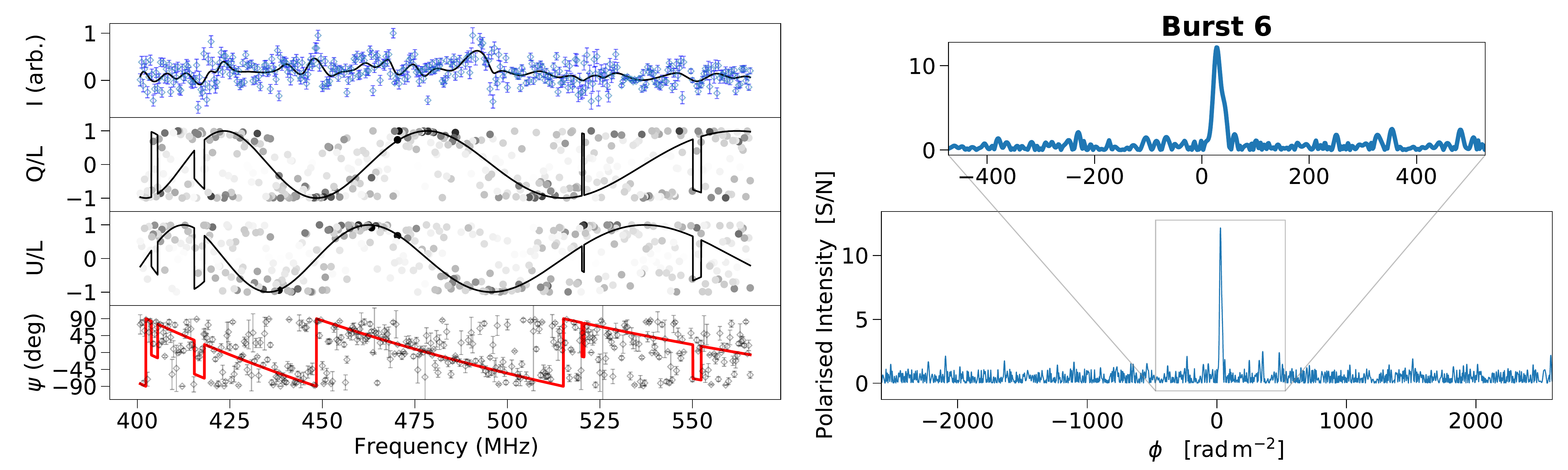}
\figsetgrpnote{}
\figsetgrpend

\figsetgrpstart
\figsetgrpnum{2.28}
\figsetgrptitle{FRB 20190208A}
\figsetplot{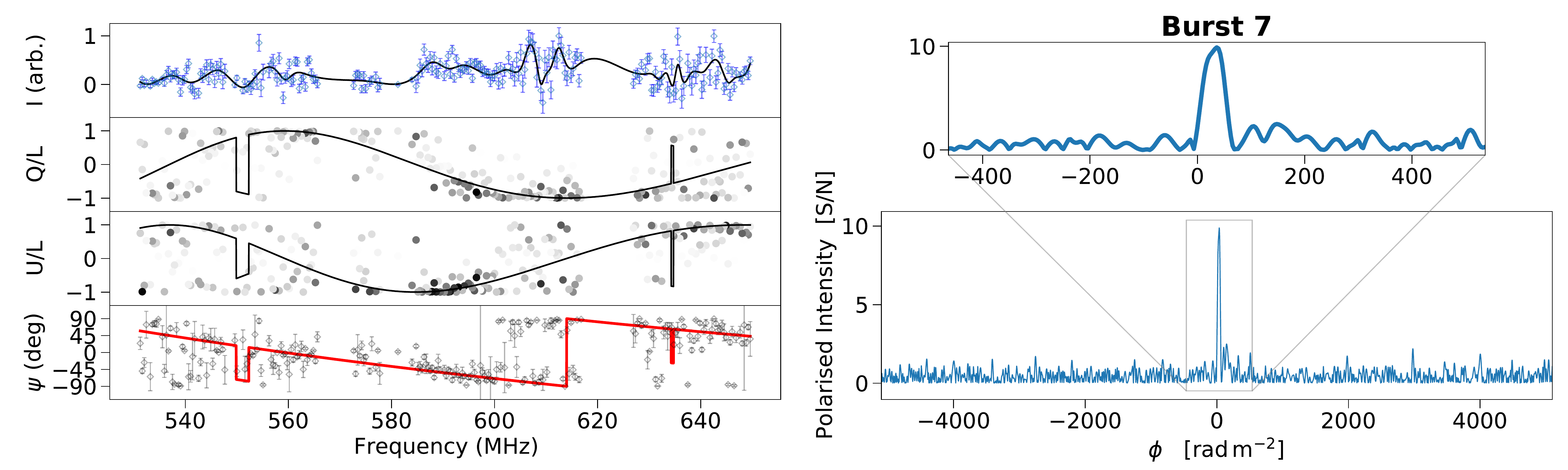}
\figsetgrpnote{}
\figsetgrpend

%%%%%%%%%%%%%%% R13

\figsetgrpstart
\figsetgrpnum{2.29}
\figsetgrptitle{FRB 20190604A}
\figsetplot{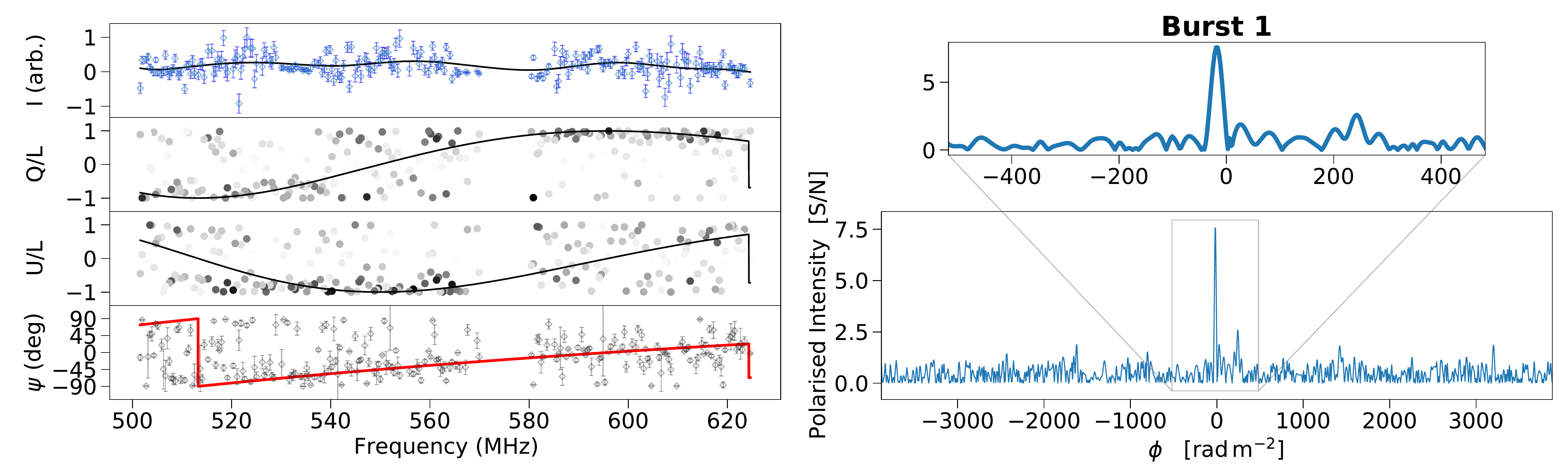}
\figsetgrpnote{}
\figsetgrpend

%%%%%%%%%%%%%%% R14
  
\figsetgrpstart
\figsetgrpnum{2.30}
\figsetgrptitle{FRB 20190213B}
\figsetplot{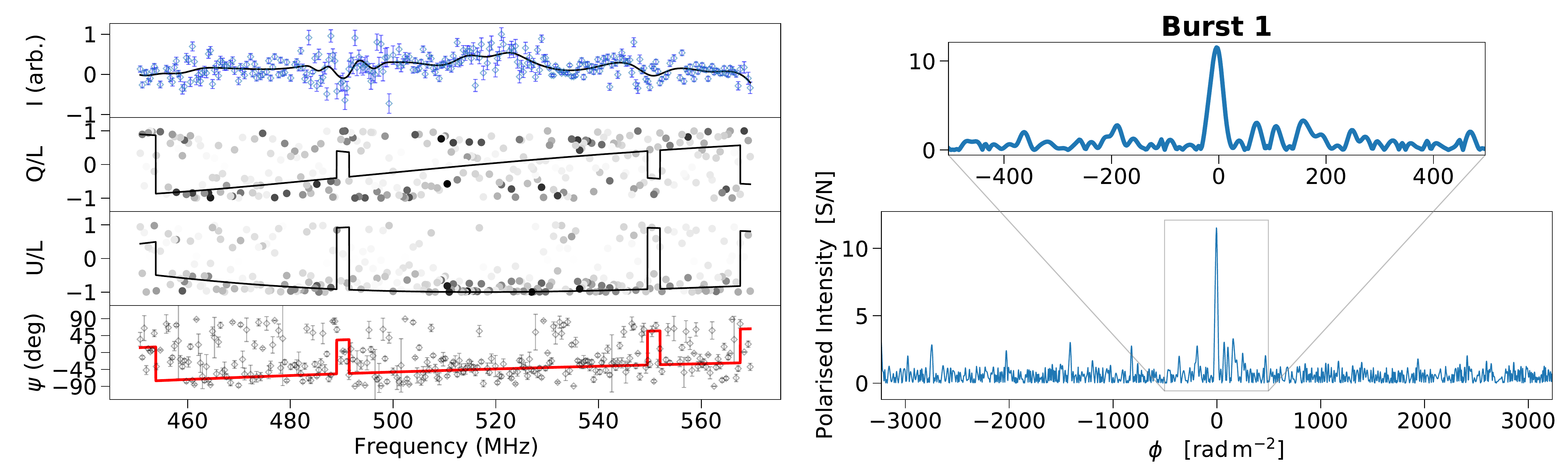}
\figsetgrpnote{}
\figsetgrpend

\figsetgrpstart
\figsetgrpnum{2.31}
\figsetgrptitle{FRB 20190213B}
\figsetplot{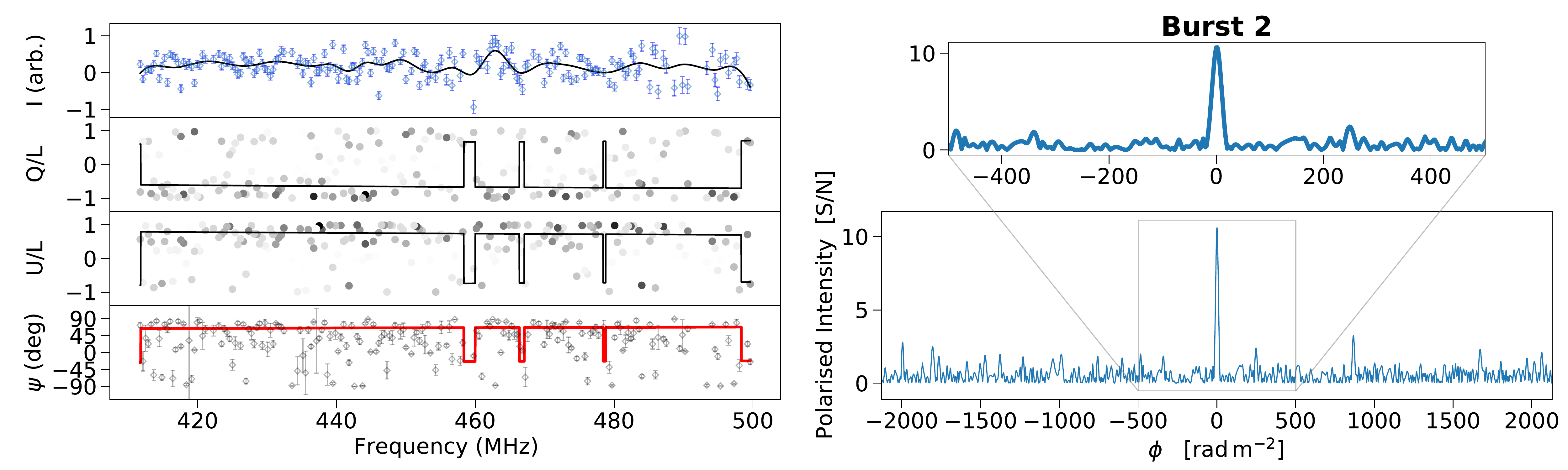}
\figsetgrpnote{}
\figsetgrpend

\figsetgrpstart
\figsetgrpnum{2.32}
\figsetgrptitle{FRB 20190213B}
\figsetplot{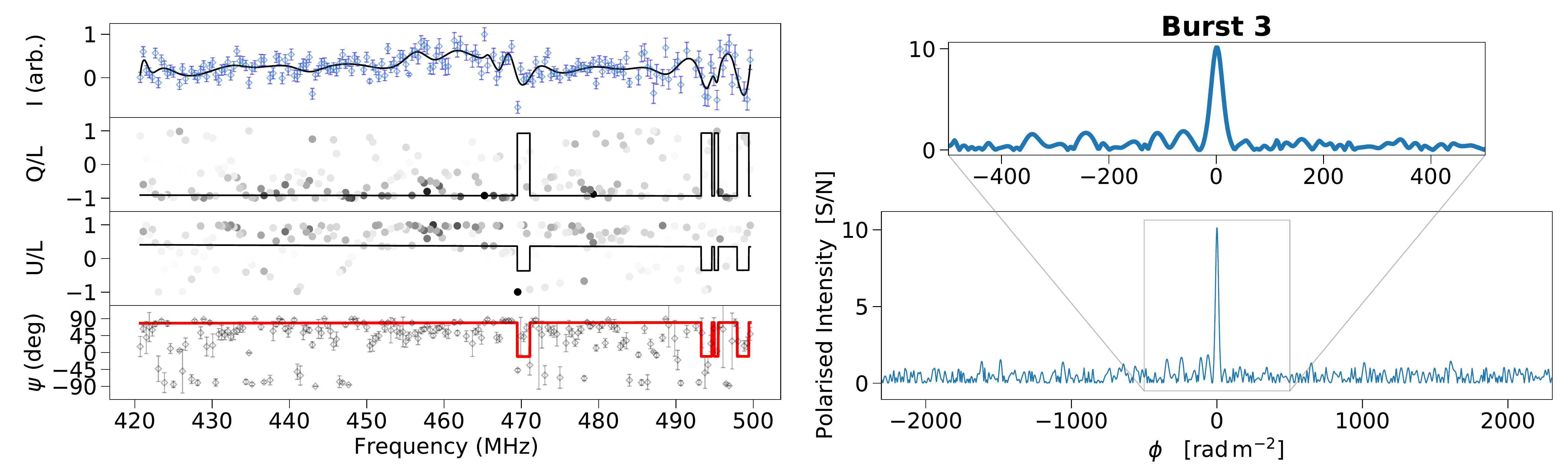}
\figsetgrpnote{}
\figsetgrpend

\figsetgrpstart
\figsetgrpnum{2.33}
\figsetgrptitle{FRB 20190213B}
\figsetplot{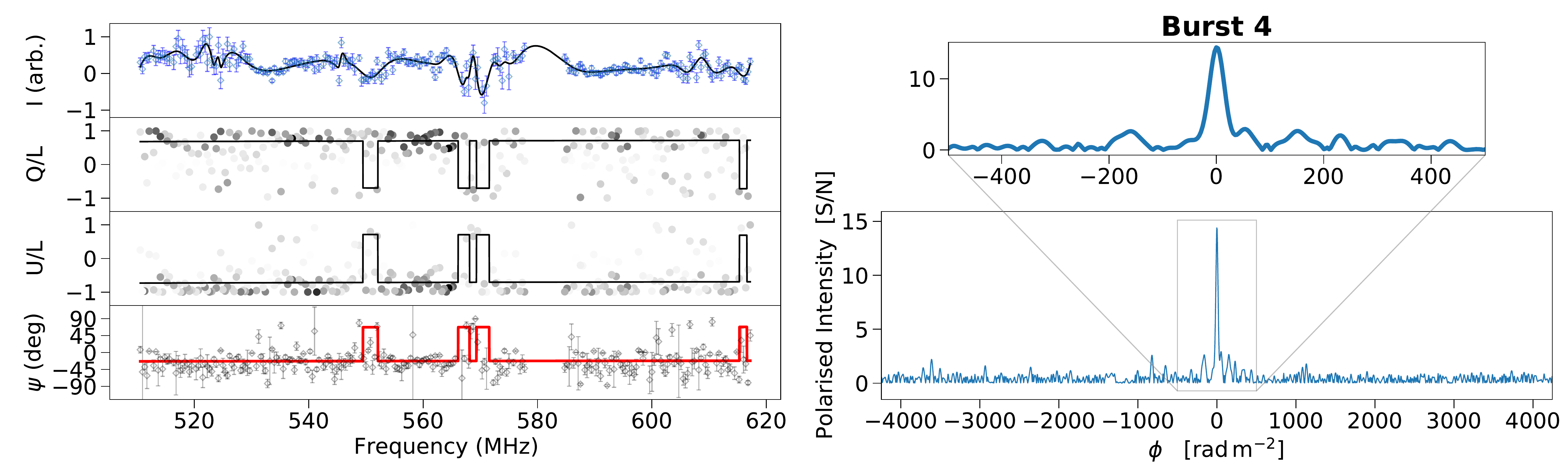}
\figsetgrpnote{}
\figsetgrpend

\figsetgrpstart
\figsetgrpnum{2.34}
\figsetgrptitle{FRB 20190213B}
\figsetplot{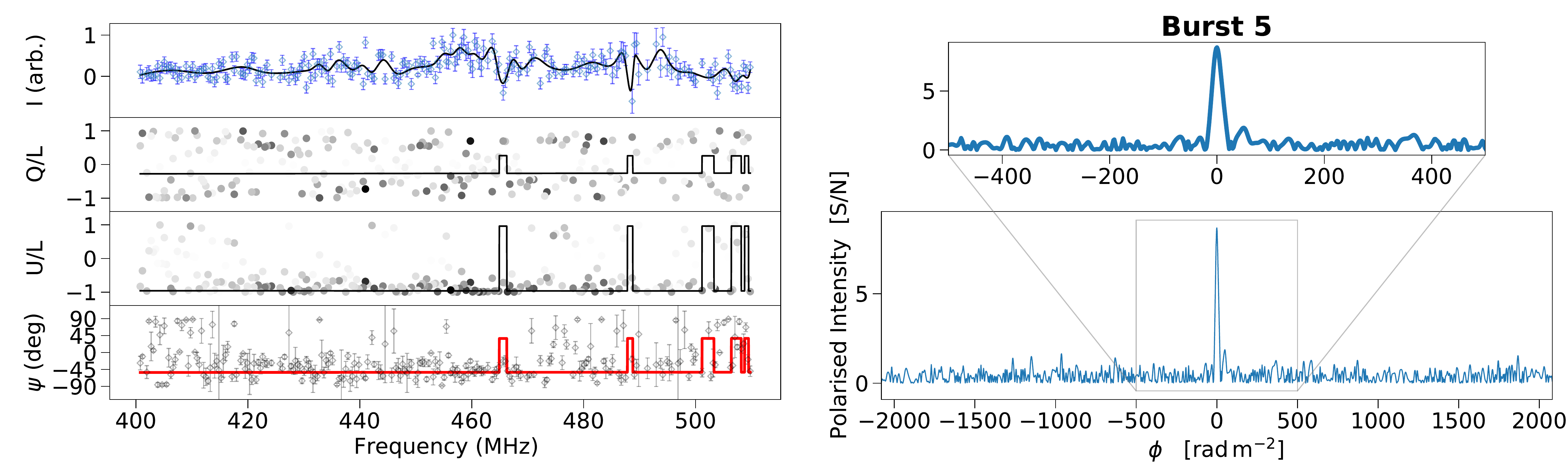}
\figsetgrpnote{}
\figsetgrpend 

%%%%%%%%%%%%%%% R16

\figsetgrpstart
\figsetgrpnum{2.35}
\figsetgrptitle{FRB 20190117A}
\figsetplot{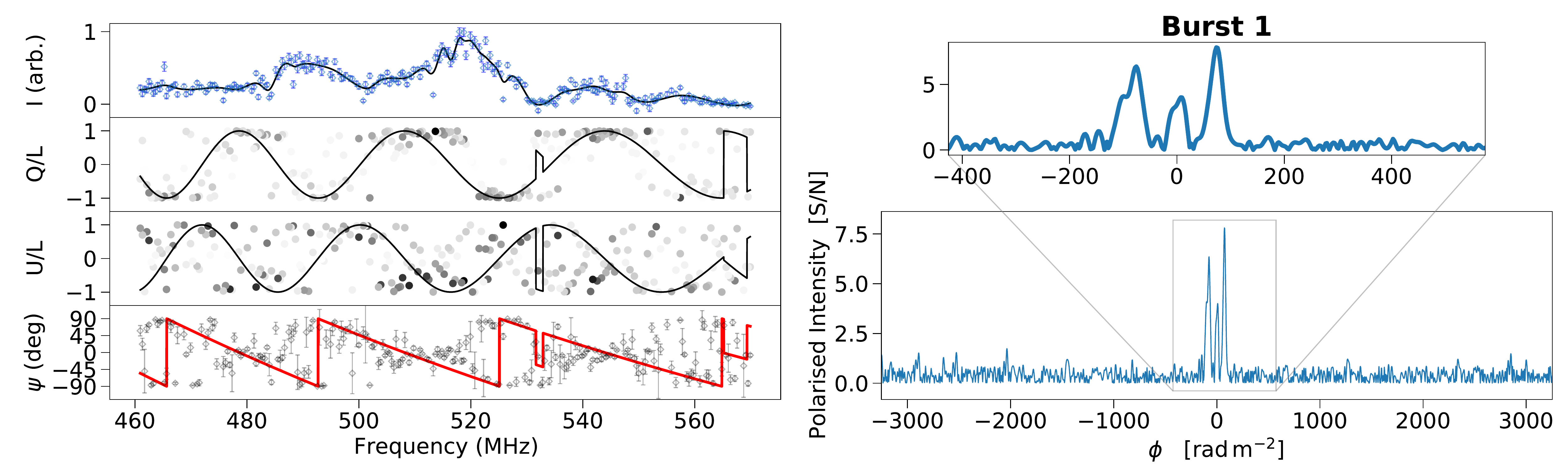}
\figsetgrpnote{}
\figsetgrpend 

\figsetgrpstart
\figsetgrpnum{2.36}
\figsetgrptitle{FRB 20190117A}
\figsetplot{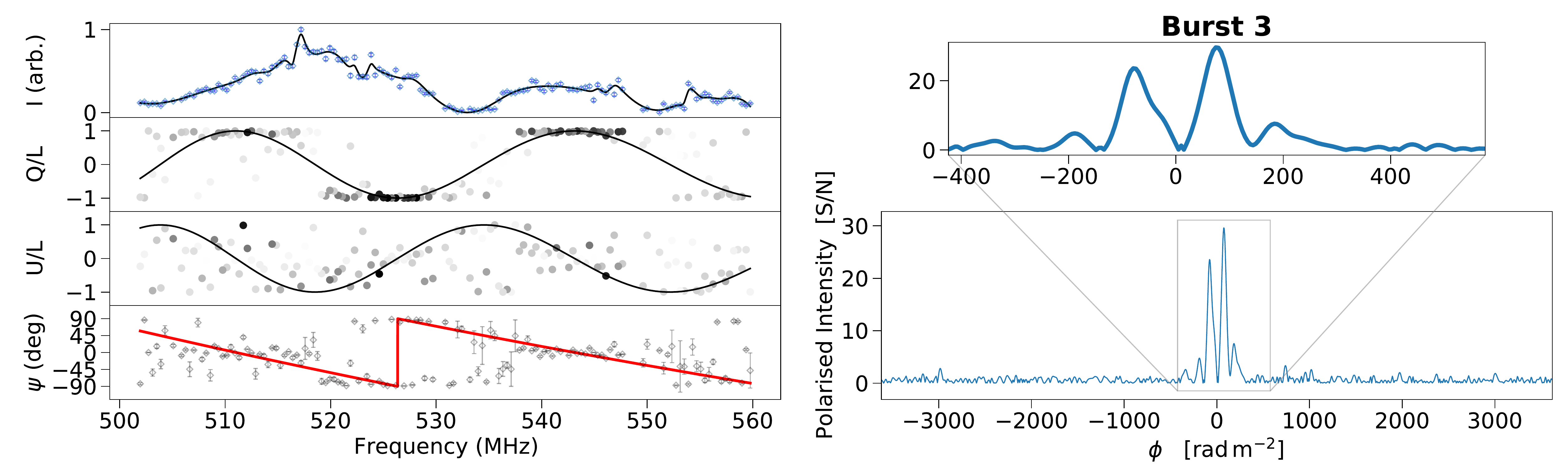}
\figsetgrpnote{}
\figsetgrpend 

%%%%%%%%%%%%%%% R18

\figsetgrpstart
\figsetgrpnum{2.37}
\figsetgrptitle{FRB 20190417A}
\figsetplot{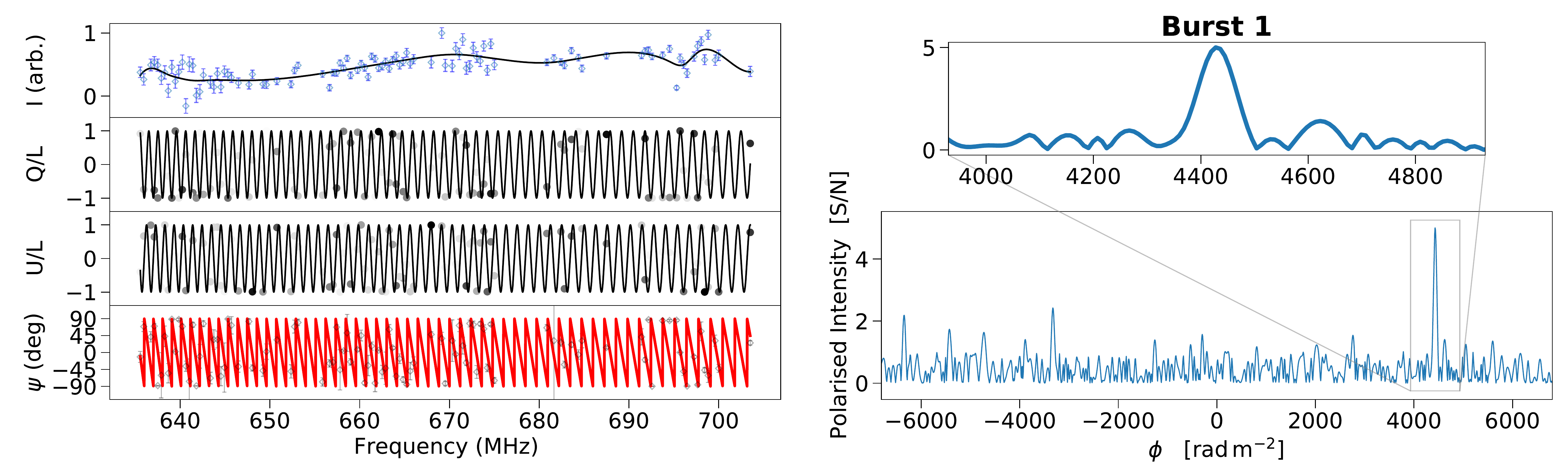}
\figsetgrpnote{}
\figsetgrpend 

\figsetend

\end{document}